\documentclass[prm,aps,twocolumn,superscriptaddress,showpacs,floatfix]{revtex4-1}

\usepackage{graphicx}
\usepackage{epstopdf}
\usepackage[colorlinks=true,linkcolor=black, citecolor=blue, urlcolor=blue, 
    unicode=true]{hyperref}
\usepackage{textcomp, gensymb}
\usepackage{booktabs}
\usepackage{dcolumn}
\usepackage{bm}
\usepackage[dvipsnames]{xcolor}
\usepackage{float}
\usepackage{placeins}

\newcommand{\liso}{Ln$_{2}$InSbO$_{7}$}

\newcommand{\LnBO}{Ln$_2$B$_2$O$_7$}
\newcommand{\LnBBO}{Ln$_2$B$_{\text{i}}^{3+}$B$_{\text{ii}}^{5+}$O$_7$}
\newcommand{\lgso}{Ln$_{2}$GaSbO$_{7}$}

\hypersetup{final}
\begin{document}

\preprint{APS/123-QED}

\title{Traversing the pyrochlore stability diagram; microwave-assisted synthesis and discovery of mixed B-site \texorpdfstring{\liso}{Ln2InSbO7}~family}

\author{Brenden R. Ortiz} 
 \thanks{This author contributed equally to this work}\email{\newline ortiz.brendenr@gmail.com}
 \affiliation{Materials Department and California Nanosystems Institute, University of California Santa Barbara, Santa Barbara, CA 93106, USA}

 \author{Paul M. Sarte}
 \thanks{This author contributed equally to this work}\email{\newline pmsarte@gmail.com}~~~
 \affiliation{Materials Department and California Nanosystems Institute, University of California Santa Barbara, Santa Barbara, CA 93106, USA}

\author{Ganesh Pokharel}
 \affiliation{Materials Department and California Nanosystems Institute, University of California Santa Barbara, Santa Barbara, CA 93106, USA}%
 
 \author{Michael Garcia}
 \affiliation{Materials Department and California Nanosystems Institute, University of California Santa Barbara, Santa Barbara, CA 93106, USA}%

\author{Marcos Marmolejo}
 \affiliation{Materials Department and California Nanosystems Institute, University of California Santa Barbara, Santa Barbara, CA 93106, USA}%

 \author{Stephen D. Wilson} \email{stephendwilson@ucsb.edu}
 \affiliation{Materials Department and California Nanosystems Institute, University of California Santa Barbara, Santa Barbara, CA 93106, USA}

\date{\today}

\begin{abstract}
The lanthanide pyrochlore oxides Ln$_2$B$_2$O$_7$ are one of the most intensely studied classes of materials within condensed matter physics, firmly centered as one of the pillars of frustrated magnetism. The extensive chemical diversity of the pyrochlores, coupled with their innate geometric frustration, enables realization of a wide array of exotic and complex magnetic ground states. Thus, the discovery of new pyrochlore compositions has been a persistent theme that continues to drive the field in exciting directions. The recent focus on the mixed $B$-site pyrochlores offers a unique route towards tuning both local coordination chemistry and sterics, while maintaining a nominally pristine magnetic sublattice. Here, we present a broad overview of the pyrochlore stability field, integrating recent synthetic efforts in mixed B-site systems with the historically established Ln$_2$B$_2$O$_7$ families. In parallel, we present the discovery and synthesis of the entire Ln$_2$InSbO$_7$ family (Ln: La, Pr, Nd, Sm, Eu, Gd, Tb, Dy, Ho, Er, Tm, Yb, Lu) located near the boundary of the pyrochlore stability field using a rapid, hybrid mechanicochemical/microwave-assisted synthesis technique.  Magnetic characterization on the entire class of compounds draws striking parallels to the stannate analogs, suggesting that these compounds may host a breadth of exotic magnetic ground states.
\end{abstract}

\maketitle

\section{Introduction}

\begin{figure}
\includegraphics[width=\linewidth]{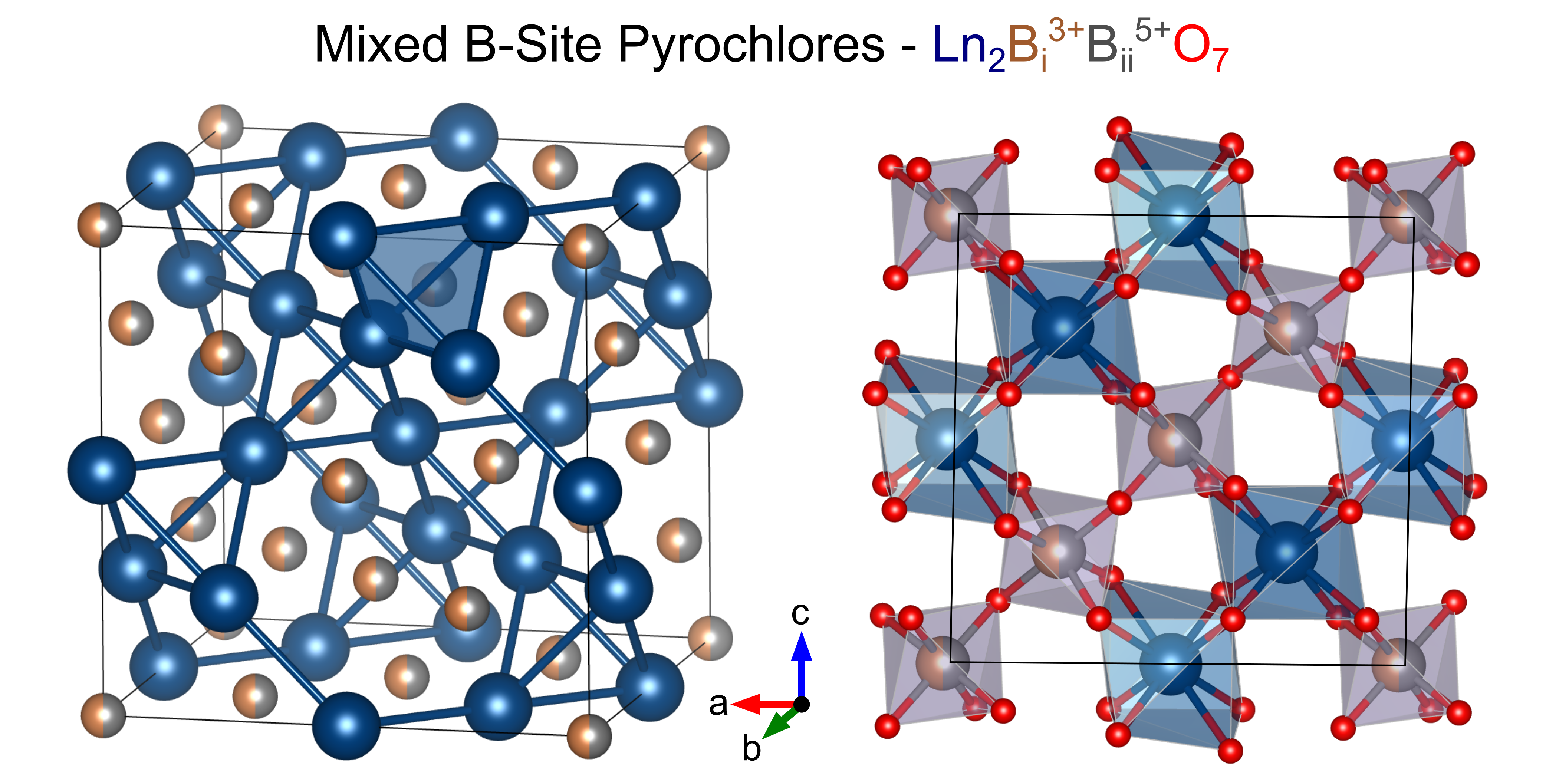}
\caption{Isometric view of the cation sublattice of the cubic $Fd\bar{3}m$ pyrochlore lattice (left), where the lanthanide sublattice consists of a 3D network of corner-sharing tetrahedra. The pyrochlore structure is an ordered superstructure variant of the defect fluorite structure, where the 8-coordinate A- and 6-coordinate B-site sites are structurally distinct (right).}
\label{fig:crystal}
\end{figure}

The cubic pyrochlore structure A$_2$B$_2$O$_7$ (Fig.~\ref{fig:crystal}) is considered as one of the canonical frustrated lattices in three dimensions~\cite{subramanian1983rare,Gardner10:82,greedan2006frustrated,rau19:10,hallas18:9}. In the case of the lanthanide pyrochlore oxides (Ln$_{2}$B$_{2}$O$_{7}$), the trivalent lanthanide Ln$^{3+}$ ions form the hallmark frustrated network of corner-sharing tetrahedra (Fig.~\ref{fig:crystal}). Even considering only the \textit{lanthanide} pyrochlore \textit{oxides}, the number of possible combinations of Ln$^{3+}$ and B$^{4+}$ is staggering~\cite{subramanian1983rare,Gardner10:82,fuentes2018critical,10.3389/fchem.2021.778140}. The flexibility to decorate the frustrated lattice with lanthanide elements of varying single-ion anisotropy, moment size, and radii -- combined with the steric and chemical flexibity offered by the B-site -- have cemented the pyrochlore family as a fruitful arena for the pursuit of the emergence of exotic and novel magnetic ground states. Experimental and theoretical studies alike have proposed the Ln$_{2}$B$_{2}$O$_{7}$ pyrochlores as a means to realize a myriad of properties ranging from unconventional long-range order~\cite{champion03:68,savary12:109,zhitomirsky12:109,wong13:88,Champion_2004,rau19:10,hallas18:9,Gardner10:82,greedan2006frustrated,li2016long,raju99:59,subramani2021comparison}, unconventional spin-glass behavior~\cite{reimers1988crystal,PhysRevLett.113.117201,PhysRevB.46.5405,greedan2006frustrated,Zhou_2008,PhysRevLett.109.247201,PhysRevB.68.174413,PhysRevLett.107.047204,ZHOU2010890,Taniguchi_2009,PhysRevB.43.3387,PhysRevLett.69.3244,PhysRevLett.78.947}, topologically-nontrivial electronic states~\cite{PhysRevLett.103.206805,PhysRevB.80.113102,pesin2010mott,otsuka2021higher,PhysRevB.89.115111,PhysRevB.83.165112,PhysRevB.82.085111,PhysRevLett.120.026801,PhysRevB.90.165119}, spin ices~\cite{PhysRevLett.79.2554,PhysRevLett.87.047205,HARRIS1998757,ramirez99:399,doi:10.1126/science.1064761,Bramwell_2020,denhertog00:84,PhysRev.102.1008,Champion_2004,PhysRevLett.95.097202,PhysRevLett.101.037204,HARRIS1998757,Melko_2004}, magnetic monopole quasiparticles~\cite{Sarte17:29,morris2009dirac,castelnovo2008magnetic,bramwell2009measurement,paulsen2016experimental}, quantum spin liquids~\cite{gao2019experimental,Benton18:121,Wen17:118,Savary17:118,sibille15:115,PhysRevB.86.104412,balents2010spin,Yb_ross2011quantum,PhysRevB.87.205130,Savary_2016,RevModPhys.89.025003}, cooperative paramagnetism~\cite{Gardner99:82,gardner2001neutron,PhysRevB.81.060408,PhysRevLett.102.237206,PhysRevLett.92.107204}, and other spin-liquid-like states~\cite{PhysRevLett.104.106407,PhysRevB.100.075125}.

Considering the rich interplay between chemistry and physics present in the pyrochlore materials, the discovery of new pyrochlore compositions has been a constant theme over the last few decades~\cite{subramanian1983rare,Gardner10:82,rau19:10,hallas18:9}. Given the vast chemical space, the community naturally sought out empirical ``rules'' to predict the relative stability of pyrochlore structures using simple chemical metrics. These approaches are reminiscent of the Goldschmidt tolerance factor for perovskites~\cite{goldschmidt1926gesetze}, and aim to provide simple guidelines and chemical intuition. Within pyrochlores, the ``radius-ratio rules'' have been the most successful, which constrain $R_{\rm{min}}$ $\leq$ $R{\rm{_{Ln^{3+}}}}$/$R{\rm{_{B^{4+}}}}$ $\leq$ $R_{\rm{max}}$~\cite{subramanian1983rare,Gardner10:82,Wiebe15:3,RR_zhou2019high}. Despite the apparent simplicity and variations in the specifics of $R_{\rm{min}}$ and $R_{\rm{max}}$, the radii-ratio rules have been applied with great success~\cite{subramanian1983rare,Gardner10:82}. Naturally, the diversity of experimentally accessible Ln$_2$B$_2$O$_7$ pyrochlore phases saw a rapid expansion in the 1960's~\cite{Wiebe15:3}, particularly with rapid technological advances in extreme synthesis conditions, particularly those at high pressure. Human ingenuity has continued to realize metastable Ln$_2$B$_2$O$_7$ phases considered outside the pyrochlore stability field, including the germanates~\cite{Hallas14:113,Hallas12:86,PhysRevLett.108.207206,li2014long,Hallas16:93} and plumbates~\cite{Hallas15:91}.

Beyond the search for the prototypical Ln$_2$B$_2$O$_7$ phases, one can also consider permutations that build upon the basic structure while maintaining the hallmark frustrated Ln$^{3+}$ sublattice. The first mention of mixed B-site pyrochlore phases emerged with Bl{\"o}te~\cite{blote1969heat}, where the six coordinate B$^{4+}$ ion was replaced by the isoelectronic combination of B$_{i}^{3+}$B$_{ii}^{5+}$. This is reminiscent of exploratory searches for more complex diamond-like semiconductors where silicon can be split iteratively into increasingly complex chemical compositions (e.g. Si, GaAs, CuGaSe$_{2}$) bound by charge balance. Unlike the diamond-like semiconductors, where the permutations reduce the global symmetry of the unit cell (Si: $Fd\bar{3}m$, GaAs: $F\bar{4}3m$), the B-site in \LnBBO~is nominally disordered, yielding an average structure with the same global symmetry as the ternary pyrochlore ($Fd\bar{3}m$). Despite the potential for a large number of additional permutations on the pyrochlore structure, systematic studies of the mixed B-site pyrochlores have been scarce so far~\cite{subramanian1983rare}. 

Part of the historical hesitation behind the mixed B-site pyrochlores comes from the negative connotation of ``disorder'' and its association with  ``dirty'' samples~\cite{martin17:7}. Recently, however, there has been a surge of renewed interest in the mixed B-site pyrochlores~\cite{gomez2021absence,YbGaSb_sarte2021dynamical,dy2gasbo7,megan_thesis,james_thesis,Mauws21:33}, driven by the emerging picture of the profound effects that disorder can play on cooperative magnetic properties. Recent experimental studies on Pr$_{2}$Zr$_{2}$O$_{7}$ suggest the possibility that disorder may in fact provide a key route in the realization of both exotic, and possibly novel magnetic ground states~\cite{Wen17:118,martin17:7}. The studies that do exist for mixed B-site pyrochlores are fascinating in their own right, and the most thorough investigations have indicated that the mixed Ga$^{3+}$Sb$^{5+}$ compositions are remarkably similar to the Ln$_2$Ti$_2$O$_7$ titanates in both structure and properties~\cite{gomez2021absence,YbGaSb_sarte2021dynamical,blote1969heat,dy2gasbo7}. As such, we postulate that the richness of the  Ln$_2$B$_2$O$_7$ phases may be mirrored in their mixed B-site counterparts, but with additional chemical degrees of freedom.

In this work we aim to provide the first systematic study of a mixed B-site family throughout the entire $f$-block in the context of the rich history of the lanthanide Ln$_2$B$_2$O$_7$ phases. We begin by providing an updated pyrochlore stability field diagram, integrating recent synthetic efforts in the metastable Ln$_2$B$_2$O$_7$ phases and mixed B-site Ln$_2$B$_{i}^{3+}$B$_{ii}^{5+}$O$_7$ pyrochlores. Simultaneously, we present our synthesis and discovery of the Ln$_2$InSbO$_7$ series of compounds. These compounds serve as a structural analog to the Ln$_2$Sn$_2$O$_7$ stannates, much as the Ln$_2$GaSbO$_7$ family corresponds to the respective titanates. The Ln$_2$InSbO$_7$ compounds straddle the pyrochlore stability window, and producing high-quality powders using classical synthesis methods is difficult~\cite{strobel2010structural}. As such, we also present a new synthesis technique -- a hybrid mechanochemical-microwave method -- for the pyrochlore oxides which can produce phase-pure crystalline material in $<$30\,min heat treatments. We discuss the magnetic properties of the entire Ln$_2$InSbO$_7$ series down to 60\,mK, analyzing the data in the context of other Ln$_2$B$_2$O$_7$ analogs and other mixed B-site pyrochlores. Our work demonstrates the rich and diverse role the mixed B-site materials can play when deciphering complex structure-property relationships in the lanthanide pyrochlores.

\section{Experimental Methods}

\subsection{Synthesis}
\label{sec:synthesis_methods}
Polycrystalline samples of the \liso~family (Ln: La, Pr, Nd, Sm, Eu, Gd, Tb, Dy, Ho, Er, Tm, Yb, Lu) were synthesized from stoichiometric amounts of the pre-dried lanthanide oxides: La$_2$O$_3$ 99.999~\% Alfa, Pr$_6$O$_{11}$ 99.99~\% Alfa, Nd$_2$O$_3$ 99.99~\% Alfa, Sm$_2$O$_3$ 99.9~\% Alfa, Eu$_2$O$_3$ 99.9~\% Alfa, Gd$_2$O$_3$ 99.99~\% Alfa, Tb$_4$O$_7$ 99.9~\% Alfa, Dy$_2$O$_3$ 99.9~\% Alfa, Ho$_2$O$_3$ 99.99~\% Alfa, Er$_2$O$_3$ 99.9~\% Alfa, Er$_2$O$_3$ 99.9~\% Alfa, Tm$_2$O$_3$ 99.9~\% Alfa, Yb$_2$O$_3$ 99.998~\% Alfa, Lu$_2$O$_3$ 99.999~\% Alfa, and the two metalloid oxides: In$_2$O$_3$ (99.9~\% Alfa) and Sb$_2$O$_5$ (99.998~\% Alfa). Two synthetic routes were investigated. 

The first corresponds to a method building on the classical heating profiles previously reported for similar pyrochlores, while integrating mechanicochemical methods. Stoichiometric mixtures of the pre-dried oxide and the two metalloid oxides were combined into a tungsten carbide ball-mill vial and milled for 60~min. The resulting powders were extracted, ground in an agate mortar, and sieved through a 50~$\mu$m sieve. These powders were loaded into 2~mL high-density alumina crucibles (CoorsTek) and annealed in a box furnace at 1250\degree C for 48~h. 
 
The second route involves abandoning classical heating profiles and instead combines mechanicochemical methods with microwave-assisted synthesis. Stoichiometric mixtures of the pre-dried oxide and the two metalloid oxides were combined into tungsten carbide ball-mill vials and milled for an initial 60~min cycle. The resulting powder was extracted, ground in an agate mortar to break up any agglomerates, and milled again for an additional 90~min. After this second cycle, 10~mL of anhydrous ethanol was added to the vial, and the sample was milled a third time for an additional 30~min. The resulting slurry was dried in a 80\degree C oven, resulting in an extremely fine, homogenous powder. 

Each precursor powder was loaded directly into a 2~mL alumina crucible (CoorsTek), which was subsequently nested within a larger 10~mL alumina crucible filled with 7~g of granular activated charcoal (DARCO 12-20 mesh, Sigma-Aldrich). The crucibles were then placed within an assembly of low density alumina foam measuring approximately 15$\times$15$\times$15~cm. Finally, the total assembly was placed in a 2.45~GHz multimode microwave oven with a maximum output of 1.2~kW (Panasonic model NN-SN651B). For samples containing large lanthanide ions (La--Gd), the alumina crucible was exposed directly to air. For smaller lanthanide ions (Tb--Lu), an alumina frit and secondary 2~mL crucible were used as an effective ``lid'' to reduce heat loss and increase the maximum temperature.

Phase purity was examined with powder x-ray diffraction (XRD) measurements at room temperature on a Panalytical Empyrean diffractometer (Cu K$_{\alpha_{1,2}}$) in standard Bragg-Brentano ($\theta$-2$\theta$) geometry. Rietveld refinements of the powder XRD patterns were performed using \texttt{TOPAS Academic} v6~\cite{Coelho}. Structural models and visualization utilized the \textsc{VESTA} software package~\cite{Momma2011}.  

\subsection{Magnetic characterization: dc \& ac magnetic susceptibility}

Measurements of the temperature- and field-dependent dc magnetization were performed on a 7~T Quantum Design Magnetic Property Measurement System (MPMS3) SQUID magnetometer in vibrating-sample magnetometry (VSM) mode. Powder of each \liso~ sample was placed in a polypropylene capsule and subsequently mounted in a Quantum Design brass holder. The zero-field-cooled (ZFC) and field-cooled (FC) DC magnetization was collected continuously in sweep mode with a ramp rate of 2~K/min in the presence of an external DC field of 10000~Oe. The ZFC isothermal dc magnetization was collected continuously in sweep mode with a ramp rate of 100~Oe/sec.  

The temperature dependence of the ac magnetization was measured on a Quantum Design 14~T Dynacool Physical Property Measurement Systems (PPMS) employing the ac susceptibility option for the dilution refrigerator (ACDR). Phase-pure powder of each member of the \liso~was cold pressed with a Carver press, and a portion of the resulting pellet with approximate dimensions of 1x1x0.5~mm was adhered to a sapphire sample mounting post with a thin layer of GE varnish. All ac measurements were collected under ZFC conditions in the absence of an external dc magnetic field.

\section{Results \& Discussion}

\subsection{Pyrochlore Stability Field}

Part of the allure of the pyrochlore lattice is the chemical diversity offered by the choice of A-site and B-site elements~\cite{subramanian1983rare,Gardner10:82,10.3389/fchem.2021.778140,talanov2021structural,trump18:9,rau19:10,hallas18:9,Wiebe15:3,RR_zhou2019high}. As established previously, the community leverages the empirically-derived ``radius-ratio rules'' to predict and synthesize new pyrochlore materials. For trivalent lanthanide oxide pyrochlores \LnBO, the relationship was previously established decades ago as 1.46 $\leq$ $R{\rm{_{Ln^{3+}}}}$/$R{\rm{_{B^{4+}}}}$ $\leq$ 1.80~\cite{subramanian1983rare}. More recently, the rules are often quoted as 1.36 $\leq$ $R{\rm{_{Ln^{3+}}}}$/$R{\rm{_{B^{4+}}}}$ $\leq$ 1.71~\cite{Gardner10:82,Wiebe15:3,RR_zhou2019high}. The pyrochlore stability rules have been applied with great success in discovery of stable and metastable \LnBO~ phases and continue to play a fundamental role in materials discovery. 

\begin{figure}
\includegraphics[width=\linewidth]{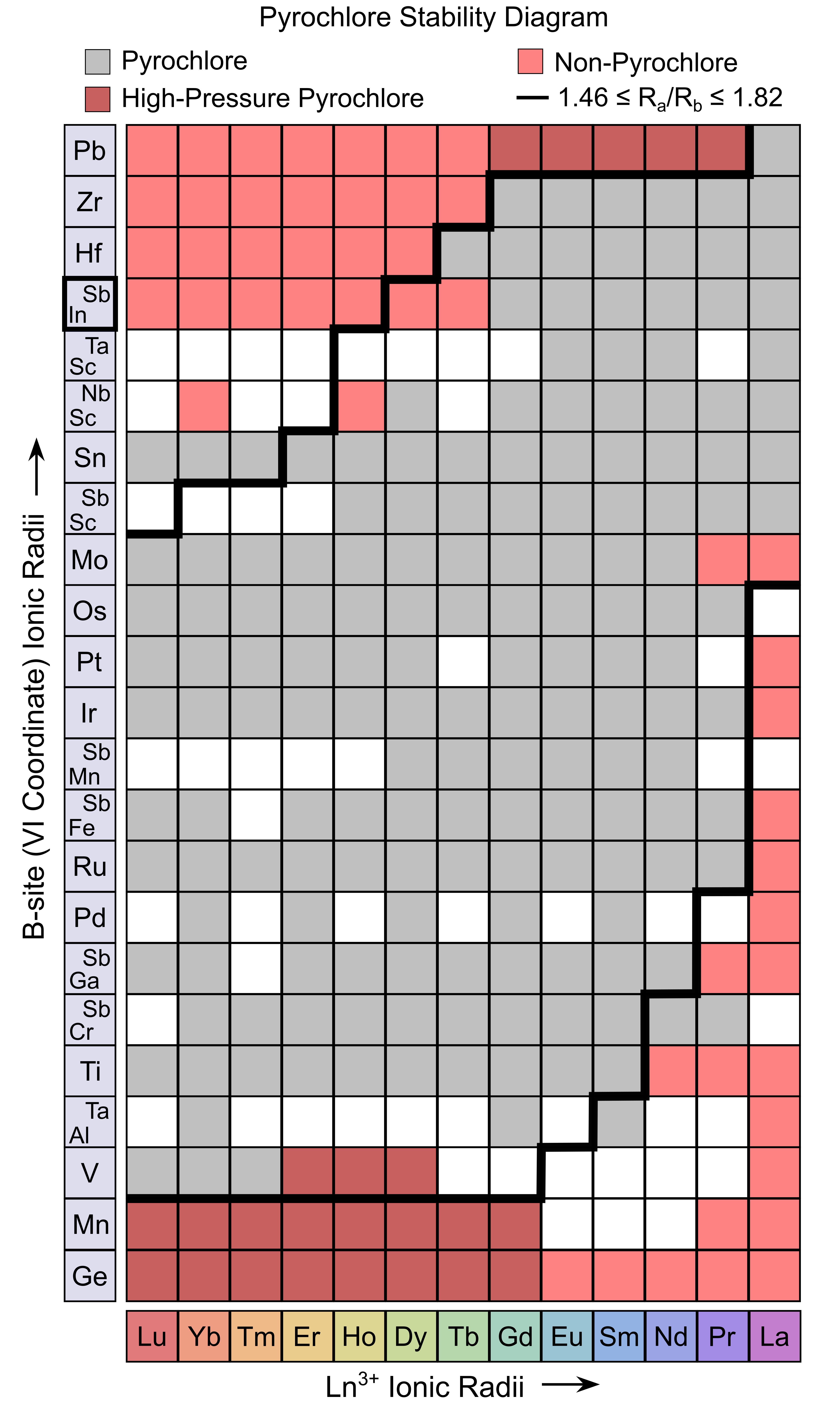}
\caption{Stability-field map for Ln$_2$B$_2$O$_7$ and \LnBBO~ lanthanide pyrochlore oxides. Known pyrochlore structures reported in the literature are superimposed with a minimized stability field (black border). ``Non-pyrochlore'' null results (red) and pyrochlores stabilized with high pressure (dark red) are distinguished from ``no data'' (white).}
\label{fig:stability}
\end{figure}

While the applicability of the ``radii-ratio rules'' within the mixed B-site systems is assumed, there has not been a exhaustive review of the B-site systems in the context of the pyrochlore stability field. Recall that mixed B-site alloys must additionally observe the charge neutrality conditions in the permutation of \LnBO~compositions to \LnBBO. As such, there are a relatively finite number of combinations that can be expected to obey both the radii-rules and charge-neutrality conditions.  Figure~\ref{fig:stability} provides a graphical review of known trivalent lanthanide pyrochlore-like oxide (gray) compositions of the form \LnBO~or \LnBBO~\cite{sleight69:8,subramanian1983rare,KLEE69:31,MICHEL74:9,perez62:7,Collongues61:70,Roth56:56,SmZr_fomina1977formation,subramanian1983rare,Gundovin75:20,KLEE69:31,Besson66:262C,SPIRIDINOV68:14,TbHf_sibille2016ms21,TbHf_shlyakhtina2006structure,strobel2010structural,shlyakhtina2014synthesis,shlyakhtina2018proton,MLZNd,SATO76:38,shlyakhtina2014synthesis,shlyakhtina2018proton,mcqueen2008frustrated,zouari2008synthesis,subramanian1983rare,Kennedy97:58,strobel2010structural,inprepOrtiz,CLARK13:203,subramanian1983rare,GREEDAN87:68,subramanian80:15,tkachenko03:39,MCCARTHY71:6,ranganathan83:52,SHAPLYGIN73:8,sleight1968new,chien1978mossbauer,subramanian1983rare,Matsuhira11:80,klicpera2020characterization,Blackrock80:72,yanagishima2001metal,LAZAREV78:13,ISMUNANDAR00:302,BERNDT15:618,Knop68:46,Montmory61:252,subramanian1983rare,Taira99:11,gaultois2015design,Kennedy95:51,Bertaut59:249,sleight1968new,subramanian1983rare,strobel2010structural,kobayashi2001nmr,gomez2021absence,blote1969heat,WHITAKER14:215,Bongers67:38,CASADO85:46,villars2017handbook,woolfson1973solid,subramanian1983rare,Knop69:47,Knop65:43,haipeng2016preparation,xiaoge2016thermal,li2009photocatalytic,subramanian1983rare,kitayama76:5,SODERHOLM82:43,troyanchuk1990preparation,YOKOKAWA92:52,SHINIKE77:12,Shimakawa99:59,Subramanian88:72,Greedan96:54,shimazaki00:19,shannon1968synthesis,BECKER1987269,Hallas14:113,Hallas12:86,PhysRevLett.108.207206,li2014long,Hallas16:93,Hallas15:91,Clark76:5}.

In our search for lanthanide pyrochlore oxides, we have chosen to maintain the nominal purity of both the Ln$^{3+}$ and oxygen sublattices, although many other variants on the pyrochlore lattice ($e.g.$ mixed anion oxy-fluoride~\cite{OF,PHRAEWPHIPHAT2018}, oxy-nitrides~\cite{PhysRevLett.113.117201}, alkali A-site~\cite{PhysRevB.95.144414}, etc.) have been reported. For graphical simplicity, compositions with a limited number ($<4$) of members reported ($e.g.$ Ln$_2$AlSbO$_7$~\cite{AlSb_che2020absence}) have been omitted. White boxes mark compositions with no reported synthetic data. Non-pyrochlore compositions are marked in red, and include: 1) phase competition with other compounds, 2) polymorph competition with fluorite-type or lower symmetry structures, or 3) persistent (substantial) phase impurity and off-stoichiometry. Special attention has been paid to pyrochlore phases that have been stabilized in the pyrochlore structure through the application of hydrostatic pressure $>$1~MPa ($e.g.$ Ln$_{2}$Ge$_{2}$O$_{7}$~\cite{Hallas14:113,Hallas12:86,PhysRevLett.108.207206,li2014long,Hallas16:93}, Ln$_{2}$Pb$_{2}$O$_{7}$~\cite{Hallas15:91}). These metastable compositions are distinguished by a dark red coloration.

Using a simple scoring algorithm, we have recalculated the limits of $R_{\text{min}}$ $\leq$ $R{\rm{_{Ln^{3+}}}}$/$R{\rm{_{B^{4+}}}}$ $\leq$ $R_{\text{max}}$. Contrary to the often quoted range [$R_{\rm{min}}$, $R_{\rm{max}}$] of [1.36\AA, 1.71\AA]~\cite{Gardner10:82,Wiebe15:3,RR_zhou2019high}, the best fit, denoted by the black boundary in Fig.~\ref{fig:stability}, was calculated to be [$R_{\rm{min}}$, $R_{\rm{max}}$] of [1.46\AA, 1.82\AA], in excellent agreement with the historical estimate provided by Subramanian \emph{et al.}~\cite{subramanian1983rare}. Our updated stability diagram suggests that the mixed B-site pyrochlores largely abide by the empirical radii rules discovered decades prior. This is interesting, as one could reasonably expect that the bond strain and disorder induced by the mixed B-site could negatively impact the energetics of the ordered pyrochlore structure.  The predictive accuracy of the model for all known compounds is $\sim$94\%. We note that the range [$R_{\rm{min}}$, $R_{\rm{max}}$] of [1.36\AA, 1.71\AA]~\cite{Gardner10:82,Wiebe15:3,RR_zhou2019high} performs substantially worse, only capturing the known pyrochlore phases with at 75\% accuracy.

Upon closer inspection of Fig.~\ref{fig:stability}, many compositions that border the edge of the stability field were noted to have mixed results reported in the literature -- where coexistance and competition of the pyrochlore and defect fluorite phases is contested~\cite{simeone2017intricate,10.3389/fchem.2021.778140}. While we have tried to survey the results independently, the binary demarcation of pyrochlore/non-pyrochlore is an oversimplification at best. The literature is rife with such results that highlight the complexity of defect formation~\cite{TbHf_sibille2017coulomb,blanchard2012does} throughout the stability field. Colloquially, our review of the literature serves as a reminder that the stability field, although convenient, is an oversimplified construct. One must respect that the defect thermodynamics and kinetics will be unique for each composition. These defect mechanisms are crucial for many key systems (e.g. Yb$_2$Ti$_2$O$_7$~\cite{YbTi_chang2012higgs,YbTi_gaudet2016gapless,YbTi_ross2012lightly,YbTi_yasui2003ferromagnetic,PhysRevB.84.172408,PhysRevB.84.174442,PhysRevB.88.134428,PhysRevB.89.184416,hallas18:9,PhysRevB.95.094407}, Tb$_2$Ti$_2$O$_7$~\cite{PhysRevB.93.144407,Gardner99:82,PhysRevB.68.180401,PhysRevB.62.6496,PhysRevB.82.100402,Takatsu_2011,PhysRevB.84.184403,PhysRevB.69.132413,PhysRevLett.112.017203,PhysRevLett.83.1854}) whose ground state properties are intimately tied with the nuances of both defects and disorder. It is important to note that such considerations must be taken into account, particularly when moving into compositions with \textit{intentional} disorder on the B-site.

\begin{figure*}
\centering
\includegraphics[width=7.05in]{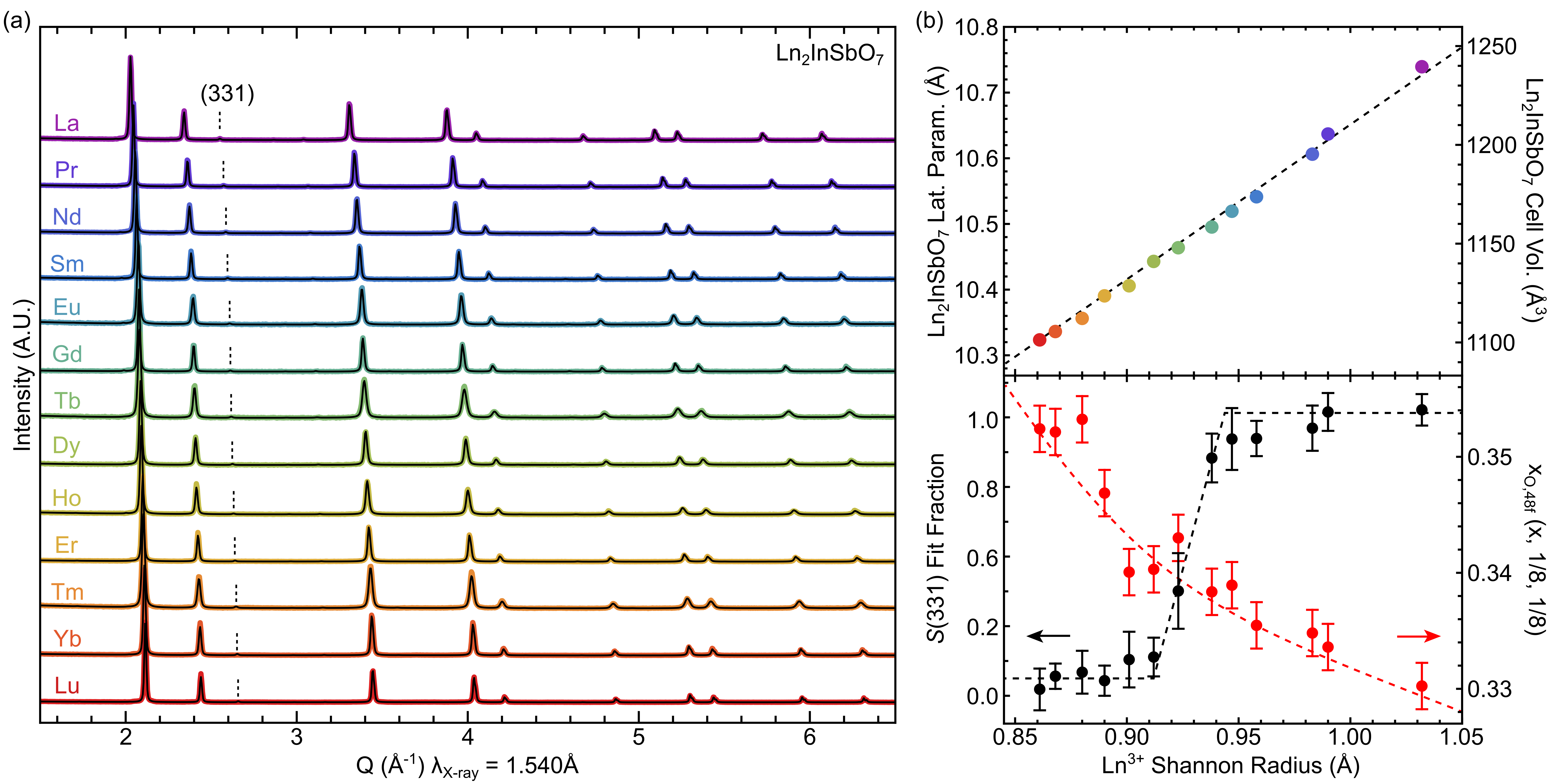}
\caption{(a) Room temperature x-ray diffraction patterns for \liso~ family. All compounds crystallize in cubic symmetry with pre-Gd compositions (La\dots Gd) displaying characteristic (311) and (331) pyrochlore superlattice reflections. (b,top) Linear dependence of the refined cubic lattice parameter and cell volume with trivalent lanthanide Shannon radius.  
(b,bottom) Deviation between predicted and observed intensity of (331) superlattice reflection as a function of composition. The suspected crossover from the pyrochlore structure for pre-Gd compositions (La\dots Gd) to a defect fluorite-like structure for post-Gd compositions (Tb\dots Yb) is mirrored by the shift of the variable $x$-coordinate of the 48$f$ oxygen. 
}
\label{fig:xray}
\end{figure*}

\subsection{\texorpdfstring{Ln$_{2}$InSbO$_{7}$}{Ln2InSbO7}: Synthesis \& Structure}

While the effective radii of the combination of InSb is approximately 0.70~\AA, nearly identical to that of Sn (0.69~\AA)~\cite{Shannon:a12967}, the discrepancy is large enough such that the InSb compositions straddle the limits of pyrochlore stability field in Fig.~\ref{fig:stability}. This is consistent with previous work by Strobel~\cite{strobel2010structural}, where only Gd$_2$InSbO$_7$ was able to be synthesized in a phase-pure form. In contrast to the case of the much smaller combination of GaSb, approximating that of Ti, preliminary work by Strobel suggested that lighter Ln$^{3+}$ cations appeared to completely destabilize the cubic pyrochlore structure, replaced by lower symmetry derivatives corresponding to numerous unidentifiable reflections present in  diffraction patterns~\cite{strobel2010structural}.

Empirical experimentation with \liso~compositions in the current study, when combined with our previous reports on the \lgso~phases, suggests that diffusion kinetics are sluggish in the mixed oxide compositions. Thus, the successful synthesis of homogeneous powders free of both impurity phases and unreacted reagents often require several cycles of classical ``shake-and-bake'' processing, consisting of multiple heating cycles with intermittent grinding. However, often these methods are insufficient to produce phase-pure powders. Such is the case for \liso~\cite{strobel2010structural,blote1969heat}, and some select \lgso~compositions~\cite{gomez2021absence,blote1969heat}, where impurities persist indefinitely. As an attempt to circumvent the limitation of slow reaction kinetics, we incorporated mechanochemical methods (see Methods \S~\ref{sec:synthesis_methods}), creating extremely fine, homogenous, glassy precursor phases.  As illustrated in Fig.~\ref{fig:xray}, in the case of La\dots Gd, a single annealing treatment (18~h) on the glassy precursor phase is sufficient to produce phase-pure powders. The successful synthesis of phase-pure \liso~cubic pyrochlore oxides suggests that prior limitations~\cite{strobel2010structural} were a consequence of the classically established processing methods commonly employed to synthesize pyrochlore oxides, and not limitations of thermodynamics.

However, \liso~compositions with small lanthanide ions (Tb\dots Yb) did not produce cubic pyrochlores, and instead exhibited diffraction patterns fraught with numerous additional reflections, reminiscent of previous exploratory work~\cite{strobel2010structural}. A combination of scanning electron microscopy (SEM) and energy dispersive spectroscopy (EDS) confirmed the heated powders remained chemically homogenous and single phase, suggesting a reduction in symmetry from cubic to monoclinic or triclinic crystal systems. We noted an extreme sensitivity of the additional peaks with anneal time and temperature, suggestive of complex crystallization kinetics in the trans-Gd compounds.

To probe the crystallization process \emph{ex situ}, we designed a series of quenching experiments to generate a crude time-temperature phase diagram. We determined that the precursor, which is largely amorphous, first crystallizes into a cubic phase within a very short time frame ($\sim$1~hr). As the anneal proceeds further, the cubic phase gradually degrades and transforms into the lower symmetry structure. Additional experiments to control atmospheric conditions and test potential volatility issues failed to stymie the decomposition of the cubic phase, suggesting that the symmetry reduction is a thermodynamic effect. In this respect, although the emergence of the cubic phase from the glassy precursor is \emph{kinetically} favored, the high symmetry phase is ultimately \emph{thermodynamically} unstable for the smaller Ln$^{3+}$ ions.

The synthetic observations summarized above are consistent with  Fig.~\ref{fig:stability}, where  \liso~compositions for early lanthanides are predicted to be cubic, while destabilizing midway through the lanthanide series.  However, the observation that the cubic phase was kinetically favored provided a means to stabilize pyrochlore-like phases for the heavier lanthanides. As described in Methods \S\ref{sec:synthesis_methods},  microwave-assisted heating~\cite{levin2019protocols,michael1991tilden,rao1999synthesis,mingos1993microwave,whittaker1994application} of the mechanicochemically synthesized amorphous precursors provided a direct experimental means of achieving the extreme heat/cooling rates required to successfully quench in the cubic \liso~phases (Fig.~\ref{fig:xray}(a)). It is worth noting that trans-Gd compounds often benefit from small amounts of excess Sb$_2$O$_5$ to compensate for increased volatility during the extreme ramp rates. We note that the microwave-assisted mechanicochemical methods also work for pyrochlore phases that are accessible through classical heating methods. Thus, the combination of microwave-assisted heating and mechanochemical methods enables the rapid synthesis of the pyrochlore oxides, on the order of minutes \emph{in lieu} of days, allowing for wide experimental surveys of chemical spaces in a short amount of time.

At first glance, it appears that the entire \liso~series can be stabilized as cubic pyrochlores. As illustrated in Fig.~\ref{fig:xray}(a), XRD confirmed that all \liso~compositions crystallize in the cubic crystal system. No impurity phases can be identified within the resolution of X-ray diffraction, and the lattice parameter and cell volume trend linearly with increasing Ln$^{3+}$ radii (Fig.~\ref{fig:xray}(b, top)), consistent with reports for other pyrochlore families~\cite{subramanian1983rare}. For lighter lanthanides (La\dots Gd), \liso~compositions are best described with the pyrochlore $Fd\bar{3}m$ structure,  possessing the characteristic (311) and (331) ``pyrochlore'' superlattice reflections~\cite{subramanian1983rare,Gardner10:82,maram2018probing,cryst6070079}. However, it is worth noting that due to the particular elements involved, these superlattice peaks in \liso~are extremely weak. For the ideal pyrochlore structure with no disorder present between the $A$- and $B$-sites, the (311) and the (331) superlattice reflections in La$_2$InSbO$_7$ are \textit{predicted} to be 0.2\% and 1.4\% of the main (222) reflection, respectively. 

In the case of trans-Gd compositions (Tb\dots Lu), assignment of the unit cell is more nuanced. Although these materials do maintain cubic symmetry (Fig.~\ref{fig:xray}(a)), the (311) and (331) superlattice reflections become vanishingly small, as illustrated in Fig.~\ref{fig:xray}(b, bottom). The black trace denotes the ``fit fraction'' of the (331) reflection, where the ``fit fraction'' is defined by:
\begin{equation}
    F = 100\% \times \left( 1-\frac{ I_{\text{hkl,Calc.}} - I_{\text{hkl,Obs.}} }{ I_{\text{hkl,Calc.}} - I_{\text{hkl,Bkg.}} } \right)
\end{equation}
\noindent where $F$ was constructed such that perfect agreement with the pyrochlore structure is defined as 100\%, and complete suppression of the (331) corresponds to 0\%.  The loss of the (331) peak as the Ln$^{3+}$ radius shrinks is mirrored by a gradual shift in the position of the  $48f$ oxygen ion to the higher symmetry point at (3/8, 1/8, 1/8). 
This shift towards a symmetric oxygen coordination is likely associated with a gradual disordering of the $A$- and $B$-sublattices~\cite{10.3389/fchem.2021.778140,POPOV2016669}. However, we caution that oxygen is notoriously difficult to fully characterize with standard X-ray diffraction. Incremental increases in disorder are consistent with recent results in the pyrochlore zirconates, where a combination of XANES, synchrotron, and neutron diffraction indicate that the defect fluorite to pyrochlore transition across the series is gradual~\cite{blanchard2012does}. However, in the case of \liso, the situation has an additional layer of complexity since the cubic phase is not the thermodynamic ground state for trans-Gd compositions. Thus, we move to conservatively classify the \liso~ compositions from (Tb...Lu) as defect fluorite phases. Unfortunately, as is the case of \liso, since the In$^{3+}$, Sb$^{5+}$, and Ln$^{3+}$ ions all share relatively similar electron densities, it is difficult to directly quantify the degree of A- and B-site disordering. However, loss of the (311) and (331) superlattice reflections is typically sufficient to suggest suppression of pyrochlore-like domains.

Despite our tentative classification of the trans-Gd \liso~ phases as defect fluorite phases, we feel it appropriate to highlight recent results that demonstrate how the transition from pyrochlore to defect fluorite is not always a simple matter. Recent work on Tb$_2$Hf$_2$O$_7$~\cite{TbHf_sibille2017coulomb} demonstrates that that the loss of the pyrochlore superlattice reflections does not \textit{necessarily} arise from cation disordering. In the case of the hafanate, a combination of neutron powder diffraction and resonant X-ray diffraction confirmed the presence of an ordered ``pyrochlore'' cation sublattice. Instead, the anionic sublattice experiences progressive disordering, marked by a high density of oxygen Frenkel defects.~\cite{TbHf_sibille2017coulomb} Reflecting on our observations in the \liso~ compounds, such an effect could conceivably be convolved in our observation of shifts in the 48$f$ oxygen position, but ultimately lies outside our experimental confidence at this time. Further experiments including neutron diffraction are underway. However, as we will demonstrate in our magnetization measurements, many of the trans-Gd compositions retain magnetic properties that are strikingly similar to their \textit{pyrochlore} stannate and titanate analogs.

Revisiting Fig.~\ref{fig:stability}, it is clear that the structural trends identified in the \liso~series are consistent with its location in the predicted pyrochlore stability field. Despite having a nearly identical effective $B$-site radius as the stannates (0.69~\AA)~\cite{Shannon:a12967}, the \liso~series (0.70~\AA) does not share the structural stability of the stannates. This striking drop in the phase stability of the pyrochlore structure is mirrored by other large mixed $B$-site pyrochlores such as Ln$_2$ScNbO$_7$~\cite{jiang2020probing,zouari2008synthesis,mcqueen2008frustrated} (0.693~\AA) and Ln$_2$ScTaO$_7$~\cite{shlyakhtina2014synthesis,shlyakhtina2018proton,SATO76:38,MLZNd} (0.693~\AA), both of whom have effective Shannon radii~\cite{Shannon:a12967} closely comparable to that of tetravalent tin. As our discussion shifts towards the magnetic properties of the \liso~family of compounds, we will aim to make comparisons between not only the stannates, but also with their heavier analogs ($e.g.$ hafanates and zirconates), and other mixed B-site pyrochlores.

\begin{figure*}
\centering
\includegraphics[width=7.05in]{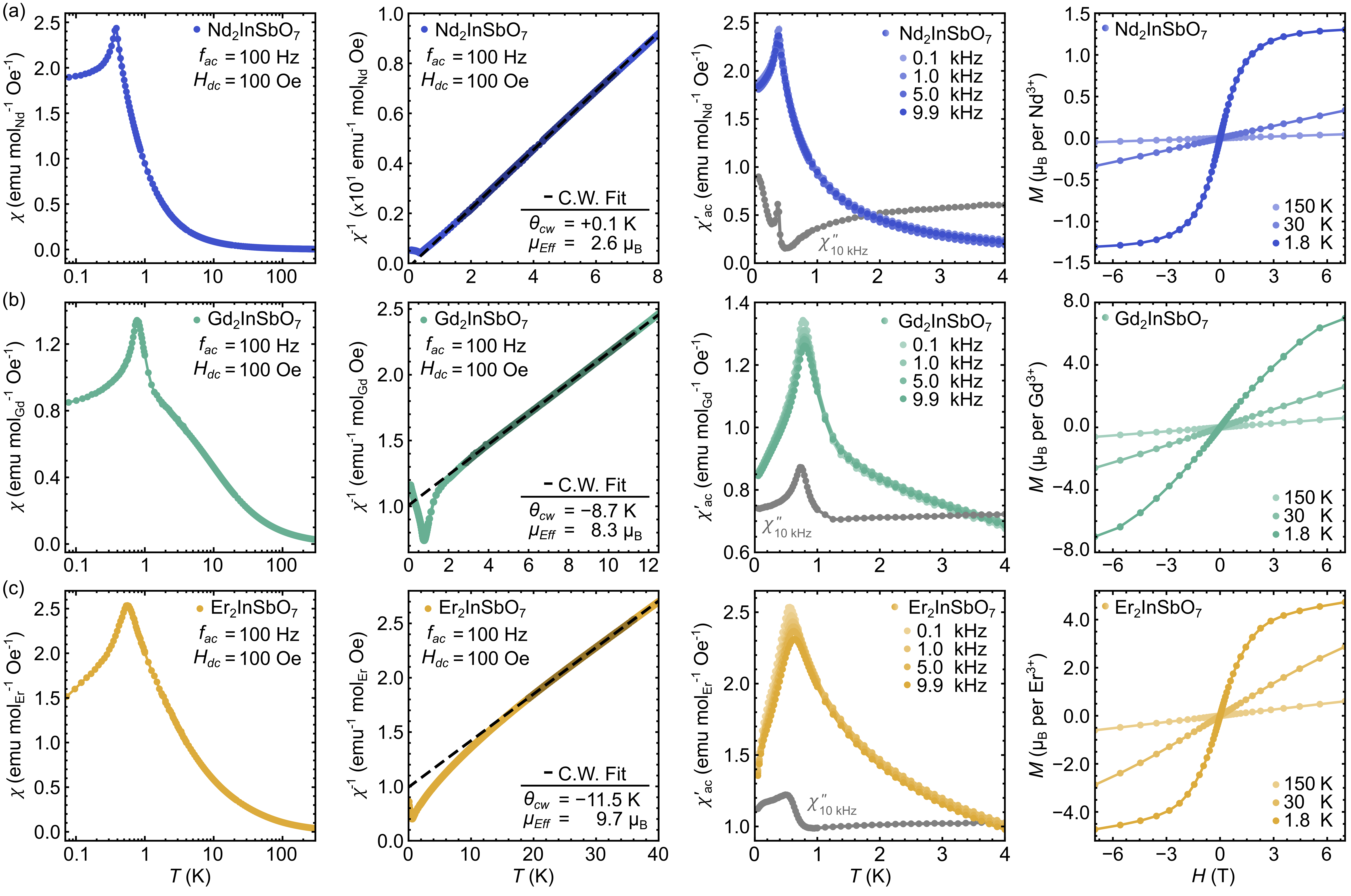}
\caption{Combined temperature dependence of the ac and dc magnetization with corresponding Curie-Weiss fits, and the isothermal dc magnetization of (a) Nd$_2$InSbO$_7$, (b) Gd$_2$InSbO$_7$, and (c) Er$_2$InSbO$_7$. For temperatures $T<4$~K, the ac susceptibility was measured at various frequencies $f_{ac}$ in the absence of a dc field. For $T>1.8$~K, the dc susceptibility was measured in an external field $H_{dc}$ of 100~Oe. 
}
\label{fig:afm}
\end{figure*}

\subsection{dc and ac magnetic susceptibility}

Drawing analogies to the stannates, members of the \liso~family have been organized into four categories based on their proposed magnetic ground state: 1) long-range antiferromagnetic order, 2) dynamic magnetic ground states, 3) dipolar spin ices, and 4) nonmagnetic singlet states. As shown below, 
despite mirroring a stability field similar to the hafanate counterparts~\cite{subramanian1983rare,Gundovin75:20,KLEE69:31,Besson66:262C,SPIRIDINOV68:14,TbHf_sibille2016ms21,TbHf_shlyakhtina2006structure}, the properties of the \liso~series are remarkably consistent with the magnetic properties of the stannates, the non-mixed $B$-site family with whom they share the closest structural properties~\cite{subramanian1983rare,Kennedy97:58}. Remarkably, even in the cases of trans-Gd compositions where the structure would be conservatively classified as defect fluorite, the transport properties of the \liso~family remain reminiscent of the behavior exhibited by their pyrochlore stannate counterparts. 

\subsubsection{Long-range antiferromagnetic order: \texorpdfstring{Nd$_{2}$InSbO$_{7}$}{Nd2InSbO7}, \texorpdfstring{Gd$_{2}$InSbO$_{7}$}{Gd2InSbO7} \& \texorpdfstring{Er$_{2}$InSbO$_{7}$}{Er2InSbO7}}

\textbf{Nd$_2$InSbO$_7$}: As illustrated in Fig.~\ref{fig:afm}, a $\lambda$-type peak at T$\rm{_{N}}$ = 0.37~K is present in the ac susceptibility of  Nd$_{2}$InSbO$_{7}$. Exhibiting no clear frequency dependence, such a peak is consistent with a transition into a long-range magnetic ordered state. As the isothermal magnetization suggests the presence of $\langle 111 \rangle$ Ising single-ion anisotropy~\cite{Bramwell99:12}, combined with the small magnitude of the Curie-Weiss temperature, our data supports the possibility that Nd$_{2}$InSbO$_{7}$ assumes non-coplanar all-in-all-out magnetic order that has been reported for all other Nd-based pyrochlores~\cite{gomez2021absence,Scheie21:104,Bertin15:92,xu15:92,Lhotel15:115,Anand17:95,Anand15:92,matsuhira13:82,tomiyasu12:81,Mauws21:33,yasui01:70,Ku18:30}. 

While possessing many of the generic features of Nd-based pyrochlores, Nd$_{2}$InSbO$_{7}$ exhibits a key distinguishing feature: a exceptionally low Curie-Weiss temperature of +0.1~K. Recently, Gomez \emph{et al.}~\cite{gomez2021absence} proposed that chemical pressure plays a crucial role in realization of moment fragmentation~\cite{brooks14:4}. The authors argue that any contraction of the lattice would decrease the value of the Curie-Weiss constant as a result of the enhancement of the antiferromagnetic Nd$^{3+}$-O$^{2-}$-Nd$^{3+}$ superexchange pathways, which would exclude moment fragmentation in favor of long-range antiferromagnetic order~\cite{Bertin15:92,Benton16:94}. Solely utilizing the lattice parameter, both T$\rm{_{N}}$ and $\theta\rm{_{CW}}$ for  Nd$_{2}$InSbO$_{7}$ compare favorably to the general trends observed among all Nd-based pyrochlores summarized by Fig.~8 in Gomez \emph{et al.}~\cite{gomez2021absence}. 

We suspect that Nd$_{2}$InSbO$_{7}$ may exhibit the same rich physics present in the ScNb~\cite{Scheie21:104,Mauws21:33}, Hf~\cite{PrHf_sibille2018experimental}, Zr~\cite{Benton16:94,petit16:12,zoghlin21:5} analogs due to its proximity to the crossover point in the sign of the Curie-Weiss temperature. Recently, Scheie \emph{et al.}~\cite{Scheie21:104} suggested that both Nd$_{2}$ScNbO$_{7}$ and Nd$_{2}$Zr$_{2}$O$_{7}$ share a common, unusual fluctuating magnetic ground state. This presents the exciting possibility that Nd$_{2}$InSbO$_{7}$ may exhibit moment fragmentation. With the presence of controlled chemical disorder, Nd$_{2}$InSbO$_{7}$ may provide additional insights into the physical origins of the strong quantum fluctuations~\cite{Benton16:94,Xu20:124,kimura13:4} underlying both the nature of moment fragmentation and the magnetic ground state present in the heavier Nd-based pyrochlores.

\textbf{Gd$_2$InSbO$_7$}: With a half-filled 4$f^{7}$ electronic configuration ($L$=0, $S$=7/2)~\cite{jensen1991}, Gd$^{3+}$ is expected to possess minimal spin anisotropy, particularly smaller than Ho$^{3+}$, Dy$^{3+}$, and Tb$^{3+}$~\cite{raju99:59,cashion1968crystal,PhysRevB.72.020409}. Combined with dominant antiferromagnetic exchange when placed on a pyrochlore lattice, Gd-based pyrochlores are excellent candidates for the experimental realization of a classical Heisenberg pyrochlore antiferromagnet and its predicted ground state degeneracy~\cite{canals98:80,canals00:61,Moessner98:80,Moessner98:58,nawa18:98,Reimers92:45,reimers91:43,villain1979insulating}.

As is the case for the other Gd$_{2}$B$_{2}$O$_{7}$ pyrochlores (Ti~\cite{Bonville_2003,cashion1968crystal,LUO2001306}, Sn~\cite{Bonville_2003,Sn_Matsuhira02:71,LUO2001306}, Sc/Nb~\cite{james_thesis}, Hf~\cite{Durand08:20}, Zr~\cite{Durand08:20}, Pb~\cite{Hallas15:91}), the presence of minimal single-ion anisotropy results in a Curie-Weiss-like magnetization response over a large temperature range. The presence of dominant antiferromagnetic interactions with a $\theta\rm{_{CW}}$=$-$8.7~K, results in the deviation of the isothermal magnetization from the Brillouin function in the low field limit before its saturation at $g_{J}J$=7~$\mu\rm{_{B}}$~\cite{Freitas11:23}.  

In stark contrast with Gd$_{2}$ScNbO$_{7}$~\cite{james_thesis}, Gd$_{2}$InSbO$_{7}$ exhibits a frequency independent $\lambda$-type anomaly at $T\rm{_{N}}$=0.78~K in the ac susceptibility, corresponding to the onset of long-range order. In the case of ScNb~\cite{james_thesis}, a Mydosh parameter of 0.020(1) places Gd$_{2}$ScNbO$_{7}$ in the conventional spin glass regime~\cite{mydosh1993spin,Zhou_2008}. The glassy nature of Gd$_{2}$ScNbO$_{7}$ is consistent with the bifurication of the ZFC/FC dc magnetization, linear low-temperature specific heat, and $\mu$SR results. Instead, Gd$_{2}$InSbO$_{7}$ has a sharp $\lambda$-type anomaly which is highly reminiscent of Gd$_{2}$Sn$_{2}$O$_{7}$~\cite{Wills06:18,Bonville_2003}. Unlike the titanate, which exhibits two closely spaced transitions~\cite{Bonville_2003,champion01:64,Stewart_2004}, the stannate assumes the $\mathbf{k}$ = (0,0,0) $\Gamma_{7}$ ``Palmer-Chalker" state~\cite{Wills06:18,Stewart08:78}. In fact, such a collinear antiferromagnetic state is theoretically predicted for a Heisenberg
antiferromagnet with dipolar interactions~\cite{PC}, and corresponds to the proposed state for Gd$_{2}$Pt$_{2}$O$_{7}$~\cite{hallas16:94}. Such a magnetic structure is much simpler to determine experimentally when compared to the proposed multi-$\mathbf{k}$ structure of the titanate that still remains the subject of debate~\cite{Stewart_2004,paddison2021suppressed,champion01:64,raju99:59,Petrenko04:70,brammall11:83,PhysRevLett.114.130601}.

The presumed long-range antiferromagnetic order of of Gd$_{2}$InSbO$_{7}$ is particularly interesting in the context of the spin-glass ScNb.~\cite{james_thesis} A generic mixed $B$-site Gd-based pyrochlores generally exhibits the three key ingredients of a spin glass: 1) frustration, 2) disorder, and 3) competing interactions. However, our conceptual intuition may be limited, particularly in the face of a recent report of the effects of sample quality in Gd$_{2}$Zr$_{2}$O$_{7}$~\cite{xu_thesis,Durand08:20}. The clear absence of glassy behavior in Gd$_{2}$InSbO$_{7}$, despite substantial (intentional) chemical disorder, demonstrates that other subtle perturbative terms such as beyond nearest neighbor exchange terms~\cite{Wills06:18,cepas4:69} may be particularly influential in the selection of its magnetic ground state.       

\textbf{Er$_2$InSbO$_7$}: As is the case for their Heisenberg Gd$^{3+}$ counterparts, the inclusion of dipolar interactions to the isotropic exchange Hamiltonian for XY pyrochlores (e.g. Er$^{3+}$) with dominant antiferromagnetic interactions is predicted to lift any extensive classical degeneracy, and uniquely select the Palmer-Chalker state~\cite{PC,guitteny2013palmer,champion03:68}. Experimentally, however, this is not the case for the majority of XY pyrochlores~\cite{dun15:92,Poole_2007,YbSn_lago2014glassy,YbSn_yaouanc2013dynamical,cai16:93,YbTi_gaudet2016gapless,YbTi_yasui2003ferromagnetic,dun15:92,Hallas16:93}. Exemplified by Er$_{2}$Ti$_{2}$O$_{7}$, where long-range magnetic order is realized \emph{via} a quantum order-by-disorder mechanism~\cite{champion03:68,savary12:109,zhitomirsky12:109,wong13:88,Champion_2004}, XY pyrochlores present a versatile experimental platform for the effects of multi-phase competition~\cite{hallas18:9}.  

The presence of multi-phase competition has been used extensively to account for key experimental signatures that are characteristic of the XY pyrochlores~\cite{hallas18:9}. These include: multiple heat capacity anomalies, suppression of the ordering temperature, unconventional spin dynamics, defect sensitivity, and pressure-dependent magnetic ground states. Recently, the application of multi-phase competition has been extended~\cite{YbCompetition_yan2017theory,jaubert15:115,PhysRevB.100.094420,Scheie20:117} further to address the vastly different magnetic ground states observed among each of the two families of XY pyrochlores. 

In Er-based pyrochlores, members with a smaller lattice parameter tend to select for long-range antiferromagnetic order in $\Gamma_{5}$~\cite{Poole_2007,dun15:92}, while their larger counterparts assume the Palmer-Chalker state at significantly lower N\'{e}el temperatures~\cite{Hallas17:119,petit2017long}. The effects of multi-phase competition becomes particularly apparent when comparing the titanate that orders at 1.23~K with $\psi_{2}$ ($\Gamma_{5}$)~\cite{champion03:68,Poole_2007,Rau16:93}, while its stannate counterpart begins to order at 108~mK in the Palmer-Chalker state~\cite{petit2017long,Yahne21:127,guitteny2013palmer}. The significant suppression of the second order magnetic transition, combined with the multiscale dynamics, is consistent with the placement of Er$_{2}$Sn$_{2}$O$_{7}$ in close proximity to the $\Gamma_{7}$/$\Gamma_{5}$ boundary~\cite{guitteny2013palmer,YbCompetition_yan2017theory,hallas18:9}. The titanate, however, exhibits little sample dependence and a robust ground state as it is located deep within $\psi_{2}$ of $\Gamma_{5}$~\cite{YbCompetition_yan2017theory,gaudet16:94}.

\begin{figure}
\centering
\includegraphics[width=\linewidth]{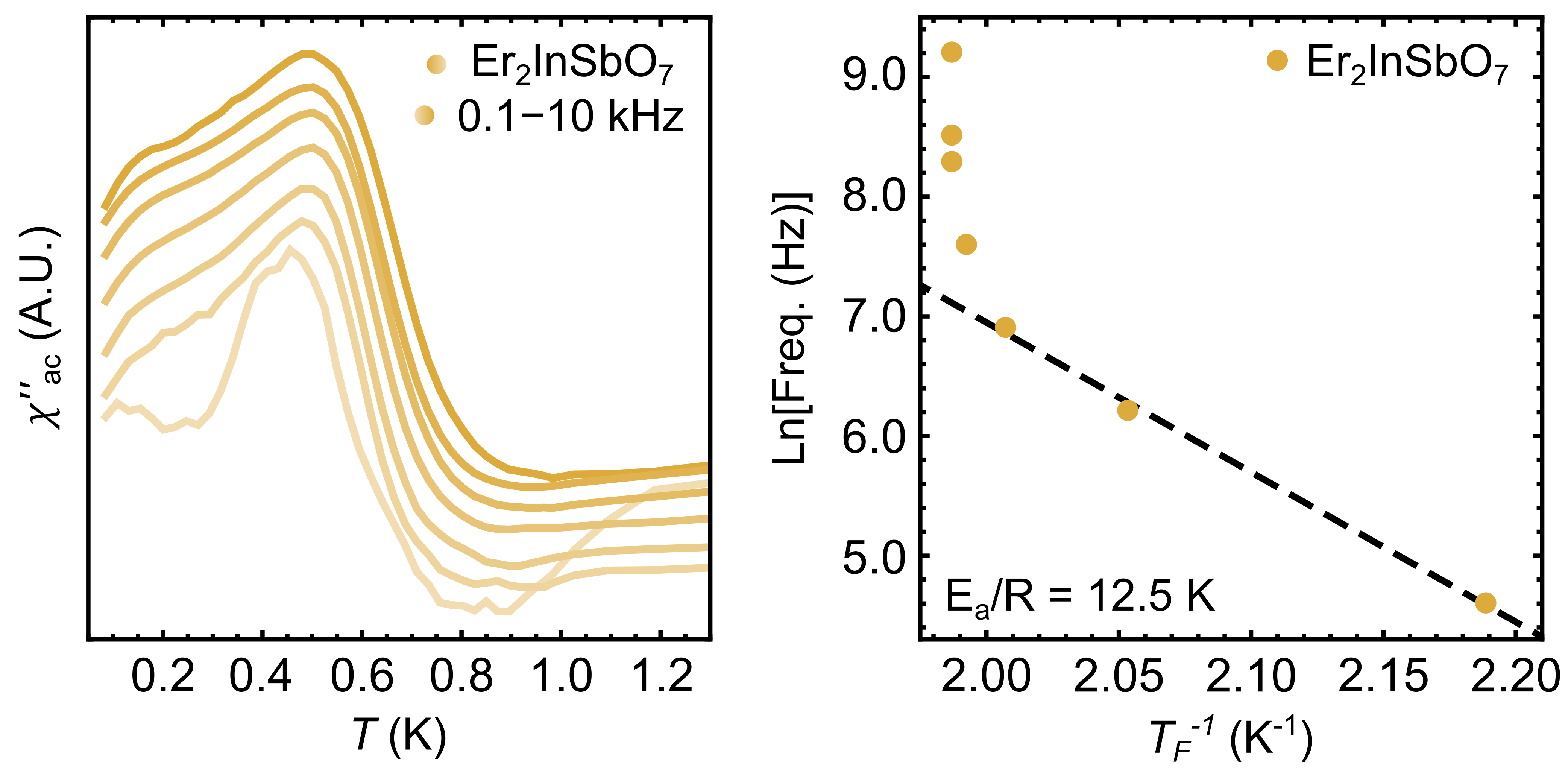}
\caption{Temperature dependence of the out-of-phase component $\chi''$ of the ac susceptibility of Er$_{2}$InSbO$_{7}$ (left) with corresponding Arrhenius analysis (right) for frequencies below 1 kHz.}
\label{fig:activationEr}
\end{figure}

As illustrated in Fig.~\ref{fig:afm}(c), both the temperature-dependent and isothermal magnetization mirrors the titanate~\cite{champion03:68,cao09:103,Bonville_2013,Petrenko_2011} and the stannate~\cite{Sarte11:23,Sn_Matsuhira02:71,EuTmSn_bondah2001magnetic,al2014thermodynamic}. Above 20~K, the magnetic susceptibility follows the Curie-Weiss law for dominant antiferromagnetic interactions ($\theta\rm{_{CW}}$=$-$11.5~K) between large Er$^{3+}$ moments ($\mu\rm{_{eff}}$=9.7~$\mu\rm{_{B}}$). The isothermal magnetization does not follow a Brillouin curve, saturating at $\sim$4.2 $\mu\rm{_{B}}$ $<<g_{J}J$. The large moment suggests a large dipolar term, and when combined with the XY single-ion anisotropy of Er$^{3+}$, a Palmer-Chalker state theoretically should be realized~\cite{PC,guitteny2013palmer,champion03:68,champion03:68,Poole_2007}. In the case of Er-based pyrochlores, B-site disorder does not preclude the assumption of long-range magnetic order. Samples produced along the Er$_{2}$Sn$_{2-x}$Ti$_{x}$O$_{7}$ solid solution retained the $\psi_{2}$ ($\Gamma_{5}$) ground state of the titanate until a quantum crossover at $x=1.7$ to $\Gamma_{7}$ of the stannate~\cite{Shirai17:96}.

In Er$_{2}$InSbO$_{7}$, the transition to a long-range ordered state is consistent with the distinct peak at $T_{N}$ = 0.56~K present in the ac susceptibility. However, unlike the Nd and Gd analogs, the prominent peak exhibits a clear frequency dependence at low frequencies. In Er$_2$Sn$_2$O$_7$, the frequency-dependence was attributed to spins located at magnetic domain boundaries~\cite{Tardif15:114,Lhotel11:107}. As illustrated in Fig.~\ref{fig:activationEr}, the frequency dependence does not follow Arrhenius-type behavior. Instead, our data exhibits extreme non-linearity for frequencies above 1 kHz. We note that this behavior may still be consistent with the stannate, as all published ac data is collected over a narrow low-frequency regime from 5-200~Hz~\cite{guitteny2013palmer,petit2017long}. We hypothesize that the non-Arrhenius behavior in Er$_{2}$InSbO$_{7}$ may be a reflection of the slow dynamics that have previously identified in Er$_{2}$Sn$_{2}$O$_{7}$~\cite{petit2017long,guitteny2013palmer}.  Ultimately, it would be interesting to determine if both the stannate and titanate analogs exhibit non-linearity in the high frequency limit. 

While Er$_{2}$InSbO$_{7}$ does exhibit clear linear behavior for lower frequencies, its activation energy of 12.5~K is over an order of magntiude larger than that of its stannate analog (0.9~K)~\cite{guitteny2013palmer,petit2017long}. Such an increase is consistent with the effects of disorder, as has been proposed in $A$-site disordered pyrochlores~\cite{snyder04:70,synder01:413,Ehlers06:73}, although it has been noted that $B$-disorder may have the opposite effect~\cite{Ke07:76}. Regardless, the distinct similarities between Er$_{2}$InSbO$_{7}$ and its stannate analogs suggest the existence of multi-scale dynamics, where slow dynamics may persist and coexist with static long-range magnetic order below T$\rm{_{N}}$.

The N\'{e}el temperature of Er$_{2}$InSbO$_{7}$ is unique among those values reported for all other Er$^{3+}$-based pyrochlores. Na\"{i}vely, with a lattice constant comparable to the stannate, it is expected that the resulting weaker exchange tensor would depress the N\'{e}el temperature of Er$_{2}$InSbO$_{7}$ towards the 0~K limit~\cite{hallas18:9}. Instead, its large T$\rm{_{N}}$=0.56~K is comparable to that of the much smaller titanate~\cite{champion03:68,cao09:103,Bonville_2013,Petrenko_2011}. Such a large N\'{e}el temperature suggests that Er$_{2}$InSbO$_{7}$ is located far from a phase boundary in the the classical ground-state phase diagram~\cite{guitteny2013palmer,YbCompetition_yan2017theory,hallas18:9}, although it important to note that without neutron diffraction, the identity of $\Gamma$ and its respective $\psi$ both remain for now an open question.    

\begin{figure*}
\centering
\includegraphics[width=7.05in]{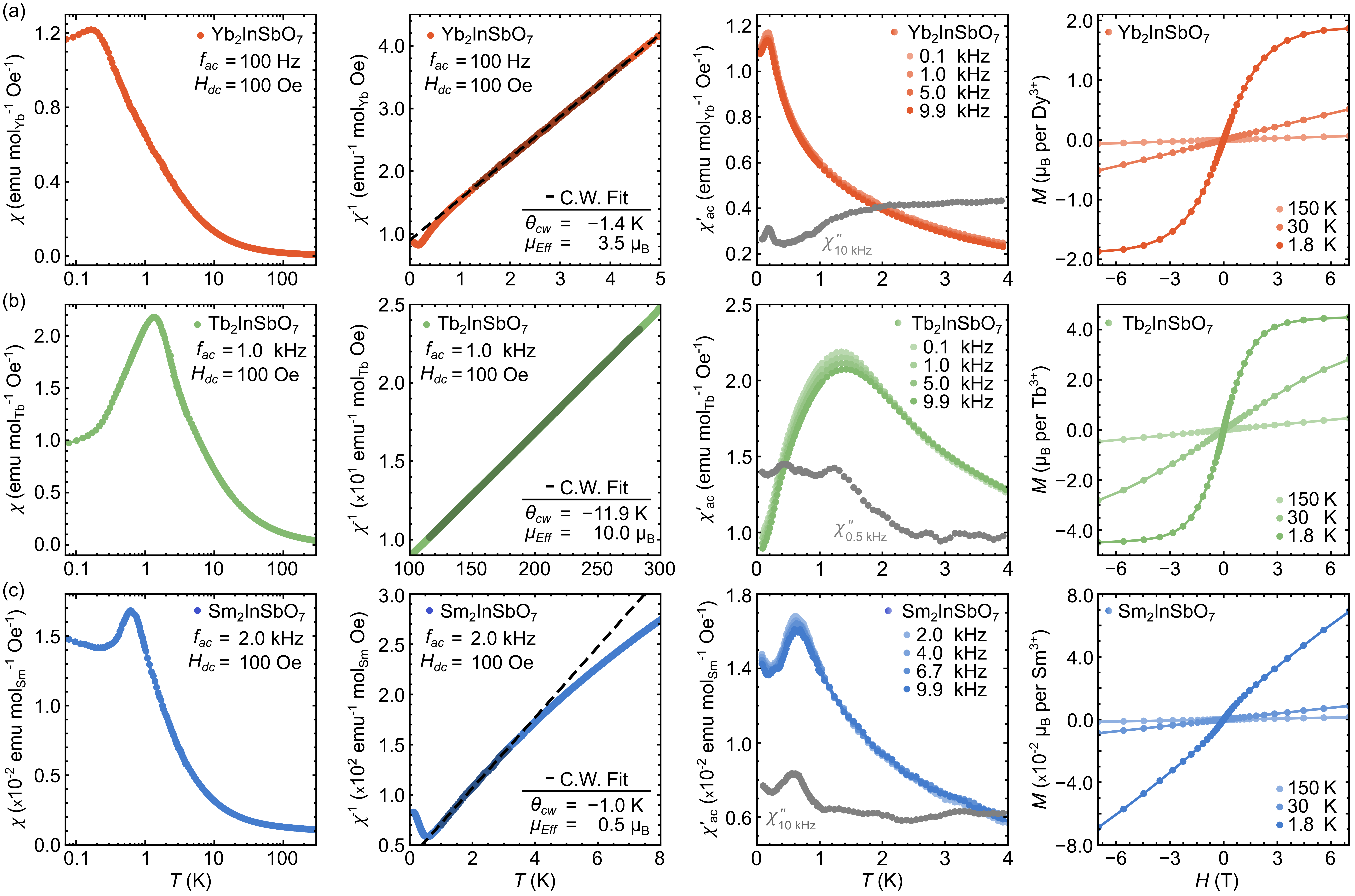}
\caption{Combined temperature dependence of the ac and dc magnetization with corresponding Curie-Weiss fits, and the isothermal dc magnetization of (a) Yb$_2$InSbO$_7$, (b) Tb$_2$InSbO$_7$, and (c) Sm$_2$InSbO$_7$. For temperatures $T<4$~K, the ac susceptibility was measured at various frequencies $f_{ac}$ in the absence of a dc field. For $T>1.8$~K, the dc susceptibility was measured in an external field $H_{dc}$ of 100~Oe. 
}
\label{fig:dynamic}
\end{figure*}

\subsubsection{Dynamic Magnetic Ground States: \texorpdfstring{Yb$_{2}$InSbO$_{7}$}{Yb2InSbO7}, \texorpdfstring{Tb$_{2}$InSbO$_{7}$}{Tb2InSbO7} \& \texorpdfstring{Sm$_{2}$InSbO$_{7}$}{Sm2InSbO7}}

\textbf{Yb$_2$InSbO$_7$}: Corresponding to the second, and arguably most prominent family of the XY pyrochlores~\cite{hallas18:9}, Yb-based pyrochlores have been subject to intense interest since the proposal of quantum spin ice behavior in Yb$_2$Ti$_2$O$_7$~\cite{Yb_ross2011quantum,Yb_gingras2014quantum,applegate12:109,petit20:117,Scheie20:117,pan2016measure}. The subsequent explosion of experimental and theoretical studies have propelled research into Yb-based magnetism in various frustrated geometries, including the  triangular~\cite{PhysRevB.98.220409,liu2018rare,PhysRevB.99.180401,bordelon2019field,PhysRevB.100.224417,PhysRevB.103.214445,PhysRevLett.115.167203,balents2010spin,paddison2017continuous} and Shastry-Sutherland lattices~\cite{SRIRAMSHASTRY19811069,PhysRevLett.110.017201,PhysRevB.83.214427,doi:10.1143/JPSJ.78.083708,liu2014antiferromagnetic,PhysRevLett.109.167202}. Fueled by the enhanced quantum fluctuations that are a result of the $J{\rm{_{eff}}}=1/2$ degrees of freedom originating from the thermally well-isolated Kramers doubet single-ion ground state~\cite{Hodges_2001,Hallas16:93,PhysRevB.92.134420,PhysRevB.70.075112,hallas18:9,YbCompetition_yan2017theory,Gardner10:82}, Yb-based magnetism has remained a fruitful arena for the search of quantum spin liquid states~\cite{hallas18:9,PhysRevB.69.064404,PhysRevLett.100.047208,PhysRevLett.108.037202,PhysRevB.86.075154,PhysRevB.86.104412,PhysRevB.87.205130,Yb_gingras2014quantum}.

As is the case for their Er$^{3+}$ counterparts, multi-phase competition in Yb-based pyrochlores is particularly influential on the magnetic properties of the selected ground state~\cite{Scheie20:117,jaubert15:115}. In addition to extremely depressed transition temperatures ($T\rm{_{N}}, T\rm{_{C}} \ll 1$~K), Yb-based pyrochlores have demonstrated particular sensitivity to sample preparation and defect formation~\cite{YbTi_ross2012lightly,YbSn_dun2014chemical,YbSn_hallas2016universal,hallas18:9,PhysRevB.95.094407,PhysRevB.84.172408,PhysRevB.88.134428,PhysRevB.89.224419,PhysRevB.87.134405,YbTi_gaudet2016gapless,bowman2019role,sala2014vacancy,PhysRevB.95.094431}. The case of Yb$_{2}$Ti$_{2}$O$_{7}$ is perhaps the most infamous. Minute changes in the Yb stoichiometry have been demonstrated to broaden, depress, or completely suppress any signature of long-range static magnetic order in heat capacity and magnetization.~\cite{YbTi_chang2012higgs,YbTi_gaudet2016gapless,YbTi_ross2012lightly,YbTi_yasui2003ferromagnetic,PhysRevB.84.172408,PhysRevB.84.174442,PhysRevB.88.134428,PhysRevB.89.184416,hallas18:9,PhysRevB.95.094407}. Interestingly enough, the application of a moderate amount of pressure~\cite{kermarrec2017ground} is sufficient to recover the $\textbf{k}=0$ splayed ferromagnetic state. Altogether, these results are consistent with the placement of the titanate near the ferromagnetic $\Gamma_{9}$/antiferromagnetic $\Gamma_{5}$ phase boundary~\cite{YbCompetition_yan2017theory,jaubert15:115,Yb_ross2011quantum}.   

Given the extreme fragility of the magnetic ground state in Yb$_2$Ti$_2$O$_7$, one would expect full suppression of the long-range order in the mixed B-site Yb$_2$GaSbO$_7$. Indeed, Yb$_{2}$GaSbO$_{7}$ does not exhibit the characteristic $\lambda$-type anomaly in its heat capacity~\cite{YbGaSb_sarte2021dynamical,blote1969heat} indicative of static long-range magnetic order in Yb$^{3+}$-based pyrochlores~\cite{YbSn_dun2013yb,YbSn_dun2014chemical,YbSn_hallas2016universal,YbSn_lago2014glassy,YbSn_yaouanc2013dynamical,YbTi_chang2012higgs,YbTi_gaudet2016gapless,YbTi_ross2012lightly,YbTi_yasui2003ferromagnetic,champion01:64,cai16:93,PhysRevLett.88.077204}. At a cursory glance this seems consistent with the deleterious effect of disorder, though more thorough investigations demonstrate that only a minority of spins freeze, with the majority remaining dynamic to base temperatures as confirmed by both inelastic neutron scattering~\cite{YbGaSb_sarte2021dynamical} and $\mu$SR spectroscopy~\cite{hodges2011magnetic}. 

In fact, strong dynamical correlations have been shown to be ubiquitous among the Yb$^{3+}$-based pyrochlores~\cite{YbSn_hallas2016universal,hallas18:9}, developing coherence at a temperature T$^{*}$~$>$~T$_{\rm{N}}$,T$_{\rm{C}}$ centered about a broad feature present in the heat capacity~\cite{PhysRevLett.88.077204,YbSn_yaouanc2013dynamical,Hallas16:93,cai16:93,YbSn_hallas2016universal}. These unconventional dynamics correspond to gapless $\mathbf{Q}$=0 excitations~\cite{YbSn_hallas2016universal,YbTi_gaudet2016gapless,YbSn_dun2013yb,Hallas17:119,petit2017long} and are remarkably robust against disorder and doping effects, regardless of the presence (or suppression) of long-range magnetic order~\cite{hallas18:9,YbSn_hallas2016universal,Sarte11:23}. Similar effects have been particularly well-documented in Tb$^{3+}$-based pyrochlores~\cite{rau19:10,PhysRevB.92.245114,PhysRevB.87.060408,doi:10.7566/JPSJ.87.064704,PhysRevB.68.180401,doi:10.1143/JPSJ.68.3802,TbHf_anand2018evidence,Gaulin15:91}.

In the case of Yb$_{2}$GaSbO$_{7}$, the application of a modest 1~T field shifts spectral weight from these low energy fluctuations into magnetic Bragg peaks corresponding to $\Gamma_{9}$ long-range magnetic order~\cite{YbGaSb_sarte2021dynamical}. Although corresponding to the zero field magnetic structure of the titanate, 
the appearance of such static magnetic order despite the presence of maximal $B$-site disorder at first is not expected, particularly when considering stuffing of the order of 1-2\% in the Yb$_{2}$Ti$_{2}$O$_{7}$ suppresses long-range ordering~\cite{hallas18:9,YbTi_ross2012lightly,PhysRevB.84.172408,PhysRevB.84.174442,PhysRevB.88.134428,PhysRevB.95.094407}. While mechanical pressure can be used to recover the magnetic order in Yb$_{2}$Ti$_{2}$O$_{7}$, the energy scale is quite disparate from the application of a 1~T field in Yb$_{2}$GaSbO$_{7}$. This is evidence that \textit{antisite} disorder (stuffing) and $B$-site disorder operate independently -- and should be experimentally discernible.        

As summarized in Fig.~\ref{fig:dynamic}(a), the low temperature magnetic properties of Yb$_{2}$InSbO$_{7}$ are reminiscent of its GaSb analog~\cite{YbGaSb_sarte2021dynamical} with comparable Curie-Weiss parameters (3.5$\mu_{\text{eff}}$/Yb$^{3+}$and $\theta\rm{_{CW}}$=$-$1.4~K). In the low temperature limit, a frequency-independent peak at 0.17~K dominates, reminiscent of both Yb$_{2}$GaSbO$_{7}$ and the relatively sharp features present in pristine samples of both the titanate~\cite{PhysRevB.95.094407,YbTi_yasui2003ferromagnetic,YbTi_gaudet2016gapless,PhysRevLett.119.127201,YbTi_yaouanc2016novel,YbTi_chang2012higgs,PhysRevB.89.184416} and stannate~\cite{YbSn_dun2013yb,YbSn_yaouanc2013dynamical,YbSn_lago2014glassy}. The energy scale is comparable to the stannate (T$\rm{_{C}}\sim$0.11--0.15~K~\cite{YbSn_dun2013yb,YbSn_yaouanc2013dynamical,YbSn_lago2014glassy}) which is proximal in terms of both average $B$-site radius~\cite{Shannon:a12967} and cell volume. The comparison between InSb:Sn is analogous to GaSb:Ti, where the smaller GaSb (0.35~K~\cite{YbGaSb_sarte2021dynamical}) mimics the respective titanate (0.26~K~\cite{YbTi_yasui2003ferromagnetic,YbTi_gaudet2016gapless,YbTi_ross2012lightly}).   

Exploiting the analogy to its GaSb analog, it is more probable that the sharp peak in both $\chi'$ and $\chi''$ corresponds to a freezing of a minority of spins instead of long-range ferromagnetic order. The striking similarities between Yb$_{2}$InSbO$_{7}$ and Yb$_{2}$GaSbO$_{7}$ are particularly noteworthy since Yb$_{2}$InSbO$_{7}$ is outside the pyrochlore stability field (Fig.~\ref{fig:stability}) and may be intermediate between the pyrochlore and defect fluorite structures. The analogous behavior of InSb and GaSb may suggest that the cation sublattice (up to a certain correlation length) remains intact, despite a highly disordered anionic sublattice, as is the case for Tb$_{2}$Hf$_{2}$O$_{7}$~\cite{TbHf_sibille2017coulomb}.  

Although it remains to be seen whether Yb$_2$InSbO$_7$ will exhibit properties consistent with Yb$_2$GaSbO$_7$ or its stannate analogs,  the interaction between long-range magnetic order and the mixed B-site pyrochlores is interesting. Given that 1-2\% of Yb disorder suppresses long-range magnetic order~\cite{hallas18:9,YbTi_ross2012lightly,PhysRevB.84.172408,PhysRevB.84.174442,PhysRevB.88.134428,PhysRevB.95.094407}, the observation that such suppression can be reversed by a modest field in Yb$_{2}$GaSbO$_{7}$~\cite{YbGaSb_sarte2021dynamical}, with its much larger structural and steric changes, suggests that $B$-site disorder may present a unique (and very distinct) chemical means of reliably suppressing long-range ferromagnetic order in Yb-based pyrochlores, while preserving the Yb$^{3+}$ sublattice and its underlying persistent dynamics~\cite{YbSn_hallas2016universal,hallas18:9}.

Despite possessing a wealth of rich and complex set of phenomena that rivals the Yb-based pyrochlores, theoretical efforts to address their Tb-based counterparts have remained an enduring challenge~\cite{rau19:10,Gardner10:82}. In contrast to the other lanthanide ions presented so far, the non-Kramers doublet of the $J$=6 free-ion ground state manifold of Tb$^{3+}$ is separated from its first excited doublet by only $\sim$1.4~meV~\cite{PhysRevB.91.224430,PhysRevB.89.134410,PhysRevB.76.184436,PhysRevB.62.6496,Gardner10:82,Gardner99:82}. Corresponding to a gap roughly an order of magnitude smaller than Yb$^{3+}$, the application of the pseudospin-1/2 model is not necessarily straightforward, necessitating the incorporation of more subtle effects, possibly even on equal footing with $\hat{\mathcal{H}}\rm{_{CEF}}$~\cite{PhysRevLett.98.157204,https://doi.org/10.48550/arxiv.0912.2957,https://doi.org/10.48550/arxiv.2009.05036}. Such subtle effects include virtual crystal field transitions, secondary corrections that introduce anistropic exchange and may potentially stabilize a quantum spin ice ground state in the cooperative paramagnet Tb$_{2}$Ti$_{2}$O$_{7}$~\cite{PhysRevLett.98.157204,PhysRevB.99.224407,https://doi.org/10.48550/arxiv.0912.2957,PhysRevB.68.172407,Molavian_2009,Yb_gingras2014quantum,PhysRevB.62.6496}. 

\textbf{Tb$_2$InSbO$_7$}: As is the case for its Yb$^{3+}$ counterpart, multi-phase competition is particularly influential on the magnetic ground state of Tb$_{2}$Ti$_{2}$O$_{7}$~\cite{rau19:10,PhysRevLett.116.217201,PhysRevLett.111.087201,PhysRevB.86.174403,Wakita_2016}. Highly reminiscent of Yb$_{2}$Ti$_{2}$O$_{7}$, the magnetic ground state of Tb$_{2}$Ti$_{2}$O$_{7}$ is extremely sample dependent~\cite{PhysRevB.93.144407,Gardner99:82,PhysRevB.68.180401,PhysRevB.62.6496,PhysRevB.82.100402,Takatsu_2011,PhysRevB.84.184403,PhysRevB.69.132413,PhysRevLett.112.017203,PhysRevLett.83.1854} with deviations in stoichometry on the level of 1-2\% yielding quadrupolar order~\cite{PhysRevB.87.060408,PhysRevLett.116.217201,doi:10.1142/S2010324715400032}, consistent with the placement of the titanate in close proximity to the boundary between quantum spin ice and quadrupolar order~\cite{rau19:10}. Contrast such behavior to that exhibited by the stannate which assumes  ``soft" (or dynamic) spin ice ordering corresponding to Tb$^{3+}$ moments oriented 13.3$^{\degree}$ relative to the local $\langle111\rangle$~\cite{TbSn_mirebeau2005ordered,TbSn_bert2006direct,TbSn_cava2011low,TbSn_chapuis2007ground,TbSn_de2006spin,TbSn_mirebeau2005ordered,TbSn_rule2007polarized,TbSn_rule2009neutron}. 

As summarized in Fig.~\ref{fig:dynamic}(b), the isothermal magnetization of Tb$_{2}$InSbO$_{7}$ is consistent with Ising single-ion anisotropy characteristic of Tb$^{3+}$-based pyrochlores~\cite{rau19:10,Yb_gingras2014quantum,PhysRevB.62.6496,PhysRevB.89.134410,Hallas14:113,TbHf_sibille2017coulomb}. The high temperature dc magnetization is qualitatively similar to both its stannate~\cite{TbSn_cava2011low,TbSn_mirebeau2005ordered,EuTmSn_bondah2001magnetic,Sn_Matsuhira02:71} and titanate~\cite{PhysRevB.62.6496,LUO2001306,Gardner99:82} analogs with comparable Curie-Weiss parameters of $\mu\rm{_{eff}}$ = 10~$\mu\rm{_{B}}$ and $\theta\rm{_{CW}}$ = $-11.9$~K. However, the sharp feature indicative of long-range order in the stannate~\cite{TbSn_cava2011low} is absent, replaced by a broad, frequency-independent feature centered about $\sim$1.5~K in $\chi'$. These properties are highly reminiscent of Tb$_{2}$SnTiO$_{7}$~\cite{Gaulin15:91}. In the titanate~\cite{LUO2001306,PhysRevB.69.132413,PhysRevB.68.180401,PhysRevB.86.020410}, stannate~\cite{TbSn_cava2011low,TbSn_mirebeau2005ordered}, and  Tb$_{2}$GaSbO$_{7}$~\cite{blote1969heat}, the broad feature has been attributed to the buildup of short-range magnetic correlations. 

\begin{figure}
\centering
\includegraphics[width=\linewidth]{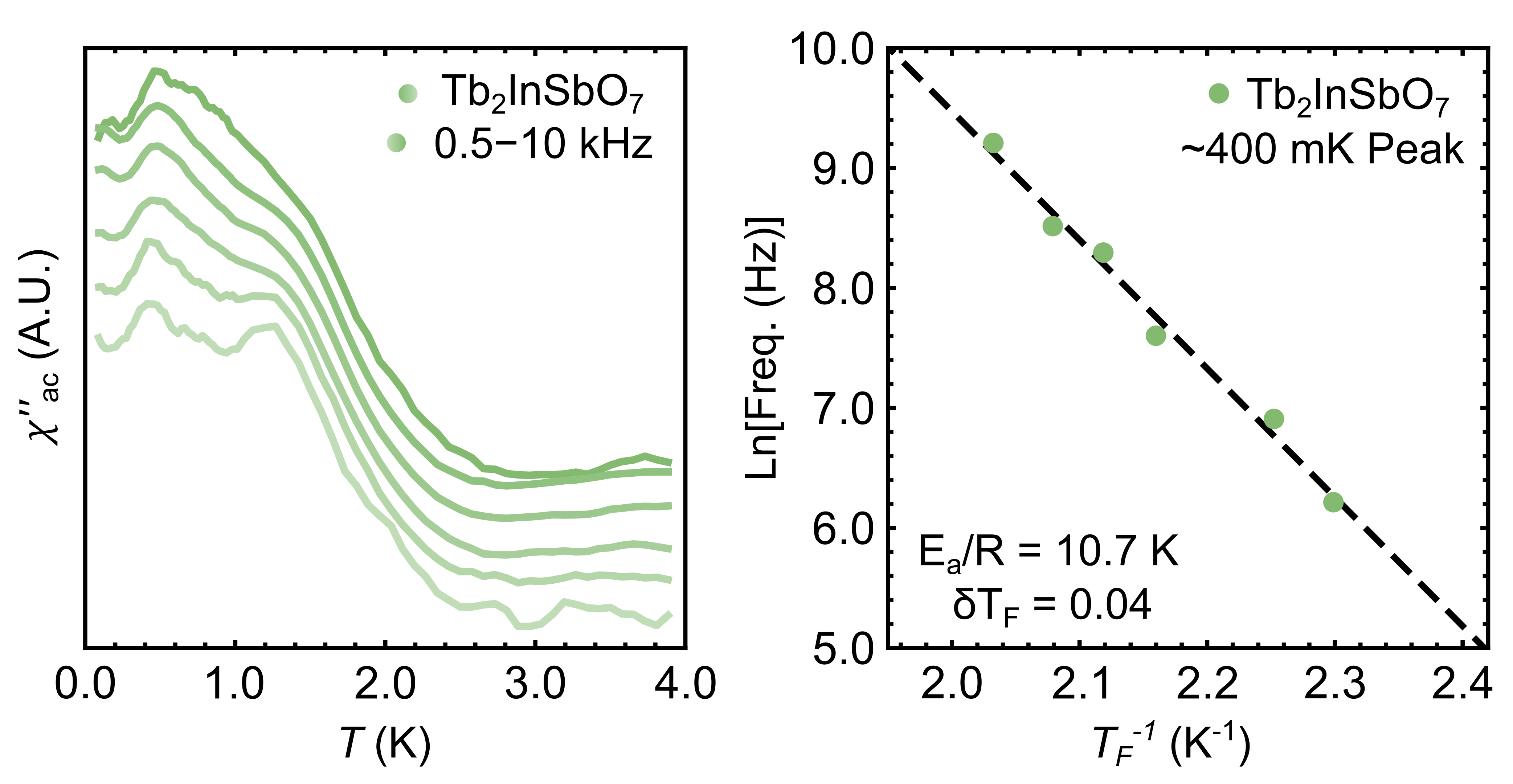}
\caption{The out-of-phase ac susceptibility ($\chi''$) in Tb$_{2}$InSbO$_{7}$ exhibits a complex spectra with multiple peaks and strong frequency-dependence (left). A subsequent Arrhenius fit to the frequency dependence of the 400~mK feature yields activation energy $E_{a}/R=10.7$~K and Mydosh parameter $\delta$T$_{\text{F}}=0.04$ (right).
} 
\label{fig:activationTb}
\end{figure}

More complexity is evident in the out-of-plane $\chi''$ ac magnetization signal (Fig. \ref{fig:activationTb}). We first observe an upturn in the magnetization around 2~K, normally associated with an increase in ferromagnetic interactions and mirrored by the broad feature in $\chi'$. A shoulder peak in $\chi''$ can be seen near 1.2~K, particularly in the low-frequency regime. This feature corresponds with the peak in the AC magnetization shown in Figure \ref{fig:dynamic}. However, unlike the pristine Tb$_2$Sn$_2$O$_7$, our data lacks the characteristic sharpness of the ferromagnetic ordering peak in the stannate, and it is difficult to extract a frequency dependence due to convolution with the rising signal. The general shape of $\chi''$ agrees \textit{remarkably well} with x=0.1 Ti-doped Tb$_2$Sn$_{2-x}$Ti$_x$O$_7$.\cite{TbSn_cava2011low} We postulate that the InSb disorder plays a similar role to the Ti-doping, destroying the long-range ferromagnetic order. An additional feature emerges around 400\,mK, a similar energy scale to a 300\,mK feature observed in the pristine Tb$_2$Sn$_2$O$_7$. However, while prior reports observed an anomalously high Mydosh parameter (0.34),\cite{TbSn_cava2011low} we estimate that the 400\,mK feature exhibits $\Delta T_f / T_f \Delta log(f) = 0.04$, an order of magnitude lower and more akin to results observed in the doped Tb$_2$Sn$_{2-x}$Ti$_x$O$_7$ samples.

Presuming that Tb$_{2}$InSbO$_{7}$ sits on the boundary between the defect fluorite and pyrochlore stability fields, the observation that Tb$_{2}$InSbO$_{7}$ exhibits behavior extremely reminiscent to lightly doped Tb$_{2}$Sn$_{2-x}$Ti$_{x}$O$_{7}$ is somewhat perplexing. However, if you consider the case of the corresponding halfnate Tb$_{2}$Hf$_{2}$O$_{7}$, where despite the presence of substantial anionic disorder (oxygen Frenkel defects), the cation sublattice remains ordered in the pyrochlore motif~\cite{TbHf_shlyakhtina2006structure,TbHf_sibille2016ms21,TbHf_sibille2017coulomb,Sn_Matsuhira02:71,subramanian1983rare}, the similarities of Tb$_{2}$InSbO$_{7}$ to its Tb$_{2}$Sn$_{2-x}$Ti$_{x}$O$_{7}$ analog can be somewhat reconciled. The question as to why $\chi''$ retains such complex structure in Tb$_{2}$InSbO$_{7}$ that is so characteristic of both the titanate and stannate, despite the known sensitivity of Tb-based pyrochlores to chemical disorder~\cite{PhysRevB.93.144407,Gardner99:82,PhysRevB.68.180401,PhysRevB.62.6496,PhysRevB.82.100402,Takatsu_2011,PhysRevB.84.184403,PhysRevB.69.132413,PhysRevLett.112.017203,PhysRevLett.83.1854}, may assist in addressing the physics that underlies the distinct sample dependence in the selection of the magnetic ground state for Tb-based pyrochlores~\cite{rau19:10}.

\begin{figure*}
\centering
\includegraphics[width=7.05in]{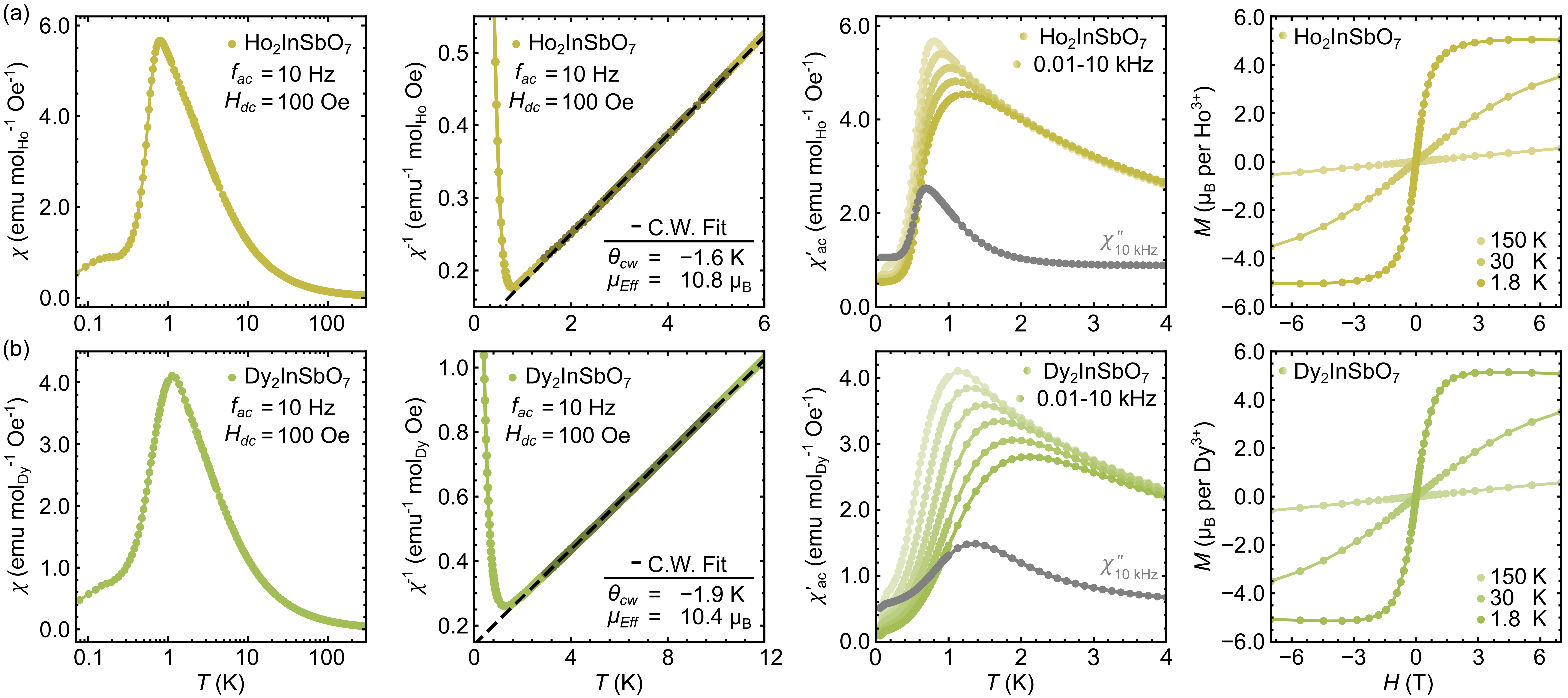}
\caption{Combined temperature dependence of the ac and dc magnetization with corresponding Curie-Weiss fits, and the isothermal dc magnetization of (a) Ho$_2$InSbO$_7$ and (b) Dy$_2$InSbO$_7$. For temperatures $T<4$~K, the ac susceptibility was measured at various frequencies $f_{ac}$ ranging from 10~Hz to 9984~Hz in the absence of a dc field. Here we show data collected for For $T>1.8$~K, the dc susceptibility was measured in an external field $H_{dc}$ of 100~Oe.}
\label{fig:dp}
\end{figure*}
Despite exhibiting a frequency independent peak at 0.62~K that compares favorably to both its titanate (T$\rm{_{N}}$ = 350~mK~\cite{Mauws18:98}) and stannate (T$\rm{_{N}}$ = 440~mK~\cite{Antonio19:99}) analogs, the peak observed in the case of Sm$_{2}$InSbO$_{7}$ is significantly broader. Although a natural suspicion is that such broadness is a reflection of poor crystallinity, this is not the case for Sm$_{2}$InSbO$_{7}$ which exhibits clear pyrochlore superstructure reflections and crystallinity comparable to all other pre-Gd InSb compounds. In fact, such behavior is highly reminiscent of its zirconate analog, which despite exhibiting excellent crystallinity with no detectable Sm/Zr antisite defects, does not exhibit a sharp $\lambda$-type anomaly in the heat capacity down to 0.1~K~\cite{xu_thesis}. Instead, Sm$_{2}$Zr$_{2}$O$_{7}$ has a broad peak centered at 500~mK corresponding to the full $Rln2$ of the ground state doublet, consistent with a buildup of short range magnetic correlations~\cite{xu_thesis,Singh08:77}. 

Such a comparison to its zirconate analog is further strengthened by clear similarities in both the temperature and field dependence of the dc magnetization. As is the case for all Sm$^{3+}$-based pyrochlores, the temperature dependence of the dc magnetization in Fig.~\ref{fig:dynamic}(c) does not exhibit a simple Curie-Weiss-like behavior due to crystal field effects~\cite{Singh08:77,Malkin10:22,Antonio19:99}. Restriction of the Curie-Weiss analysis to a narrow temperature range yields a Curie-Weiss temperature $\theta\rm{_{CW}}$ of $-1.0$~K and an effective paramagnetic moment $\mu\rm{_{eff}}$ of 0.5~$\mu\rm{_{B}}$/Sm$^{3+}$, comparing favorably with those values recently reported for the zirconate~\cite{xu_thesis} ($\theta\rm{_{CW}}$=$-1.27$~K and $\mu\rm{_{eff}}$=0.39~$\mu\rm{_{B}}$/Sm$^{3+}$). As is the case for the zirconate, the isothermal magnetization does not saturate up to $\mu_{o}H{\rm{_{ext}}}\sim$ 7~T, instead the magnetization is linear, consistent with a pseudospin-1/2 model assuming Ising anisotropy with a $g$ factor significantly less than 1~\cite{xu_thesis,Singh08:77}. Such single-ion anisotropy is consistent with the $|\pm 3/2\rangle$ dipolar-octupolar doublet previously reported for the stannate~\cite{Antonio19:99}, titanate~\cite{Mauws18:98,Antonio19:99}, and zirconate~\cite{xu_thesis}. 

Possessing a moment over a magnitude smaller relative to the dipolar spin ice candidates Ho$^{3+}$ and Dy$^{3+}$, the corresponding reduction in the dipolar term that is characteristic of all Sm$^{3+}$-based pyrochlores~\cite{Mauws18:98,Antonio19:99,Singh08:77,EuTmSn_bondah2001magnetic,Donnerer16:117}, suggesting that exchange will be particularly influential in the magnetic ground state of Sm$_{2}$InSbO$_{7}$. When combined with the slightly larger lattice parameter, there is a possibility that quantum fluctuations may be strong enough to prohibit the assumption of the long-range antiferromagnetic all-in-all-out magnetic order that usually accompanies antiferromagnetically coupled Ising moments on the pyrochlore lattice~\cite{gomez2021absence,denhertog00:84,guruciaga14:90,Ikeda08:77,Bramwell_1998,opherden17:95,sibille15:115}. As is the case for the zirconate~\cite{xu_thesis}, strong dynamics may persist in Sm$_{2}$InSbO$_{7}$ down to the lowest measurable temperatures, placing Sm$_{2}$InSbO$_{7}$ as a potential quantum spin ice candidate.

\subsubsection{Dipolar spin ices: \texorpdfstring{Ho$_{2}$InSbO$_{7}$}{Ho2InSbO7} \& \texorpdfstring{Dy$_{2}$InSbO$_{7}$}{Dy2InSbO7}}

The discovery of the the dipolar (``classical") spin ice state in both Ho$_{2}$Ti$_{2}$O$_{7}$~\cite{PhysRevLett.79.2554,PhysRevLett.87.047205,HARRIS1998757} and Dy$_{2}$Ti$_{2}$O$_{7}$~\cite{ramirez99:399} spurred a flurry of interest in lanthanide pyrochlore oxides that has spanned the past two decades~\cite{Gardner10:82,Bramwell_2020,PhysRevLett.121.067202,hallas18:9,rau19:10,doi:10.1142/8676}. When decorated with large Ho$^{3+}$ or Dy$^{3+}$ moments that exhibit local $\langle111\rangle$ Ising single-ion anisotropy amd net effective ferromagnetic exchange,\cite{doi:10.1063/1.372565,PhysRevB.65.054410,blote1969heat,doi:10.1063/1.1663909,doi:10.1063/1.372565,JANA20027} Ho$_{2}$Ti$_{2}$O$_{7}$ and Dy$_{2}$Ti$_{2}$O$_{7}$ exhibit the macroscopically degenerate ``two-in-two-out" motif fundamental to the dipolar spin ice~\cite{doi:10.1126/science.1064761,Bramwell_2020,denhertog00:84,PhysRev.102.1008,Champion_2004,PhysRevLett.95.097202,PhysRevLett.101.037204,HARRIS1998757,Melko_2004}.

\begin{figure}
\centering
\includegraphics[width=\linewidth]{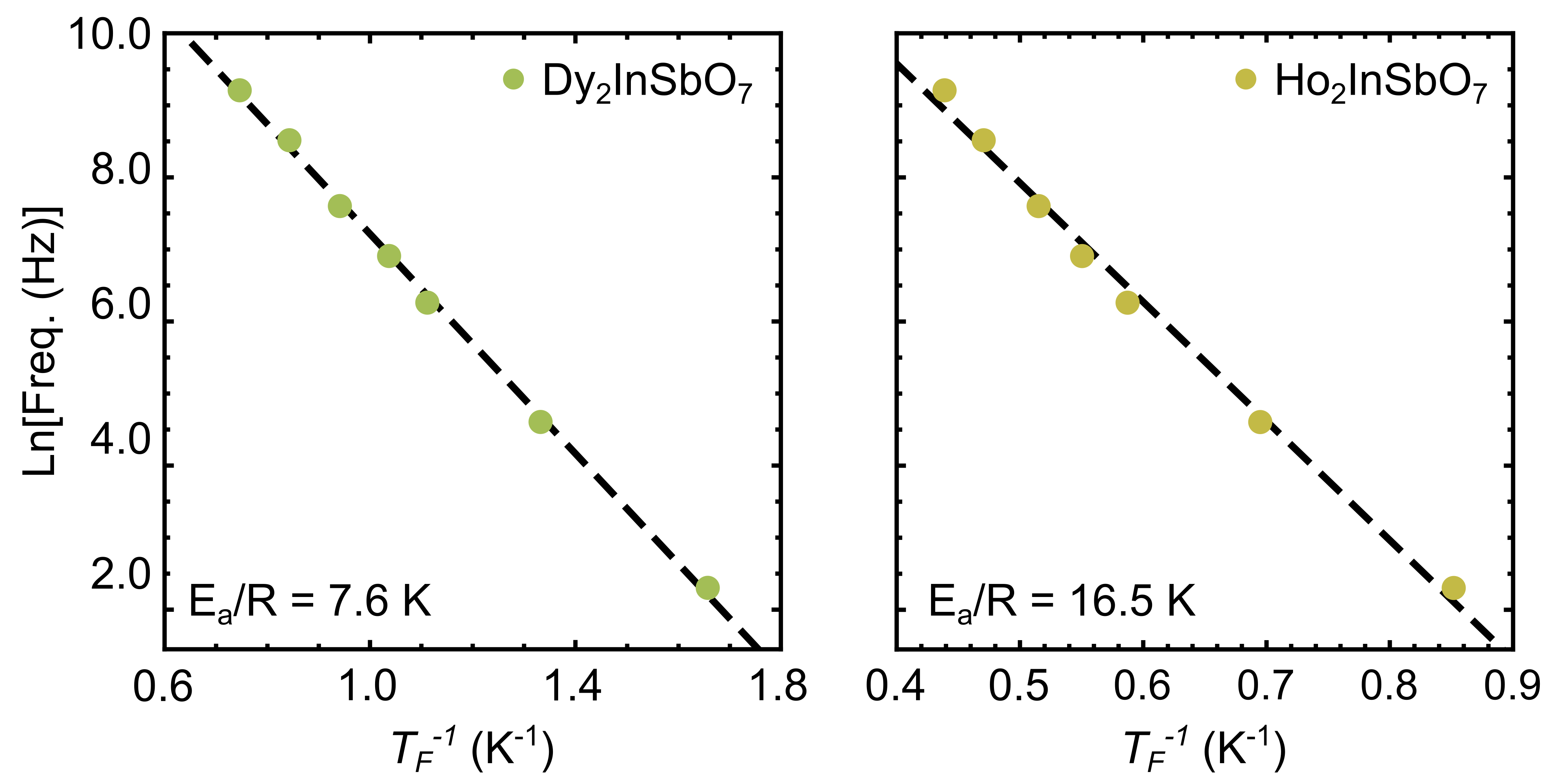}
\caption{Arrhenius fit of the frequency dependence of the freezing transition in Dy$_2$InSbO$_7$ (left) and Ho$_2$InSbO$_7$ (right), yielding activation energies of $E_a/R = 7.6$~K and 16.5~K, respectively, comparable to their stannate and titanate analogs.}
\label{fig:activation}
\end{figure}

As illustrated in Fig.~\ref{fig:dp}, both Ho$_{2}$InSbO$_{7}$ and Dy$_{2}$InSbO$_{7}$ exhibit  quantitative behavior highly reminiscent of both their titanate~\cite{Matsuhira00:12,Wiebe15:3,Bramwell99:12,PhysRevLett.79.2554,ramirez99:399,PhysRevLett.83.1854,Petrenko_2011,PhysRevB.70.134408,cashion1968crystal} and stannate~\cite{Sn_Matsuhira02:71,EuTmSn_bondah2001magnetic,Matsuhira00:12,Wiebe15:3,Bramwell99:12,PhysRevB.65.144421} counterparts. The isothermal magnetization saturates to half of the theoretical value, consistent with a pseudospin-1/2 model
assuming Ising anisotropy~\cite{Bramwell99:12}. Temperature-dependent susceptibility measurements confirm the clear absence of long-range magnetic order. Above 3~K, both systems exhibit Curie-Weiss-like behavior with parameters similar to those reported for the titanate and stannate. Frequency-dependent maxima at 0.80~K and 1.12~K appear in the ac susceptibility of Ho$_{2}$InSbO$_{7}$ and Dy$_{2}$InSbO$_{7}$, respectively, comparable to those temperatures reported for their stannate~\cite{Matsuhira00:12,Matsuhira00:12} and titanate~\cite{Matsuhira01:13,Matsuhira00:12,PhysRevLett.91.107201,snyder2001spin,Ehlers02:15} counterparts. Such behavior is also consistent with other Ho- and Dy-based spin ices, where frequency dependence is indicative of local spins freezing into the spin ice state~\cite{PrSn_zhou2008dynamic,Gardner_2011,Ehlers04:16,Ehlers02:15,PhysRevLett.91.107201,Hallas12:86,PhysRevLett.108.207206,PhysRevB.66.064432,snyder2001spin,Matsuhira01:13,Matsuhira00:12}. An Arrhenius analysis of the frequency-dependent ac susceptibility (Fig.~\ref{fig:activation}) yielded activation energies $E_{a}/R$ of 16.5~K and 7.6~K for Ho$_{2}$InSbO$_{7}$ and Dy$_{2}$InSbO$_{7}$, respectively, that are comparable to those values reported~\cite{Matsuhira_2011,PhysRevB.69.064414,PhysRevB.85.020410,PhysRevB.69.064414,PhysRevB.83.094424,Matsuhira01:13,megan_thesis} for the corresponding stannate and titanate systems.

\begin{figure*}
\centering
\includegraphics[width=7.05in]{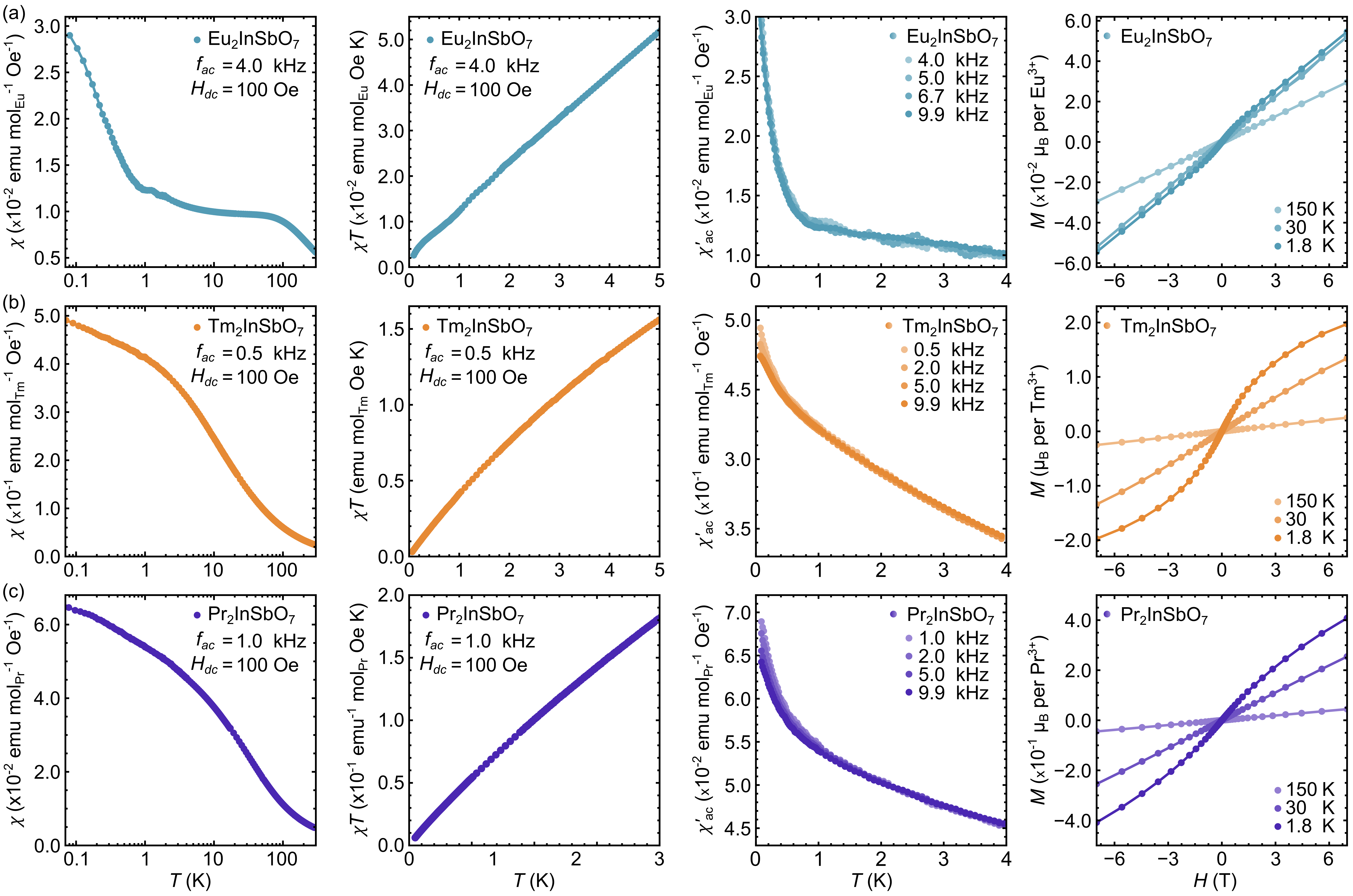}
\caption{Combined temperature dependence of the ac and dc magnetization, and the isothermal dc magnetization of (a) Eu$_2$InSbO$_7$,  (b) Tm$_2$InSbO$_7$, and (c) Pr$_2$InSbO$_7$. For temperatures $T<4$~K, the ac susceptibility was measured at various frequencies $f_{ac}$ in the absence of a dc field. For $T>1.8$~K, the dc susceptibility was measured in an external field $H_{dc}$ of 100~Oe.   
}
\label{fig:nm}
\end{figure*}

The realization of a dipolar spin ice for two trans-Gd members of the Ln$_{2}$InSbO$_{7}$ family is not completely unexpected. Early studies on ``stuffed" spin ice Ho$_{2}$(Ti$_{2-x}$Ho$_{x}$)O$_{7-x/2}$ ($0 \leq x \leq 2/3$) demonstrated that the spin ice state is remarkably robust against significant amounts of disorder~\cite{lau2006zero,Aldus_2013,LAU20063126,Zhou_2007,PhysRevB.77.052404}. The resistance to local disorder appears to persist up to the defect fluorite limit, where both $A$ and $B$-sites become indistinguishable. The presence of antisite disorder naturally accounts for both the negative Weiss temperature and reduced activation energy barrier~\cite{Ke07:76,lau2006zero}. However there are noteworthy differences between the Ln$_{2}$InSbO$_{7}$ systems and the ``stuffed'' spin ices.  In particular, the magnitude of the normalized values of both $\chi'$ and $\chi''$ are much larger than their stuffed spin ice counterparts~\cite{lau2006zero}. The freezing transitions of the InSb pyrochlores are also much sharper with minimal temperature suppression, reminiscent of their pure stannate~\cite{Matsuhira00:12,Matsuhira00:12} and titanate~\cite{Matsuhira01:13,Matsuhira00:12,PhysRevLett.91.107201,snyder2001spin,Ehlers02:15} analogs.

Recently it was shown that in many Dy-based pyrochlores, anti-site disorder between the A- and B-sites may destroy the spin ice state. This effect was observed for the ``stuffed" spin ice ${\text{Dy}}_{2}({\text{Dy}}_{x}{\text{Ti}}_{2\ensuremath{-}x}){\text{O}}_{7\ensuremath{-}x/2}$~\cite{PhysRevB.77.144412} and Dy$_{2}$Zr$_{2}$O$_{7}$, the latter of which hosts a disordered spin liquid-like magnetic ground state~\cite{Ramon19:99}. Recall that Dy$_2$InSbO$_7$ is initially considered to be within the defect fluorite regime of the stability diagram, naively characterized by gradual disordering of the A- and B-sites. In this sense, the prototypical spin-ice behavior seen in Figure \ref{fig:dp} is surprising. Potentially both Ho$_{2}$InSbO$_{7}$ and Dy$_{2}$InSbO$_{7}$ may exhibit the same anionic sublattice disorder at Tb$_{2}$Hf$_{2}$O$_{7}$~\cite{TbHf_sibille2017coulomb}, which would preserve the Dy$^{3+}$ and Ho$^{3+}$ cation sublattice and the requisite spin-ice state. Consistent with this thesis, the spin state has been shown to remarkably robust against disorder isolated to the non-magnetic $B$-site sublattice. Examples include the doped systems Ho$_{2}$(Ti$_{2-x}$Ho$_{x}$)O$_{7-x/2}$ ($x \lesssim$ 0.3)~\cite{lau2006zero}, Dy$_{2}$Sn$_{2-x}$Sb$_{x}$O$_{7+x/2}$~\cite{Ke07:76}, and Dy$_{2}$Ti$_{2-x}$Fe$_{x}$O$_{7}$~\cite{LIU2014107}. Similar effects are observed in the mixed B-site analogs Dy$_{2}$ScNbO$_{7}$~\cite{Ke07:76,megan_thesis},  Dy$_{2}$GaSbO$_{7}$~\cite{dy2gasbo7}, and Dy$_{2}$FeSbO$_{7}$~\cite{NANDI2017318} as well.


As is the case for all trans-Gd members, additional measurements, ($e.g.$ neutron diffraction and PDF, heat capacity, $etc.$) will be necessary to distinguish between which particular scenario: ``stuffed" ($A^{3+}_{B^{4+}}$ anti-site) or $B$-site ($B_{i}B_{ii}$) disordered spin ice applies to both the Ho$^{3+}$ and Dy$^{3+}$ members.  

\subsubsection{Nonmagnetic Singlet Ground States: 
\texorpdfstring{Tm$_{2}$InSbO$_{7}$}{Eu2InSbO7}, \texorpdfstring{Eu$_{2}$InSbO$_{7}$}{Tm2InSbO7} \& \texorpdfstring{ Pr$_{2}$InSbO$_{7}$}{Pr2InSbO7}}

\textbf{Eu$_2$InSbO$_7$}: Both Eu- and Tm-based pyrochlores host trivial non-magnetic spin singlet ground states~\cite{chien78:18,EuTmSn_bondah2001magnetic,PAL18:462,DASGUPTA2007347,TmTi_zinkin1996lifting}. In the case of Eu$^{3+}$, the combination of the 4$f^{6}$ electronic configuration, strong spin-orbit coupling, and $D_{3d}$ point group symmetry yields a non-magnetic $^{7}F_{0}$ ground state~\cite{PAL18:462,DASGUPTA2007347,CHATTOPADHYAY20046}. With a gap of over 100~K to its first excited magnetic state $^{7}F_{1}$, the Van Vleck term dominates the low temperature magnetism of Eu$^{3+}$ pyrochlores~\cite{EuTmSn_bondah2001magnetic,strobel2010structural,SHEETAL2022169255}. The dominance of the Van Vleck term for Eu$_{2}$InSbO$_{7}$ is consistent with the temperature independence of the magnetic susceptibility over a large temperature range below 100~K (Fig.~\ref{fig:nm}(a)), in stark contrast to the behavior predicted by the Curie-Weiss law. As was the case for the titanate~\cite{PAL18:462,DASGUPTA2007347} and stannate~\cite{EuTmSn_bondah2001magnetic} 
analogs, the isothermal magnetization does not saturate but continues linearly up to 7~T, consistent with antiferromagnetic exchange~\cite{PAL18:462}. The increase in the magnetization in the low temperature limit can be attributed to the presence of small amounts of magnetic impurities that eventually freeze.

\textbf{Tm$_2$InSbO$_7$}: In the case of Tm$^{3+}$, the ground state is a non-Kramers doublet that is weakly split due to crystal field effects -- resulting in a non-magnetic singlet state at sufficiently low temepratures.~\cite{EuTmSn_bondah2001magnetic,blote1969heat,TmTi_zinkin1996lifting,Taira02:12,Lea62:23}. 
As is the case for both the stannate~\cite{EuTmSn_bondah2001magnetic} and titanate~\cite{TmTi_zinkin1996lifting} analogs, the isothermal magnetization of Tm$_{2}$InSbO$_{7}$ does not saturate, with values over an order of magnitude smaller than its predicted saturation magnetization~\cite{jensen1991}. The temperature dependence of the magnetic susceptibility illustrated in Fig.~\ref{fig:nm}(b) does not exhibit Curie-Weiss-like behavior, instead it is reminiscent of the behavior predicted for a singlet–triplet level scheme~\cite{TmTi_zinkin1996lifting,LandoltBornstein1992}. The slight upturn and resulting non-linearity at the low temperature limit is attributed to the presence of small amounts of magnetic impurities. The frequency dependence of the ac susceptibility suggests that these impurities freeze in the low temperature limit.

\textbf{Pr$_2$InSbO$_7$}: In contrast to their Tm$^{3+}$ and Eu$^{3+}$ counterparts, Pr-based pyrochlores can host a plethora of exotic magnetic ground states including the possibility of the experimental realization of quantum spin ice in the stannate~\cite{PrSn_zhou2008dynamic,Sarte17:29}, hafanate~\cite{PrHf_sibille2016candidate,PrHf_sibille2018experimental,PrHf_anand2016physical}, and zirconate~\cite{kimura13:4,martin17:7,Wen17:118,Pettit16:94}. Pr$_{2}$Zr$_{2}$O$_{7}$ in particular has garnered significant attention as theoretical and experimental treatments have identified $J_{zz}>$ 0 that is much larger than both $|J_{\pm}|$ and $|J_{\pm\pm}|$, placing Pr$_{2}$Zr$_{2}$O$_{7}$ near the QSI limit~\cite{kimura13:4,Pettit16:94,martin17:7}. In these systems, the crystalline electric field acting on the $J=4$ manifold yields a non-Kramers doublet that is well-thermally isolated ($\Delta >$ 10~meV), and thus can be projected onto an effective $J$=1/2 description~\cite{kimura13:4,PrSn_zhou2008dynamic,Savary17:118,Bonville16:94}. Thus, Pr$^{3+}$-based pyrochlores can provide an experimental platform to investigate the complex, and often subtle interplay that result from geometric frustration, strongly correlated electrons, and orbital degrees of freedom~\cite{Onoda11:83}.

However, the zirconate is also fraught with substantial chemical and structural disorder~\cite{Wen17:118,martin17:7,trump18:9}. Though the physical origin of the disorder is still a subject of debate, ($e.g.$ lattice strain~\cite{martin17:7}, Pr$^{3+}$ off-centering~\cite{KOOHPAYEH2014291}) the effect is to induce a temperature-independent continuum of scattering~\cite{Wen17:118}. The profound influence of structural disorder in the zirconate is intuitive, considering that Pr$^{3+}$ is a non-Kramers ion. It was shown that any perturbation away from ideal $D_{3d}$ site symmetry can be mapped onto the introduction of transverse fields onto the pseudospin manifold~\cite{Wen17:118,martin17:7}. Although such a strong random transverse theoretical treatment was shown to be consistent with experimental observations, the role of exchange disorder plays remains an open question~\cite{rau19:10}. Theoretical attempts to address such a question have only focused on both extremes. In the case of strong exchange and weak exchange disorder, a disordered variant of a QSI is expected, while in the case of the exchange being significantly weaker than disorder, one would expect a frozen quadrupolar gas~\cite{Savary17:118,Benton18:121}. Instead, the intermediate regime, where the strength of the exchange and exchange disorder are comparable, has remained largely unexplored and represents an open question of great interest~\cite{rau19:10}. 

At a cursory level, Pr$_{2}$InSbO$_{7}$ presents an opportunity to address this intermediate regime. Its significantly smaller lattice parameter relative to the zirconate should result in an increase in strength of the Pr$^{3+}$-O$^{2-}$-Pr$^{3+}$ superexchange pathways. Instead, as illustrated in Fig.~\ref{fig:nm}(a), both dc and ac magnetization results suggest a trivial non-magnetic singlet state resembling that of Tm$_{2}$InSbO$_{7}$. The absolute magnitude of the isothermal magnetization is suppressed, and combined with the lack of saturation is inconsistent with a pseudospin-1/2 model assuming Ising anisotropy. Further, many quantum spin ice candidates~\cite{PrSn_zhou2008dynamic} exhibit a weakly frequency-dependent feature at base temperatures -- but this feature is replaced with a Curie-like tail in Pr$_{2}$InSbO$_{7}$ that may be attributed to the freezing of small amounts of impurities.

A natural conclusion would be that the splitting of the ground state doublet into a non-magnetic singlet state was a result of maximal $B$-site disorder, in strong contrast with the minimal levels of disorder in the zirconate~\cite{Wen17:118,martin17:7,KOOHPAYEH2014291}. However, other mixed $B$-site 
Pr-based pyrochlores such as Pr$_{2}$Sn$_{2-x}$Ti$_{x}$O$_{7}$~\cite{Sarte17:29} and Pr$_{2}$ScNbO$_{7}$~\cite{private} have been identified as spin liquid-like candidates. It is possible that the various magnetic ground states stem from the deviation between the Shannon radii~\cite{Shannon:a12967} of In$^{3+}$ and Sb$^{5+}$ ($\delta$=0.2), which is substantially larger than that of Pr$_{2}$Sn$_{2-x}$Ti$_{x}$O$_{7}$ ($\delta$=0.085) or Pr$_{2}$ScNbO$_{7}$ ($\delta$=0.105). The role of the deviation $\delta$ in the selection of the magnetic ground state is a promising avenue for future investigation, with direct experimental routes in the form of the plethora of mixed $B$-site combinations currently synthetically available, each with differing values of $\delta$.

\section{Conclusion}
With a wide array of possible compositions and the hallmark frustrated lattice, mixed-site pyrochlores allow access to exotic properties ranging from moment fragmentation, quantum spin ices, spin liquids, magnetic monopoles, and many others~\cite{Gardner10:82,Wiebe15:3,hallas18:9,rau19:10}. In this work we analyzed the known stability field for the lanthanide oxide pyrochlores Ln$_2$B$_2$O$_7$. We merged recent synthetic results for the mixed B-site compositions with the previously established families, providing a broad review of the existing Ln-B-O compounds. We find the slightly modified ``radius-ratio rule'' of $1.43$ $\leq$ $R{\rm{_{Ln^{3+}}}}$/$R{\rm{_{B^{4+}}}}$ $\leq$ $1.82$ to perform across the entire family with a predictive accuracy of $\sim$94\%.

In tandem with our review of pyrochlore compositions, we presented the phase-pure, rapid synthesis of the entire Ln$_2$InSbO$_7$ mixed B-site family. Our synthetic improvements enable synthesis of the entire family, improving upon prior reports wherein only the Gd and Nd variants were produced with sufficient purity~\cite{strobel2010structural}. Furthermore, we discuss how microwave-heating methods can stabilize the cubic phase throughout the entire family, despite small Ln$^{3+}$ (Lu...Tb) showing the proclivity to crystallize in low symmetry (monoclinic/triclinic) forms. We have provided empirical evidence that the InSb family may transition towards the defect fluorite phase for small Ln$^{3+}$ (Lu...Tb) while compounds with Ln$^{3+}$ (La...Gd) are consistent with the pyrochlore $Fd\bar{3}m$ structure. 

This work additionally presents a broad survey of the magnetic properties throughout the entire Ln$_2$InSbO$_7$ mixed B-site family. We provide analysis and comparisons to the relevant Ln$_2$B$_2$O$_7$ analogs, examining how the magnetic properties of the InSb compounds compare to the parent structures. We find that Nd$_2$InSbO$_7$, Gd$_2$InSbO$_7$, and Er$_2$InSbO$_7$ exhibit properties consistent with the antiferromagnetic ordering typical of their analogous stannates. Yb$_2$InSbO$_7$, Tb$_2$InSbO$_7$, and Sm$_2$InSbO$_7$ show properties consistent with a dynamic ground state. Ho$_2$InSbO$_7$ and Dy$_2$InSbO$_7$ exhibit behavior consistent with the canonical dipolar spin ices, with energetics consistent with their stannate equivalents. Finally, Eu$_2$InSbO$_7$, Tm$_2$InSbO$_7$, and Pr$_2$InSbO$_7$ host non-magnetic singlet ground states. Such behavior is not unexpected for Eu$_2$InSbO$_7$ and Tm$_2$InSbO$_7$ due to their local crystal field configurations. However, in the case of Pr$^{3+}$, the non-magnetic ground state appears to arise from a splitting of the non-Kramers ground state doublet due to the large Shannon radii contrast between In$^{3+}$(VI) and Sb$^{5+}$(VI).

Our survey of the existing lanthanide pyrochlore oxides combined with the magnetic characterization of the Ln$_2$InSbO$_7$ family provides a fresh perspective on the properties of mixed B-site pyrochlores. The chemical diversity of the mixed B-site systems have become a highlight of recent pyrochlore research, as the community reexamines the influence of disorder and local sterics on the realization of complex quantum phenomenon in the pyrochlore phases. Striking similarities between mixed B-site systems and the analogous mono B-site pyrochlores continues to emerge, forcing us to reevaluate the presumed deleterious role of disorder. Instead, chemical disorder becomes a tool -- enabling controlled local chemical changes while maintaining a pristine Ln$^{3+}$ magnetic sublattice.

\section{Acknowledgments}

The authors acknowledge fruitful conversations with A.~A.~Aczel, J.~A.~M.~Paddison, C.~R.~Wiebe, B.~A.~Frandsen, and A.~Krajewska. B.R.O. and P.M.S.~acknowledge financial support from the University of California, Santa Barbara, through the Elings Fellowship. This work was supported by DOE, Office of Science, Basic Energy Sciences under Award DE-SC0017752 (S.D.W., P.M.S., B.R.O.)  This work used facilities supported via the UC Santa Barbara NSF Quantum Foundry funded via the Q-AMASE-i program under award DMR-1906325. The research made use of the shared experimental facilities of the NSF Materials Research Science and Engineering Center at UC Santa Barbara
(DMR- 1720256). The UC Santa Barbara MRSEC is a member of the Materials Research Facilities Network~\cite{MRL}.

\bibliography{MicrowaveLn2InSbO7}

\providecommand{\noopsort}[1]{}\providecommand{\singleletter}[1]{#1}%
\begin{thebibliography}{389}%
\makeatletter
\providecommand \@ifxundefined [1]{%
 \@ifx{#1\undefined}
}%
\providecommand \@ifnum [1]{%
 \ifnum #1\expandafter \@firstoftwo
 \else \expandafter \@secondoftwo
 \fi
}%
\providecommand \@ifx [1]{%
 \ifx #1\expandafter \@firstoftwo
 \else \expandafter \@secondoftwo
 \fi
}%
\providecommand \natexlab [1]{#1}%
\providecommand \enquote  [1]{``#1''}%
\providecommand \bibnamefont  [1]{#1}%
\providecommand \bibfnamefont [1]{#1}%
\providecommand \citenamefont [1]{#1}%
\providecommand \href@noop [0]{\@secondoftwo}%
\providecommand \href [0]{\begingroup \@sanitize@url \@href}%
\providecommand \@href[1]{\@@startlink{#1}\@@href}%
\providecommand \@@href[1]{\endgroup#1\@@endlink}%
\providecommand \@sanitize@url [0]{\catcode `\\12\catcode `\$12\catcode
  `\&12\catcode `\#12\catcode `\^12\catcode `\_12\catcode `\%12\relax}%
\providecommand \@@startlink[1]{}%
\providecommand \@@endlink[0]{}%
\providecommand \url  [0]{\begingroup\@sanitize@url \@url }%
\providecommand \@url [1]{\endgroup\@href {#1}{\urlprefix }}%
\providecommand \urlprefix  [0]{URL }%
\providecommand \Eprint [0]{\href }%
\providecommand \doibase [0]{http://dx.doi.org/}%
\providecommand \selectlanguage [0]{\@gobble}%
\providecommand \bibinfo  [0]{\@secondoftwo}%
\providecommand \bibfield  [0]{\@secondoftwo}%
\providecommand \translation [1]{[#1]}%
\providecommand \BibitemOpen [0]{}%
\providecommand \bibitemStop [0]{}%
\providecommand \bibitemNoStop [0]{.\EOS\space}%
\providecommand \EOS [0]{\spacefactor3000\relax}%
\providecommand \BibitemShut  [1]{\csname bibitem#1\endcsname}%
\let\auto@bib@innerbib\@empty
\bibitem [{\citenamefont {Subramanian}\ \emph {et~al.}(1983)\citenamefont
  {Subramanian}, \citenamefont {Aravamudan},\ and\ \citenamefont
  {Subba~Rao}}]{subramanian1983rare}%
  \BibitemOpen
  \bibfield  {author} {\bibinfo {author} {\bibfnamefont {M.~A.}\ \bibnamefont
  {Subramanian}}, \bibinfo {author} {\bibfnamefont {G.}~\bibnamefont
  {Aravamudan}}, \ and\ \bibinfo {author} {\bibfnamefont {G.~V.}\ \bibnamefont
  {Subba~Rao}},\ }\href {\doibase
  https://doi.org/10.1016/S0168-1273(05)80018-2} {\bibfield  {journal}
  {\bibinfo  {journal} {Prog. Solid State Chem.}\ }\textbf {\bibinfo {volume}
  {15}},\ \bibinfo {pages} {55} (\bibinfo {year} {1983})}\BibitemShut {NoStop}%
\bibitem [{\citenamefont {Gardner}\ \emph {et~al.}(2010)\citenamefont
  {Gardner}, \citenamefont {Gingras},\ and\ \citenamefont
  {Greedan}}]{Gardner10:82}%
  \BibitemOpen
  \bibfield  {author} {\bibinfo {author} {\bibfnamefont {J.~S.}\ \bibnamefont
  {Gardner}}, \bibinfo {author} {\bibfnamefont {M.~J.~P.}\ \bibnamefont
  {Gingras}}, \ and\ \bibinfo {author} {\bibfnamefont {J.~E.}\ \bibnamefont
  {Greedan}},\ }\href {\doibase 10.1103/RevModPhys.82.53} {\bibfield  {journal}
  {\bibinfo  {journal} {Rev. Mod. Phys.}\ }\textbf {\bibinfo {volume} {82}},\
  \bibinfo {pages} {53} (\bibinfo {year} {2010})}\BibitemShut {NoStop}%
\bibitem [{\citenamefont {Greedan}(2006)}]{greedan2006frustrated}%
  \BibitemOpen
  \bibfield  {author} {\bibinfo {author} {\bibfnamefont {J.~E.}\ \bibnamefont
  {Greedan}},\ }\href {\doibase https://doi.org/10.1016/j.jallcom.2004.12.084}
  {\bibfield  {journal} {\bibinfo  {journal} {J. Alloys Compd.}\ }\textbf
  {\bibinfo {volume} {408}},\ \bibinfo {pages} {444} (\bibinfo {year}
  {2006})}\BibitemShut {NoStop}%
\bibitem [{\citenamefont {Rau}\ and\ \citenamefont {Gingras}(2019)}]{rau19:10}%
  \BibitemOpen
  \bibfield  {author} {\bibinfo {author} {\bibfnamefont {J.~G.}\ \bibnamefont
  {Rau}}\ and\ \bibinfo {author} {\bibfnamefont {M.~J.}\ \bibnamefont
  {Gingras}},\ }\href {\doibase 10.1146/annurev-conmatphys-022317-110520}
  {\bibfield  {journal} {\bibinfo  {journal} {Annu. Rev. Condens. Matter
  Phys.}\ }\textbf {\bibinfo {volume} {10}},\ \bibinfo {pages} {357} (\bibinfo
  {year} {2019})}\BibitemShut {NoStop}%
\bibitem [{\citenamefont {Hallas}\ \emph {et~al.}(2018)\citenamefont {Hallas},
  \citenamefont {Gaudet},\ and\ \citenamefont {Gaulin}}]{hallas18:9}%
  \BibitemOpen
  \bibfield  {author} {\bibinfo {author} {\bibfnamefont {A.~M.}\ \bibnamefont
  {Hallas}}, \bibinfo {author} {\bibfnamefont {J.}~\bibnamefont {Gaudet}}, \
  and\ \bibinfo {author} {\bibfnamefont {B.~D.}\ \bibnamefont {Gaulin}},\
  }\href {\doibase 10.1146/annurev-conmatphys-031016-025218} {\bibfield
  {journal} {\bibinfo  {journal} {Annu. Rev. Condens. Matter Phys.}\ }\textbf
  {\bibinfo {volume} {9}},\ \bibinfo {pages} {105} (\bibinfo {year}
  {2018})}\BibitemShut {NoStop}%
\bibitem [{\citenamefont {Fuentes}\ \emph {et~al.}(2018)\citenamefont
  {Fuentes}, \citenamefont {Montemayor}, \citenamefont {Maczka}, \citenamefont
  {Lang}, \citenamefont {Ewing},\ and\ \citenamefont
  {Amador}}]{fuentes2018critical}%
  \BibitemOpen
  \bibfield  {author} {\bibinfo {author} {\bibfnamefont {A.~F.}\ \bibnamefont
  {Fuentes}}, \bibinfo {author} {\bibfnamefont {S.~M.}\ \bibnamefont
  {Montemayor}}, \bibinfo {author} {\bibfnamefont {M.}~\bibnamefont {Maczka}},
  \bibinfo {author} {\bibfnamefont {M.}~\bibnamefont {Lang}}, \bibinfo {author}
  {\bibfnamefont {R.~C.}\ \bibnamefont {Ewing}}, \ and\ \bibinfo {author}
  {\bibfnamefont {U.}~\bibnamefont {Amador}},\ }\href {\doibase
  https://doi.org/10.1021/acs.inorgchem.8b01665} {\bibfield  {journal}
  {\bibinfo  {journal} {Inorg. Chem.}\ }\textbf {\bibinfo {volume} {57}},\
  \bibinfo {pages} {12093} (\bibinfo {year} {2018})}\BibitemShut {NoStop}%
\bibitem [{\citenamefont {Lumpkin}\ and\ \citenamefont
  {Aughterson}(2021)}]{10.3389/fchem.2021.778140}%
  \BibitemOpen
  \bibfield  {author} {\bibinfo {author} {\bibfnamefont {G.~R.}\ \bibnamefont
  {Lumpkin}}\ and\ \bibinfo {author} {\bibfnamefont {R.~D.}\ \bibnamefont
  {Aughterson}},\ }\href {\doibase 10.3389/fchem.2021.778140} {\bibfield
  {journal} {\bibinfo  {journal} {Front. Chem.}\ }\textbf {\bibinfo {volume}
  {9}} (\bibinfo {year} {2021}),\ 10.3389/fchem.2021.778140}\BibitemShut
  {NoStop}%
\bibitem [{\citenamefont {Champion}\ \emph {et~al.}(2003)\citenamefont
  {Champion}, \citenamefont {Harris}, \citenamefont {Holdsworth}, \citenamefont
  {Wills}, \citenamefont {Balakrishnan}, \citenamefont {Bramwell},
  \citenamefont {\ifmmode \check{C}\else \v{C}\fi{}i\ifmmode~\check{z}\else
  \v{z}\fi{}m\'ar}, \citenamefont {Fennell}, \citenamefont {Gardner},
  \citenamefont {Lago}, \citenamefont {McMorrow}, \citenamefont
  {Orend\'a\ifmmode~\check{c}\else \v{c}\fi{}}, \citenamefont
  {Orend\'a\ifmmode~\check{c}\else \v{c}\fi{}ov\'a}, \citenamefont {Paul},
  \citenamefont {Smith}, \citenamefont {Telling},\ and\ \citenamefont
  {Wildes}}]{champion03:68}%
  \BibitemOpen
  \bibfield  {author} {\bibinfo {author} {\bibfnamefont {J.~D.~M.}\
  \bibnamefont {Champion}}, \bibinfo {author} {\bibfnamefont {M.~J.}\
  \bibnamefont {Harris}}, \bibinfo {author} {\bibfnamefont {P.~C.~W.}\
  \bibnamefont {Holdsworth}}, \bibinfo {author} {\bibfnamefont {A.~S.}\
  \bibnamefont {Wills}}, \bibinfo {author} {\bibfnamefont {G.}~\bibnamefont
  {Balakrishnan}}, \bibinfo {author} {\bibfnamefont {S.~T.}\ \bibnamefont
  {Bramwell}}, \bibinfo {author} {\bibfnamefont {E.}~\bibnamefont {\ifmmode
  \check{C}\else \v{C}\fi{}i\ifmmode~\check{z}\else \v{z}\fi{}m\'ar}}, \bibinfo
  {author} {\bibfnamefont {T.}~\bibnamefont {Fennell}}, \bibinfo {author}
  {\bibfnamefont {J.~S.}\ \bibnamefont {Gardner}}, \bibinfo {author}
  {\bibfnamefont {J.}~\bibnamefont {Lago}}, \bibinfo {author} {\bibfnamefont
  {D.~F.}\ \bibnamefont {McMorrow}}, \bibinfo {author} {\bibfnamefont
  {M.}~\bibnamefont {Orend\'a\ifmmode~\check{c}\else \v{c}\fi{}}}, \bibinfo
  {author} {\bibfnamefont {A.}~\bibnamefont {Orend\'a\ifmmode~\check{c}\else
  \v{c}\fi{}ov\'a}}, \bibinfo {author} {\bibfnamefont {D.~M.}\ \bibnamefont
  {Paul}}, \bibinfo {author} {\bibfnamefont {R.~I.}\ \bibnamefont {Smith}},
  \bibinfo {author} {\bibfnamefont {M.~T.~F.}\ \bibnamefont {Telling}}, \ and\
  \bibinfo {author} {\bibfnamefont {A.}~\bibnamefont {Wildes}},\ }\href
  {\doibase 10.1103/PhysRevB.68.020401} {\bibfield  {journal} {\bibinfo
  {journal} {Phys. Rev. B}\ }\textbf {\bibinfo {volume} {68}},\ \bibinfo
  {pages} {020401} (\bibinfo {year} {2003})}\BibitemShut {NoStop}%
\bibitem [{\citenamefont {Savary}\ \emph {et~al.}(2012)\citenamefont {Savary},
  \citenamefont {Ross}, \citenamefont {Gaulin}, \citenamefont {Ruff},\ and\
  \citenamefont {Balents}}]{savary12:109}%
  \BibitemOpen
  \bibfield  {author} {\bibinfo {author} {\bibfnamefont {L.}~\bibnamefont
  {Savary}}, \bibinfo {author} {\bibfnamefont {K.~A.}\ \bibnamefont {Ross}},
  \bibinfo {author} {\bibfnamefont {B.~D.}\ \bibnamefont {Gaulin}}, \bibinfo
  {author} {\bibfnamefont {J.~P.~C.}\ \bibnamefont {Ruff}}, \ and\ \bibinfo
  {author} {\bibfnamefont {L.}~\bibnamefont {Balents}},\ }\href {\doibase
  10.1103/PhysRevLett.109.167201} {\bibfield  {journal} {\bibinfo  {journal}
  {Phys. Rev. Lett.}\ }\textbf {\bibinfo {volume} {109}},\ \bibinfo {pages}
  {167201} (\bibinfo {year} {2012})}\BibitemShut {NoStop}%
\bibitem [{\citenamefont {Zhitomirsky}\ \emph {et~al.}(2012)\citenamefont
  {Zhitomirsky}, \citenamefont {Gvozdikova}, \citenamefont {Holdsworth},\ and\
  \citenamefont {Moessner}}]{zhitomirsky12:109}%
  \BibitemOpen
  \bibfield  {author} {\bibinfo {author} {\bibfnamefont {M.~E.}\ \bibnamefont
  {Zhitomirsky}}, \bibinfo {author} {\bibfnamefont {M.~V.}\ \bibnamefont
  {Gvozdikova}}, \bibinfo {author} {\bibfnamefont {P.~C.~W.}\ \bibnamefont
  {Holdsworth}}, \ and\ \bibinfo {author} {\bibfnamefont {R.}~\bibnamefont
  {Moessner}},\ }\href {\doibase 10.1103/PhysRevLett.109.077204} {\bibfield
  {journal} {\bibinfo  {journal} {Phys. Rev. Lett.}\ }\textbf {\bibinfo
  {volume} {109}},\ \bibinfo {pages} {077204} (\bibinfo {year}
  {2012})}\BibitemShut {NoStop}%
\bibitem [{\citenamefont {Wong}\ \emph {et~al.}(2013)\citenamefont {Wong},
  \citenamefont {Hao},\ and\ \citenamefont {Gingras}}]{wong13:88}%
  \BibitemOpen
  \bibfield  {author} {\bibinfo {author} {\bibfnamefont {A.~W.~C.}\
  \bibnamefont {Wong}}, \bibinfo {author} {\bibfnamefont {Z.}~\bibnamefont
  {Hao}}, \ and\ \bibinfo {author} {\bibfnamefont {M.~J.~P.}\ \bibnamefont
  {Gingras}},\ }\href {\doibase 10.1103/PhysRevB.88.144402} {\bibfield
  {journal} {\bibinfo  {journal} {Phys. Rev. B}\ }\textbf {\bibinfo {volume}
  {88}},\ \bibinfo {pages} {144402} (\bibinfo {year} {2013})}\BibitemShut
  {NoStop}%
\bibitem [{\citenamefont {Champion}\ and\ \citenamefont
  {Holdsworth}(2004)}]{Champion_2004}%
  \BibitemOpen
  \bibfield  {author} {\bibinfo {author} {\bibfnamefont {J.~D.~M.}\
  \bibnamefont {Champion}}\ and\ \bibinfo {author} {\bibfnamefont {P.~C.~W.}\
  \bibnamefont {Holdsworth}},\ }\href {\doibase 10.1088/0953-8984/16/11/013}
  {\bibfield  {journal} {\bibinfo  {journal} {J. Phys.: Condens. Matter}\
  }\textbf {\bibinfo {volume} {16}},\ \bibinfo {pages} {S665} (\bibinfo {year}
  {2004})}\BibitemShut {NoStop}%
\bibitem [{\citenamefont {Li}\ \emph {et~al.}(2016)\citenamefont {Li},
  \citenamefont {Cai}, \citenamefont {Cui}, \citenamefont {Lin}, \citenamefont
  {Dun}, \citenamefont {Matsubayashi}, \citenamefont {Uwatoko}, \citenamefont
  {Sato}, \citenamefont {Kawae}, \citenamefont {Lv}, \citenamefont {Jin},
  \citenamefont {Zhou}, \citenamefont {Goodenough}, \citenamefont {Zhou},\ and\
  \citenamefont {Cheng}}]{li2016long}%
  \BibitemOpen
  \bibfield  {author} {\bibinfo {author} {\bibfnamefont {X.}~\bibnamefont
  {Li}}, \bibinfo {author} {\bibfnamefont {Y.~Q.}\ \bibnamefont {Cai}},
  \bibinfo {author} {\bibfnamefont {Q.}~\bibnamefont {Cui}}, \bibinfo {author}
  {\bibfnamefont {C.~J.}\ \bibnamefont {Lin}}, \bibinfo {author} {\bibfnamefont
  {Z.~L.}\ \bibnamefont {Dun}}, \bibinfo {author} {\bibfnamefont
  {K.}~\bibnamefont {Matsubayashi}}, \bibinfo {author} {\bibfnamefont
  {Y.}~\bibnamefont {Uwatoko}}, \bibinfo {author} {\bibfnamefont
  {Y.}~\bibnamefont {Sato}}, \bibinfo {author} {\bibfnamefont {T.}~\bibnamefont
  {Kawae}}, \bibinfo {author} {\bibfnamefont {S.~J.}\ \bibnamefont {Lv}},
  \bibinfo {author} {\bibfnamefont {C.~Q.}\ \bibnamefont {Jin}}, \bibinfo
  {author} {\bibfnamefont {J.-S.}\ \bibnamefont {Zhou}}, \bibinfo {author}
  {\bibfnamefont {J.~B.}\ \bibnamefont {Goodenough}}, \bibinfo {author}
  {\bibfnamefont {H.~D.}\ \bibnamefont {Zhou}}, \ and\ \bibinfo {author}
  {\bibfnamefont {J.-G.}\ \bibnamefont {Cheng}},\ }\href {\doibase
  10.1103/PhysRevB.94.214429} {\bibfield  {journal} {\bibinfo  {journal} {Phys.
  Rev. B}\ }\textbf {\bibinfo {volume} {94}},\ \bibinfo {pages} {214429}
  (\bibinfo {year} {2016})}\BibitemShut {NoStop}%
\bibitem [{\citenamefont {Raju}\ \emph {et~al.}(1999)\citenamefont {Raju},
  \citenamefont {Dion}, \citenamefont {Gingras}, \citenamefont {Mason},\ and\
  \citenamefont {Greedan}}]{raju99:59}%
  \BibitemOpen
  \bibfield  {author} {\bibinfo {author} {\bibfnamefont {N.~P.}\ \bibnamefont
  {Raju}}, \bibinfo {author} {\bibfnamefont {M.}~\bibnamefont {Dion}}, \bibinfo
  {author} {\bibfnamefont {M.~J.~P.}\ \bibnamefont {Gingras}}, \bibinfo
  {author} {\bibfnamefont {T.~E.}\ \bibnamefont {Mason}}, \ and\ \bibinfo
  {author} {\bibfnamefont {J.~E.}\ \bibnamefont {Greedan}},\ }\href {\doibase
  10.1103/PhysRevB.59.14489} {\bibfield  {journal} {\bibinfo  {journal} {Phys.
  Rev. B}\ }\textbf {\bibinfo {volume} {59}},\ \bibinfo {pages} {14489}
  (\bibinfo {year} {1999})}\BibitemShut {NoStop}%
\bibitem [{\citenamefont {Subramani}\ \emph {et~al.}(2021)\citenamefont
  {Subramani}, \citenamefont {Voskanyan}, \citenamefont {Jayanthi},
  \citenamefont {Abramchuk},\ and\ \citenamefont
  {Navrotsky}}]{subramani2021comparison}%
  \BibitemOpen
  \bibfield  {author} {\bibinfo {author} {\bibfnamefont {T.}~\bibnamefont
  {Subramani}}, \bibinfo {author} {\bibfnamefont {A.}~\bibnamefont
  {Voskanyan}}, \bibinfo {author} {\bibfnamefont {K.}~\bibnamefont {Jayanthi}},
  \bibinfo {author} {\bibfnamefont {M.}~\bibnamefont {Abramchuk}}, \ and\
  \bibinfo {author} {\bibfnamefont {A.}~\bibnamefont {Navrotsky}},\ }\href
  {\doibase 10.3389/fchem.2021.719169} {\bibfield  {journal} {\bibinfo
  {journal} {Front. Chem.}\ }\textbf {\bibinfo {volume} {9}} (\bibinfo {year}
  {2021}),\ 10.3389/fchem.2021.719169}\BibitemShut {NoStop}%
\bibitem [{\citenamefont {Reimers}\ \emph {et~al.}(1988)\citenamefont
  {Reimers}, \citenamefont {Greedan},\ and\ \citenamefont
  {Sato}}]{reimers1988crystal}%
  \BibitemOpen
  \bibfield  {author} {\bibinfo {author} {\bibfnamefont {J.~N.}\ \bibnamefont
  {Reimers}}, \bibinfo {author} {\bibfnamefont {J.~E.}\ \bibnamefont
  {Greedan}}, \ and\ \bibinfo {author} {\bibfnamefont {M.}~\bibnamefont
  {Sato}},\ }\href {\doibase https://doi.org/10.1016/0022-4596(88)90042-4}
  {\bibfield  {journal} {\bibinfo  {journal} {J. Solid State Chem.}\ }\textbf
  {\bibinfo {volume} {72}},\ \bibinfo {pages} {390} (\bibinfo {year}
  {1988})}\BibitemShut {NoStop}%
\bibitem [{\citenamefont {Clark}\ \emph {et~al.}(2014)\citenamefont {Clark},
  \citenamefont {Nilsen}, \citenamefont {Kermarrec}, \citenamefont {Ehlers},
  \citenamefont {Knight}, \citenamefont {Harrison}, \citenamefont {Attfield},\
  and\ \citenamefont {Gaulin}}]{PhysRevLett.113.117201}%
  \BibitemOpen
  \bibfield  {author} {\bibinfo {author} {\bibfnamefont {L.}~\bibnamefont
  {Clark}}, \bibinfo {author} {\bibfnamefont {G.~J.}\ \bibnamefont {Nilsen}},
  \bibinfo {author} {\bibfnamefont {E.}~\bibnamefont {Kermarrec}}, \bibinfo
  {author} {\bibfnamefont {G.}~\bibnamefont {Ehlers}}, \bibinfo {author}
  {\bibfnamefont {K.~S.}\ \bibnamefont {Knight}}, \bibinfo {author}
  {\bibfnamefont {A.}~\bibnamefont {Harrison}}, \bibinfo {author}
  {\bibfnamefont {J.~P.}\ \bibnamefont {Attfield}}, \ and\ \bibinfo {author}
  {\bibfnamefont {B.~D.}\ \bibnamefont {Gaulin}},\ }\href {\doibase
  10.1103/PhysRevLett.113.117201} {\bibfield  {journal} {\bibinfo  {journal}
  {Phys. Rev. Lett.}\ }\textbf {\bibinfo {volume} {113}},\ \bibinfo {pages}
  {117201} (\bibinfo {year} {2014})}\BibitemShut {NoStop}%
\bibitem [{\citenamefont {Raju}\ \emph {et~al.}(1992)\citenamefont {Raju},
  \citenamefont {Gmelin},\ and\ \citenamefont {Kremer}}]{PhysRevB.46.5405}%
  \BibitemOpen
  \bibfield  {author} {\bibinfo {author} {\bibfnamefont {N.~P.}\ \bibnamefont
  {Raju}}, \bibinfo {author} {\bibfnamefont {E.}~\bibnamefont {Gmelin}}, \ and\
  \bibinfo {author} {\bibfnamefont {R.~K.}\ \bibnamefont {Kremer}},\ }\href
  {\doibase 10.1103/PhysRevB.46.5405} {\bibfield  {journal} {\bibinfo
  {journal} {Phys. Rev. B}\ }\textbf {\bibinfo {volume} {46}},\ \bibinfo
  {pages} {5405} (\bibinfo {year} {1992})}\BibitemShut {NoStop}%
\bibitem [{\citenamefont {Zhou}\ \emph
  {et~al.}(2008{\natexlab{a}})\citenamefont {Zhou}, \citenamefont {Wiebe},
  \citenamefont {Harter}, \citenamefont {Dalal},\ and\ \citenamefont
  {Gardner}}]{Zhou_2008}%
  \BibitemOpen
  \bibfield  {author} {\bibinfo {author} {\bibfnamefont {H.~D.}\ \bibnamefont
  {Zhou}}, \bibinfo {author} {\bibfnamefont {C.~R.}\ \bibnamefont {Wiebe}},
  \bibinfo {author} {\bibfnamefont {A.}~\bibnamefont {Harter}}, \bibinfo
  {author} {\bibfnamefont {N.~S.}\ \bibnamefont {Dalal}}, \ and\ \bibinfo
  {author} {\bibfnamefont {J.~S.}\ \bibnamefont {Gardner}},\ }\href {\doibase
  10.1088/0953-8984/20/32/325201} {\bibfield  {journal} {\bibinfo  {journal}
  {J. Phys.: Condens. Matter}\ }\textbf {\bibinfo {volume} {20}},\ \bibinfo
  {pages} {325201} (\bibinfo {year} {2008}{\natexlab{a}})}\BibitemShut
  {NoStop}%
\bibitem [{\citenamefont {Singh}\ and\ \citenamefont
  {Lee}(2012)}]{PhysRevLett.109.247201}%
  \BibitemOpen
  \bibfield  {author} {\bibinfo {author} {\bibfnamefont {D.~K.}\ \bibnamefont
  {Singh}}\ and\ \bibinfo {author} {\bibfnamefont {Y.~S.}\ \bibnamefont
  {Lee}},\ }\href {\doibase 10.1103/PhysRevLett.109.247201} {\bibfield
  {journal} {\bibinfo  {journal} {Phys. Rev. Lett.}\ }\textbf {\bibinfo
  {volume} {109}},\ \bibinfo {pages} {247201} (\bibinfo {year}
  {2012})}\BibitemShut {NoStop}%
\bibitem [{\citenamefont {Jana}\ \emph {et~al.}(2003)\citenamefont {Jana},
  \citenamefont {Sakai}, \citenamefont {Higashinaka}, \citenamefont {Fukazawa},
  \citenamefont {Maeno}, \citenamefont {Dasgupta},\ and\ \citenamefont
  {Ghosh}}]{PhysRevB.68.174413}%
  \BibitemOpen
  \bibfield  {author} {\bibinfo {author} {\bibfnamefont {Y.~M.}\ \bibnamefont
  {Jana}}, \bibinfo {author} {\bibfnamefont {O.}~\bibnamefont {Sakai}},
  \bibinfo {author} {\bibfnamefont {R.}~\bibnamefont {Higashinaka}}, \bibinfo
  {author} {\bibfnamefont {H.}~\bibnamefont {Fukazawa}}, \bibinfo {author}
  {\bibfnamefont {Y.}~\bibnamefont {Maeno}}, \bibinfo {author} {\bibfnamefont
  {P.}~\bibnamefont {Dasgupta}}, \ and\ \bibinfo {author} {\bibfnamefont
  {D.}~\bibnamefont {Ghosh}},\ }\href {\doibase 10.1103/PhysRevB.68.174413}
  {\bibfield  {journal} {\bibinfo  {journal} {Phys. Rev. B}\ }\textbf {\bibinfo
  {volume} {68}},\ \bibinfo {pages} {174413} (\bibinfo {year}
  {2003})}\BibitemShut {NoStop}%
\bibitem [{\citenamefont {Shinaoka}\ \emph {et~al.}(2011)\citenamefont
  {Shinaoka}, \citenamefont {Tomita},\ and\ \citenamefont
  {Motome}}]{PhysRevLett.107.047204}%
  \BibitemOpen
  \bibfield  {author} {\bibinfo {author} {\bibfnamefont {H.}~\bibnamefont
  {Shinaoka}}, \bibinfo {author} {\bibfnamefont {Y.}~\bibnamefont {Tomita}}, \
  and\ \bibinfo {author} {\bibfnamefont {Y.}~\bibnamefont {Motome}},\ }\href
  {\doibase 10.1103/PhysRevLett.107.047204} {\bibfield  {journal} {\bibinfo
  {journal} {Phys. Rev. Lett.}\ }\textbf {\bibinfo {volume} {107}},\ \bibinfo
  {pages} {047204} (\bibinfo {year} {2011})}\BibitemShut {NoStop}%
\bibitem [{\citenamefont {Zhou}\ \emph {et~al.}(2010)\citenamefont {Zhou},
  \citenamefont {Wiebe}, \citenamefont {Janik}, \citenamefont {Vogt},
  \citenamefont {Harter}, \citenamefont {Dalal},\ and\ \citenamefont
  {Gardner}}]{ZHOU2010890}%
  \BibitemOpen
  \bibfield  {author} {\bibinfo {author} {\bibfnamefont {H.}~\bibnamefont
  {Zhou}}, \bibinfo {author} {\bibfnamefont {C.}~\bibnamefont {Wiebe}},
  \bibinfo {author} {\bibfnamefont {J.}~\bibnamefont {Janik}}, \bibinfo
  {author} {\bibfnamefont {B.}~\bibnamefont {Vogt}}, \bibinfo {author}
  {\bibfnamefont {A.}~\bibnamefont {Harter}}, \bibinfo {author} {\bibfnamefont
  {N.}~\bibnamefont {Dalal}}, \ and\ \bibinfo {author} {\bibfnamefont
  {J.}~\bibnamefont {Gardner}},\ }\href {\doibase
  https://doi.org/10.1016/j.jssc.2010.01.025} {\bibfield  {journal} {\bibinfo
  {journal} {J. Solid State Chem.}\ }\textbf {\bibinfo {volume} {183}},\
  \bibinfo {pages} {890} (\bibinfo {year} {2010})}\BibitemShut {NoStop}%
\bibitem [{\citenamefont {Taniguchi}\ \emph {et~al.}(2009)\citenamefont
  {Taniguchi}, \citenamefont {Munenaka},\ and\ \citenamefont
  {Sato}}]{Taniguchi_2009}%
  \BibitemOpen
  \bibfield  {author} {\bibinfo {author} {\bibfnamefont {T.}~\bibnamefont
  {Taniguchi}}, \bibinfo {author} {\bibfnamefont {T.}~\bibnamefont {Munenaka}},
  \ and\ \bibinfo {author} {\bibfnamefont {H.}~\bibnamefont {Sato}},\ }\href
  {\doibase 10.1088/1742-6596/145/1/012017} {\bibfield  {journal} {\bibinfo
  {journal} {J. Phys.: Conf. Ser.}\ }\textbf {\bibinfo {volume} {145}},\
  \bibinfo {pages} {012017} (\bibinfo {year} {2009})}\BibitemShut {NoStop}%
\bibitem [{\citenamefont {Reimers}\ \emph
  {et~al.}(1991{\natexlab{a}})\citenamefont {Reimers}, \citenamefont {Greedan},
  \citenamefont {Kremer}, \citenamefont {Gmelin},\ and\ \citenamefont
  {Subramanian}}]{PhysRevB.43.3387}%
  \BibitemOpen
  \bibfield  {author} {\bibinfo {author} {\bibfnamefont {J.~N.}\ \bibnamefont
  {Reimers}}, \bibinfo {author} {\bibfnamefont {J.~E.}\ \bibnamefont
  {Greedan}}, \bibinfo {author} {\bibfnamefont {R.~K.}\ \bibnamefont {Kremer}},
  \bibinfo {author} {\bibfnamefont {E.}~\bibnamefont {Gmelin}}, \ and\ \bibinfo
  {author} {\bibfnamefont {M.~A.}\ \bibnamefont {Subramanian}},\ }\href
  {\doibase 10.1103/PhysRevB.43.3387} {\bibfield  {journal} {\bibinfo
  {journal} {Phys. Rev. B}\ }\textbf {\bibinfo {volume} {43}},\ \bibinfo
  {pages} {3387} (\bibinfo {year} {1991}{\natexlab{a}})}\BibitemShut {NoStop}%
\bibitem [{\citenamefont {Gaulin}\ \emph {et~al.}(1992)\citenamefont {Gaulin},
  \citenamefont {Reimers}, \citenamefont {Mason}, \citenamefont {Greedan},\
  and\ \citenamefont {Tun}}]{PhysRevLett.69.3244}%
  \BibitemOpen
  \bibfield  {author} {\bibinfo {author} {\bibfnamefont {B.~D.}\ \bibnamefont
  {Gaulin}}, \bibinfo {author} {\bibfnamefont {J.~N.}\ \bibnamefont {Reimers}},
  \bibinfo {author} {\bibfnamefont {T.~E.}\ \bibnamefont {Mason}}, \bibinfo
  {author} {\bibfnamefont {J.~E.}\ \bibnamefont {Greedan}}, \ and\ \bibinfo
  {author} {\bibfnamefont {Z.}~\bibnamefont {Tun}},\ }\href {\doibase
  10.1103/PhysRevLett.69.3244} {\bibfield  {journal} {\bibinfo  {journal}
  {Phys. Rev. Lett.}\ }\textbf {\bibinfo {volume} {69}},\ \bibinfo {pages}
  {3244} (\bibinfo {year} {1992})}\BibitemShut {NoStop}%
\bibitem [{\citenamefont {Gingras}\ \emph {et~al.}(1997)\citenamefont
  {Gingras}, \citenamefont {Stager}, \citenamefont {Raju}, \citenamefont
  {Gaulin},\ and\ \citenamefont {Greedan}}]{PhysRevLett.78.947}%
  \BibitemOpen
  \bibfield  {author} {\bibinfo {author} {\bibfnamefont {M.~J.~P.}\
  \bibnamefont {Gingras}}, \bibinfo {author} {\bibfnamefont {C.~V.}\
  \bibnamefont {Stager}}, \bibinfo {author} {\bibfnamefont {N.~P.}\
  \bibnamefont {Raju}}, \bibinfo {author} {\bibfnamefont {B.~D.}\ \bibnamefont
  {Gaulin}}, \ and\ \bibinfo {author} {\bibfnamefont {J.~E.}\ \bibnamefont
  {Greedan}},\ }\href {\doibase 10.1103/PhysRevLett.78.947} {\bibfield
  {journal} {\bibinfo  {journal} {Phys. Rev. Lett.}\ }\textbf {\bibinfo
  {volume} {78}},\ \bibinfo {pages} {947} (\bibinfo {year} {1997})}\BibitemShut
  {NoStop}%
\bibitem [{\citenamefont {Guo}\ and\ \citenamefont
  {Franz}(2009{\natexlab{a}})}]{PhysRevLett.103.206805}%
  \BibitemOpen
  \bibfield  {author} {\bibinfo {author} {\bibfnamefont {H.-M.}\ \bibnamefont
  {Guo}}\ and\ \bibinfo {author} {\bibfnamefont {M.}~\bibnamefont {Franz}},\
  }\href {\doibase 10.1103/PhysRevLett.103.206805} {\bibfield  {journal}
  {\bibinfo  {journal} {Phys. Rev. Lett.}\ }\textbf {\bibinfo {volume} {103}},\
  \bibinfo {pages} {206805} (\bibinfo {year} {2009}{\natexlab{a}})}\BibitemShut
  {NoStop}%
\bibitem [{\citenamefont {Guo}\ and\ \citenamefont
  {Franz}(2009{\natexlab{b}})}]{PhysRevB.80.113102}%
  \BibitemOpen
  \bibfield  {author} {\bibinfo {author} {\bibfnamefont {H.-M.}\ \bibnamefont
  {Guo}}\ and\ \bibinfo {author} {\bibfnamefont {M.}~\bibnamefont {Franz}},\
  }\href {\doibase 10.1103/PhysRevB.80.113102} {\bibfield  {journal} {\bibinfo
  {journal} {Phys. Rev. B}\ }\textbf {\bibinfo {volume} {80}},\ \bibinfo
  {pages} {113102} (\bibinfo {year} {2009}{\natexlab{b}})}\BibitemShut
  {NoStop}%
\bibitem [{\citenamefont {Pesin}\ and\ \citenamefont
  {Balents}(2010)}]{pesin2010mott}%
  \BibitemOpen
  \bibfield  {author} {\bibinfo {author} {\bibfnamefont {D.}~\bibnamefont
  {Pesin}}\ and\ \bibinfo {author} {\bibfnamefont {L.}~\bibnamefont
  {Balents}},\ }\href {\doibase https://doi.org/10.1038/nphys1606} {\bibfield
  {journal} {\bibinfo  {journal} {Nat. Phys.}\ }\textbf {\bibinfo {volume}
  {6}},\ \bibinfo {pages} {376} (\bibinfo {year} {2010})}\BibitemShut {NoStop}%
\bibitem [{\citenamefont {Otsuka}\ \emph {et~al.}(2021)\citenamefont {Otsuka},
  \citenamefont {Yoshida}, \citenamefont {Kudo}, \citenamefont {Yunoki},\ and\
  \citenamefont {Hatsugai}}]{otsuka2021higher}%
  \BibitemOpen
  \bibfield  {author} {\bibinfo {author} {\bibfnamefont {Y.}~\bibnamefont
  {Otsuka}}, \bibinfo {author} {\bibfnamefont {T.}~\bibnamefont {Yoshida}},
  \bibinfo {author} {\bibfnamefont {K.}~\bibnamefont {Kudo}}, \bibinfo {author}
  {\bibfnamefont {S.}~\bibnamefont {Yunoki}}, \ and\ \bibinfo {author}
  {\bibfnamefont {Y.}~\bibnamefont {Hatsugai}},\ }\href {\doibase
  https://doi.org/10.1038/s41598-021-99213-z} {\bibfield  {journal} {\bibinfo
  {journal} {Sci. Rep.}\ }\textbf {\bibinfo {volume} {11}},\ \bibinfo {pages}
  {1} (\bibinfo {year} {2021})}\BibitemShut {NoStop}%
\bibitem [{\citenamefont {Hozoi}\ \emph {et~al.}(2014)\citenamefont {Hozoi},
  \citenamefont {Gretarsson}, \citenamefont {Clancy}, \citenamefont {Jeon},
  \citenamefont {Lee}, \citenamefont {Kim}, \citenamefont {Yushankhai},
  \citenamefont {Fulde}, \citenamefont {Casa}, \citenamefont {Gog},
  \citenamefont {Kim}, \citenamefont {Said}, \citenamefont {Upton},
  \citenamefont {Kim},\ and\ \citenamefont {van~den
  Brink}}]{PhysRevB.89.115111}%
  \BibitemOpen
  \bibfield  {author} {\bibinfo {author} {\bibfnamefont {L.}~\bibnamefont
  {Hozoi}}, \bibinfo {author} {\bibfnamefont {H.}~\bibnamefont {Gretarsson}},
  \bibinfo {author} {\bibfnamefont {J.~P.}\ \bibnamefont {Clancy}}, \bibinfo
  {author} {\bibfnamefont {B.-G.}\ \bibnamefont {Jeon}}, \bibinfo {author}
  {\bibfnamefont {B.}~\bibnamefont {Lee}}, \bibinfo {author} {\bibfnamefont
  {K.~H.}\ \bibnamefont {Kim}}, \bibinfo {author} {\bibfnamefont
  {V.}~\bibnamefont {Yushankhai}}, \bibinfo {author} {\bibfnamefont
  {P.}~\bibnamefont {Fulde}}, \bibinfo {author} {\bibfnamefont
  {D.}~\bibnamefont {Casa}}, \bibinfo {author} {\bibfnamefont {T.}~\bibnamefont
  {Gog}}, \bibinfo {author} {\bibfnamefont {J.}~\bibnamefont {Kim}}, \bibinfo
  {author} {\bibfnamefont {A.~H.}\ \bibnamefont {Said}}, \bibinfo {author}
  {\bibfnamefont {M.~H.}\ \bibnamefont {Upton}}, \bibinfo {author}
  {\bibfnamefont {Y.-J.}\ \bibnamefont {Kim}}, \ and\ \bibinfo {author}
  {\bibfnamefont {J.}~\bibnamefont {van~den Brink}},\ }\href {\doibase
  10.1103/PhysRevB.89.115111} {\bibfield  {journal} {\bibinfo  {journal} {Phys.
  Rev. B}\ }\textbf {\bibinfo {volume} {89}},\ \bibinfo {pages} {115111}
  (\bibinfo {year} {2014})}\BibitemShut {NoStop}%
\bibitem [{\citenamefont {Kargarian}\ \emph {et~al.}(2011)\citenamefont
  {Kargarian}, \citenamefont {Wen},\ and\ \citenamefont
  {Fiete}}]{PhysRevB.83.165112}%
  \BibitemOpen
  \bibfield  {author} {\bibinfo {author} {\bibfnamefont {M.}~\bibnamefont
  {Kargarian}}, \bibinfo {author} {\bibfnamefont {J.}~\bibnamefont {Wen}}, \
  and\ \bibinfo {author} {\bibfnamefont {G.~A.}\ \bibnamefont {Fiete}},\ }\href
  {\doibase 10.1103/PhysRevB.83.165112} {\bibfield  {journal} {\bibinfo
  {journal} {Phys. Rev. B}\ }\textbf {\bibinfo {volume} {83}},\ \bibinfo
  {pages} {165112} (\bibinfo {year} {2011})}\BibitemShut {NoStop}%
\bibitem [{\citenamefont {Yang}\ and\ \citenamefont
  {Kim}(2010)}]{PhysRevB.82.085111}%
  \BibitemOpen
  \bibfield  {author} {\bibinfo {author} {\bibfnamefont {B.-J.}\ \bibnamefont
  {Yang}}\ and\ \bibinfo {author} {\bibfnamefont {Y.~B.}\ \bibnamefont {Kim}},\
  }\href {\doibase 10.1103/PhysRevB.82.085111} {\bibfield  {journal} {\bibinfo
  {journal} {Phys. Rev. B}\ }\textbf {\bibinfo {volume} {82}},\ \bibinfo
  {pages} {085111} (\bibinfo {year} {2010})}\BibitemShut {NoStop}%
\bibitem [{\citenamefont {Ezawa}(2018)}]{PhysRevLett.120.026801}%
  \BibitemOpen
  \bibfield  {author} {\bibinfo {author} {\bibfnamefont {M.}~\bibnamefont
  {Ezawa}},\ }\href {\doibase 10.1103/PhysRevLett.120.026801} {\bibfield
  {journal} {\bibinfo  {journal} {Phys. Rev. Lett.}\ }\textbf {\bibinfo
  {volume} {120}},\ \bibinfo {pages} {026801} (\bibinfo {year}
  {2018})}\BibitemShut {NoStop}%
\bibitem [{\citenamefont {Shinaoka}\ \emph {et~al.}(2014)\citenamefont
  {Shinaoka}, \citenamefont {Tomita},\ and\ \citenamefont
  {Motome}}]{PhysRevB.90.165119}%
  \BibitemOpen
  \bibfield  {author} {\bibinfo {author} {\bibfnamefont {H.}~\bibnamefont
  {Shinaoka}}, \bibinfo {author} {\bibfnamefont {Y.}~\bibnamefont {Tomita}}, \
  and\ \bibinfo {author} {\bibfnamefont {Y.}~\bibnamefont {Motome}},\ }\href
  {\doibase 10.1103/PhysRevB.90.165119} {\bibfield  {journal} {\bibinfo
  {journal} {Phys. Rev. B}\ }\textbf {\bibinfo {volume} {90}},\ \bibinfo
  {pages} {165119} (\bibinfo {year} {2014})}\BibitemShut {NoStop}%
\bibitem [{\citenamefont {Harris}\ \emph {et~al.}(1997)\citenamefont {Harris},
  \citenamefont {Bramwell}, \citenamefont {McMorrow}, \citenamefont {Zeiske},\
  and\ \citenamefont {Godfrey}}]{PhysRevLett.79.2554}%
  \BibitemOpen
  \bibfield  {author} {\bibinfo {author} {\bibfnamefont {M.~J.}\ \bibnamefont
  {Harris}}, \bibinfo {author} {\bibfnamefont {S.~T.}\ \bibnamefont
  {Bramwell}}, \bibinfo {author} {\bibfnamefont {D.~F.}\ \bibnamefont
  {McMorrow}}, \bibinfo {author} {\bibfnamefont {T.}~\bibnamefont {Zeiske}}, \
  and\ \bibinfo {author} {\bibfnamefont {K.~W.}\ \bibnamefont {Godfrey}},\
  }\href {\doibase 10.1103/PhysRevLett.79.2554} {\bibfield  {journal} {\bibinfo
   {journal} {Phys. Rev. Lett.}\ }\textbf {\bibinfo {volume} {79}},\ \bibinfo
  {pages} {2554} (\bibinfo {year} {1997})}\BibitemShut {NoStop}%
\bibitem [{\citenamefont {Bramwell}\ \emph {et~al.}(2001)\citenamefont
  {Bramwell}, \citenamefont {Harris}, \citenamefont {den Hertog}, \citenamefont
  {Gingras}, \citenamefont {Gardner}, \citenamefont {McMorrow}, \citenamefont
  {Wildes}, \citenamefont {Cornelius}, \citenamefont {Champion}, \citenamefont
  {Melko},\ and\ \citenamefont {Fennell}}]{PhysRevLett.87.047205}%
  \BibitemOpen
  \bibfield  {author} {\bibinfo {author} {\bibfnamefont {S.~T.}\ \bibnamefont
  {Bramwell}}, \bibinfo {author} {\bibfnamefont {M.~J.}\ \bibnamefont
  {Harris}}, \bibinfo {author} {\bibfnamefont {B.~C.}\ \bibnamefont {den
  Hertog}}, \bibinfo {author} {\bibfnamefont {M.~J.~P.}\ \bibnamefont
  {Gingras}}, \bibinfo {author} {\bibfnamefont {J.~S.}\ \bibnamefont
  {Gardner}}, \bibinfo {author} {\bibfnamefont {D.~F.}\ \bibnamefont
  {McMorrow}}, \bibinfo {author} {\bibfnamefont {A.~R.}\ \bibnamefont
  {Wildes}}, \bibinfo {author} {\bibfnamefont {A.~L.}\ \bibnamefont
  {Cornelius}}, \bibinfo {author} {\bibfnamefont {J.~D.~M.}\ \bibnamefont
  {Champion}}, \bibinfo {author} {\bibfnamefont {R.~G.}\ \bibnamefont {Melko}},
  \ and\ \bibinfo {author} {\bibfnamefont {T.}~\bibnamefont {Fennell}},\ }\href
  {\doibase 10.1103/PhysRevLett.87.047205} {\bibfield  {journal} {\bibinfo
  {journal} {Phys. Rev. Lett.}\ }\textbf {\bibinfo {volume} {87}},\ \bibinfo
  {pages} {047205} (\bibinfo {year} {2001})}\BibitemShut {NoStop}%
\bibitem [{\citenamefont {Harris}\ \emph {et~al.}(1998)\citenamefont {Harris},
  \citenamefont {Bramwell}, \citenamefont {Zeiske}, \citenamefont {McMorrow},\
  and\ \citenamefont {King}}]{HARRIS1998757}%
  \BibitemOpen
  \bibfield  {author} {\bibinfo {author} {\bibfnamefont {M.}~\bibnamefont
  {Harris}}, \bibinfo {author} {\bibfnamefont {S.}~\bibnamefont {Bramwell}},
  \bibinfo {author} {\bibfnamefont {T.}~\bibnamefont {Zeiske}}, \bibinfo
  {author} {\bibfnamefont {D.}~\bibnamefont {McMorrow}}, \ and\ \bibinfo
  {author} {\bibfnamefont {P.}~\bibnamefont {King}},\ }\href {\doibase
  https://doi.org/10.1016/S0304-8853(97)00796-8} {\bibfield  {journal}
  {\bibinfo  {journal} {J. Magn. Magn. Mater.}\ }\textbf {\bibinfo {volume}
  {177-181}},\ \bibinfo {pages} {757} (\bibinfo {year} {1998})},\ \bibinfo
  {note} {international Conference on Magnetism (Part II)}\BibitemShut
  {NoStop}%
\bibitem [{\citenamefont {Ramirez}\ \emph {et~al.}(1999)\citenamefont
  {Ramirez}, \citenamefont {Hayashi}, \citenamefont {Cava}, \citenamefont
  {Siddharthan},\ and\ \citenamefont {Shastry}}]{ramirez99:399}%
  \BibitemOpen
  \bibfield  {author} {\bibinfo {author} {\bibfnamefont {A.~P.}\ \bibnamefont
  {Ramirez}}, \bibinfo {author} {\bibfnamefont {A.}~\bibnamefont {Hayashi}},
  \bibinfo {author} {\bibfnamefont {R.~J.}\ \bibnamefont {Cava}}, \bibinfo
  {author} {\bibfnamefont {R.}~\bibnamefont {Siddharthan}}, \ and\ \bibinfo
  {author} {\bibfnamefont {B.}~\bibnamefont {Shastry}},\ }\href {\doibase
  https://doi.org/10.1038/20619} {\bibfield  {journal} {\bibinfo  {journal}
  {Nature}\ }\textbf {\bibinfo {volume} {399}},\ \bibinfo {pages} {333}
  (\bibinfo {year} {1999})}\BibitemShut {NoStop}%
\bibitem [{\citenamefont {Bramwell}\ and\ \citenamefont
  {Gingras}(2001)}]{doi:10.1126/science.1064761}%
  \BibitemOpen
  \bibfield  {author} {\bibinfo {author} {\bibfnamefont {S.~T.}\ \bibnamefont
  {Bramwell}}\ and\ \bibinfo {author} {\bibfnamefont {M.~J.~P.}\ \bibnamefont
  {Gingras}},\ }\href {\doibase 10.1126/science.1064761} {\bibfield  {journal}
  {\bibinfo  {journal} {Science}\ }\textbf {\bibinfo {volume} {294}},\ \bibinfo
  {pages} {1495} (\bibinfo {year} {2001})}\BibitemShut {NoStop}%
\bibitem [{\citenamefont {Bramwell}\ and\ \citenamefont
  {Harris}(2020)}]{Bramwell_2020}%
  \BibitemOpen
  \bibfield  {author} {\bibinfo {author} {\bibfnamefont {S.~T.}\ \bibnamefont
  {Bramwell}}\ and\ \bibinfo {author} {\bibfnamefont {M.~J.}\ \bibnamefont
  {Harris}},\ }\href {\doibase 10.1088/1361-648x/ab8423} {\bibfield  {journal}
  {\bibinfo  {journal} {J. Condens. Matter Phys.}\ }\textbf {\bibinfo {volume}
  {32}},\ \bibinfo {pages} {374010} (\bibinfo {year} {2020})}\BibitemShut
  {NoStop}%
\bibitem [{\citenamefont {den Hertog}\ and\ \citenamefont
  {Gingras}(2000)}]{denhertog00:84}%
  \BibitemOpen
  \bibfield  {author} {\bibinfo {author} {\bibfnamefont {B.~C.}\ \bibnamefont
  {den Hertog}}\ and\ \bibinfo {author} {\bibfnamefont {M.~J.~P.}\ \bibnamefont
  {Gingras}},\ }\href {\doibase 10.1103/PhysRevLett.84.3430} {\bibfield
  {journal} {\bibinfo  {journal} {Phys. Rev. Lett.}\ }\textbf {\bibinfo
  {volume} {84}},\ \bibinfo {pages} {3430} (\bibinfo {year}
  {2000})}\BibitemShut {NoStop}%
\bibitem [{\citenamefont {Anderson}(1956)}]{PhysRev.102.1008}%
  \BibitemOpen
  \bibfield  {author} {\bibinfo {author} {\bibfnamefont {P.~W.}\ \bibnamefont
  {Anderson}},\ }\href {\doibase 10.1103/PhysRev.102.1008} {\bibfield
  {journal} {\bibinfo  {journal} {Phys. Rev.}\ }\textbf {\bibinfo {volume}
  {102}},\ \bibinfo {pages} {1008} (\bibinfo {year} {1956})}\BibitemShut
  {NoStop}%
\bibitem [{\citenamefont {Ruff}\ \emph {et~al.}(2005)\citenamefont {Ruff},
  \citenamefont {Melko},\ and\ \citenamefont
  {Gingras}}]{PhysRevLett.95.097202}%
  \BibitemOpen
  \bibfield  {author} {\bibinfo {author} {\bibfnamefont {J.~P.~C.}\
  \bibnamefont {Ruff}}, \bibinfo {author} {\bibfnamefont {R.~G.}\ \bibnamefont
  {Melko}}, \ and\ \bibinfo {author} {\bibfnamefont {M.~J.~P.}\ \bibnamefont
  {Gingras}},\ }\href {\doibase 10.1103/PhysRevLett.95.097202} {\bibfield
  {journal} {\bibinfo  {journal} {Phys. Rev. Lett.}\ }\textbf {\bibinfo
  {volume} {95}},\ \bibinfo {pages} {097202} (\bibinfo {year}
  {2005})}\BibitemShut {NoStop}%
\bibitem [{\citenamefont {Yavors'kii}\ \emph {et~al.}(2008)\citenamefont
  {Yavors'kii}, \citenamefont {Fennell}, \citenamefont {Gingras},\ and\
  \citenamefont {Bramwell}}]{PhysRevLett.101.037204}%
  \BibitemOpen
  \bibfield  {author} {\bibinfo {author} {\bibfnamefont {T.}~\bibnamefont
  {Yavors'kii}}, \bibinfo {author} {\bibfnamefont {T.}~\bibnamefont {Fennell}},
  \bibinfo {author} {\bibfnamefont {M.~J.~P.}\ \bibnamefont {Gingras}}, \ and\
  \bibinfo {author} {\bibfnamefont {S.~T.}\ \bibnamefont {Bramwell}},\ }\href
  {\doibase 10.1103/PhysRevLett.101.037204} {\bibfield  {journal} {\bibinfo
  {journal} {Phys. Rev. Lett.}\ }\textbf {\bibinfo {volume} {101}},\ \bibinfo
  {pages} {037204} (\bibinfo {year} {2008})}\BibitemShut {NoStop}%
\bibitem [{\citenamefont {Melko}\ and\ \citenamefont
  {Gingras}(2004)}]{Melko_2004}%
  \BibitemOpen
  \bibfield  {author} {\bibinfo {author} {\bibfnamefont {R.~G.}\ \bibnamefont
  {Melko}}\ and\ \bibinfo {author} {\bibfnamefont {M.~J.~P.}\ \bibnamefont
  {Gingras}},\ }\href {\doibase 10.1088/0953-8984/16/43/r02} {\bibfield
  {journal} {\bibinfo  {journal} {J. Condens. Matter Phys.}\ }\textbf {\bibinfo
  {volume} {16}},\ \bibinfo {pages} {R1277} (\bibinfo {year}
  {2004})}\BibitemShut {NoStop}%
\bibitem [{\citenamefont {Sarte}\ \emph {et~al.}(2017)\citenamefont {Sarte},
  \citenamefont {Aczel}, \citenamefont {Ehlers}, \citenamefont {Stock},
  \citenamefont {Gaulin}, \citenamefont {Mauws}, \citenamefont {Stone},
  \citenamefont {Calder}, \citenamefont {Nagler}, \citenamefont {Hollett},
  \citenamefont {Zhou}, \citenamefont {Gardner}, \citenamefont {Attfield},\
  and\ \citenamefont {Wiebe}}]{Sarte17:29}%
  \BibitemOpen
  \bibfield  {author} {\bibinfo {author} {\bibfnamefont {P.~M.}\ \bibnamefont
  {Sarte}}, \bibinfo {author} {\bibfnamefont {A.~A.}\ \bibnamefont {Aczel}},
  \bibinfo {author} {\bibfnamefont {G.}~\bibnamefont {Ehlers}}, \bibinfo
  {author} {\bibfnamefont {C.}~\bibnamefont {Stock}}, \bibinfo {author}
  {\bibfnamefont {B.~D.}\ \bibnamefont {Gaulin}}, \bibinfo {author}
  {\bibfnamefont {C.}~\bibnamefont {Mauws}}, \bibinfo {author} {\bibfnamefont
  {M.~B.}\ \bibnamefont {Stone}}, \bibinfo {author} {\bibfnamefont
  {S.}~\bibnamefont {Calder}}, \bibinfo {author} {\bibfnamefont {S.~E.}\
  \bibnamefont {Nagler}}, \bibinfo {author} {\bibfnamefont {J.~W.}\
  \bibnamefont {Hollett}}, \bibinfo {author} {\bibfnamefont {H.~D.}\
  \bibnamefont {Zhou}}, \bibinfo {author} {\bibfnamefont {J.~S.}\ \bibnamefont
  {Gardner}}, \bibinfo {author} {\bibfnamefont {J.~P.}\ \bibnamefont
  {Attfield}}, \ and\ \bibinfo {author} {\bibfnamefont {C.~R.}\ \bibnamefont
  {Wiebe}},\ }\href {\doibase 10.1088/1361-648x/aa8ec2} {\bibfield  {journal}
  {\bibinfo  {journal} {J. Phys. Condens. Matter}\ }\textbf {\bibinfo {volume}
  {29}},\ \bibinfo {pages} {45LT01} (\bibinfo {year} {2017})}\BibitemShut
  {NoStop}%
\bibitem [{\citenamefont {Morris}\ \emph {et~al.}(2009)\citenamefont {Morris},
  \citenamefont {Tennant}, \citenamefont {Grigera}, \citenamefont {Klemke},
  \citenamefont {Castelnovo}, \citenamefont {Moessner}, \citenamefont
  {Czternasty}, \citenamefont {Meissner}, \citenamefont {Rule}, \citenamefont
  {Hoffmann}, \citenamefont {Kiefer}, \citenamefont {Gerischer}, \citenamefont
  {Slobinsky},\ and\ \citenamefont {Perry}}]{morris2009dirac}%
  \BibitemOpen
  \bibfield  {author} {\bibinfo {author} {\bibfnamefont {D.~J.~P.}\
  \bibnamefont {Morris}}, \bibinfo {author} {\bibfnamefont {D.~A.}\
  \bibnamefont {Tennant}}, \bibinfo {author} {\bibfnamefont {S.}~\bibnamefont
  {Grigera}}, \bibinfo {author} {\bibfnamefont {B.}~\bibnamefont {Klemke}},
  \bibinfo {author} {\bibfnamefont {C.}~\bibnamefont {Castelnovo}}, \bibinfo
  {author} {\bibfnamefont {R.}~\bibnamefont {Moessner}}, \bibinfo {author}
  {\bibfnamefont {C.}~\bibnamefont {Czternasty}}, \bibinfo {author}
  {\bibfnamefont {M.}~\bibnamefont {Meissner}}, \bibinfo {author}
  {\bibfnamefont {K.}~\bibnamefont {Rule}}, \bibinfo {author} {\bibfnamefont
  {J.-U.}\ \bibnamefont {Hoffmann}}, \bibinfo {author} {\bibfnamefont
  {K.}~\bibnamefont {Kiefer}}, \bibinfo {author} {\bibfnamefont
  {S.}~\bibnamefont {Gerischer}}, \bibinfo {author} {\bibfnamefont
  {D.}~\bibnamefont {Slobinsky}}, \ and\ \bibinfo {author} {\bibfnamefont
  {R.~S.}\ \bibnamefont {Perry}},\ }\href {\doibase 10.1126/science.1178868}
  {\bibfield  {journal} {\bibinfo  {journal} {Science}\ }\textbf {\bibinfo
  {volume} {326}},\ \bibinfo {pages} {411} (\bibinfo {year}
  {2009})}\BibitemShut {NoStop}%
\bibitem [{\citenamefont {Castelnovo}\ \emph {et~al.}(2008)\citenamefont
  {Castelnovo}, \citenamefont {Moessner},\ and\ \citenamefont
  {Sondhi}}]{castelnovo2008magnetic}%
  \BibitemOpen
  \bibfield  {author} {\bibinfo {author} {\bibfnamefont {C.}~\bibnamefont
  {Castelnovo}}, \bibinfo {author} {\bibfnamefont {R.}~\bibnamefont
  {Moessner}}, \ and\ \bibinfo {author} {\bibfnamefont {S.~L.}\ \bibnamefont
  {Sondhi}},\ }\href {\doibase https://doi.org/10.1038/nature06433} {\bibfield
  {journal} {\bibinfo  {journal} {Nature}\ }\textbf {\bibinfo {volume} {451}},\
  \bibinfo {pages} {42} (\bibinfo {year} {2008})}\BibitemShut {NoStop}%
\bibitem [{\citenamefont {Bramwell}\ \emph {et~al.}(2009)\citenamefont
  {Bramwell}, \citenamefont {Giblin}, \citenamefont {Calder}, \citenamefont
  {Aldus}, \citenamefont {Prabhakaran},\ and\ \citenamefont
  {Fennell}}]{bramwell2009measurement}%
  \BibitemOpen
  \bibfield  {author} {\bibinfo {author} {\bibfnamefont {S.~T.}\ \bibnamefont
  {Bramwell}}, \bibinfo {author} {\bibfnamefont {S.~R.}\ \bibnamefont
  {Giblin}}, \bibinfo {author} {\bibfnamefont {S.}~\bibnamefont {Calder}},
  \bibinfo {author} {\bibfnamefont {R.}~\bibnamefont {Aldus}}, \bibinfo
  {author} {\bibfnamefont {D.}~\bibnamefont {Prabhakaran}}, \ and\ \bibinfo
  {author} {\bibfnamefont {T.}~\bibnamefont {Fennell}},\ }\href {\doibase
  https://doi.org/10.1038/nature08500} {\bibfield  {journal} {\bibinfo
  {journal} {Nature}\ }\textbf {\bibinfo {volume} {461}},\ \bibinfo {pages}
  {956} (\bibinfo {year} {2009})}\BibitemShut {NoStop}%
\bibitem [{\citenamefont {Paulsen}\ \emph {et~al.}(2016)\citenamefont
  {Paulsen}, \citenamefont {Giblin}, \citenamefont {Lhotel}, \citenamefont
  {Prabhakaran}, \citenamefont {Balakrishnan}, \citenamefont {Matsuhira},\ and\
  \citenamefont {Bramwell}}]{paulsen2016experimental}%
  \BibitemOpen
  \bibfield  {author} {\bibinfo {author} {\bibfnamefont {C.}~\bibnamefont
  {Paulsen}}, \bibinfo {author} {\bibfnamefont {S.~R.}\ \bibnamefont {Giblin}},
  \bibinfo {author} {\bibfnamefont {E.}~\bibnamefont {Lhotel}}, \bibinfo
  {author} {\bibfnamefont {D.}~\bibnamefont {Prabhakaran}}, \bibinfo {author}
  {\bibfnamefont {G.}~\bibnamefont {Balakrishnan}}, \bibinfo {author}
  {\bibfnamefont {K.}~\bibnamefont {Matsuhira}}, \ and\ \bibinfo {author}
  {\bibfnamefont {S.~T.}\ \bibnamefont {Bramwell}},\ }\href {\doibase
  https://doi.org/10.1038/nphys3704} {\bibfield  {journal} {\bibinfo  {journal}
  {Nat. Phys.}\ }\textbf {\bibinfo {volume} {12}},\ \bibinfo {pages} {661}
  (\bibinfo {year} {2016})}\BibitemShut {NoStop}%
\bibitem [{\citenamefont {Gao}\ \emph {et~al.}(2019)\citenamefont {Gao},
  \citenamefont {Chen}, \citenamefont {Tam}, \citenamefont {Huang},
  \citenamefont {Sasmal}, \citenamefont {Adroja}, \citenamefont {Ye},
  \citenamefont {Cao}, \citenamefont {Sala}, \citenamefont {Stone},
  \citenamefont {Baines}, \citenamefont {Verezhak}, \citenamefont {Hu},
  \citenamefont {J.-H.}, \citenamefont {Xu}, \citenamefont {Cheong},
  \citenamefont {Nallaiyan}, \citenamefont {Spagna}, \citenamefont {Maple},
  \citenamefont {Nevidomskyy}, \citenamefont {Morosan},\ and\ \citenamefont
  {Dai}}]{gao2019experimental}%
  \BibitemOpen
  \bibfield  {author} {\bibinfo {author} {\bibfnamefont {B.}~\bibnamefont
  {Gao}}, \bibinfo {author} {\bibfnamefont {T.}~\bibnamefont {Chen}}, \bibinfo
  {author} {\bibfnamefont {D.~W.}\ \bibnamefont {Tam}}, \bibinfo {author}
  {\bibfnamefont {C.-L.}\ \bibnamefont {Huang}}, \bibinfo {author}
  {\bibfnamefont {K.}~\bibnamefont {Sasmal}}, \bibinfo {author} {\bibfnamefont
  {D.~T.}\ \bibnamefont {Adroja}}, \bibinfo {author} {\bibfnamefont
  {F.}~\bibnamefont {Ye}}, \bibinfo {author} {\bibfnamefont {H.}~\bibnamefont
  {Cao}}, \bibinfo {author} {\bibfnamefont {G.}~\bibnamefont {Sala}}, \bibinfo
  {author} {\bibfnamefont {M.~B.}\ \bibnamefont {Stone}}, \bibinfo {author}
  {\bibfnamefont {C.}~\bibnamefont {Baines}}, \bibinfo {author} {\bibfnamefont
  {J.~A.~T.}\ \bibnamefont {Verezhak}}, \bibinfo {author} {\bibfnamefont
  {H.}~\bibnamefont {Hu}}, \bibinfo {author} {\bibfnamefont {C.}~\bibnamefont
  {J.-H.}}, \bibinfo {author} {\bibfnamefont {X.}~\bibnamefont {Xu}}, \bibinfo
  {author} {\bibfnamefont {S.-W.}\ \bibnamefont {Cheong}}, \bibinfo {author}
  {\bibfnamefont {M.}~\bibnamefont {Nallaiyan}}, \bibinfo {author}
  {\bibfnamefont {S.}~\bibnamefont {Spagna}}, \bibinfo {author} {\bibfnamefont
  {M.~B.}\ \bibnamefont {Maple}}, \bibinfo {author} {\bibfnamefont {A.~H.}\
  \bibnamefont {Nevidomskyy}}, \bibinfo {author} {\bibfnamefont
  {G.}~\bibnamefont {Morosan}, \bibfnamefont {E.~Chen}}, \ and\ \bibinfo
  {author} {\bibfnamefont {P.}~\bibnamefont {Dai}},\ }\href {\doibase
  https://doi.org/10.1038/s41567-019-0577-6} {\bibfield  {journal} {\bibinfo
  {journal} {Nat. Phys.}\ }\textbf {\bibinfo {volume} {15}},\ \bibinfo {pages}
  {1052} (\bibinfo {year} {2019})}\BibitemShut {NoStop}%
\bibitem [{\citenamefont {Benton}(2018)}]{Benton18:121}%
  \BibitemOpen
  \bibfield  {author} {\bibinfo {author} {\bibfnamefont {O.}~\bibnamefont
  {Benton}},\ }\href {\doibase 10.1103/PhysRevLett.121.037203} {\bibfield
  {journal} {\bibinfo  {journal} {Phys. Rev. Lett.}\ }\textbf {\bibinfo
  {volume} {121}},\ \bibinfo {pages} {037203} (\bibinfo {year}
  {2018})}\BibitemShut {NoStop}%
\bibitem [{\citenamefont {Wen}\ \emph {et~al.}(2017)\citenamefont {Wen},
  \citenamefont {Koohpayeh}, \citenamefont {Ross}, \citenamefont {Trump},
  \citenamefont {McQueen}, \citenamefont {Kimura}, \citenamefont {Nakatsuji},
  \citenamefont {Qiu}, \citenamefont {Pajerowski}, \citenamefont {Copley},\
  and\ \citenamefont {Broholm}}]{Wen17:118}%
  \BibitemOpen
  \bibfield  {author} {\bibinfo {author} {\bibfnamefont {J.-J.}\ \bibnamefont
  {Wen}}, \bibinfo {author} {\bibfnamefont {S.~M.}\ \bibnamefont {Koohpayeh}},
  \bibinfo {author} {\bibfnamefont {K.~A.}\ \bibnamefont {Ross}}, \bibinfo
  {author} {\bibfnamefont {B.~A.}\ \bibnamefont {Trump}}, \bibinfo {author}
  {\bibfnamefont {T.~M.}\ \bibnamefont {McQueen}}, \bibinfo {author}
  {\bibfnamefont {K.}~\bibnamefont {Kimura}}, \bibinfo {author} {\bibfnamefont
  {S.}~\bibnamefont {Nakatsuji}}, \bibinfo {author} {\bibfnamefont
  {Y.}~\bibnamefont {Qiu}}, \bibinfo {author} {\bibfnamefont {D.~M.}\
  \bibnamefont {Pajerowski}}, \bibinfo {author} {\bibfnamefont {J.~R.~D.}\
  \bibnamefont {Copley}}, \ and\ \bibinfo {author} {\bibfnamefont {C.~L.}\
  \bibnamefont {Broholm}},\ }\href {\doibase 10.1103/PhysRevLett.118.107206}
  {\bibfield  {journal} {\bibinfo  {journal} {Phys. Rev. Lett.}\ }\textbf
  {\bibinfo {volume} {118}},\ \bibinfo {pages} {107206} (\bibinfo {year}
  {2017})}\BibitemShut {NoStop}%
\bibitem [{\citenamefont {Savary}\ and\ \citenamefont
  {Balents}(2017)}]{Savary17:118}%
  \BibitemOpen
  \bibfield  {author} {\bibinfo {author} {\bibfnamefont {L.}~\bibnamefont
  {Savary}}\ and\ \bibinfo {author} {\bibfnamefont {L.}~\bibnamefont
  {Balents}},\ }\href {\doibase 10.1103/PhysRevLett.118.087203} {\bibfield
  {journal} {\bibinfo  {journal} {Phys. Rev. Lett.}\ }\textbf {\bibinfo
  {volume} {118}},\ \bibinfo {pages} {087203} (\bibinfo {year}
  {2017})}\BibitemShut {NoStop}%
\bibitem [{\citenamefont {Sibille}\ \emph {et~al.}(2015)\citenamefont
  {Sibille}, \citenamefont {Lhotel}, \citenamefont {Pomjakushin}, \citenamefont
  {Baines}, \citenamefont {Fennell},\ and\ \citenamefont
  {Kenzelmann}}]{sibille15:115}%
  \BibitemOpen
  \bibfield  {author} {\bibinfo {author} {\bibfnamefont {R.}~\bibnamefont
  {Sibille}}, \bibinfo {author} {\bibfnamefont {E.}~\bibnamefont {Lhotel}},
  \bibinfo {author} {\bibfnamefont {V.}~\bibnamefont {Pomjakushin}}, \bibinfo
  {author} {\bibfnamefont {C.}~\bibnamefont {Baines}}, \bibinfo {author}
  {\bibfnamefont {T.}~\bibnamefont {Fennell}}, \ and\ \bibinfo {author}
  {\bibfnamefont {M.}~\bibnamefont {Kenzelmann}},\ }\href {\doibase
  10.1103/PhysRevLett.115.097202} {\bibfield  {journal} {\bibinfo  {journal}
  {Phys. Rev. Lett.}\ }\textbf {\bibinfo {volume} {115}},\ \bibinfo {pages}
  {097202} (\bibinfo {year} {2015})}\BibitemShut {NoStop}%
\bibitem [{\citenamefont {Lee}\ \emph {et~al.}(2012)\citenamefont {Lee},
  \citenamefont {Onoda},\ and\ \citenamefont {Balents}}]{PhysRevB.86.104412}%
  \BibitemOpen
  \bibfield  {author} {\bibinfo {author} {\bibfnamefont {S.}~\bibnamefont
  {Lee}}, \bibinfo {author} {\bibfnamefont {S.}~\bibnamefont {Onoda}}, \ and\
  \bibinfo {author} {\bibfnamefont {L.}~\bibnamefont {Balents}},\ }\href
  {\doibase 10.1103/PhysRevB.86.104412} {\bibfield  {journal} {\bibinfo
  {journal} {Phys. Rev. B}\ }\textbf {\bibinfo {volume} {86}},\ \bibinfo
  {pages} {104412} (\bibinfo {year} {2012})}\BibitemShut {NoStop}%
\bibitem [{\citenamefont {Balents}(2010)}]{balents2010spin}%
  \BibitemOpen
  \bibfield  {author} {\bibinfo {author} {\bibfnamefont {L.}~\bibnamefont
  {Balents}},\ }\href {\doibase https://doi.org/10.1038/nature08917} {\bibfield
   {journal} {\bibinfo  {journal} {Nature}\ }\textbf {\bibinfo {volume}
  {464}},\ \bibinfo {pages} {199} (\bibinfo {year} {2010})}\BibitemShut
  {NoStop}%
\bibitem [{\citenamefont {Ross}\ \emph
  {et~al.}(2011{\natexlab{a}})\citenamefont {Ross}, \citenamefont {Savary},
  \citenamefont {Gaulin},\ and\ \citenamefont {Balents}}]{Yb_ross2011quantum}%
  \BibitemOpen
  \bibfield  {author} {\bibinfo {author} {\bibfnamefont {K.~A.}\ \bibnamefont
  {Ross}}, \bibinfo {author} {\bibfnamefont {L.}~\bibnamefont {Savary}},
  \bibinfo {author} {\bibfnamefont {B.~D.}\ \bibnamefont {Gaulin}}, \ and\
  \bibinfo {author} {\bibfnamefont {L.}~\bibnamefont {Balents}},\ }\href
  {\doibase 10.1103/PhysRevX.1.021002} {\bibfield  {journal} {\bibinfo
  {journal} {Phys. Rev. X}\ }\textbf {\bibinfo {volume} {1}},\ \bibinfo {pages}
  {021002} (\bibinfo {year} {2011}{\natexlab{a}})}\BibitemShut {NoStop}%
\bibitem [{\citenamefont {Savary}\ and\ \citenamefont
  {Balents}(2013)}]{PhysRevB.87.205130}%
  \BibitemOpen
  \bibfield  {author} {\bibinfo {author} {\bibfnamefont {L.}~\bibnamefont
  {Savary}}\ and\ \bibinfo {author} {\bibfnamefont {L.}~\bibnamefont
  {Balents}},\ }\href {\doibase 10.1103/PhysRevB.87.205130} {\bibfield
  {journal} {\bibinfo  {journal} {Phys. Rev. B}\ }\textbf {\bibinfo {volume}
  {87}},\ \bibinfo {pages} {205130} (\bibinfo {year} {2013})}\BibitemShut
  {NoStop}%
\bibitem [{\citenamefont {Savary}\ and\ \citenamefont
  {Balents}(2016)}]{Savary_2016}%
  \BibitemOpen
  \bibfield  {author} {\bibinfo {author} {\bibfnamefont {L.}~\bibnamefont
  {Savary}}\ and\ \bibinfo {author} {\bibfnamefont {L.}~\bibnamefont
  {Balents}},\ }\href {\doibase 10.1088/0034-4885/80/1/016502} {\bibfield
  {journal} {\bibinfo  {journal} {Rep. Prog. Phys.}\ }\textbf {\bibinfo
  {volume} {80}},\ \bibinfo {pages} {016502} (\bibinfo {year}
  {2016})}\BibitemShut {NoStop}%
\bibitem [{\citenamefont {Zhou}\ \emph {et~al.}(2017)\citenamefont {Zhou},
  \citenamefont {Kanoda},\ and\ \citenamefont {Ng}}]{RevModPhys.89.025003}%
  \BibitemOpen
  \bibfield  {author} {\bibinfo {author} {\bibfnamefont {Y.}~\bibnamefont
  {Zhou}}, \bibinfo {author} {\bibfnamefont {K.}~\bibnamefont {Kanoda}}, \ and\
  \bibinfo {author} {\bibfnamefont {T.-K.}\ \bibnamefont {Ng}},\ }\href
  {\doibase 10.1103/RevModPhys.89.025003} {\bibfield  {journal} {\bibinfo
  {journal} {Rev. Mod. Phys.}\ }\textbf {\bibinfo {volume} {89}},\ \bibinfo
  {pages} {025003} (\bibinfo {year} {2017})}\BibitemShut {NoStop}%
\bibitem [{\citenamefont {Gardner}\ \emph {et~al.}(1999)\citenamefont
  {Gardner}, \citenamefont {Dunsiger}, \citenamefont {Gaulin}, \citenamefont
  {Gingras}, \citenamefont {Greedan}, \citenamefont {Kiefl}, \citenamefont
  {Lumsden}, \citenamefont {MacFarlane}, \citenamefont {Raju}, \citenamefont
  {Sonier}, \citenamefont {Swainson},\ and\ \citenamefont
  {Tun}}]{Gardner99:82}%
  \BibitemOpen
  \bibfield  {author} {\bibinfo {author} {\bibfnamefont {J.~S.}\ \bibnamefont
  {Gardner}}, \bibinfo {author} {\bibfnamefont {S.~R.}\ \bibnamefont
  {Dunsiger}}, \bibinfo {author} {\bibfnamefont {B.~D.}\ \bibnamefont
  {Gaulin}}, \bibinfo {author} {\bibfnamefont {M.~J.~P.}\ \bibnamefont
  {Gingras}}, \bibinfo {author} {\bibfnamefont {J.~E.}\ \bibnamefont
  {Greedan}}, \bibinfo {author} {\bibfnamefont {R.~F.}\ \bibnamefont {Kiefl}},
  \bibinfo {author} {\bibfnamefont {M.~D.}\ \bibnamefont {Lumsden}}, \bibinfo
  {author} {\bibfnamefont {W.~A.}\ \bibnamefont {MacFarlane}}, \bibinfo
  {author} {\bibfnamefont {N.~P.}\ \bibnamefont {Raju}}, \bibinfo {author}
  {\bibfnamefont {J.~E.}\ \bibnamefont {Sonier}}, \bibinfo {author}
  {\bibfnamefont {I.}~\bibnamefont {Swainson}}, \ and\ \bibinfo {author}
  {\bibfnamefont {Z.}~\bibnamefont {Tun}},\ }\href {\doibase
  10.1103/PhysRevLett.82.1012} {\bibfield  {journal} {\bibinfo  {journal}
  {Phys. Rev. Lett.}\ }\textbf {\bibinfo {volume} {82}},\ \bibinfo {pages}
  {1012} (\bibinfo {year} {1999})}\BibitemShut {NoStop}%
\bibitem [{\citenamefont {Gardner}\ \emph {et~al.}(2001)\citenamefont
  {Gardner}, \citenamefont {Gaulin}, \citenamefont {Berlinsky}, \citenamefont
  {Waldron}, \citenamefont {Dunsiger}, \citenamefont {Raju},\ and\
  \citenamefont {Greedan}}]{gardner2001neutron}%
  \BibitemOpen
  \bibfield  {author} {\bibinfo {author} {\bibfnamefont {J.~S.}\ \bibnamefont
  {Gardner}}, \bibinfo {author} {\bibfnamefont {B.~D.}\ \bibnamefont {Gaulin}},
  \bibinfo {author} {\bibfnamefont {A.~J.}\ \bibnamefont {Berlinsky}}, \bibinfo
  {author} {\bibfnamefont {P.}~\bibnamefont {Waldron}}, \bibinfo {author}
  {\bibfnamefont {S.~R.}\ \bibnamefont {Dunsiger}}, \bibinfo {author}
  {\bibfnamefont {N.~P.}\ \bibnamefont {Raju}}, \ and\ \bibinfo {author}
  {\bibfnamefont {J.~E.}\ \bibnamefont {Greedan}},\ }\href {\doibase
  10.1103/PhysRevB.64.224416} {\bibfield  {journal} {\bibinfo  {journal} {Phys.
  Rev. B}\ }\textbf {\bibinfo {volume} {64}},\ \bibinfo {pages} {224416}
  (\bibinfo {year} {2001})}\BibitemShut {NoStop}%
\bibitem [{\citenamefont {Ueland}\ \emph {et~al.}(2010)\citenamefont {Ueland},
  \citenamefont {Gardner}, \citenamefont {Williams}, \citenamefont {Dahlberg},
  \citenamefont {Kim}, \citenamefont {Qiu}, \citenamefont {Copley},
  \citenamefont {Schiffer},\ and\ \citenamefont {Cava}}]{PhysRevB.81.060408}%
  \BibitemOpen
  \bibfield  {author} {\bibinfo {author} {\bibfnamefont {B.~G.}\ \bibnamefont
  {Ueland}}, \bibinfo {author} {\bibfnamefont {J.~S.}\ \bibnamefont {Gardner}},
  \bibinfo {author} {\bibfnamefont {A.~J.}\ \bibnamefont {Williams}}, \bibinfo
  {author} {\bibfnamefont {M.~L.}\ \bibnamefont {Dahlberg}}, \bibinfo {author}
  {\bibfnamefont {J.~G.}\ \bibnamefont {Kim}}, \bibinfo {author} {\bibfnamefont
  {Y.}~\bibnamefont {Qiu}}, \bibinfo {author} {\bibfnamefont {J.~R.~D.}\
  \bibnamefont {Copley}}, \bibinfo {author} {\bibfnamefont {P.}~\bibnamefont
  {Schiffer}}, \ and\ \bibinfo {author} {\bibfnamefont {R.~J.}\ \bibnamefont
  {Cava}},\ }\href {\doibase 10.1103/PhysRevB.81.060408} {\bibfield  {journal}
  {\bibinfo  {journal} {Phys. Rev. B}\ }\textbf {\bibinfo {volume} {81}},\
  \bibinfo {pages} {060408} (\bibinfo {year} {2010})}\BibitemShut {NoStop}%
\bibitem [{\citenamefont {Conlon}\ and\ \citenamefont
  {Chalker}(2009)}]{PhysRevLett.102.237206}%
  \BibitemOpen
  \bibfield  {author} {\bibinfo {author} {\bibfnamefont {P.~H.}\ \bibnamefont
  {Conlon}}\ and\ \bibinfo {author} {\bibfnamefont {J.~T.}\ \bibnamefont
  {Chalker}},\ }\href {\doibase 10.1103/PhysRevLett.102.237206} {\bibfield
  {journal} {\bibinfo  {journal} {Phys. Rev. Lett.}\ }\textbf {\bibinfo
  {volume} {102}},\ \bibinfo {pages} {237206} (\bibinfo {year}
  {2009})}\BibitemShut {NoStop}%
\bibitem [{\citenamefont {Keren}\ \emph {et~al.}(2004)\citenamefont {Keren},
  \citenamefont {Gardner}, \citenamefont {Ehlers}, \citenamefont {Fukaya},
  \citenamefont {Segal},\ and\ \citenamefont {Uemura}}]{PhysRevLett.92.107204}%
  \BibitemOpen
  \bibfield  {author} {\bibinfo {author} {\bibfnamefont {A.}~\bibnamefont
  {Keren}}, \bibinfo {author} {\bibfnamefont {J.~S.}\ \bibnamefont {Gardner}},
  \bibinfo {author} {\bibfnamefont {G.}~\bibnamefont {Ehlers}}, \bibinfo
  {author} {\bibfnamefont {A.}~\bibnamefont {Fukaya}}, \bibinfo {author}
  {\bibfnamefont {E.}~\bibnamefont {Segal}}, \ and\ \bibinfo {author}
  {\bibfnamefont {Y.~J.}\ \bibnamefont {Uemura}},\ }\href {\doibase
  10.1103/PhysRevLett.92.107204} {\bibfield  {journal} {\bibinfo  {journal}
  {Phys. Rev. Lett.}\ }\textbf {\bibinfo {volume} {92}},\ \bibinfo {pages}
  {107204} (\bibinfo {year} {2004})}\BibitemShut {NoStop}%
\bibitem [{\citenamefont {Motome}\ and\ \citenamefont
  {Furukawa}(2010)}]{PhysRevLett.104.106407}%
  \BibitemOpen
  \bibfield  {author} {\bibinfo {author} {\bibfnamefont {Y.}~\bibnamefont
  {Motome}}\ and\ \bibinfo {author} {\bibfnamefont {N.}~\bibnamefont
  {Furukawa}},\ }\href {\doibase 10.1103/PhysRevLett.104.106407} {\bibfield
  {journal} {\bibinfo  {journal} {Phys. Rev. Lett.}\ }\textbf {\bibinfo
  {volume} {104}},\ \bibinfo {pages} {106407} (\bibinfo {year}
  {2010})}\BibitemShut {NoStop}%
\bibitem [{\citenamefont {Liu}\ \emph {et~al.}(2019{\natexlab{a}})\citenamefont
  {Liu}, \citenamefont {Hal\'asz},\ and\ \citenamefont
  {Balents}}]{PhysRevB.100.075125}%
  \BibitemOpen
  \bibfield  {author} {\bibinfo {author} {\bibfnamefont {C.}~\bibnamefont
  {Liu}}, \bibinfo {author} {\bibfnamefont {G.~B.}\ \bibnamefont {Hal\'asz}}, \
  and\ \bibinfo {author} {\bibfnamefont {L.}~\bibnamefont {Balents}},\ }\href
  {\doibase 10.1103/PhysRevB.100.075125} {\bibfield  {journal} {\bibinfo
  {journal} {Phys. Rev. B}\ }\textbf {\bibinfo {volume} {100}},\ \bibinfo
  {pages} {075125} (\bibinfo {year} {2019}{\natexlab{a}})}\BibitemShut
  {NoStop}%
\bibitem [{\citenamefont {Goldschmidt}(1926)}]{goldschmidt1926gesetze}%
  \BibitemOpen
  \bibfield  {author} {\bibinfo {author} {\bibfnamefont {V.~M.}\ \bibnamefont
  {Goldschmidt}},\ }\href {\doibase 10.1007/BF01507527} {\bibfield  {journal}
  {\bibinfo  {journal} {Sci. Nat.}\ }\textbf {\bibinfo {volume} {14}},\
  \bibinfo {pages} {477} (\bibinfo {year} {1926})}\BibitemShut {NoStop}%
\bibitem [{\citenamefont {Wiebe}\ and\ \citenamefont
  {Hallas}(2015)}]{Wiebe15:3}%
  \BibitemOpen
  \bibfield  {author} {\bibinfo {author} {\bibfnamefont {C.}~\bibnamefont
  {Wiebe}}\ and\ \bibinfo {author} {\bibfnamefont {A.}~\bibnamefont {Hallas}},\
  }\href {\doibase 10.1063/1.4916020} {\bibfield  {journal} {\bibinfo
  {journal} {APL materials}\ }\textbf {\bibinfo {volume} {3}},\ \bibinfo
  {pages} {041519} (\bibinfo {year} {2015})}\BibitemShut {NoStop}%
\bibitem [{\citenamefont {Zhou}\ and\ \citenamefont
  {Wiebe}(2019)}]{RR_zhou2019high}%
  \BibitemOpen
  \bibfield  {author} {\bibinfo {author} {\bibfnamefont {H.}~\bibnamefont
  {Zhou}}\ and\ \bibinfo {author} {\bibfnamefont {C.~R.}\ \bibnamefont
  {Wiebe}},\ }\href {\doibase 10.3390/inorganics7040049} {\bibfield  {journal}
  {\bibinfo  {journal} {Inorganics}\ }\textbf {\bibinfo {volume} {7}},\
  \bibinfo {pages} {49} (\bibinfo {year} {2019})}\BibitemShut {NoStop}%
\bibitem [{\citenamefont {Hallas}\ \emph {et~al.}(2014)\citenamefont {Hallas},
  \citenamefont {Cheng}, \citenamefont {Arevalo-Lopez}, \citenamefont
  {Silverstein}, \citenamefont {Su}, \citenamefont {Sarte}, \citenamefont
  {Zhou}, \citenamefont {Choi}, \citenamefont {Attfield}, \citenamefont
  {Luke},\ and\ \citenamefont {Wiebe}}]{Hallas14:113}%
  \BibitemOpen
  \bibfield  {author} {\bibinfo {author} {\bibfnamefont {A.~M.}\ \bibnamefont
  {Hallas}}, \bibinfo {author} {\bibfnamefont {J.~G.}\ \bibnamefont {Cheng}},
  \bibinfo {author} {\bibfnamefont {A.~M.}\ \bibnamefont {Arevalo-Lopez}},
  \bibinfo {author} {\bibfnamefont {H.~J.}\ \bibnamefont {Silverstein}},
  \bibinfo {author} {\bibfnamefont {Y.}~\bibnamefont {Su}}, \bibinfo {author}
  {\bibfnamefont {P.~M.}\ \bibnamefont {Sarte}}, \bibinfo {author}
  {\bibfnamefont {H.~D.}\ \bibnamefont {Zhou}}, \bibinfo {author}
  {\bibfnamefont {E.~S.}\ \bibnamefont {Choi}}, \bibinfo {author}
  {\bibfnamefont {J.~P.}\ \bibnamefont {Attfield}}, \bibinfo {author}
  {\bibfnamefont {G.~M.}\ \bibnamefont {Luke}}, \ and\ \bibinfo {author}
  {\bibfnamefont {C.~R.}\ \bibnamefont {Wiebe}},\ }\href {\doibase
  10.1103/PhysRevLett.113.267205} {\bibfield  {journal} {\bibinfo  {journal}
  {Phys. Rev. Lett.}\ }\textbf {\bibinfo {volume} {113}},\ \bibinfo {pages}
  {267205} (\bibinfo {year} {2014})}\BibitemShut {NoStop}%
\bibitem [{\citenamefont {Hallas}\ \emph {et~al.}(2012)\citenamefont {Hallas},
  \citenamefont {Paddison}, \citenamefont {Silverstein}, \citenamefont
  {Goodwin}, \citenamefont {Stewart}, \citenamefont {Wildes}, \citenamefont
  {Cheng}, \citenamefont {Zhou}, \citenamefont {Goodenough}, \citenamefont
  {Choi}, \citenamefont {Ehlers}, \citenamefont {Gardner}, \citenamefont
  {Wiebe},\ and\ \citenamefont {Zhou}}]{Hallas12:86}%
  \BibitemOpen
  \bibfield  {author} {\bibinfo {author} {\bibfnamefont {A.~M.}\ \bibnamefont
  {Hallas}}, \bibinfo {author} {\bibfnamefont {J.~A.~M.}\ \bibnamefont
  {Paddison}}, \bibinfo {author} {\bibfnamefont {H.~J.}\ \bibnamefont
  {Silverstein}}, \bibinfo {author} {\bibfnamefont {A.~L.}\ \bibnamefont
  {Goodwin}}, \bibinfo {author} {\bibfnamefont {J.~R.}\ \bibnamefont
  {Stewart}}, \bibinfo {author} {\bibfnamefont {A.~R.}\ \bibnamefont {Wildes}},
  \bibinfo {author} {\bibfnamefont {J.~G.}\ \bibnamefont {Cheng}}, \bibinfo
  {author} {\bibfnamefont {J.~S.}\ \bibnamefont {Zhou}}, \bibinfo {author}
  {\bibfnamefont {J.~B.}\ \bibnamefont {Goodenough}}, \bibinfo {author}
  {\bibfnamefont {E.~S.}\ \bibnamefont {Choi}}, \bibinfo {author}
  {\bibfnamefont {G.}~\bibnamefont {Ehlers}}, \bibinfo {author} {\bibfnamefont
  {J.~S.}\ \bibnamefont {Gardner}}, \bibinfo {author} {\bibfnamefont {C.~R.}\
  \bibnamefont {Wiebe}}, \ and\ \bibinfo {author} {\bibfnamefont {H.~D.}\
  \bibnamefont {Zhou}},\ }\href {\doibase 10.1103/PhysRevB.86.134431}
  {\bibfield  {journal} {\bibinfo  {journal} {Phys. Rev. B}\ }\textbf {\bibinfo
  {volume} {86}},\ \bibinfo {pages} {134431} (\bibinfo {year}
  {2012})}\BibitemShut {NoStop}%
\bibitem [{\citenamefont {Zhou}\ \emph {et~al.}(2012)\citenamefont {Zhou},
  \citenamefont {Cheng}, \citenamefont {Hallas}, \citenamefont {Wiebe},
  \citenamefont {Li}, \citenamefont {Balicas}, \citenamefont {Zhou},
  \citenamefont {Goodenough}, \citenamefont {Gardner},\ and\ \citenamefont
  {Choi}}]{PhysRevLett.108.207206}%
  \BibitemOpen
  \bibfield  {author} {\bibinfo {author} {\bibfnamefont {H.~D.}\ \bibnamefont
  {Zhou}}, \bibinfo {author} {\bibfnamefont {J.~G.}\ \bibnamefont {Cheng}},
  \bibinfo {author} {\bibfnamefont {A.~M.}\ \bibnamefont {Hallas}}, \bibinfo
  {author} {\bibfnamefont {C.~R.}\ \bibnamefont {Wiebe}}, \bibinfo {author}
  {\bibfnamefont {G.}~\bibnamefont {Li}}, \bibinfo {author} {\bibfnamefont
  {L.}~\bibnamefont {Balicas}}, \bibinfo {author} {\bibfnamefont {J.~S.}\
  \bibnamefont {Zhou}}, \bibinfo {author} {\bibfnamefont {J.~B.}\ \bibnamefont
  {Goodenough}}, \bibinfo {author} {\bibfnamefont {J.~S.}\ \bibnamefont
  {Gardner}}, \ and\ \bibinfo {author} {\bibfnamefont {E.~S.}\ \bibnamefont
  {Choi}},\ }\href {\doibase 10.1103/PhysRevLett.108.207206} {\bibfield
  {journal} {\bibinfo  {journal} {Phys. Rev. Lett.}\ }\textbf {\bibinfo
  {volume} {108}},\ \bibinfo {pages} {207206} (\bibinfo {year}
  {2012})}\BibitemShut {NoStop}%
\bibitem [{\citenamefont {Li}\ \emph {et~al.}(2014)\citenamefont {Li},
  \citenamefont {Li}, \citenamefont {Matsubayashi}, \citenamefont {Sato},
  \citenamefont {Jin}, \citenamefont {Uwatoko}, \citenamefont {Kawae},
  \citenamefont {Hallas}, \citenamefont {Wiebe}, \citenamefont {Arevalo-Lopez},
  \citenamefont {Attfield}, \citenamefont {Gardner}, \citenamefont {Freitas},
  \citenamefont {Zhou},\ and\ \citenamefont {Cheng}}]{li2014long}%
  \BibitemOpen
  \bibfield  {author} {\bibinfo {author} {\bibfnamefont {X.}~\bibnamefont
  {Li}}, \bibinfo {author} {\bibfnamefont {W.~M.}\ \bibnamefont {Li}}, \bibinfo
  {author} {\bibfnamefont {K.}~\bibnamefont {Matsubayashi}}, \bibinfo {author}
  {\bibfnamefont {Y.}~\bibnamefont {Sato}}, \bibinfo {author} {\bibfnamefont
  {C.~Q.}\ \bibnamefont {Jin}}, \bibinfo {author} {\bibfnamefont
  {Y.}~\bibnamefont {Uwatoko}}, \bibinfo {author} {\bibfnamefont
  {T.}~\bibnamefont {Kawae}}, \bibinfo {author} {\bibfnamefont {A.~M.}\
  \bibnamefont {Hallas}}, \bibinfo {author} {\bibfnamefont {C.~R.}\
  \bibnamefont {Wiebe}}, \bibinfo {author} {\bibfnamefont {A.~M.}\ \bibnamefont
  {Arevalo-Lopez}}, \bibinfo {author} {\bibfnamefont {J.~P.}\ \bibnamefont
  {Attfield}}, \bibinfo {author} {\bibfnamefont {J.~S.}\ \bibnamefont
  {Gardner}}, \bibinfo {author} {\bibfnamefont {R.~S.}\ \bibnamefont
  {Freitas}}, \bibinfo {author} {\bibfnamefont {H.~D.}\ \bibnamefont {Zhou}}, \
  and\ \bibinfo {author} {\bibfnamefont {J.-G.}\ \bibnamefont {Cheng}},\ }\href
  {\doibase 10.1103/PhysRevB.89.064409} {\bibfield  {journal} {\bibinfo
  {journal} {Phys. Rev. B}\ }\textbf {\bibinfo {volume} {89}},\ \bibinfo
  {pages} {064409} (\bibinfo {year} {2014})}\BibitemShut {NoStop}%
\bibitem [{\citenamefont {Hallas}\ \emph
  {et~al.}(2016{\natexlab{a}})\citenamefont {Hallas}, \citenamefont {Gaudet},
  \citenamefont {Wilson}, \citenamefont {Munsie}, \citenamefont {Aczel},
  \citenamefont {Stone}, \citenamefont {Freitas}, \citenamefont
  {Arevalo-Lopez}, \citenamefont {Attfield}, \citenamefont {Tachibana},
  \citenamefont {Wiebe}, \citenamefont {Luke},\ and\ \citenamefont
  {Gaulin}}]{Hallas16:93}%
  \BibitemOpen
  \bibfield  {author} {\bibinfo {author} {\bibfnamefont {A.~M.}\ \bibnamefont
  {Hallas}}, \bibinfo {author} {\bibfnamefont {J.}~\bibnamefont {Gaudet}},
  \bibinfo {author} {\bibfnamefont {M.~N.}\ \bibnamefont {Wilson}}, \bibinfo
  {author} {\bibfnamefont {T.~J.}\ \bibnamefont {Munsie}}, \bibinfo {author}
  {\bibfnamefont {A.~A.}\ \bibnamefont {Aczel}}, \bibinfo {author}
  {\bibfnamefont {M.~B.}\ \bibnamefont {Stone}}, \bibinfo {author}
  {\bibfnamefont {R.~S.}\ \bibnamefont {Freitas}}, \bibinfo {author}
  {\bibfnamefont {A.~M.}\ \bibnamefont {Arevalo-Lopez}}, \bibinfo {author}
  {\bibfnamefont {J.~P.}\ \bibnamefont {Attfield}}, \bibinfo {author}
  {\bibfnamefont {M.}~\bibnamefont {Tachibana}}, \bibinfo {author}
  {\bibfnamefont {C.~R.}\ \bibnamefont {Wiebe}}, \bibinfo {author}
  {\bibfnamefont {G.~M.}\ \bibnamefont {Luke}}, \ and\ \bibinfo {author}
  {\bibfnamefont {B.~D.}\ \bibnamefont {Gaulin}},\ }\href {\doibase
  10.1103/PhysRevB.93.104405} {\bibfield  {journal} {\bibinfo  {journal} {Phys.
  Rev. B}\ }\textbf {\bibinfo {volume} {93}},\ \bibinfo {pages} {104405}
  (\bibinfo {year} {2016}{\natexlab{a}})}\BibitemShut {NoStop}%
\bibitem [{\citenamefont {Hallas}\ \emph {et~al.}(2015)\citenamefont {Hallas},
  \citenamefont {Arevalo-Lopez}, \citenamefont {Sharma}, \citenamefont
  {Munsie}, \citenamefont {Attfield}, \citenamefont {Wiebe},\ and\
  \citenamefont {Luke}}]{Hallas15:91}%
  \BibitemOpen
  \bibfield  {author} {\bibinfo {author} {\bibfnamefont {A.~M.}\ \bibnamefont
  {Hallas}}, \bibinfo {author} {\bibfnamefont {A.~M.}\ \bibnamefont
  {Arevalo-Lopez}}, \bibinfo {author} {\bibfnamefont {A.~Z.}\ \bibnamefont
  {Sharma}}, \bibinfo {author} {\bibfnamefont {T.}~\bibnamefont {Munsie}},
  \bibinfo {author} {\bibfnamefont {J.~P.}\ \bibnamefont {Attfield}}, \bibinfo
  {author} {\bibfnamefont {C.~R.}\ \bibnamefont {Wiebe}}, \ and\ \bibinfo
  {author} {\bibfnamefont {G.~M.}\ \bibnamefont {Luke}},\ }\href {\doibase
  10.1103/PhysRevB.91.104417} {\bibfield  {journal} {\bibinfo  {journal} {Phys.
  Rev. B}\ }\textbf {\bibinfo {volume} {91}},\ \bibinfo {pages} {104417}
  (\bibinfo {year} {2015})}\BibitemShut {NoStop}%
\bibitem [{\citenamefont {Bl{\"o}te}\ \emph {et~al.}(1969)\citenamefont
  {Bl{\"o}te}, \citenamefont {Wielinga},\ and\ \citenamefont
  {Huiskamp}}]{blote1969heat}%
  \BibitemOpen
  \bibfield  {author} {\bibinfo {author} {\bibfnamefont {H.~W.~J.}\
  \bibnamefont {Bl{\"o}te}}, \bibinfo {author} {\bibfnamefont {R.~F.}\
  \bibnamefont {Wielinga}}, \ and\ \bibinfo {author} {\bibfnamefont {W.~J.}\
  \bibnamefont {Huiskamp}},\ }\href {\doibase
  https://doi.org/10.1016/0031-8914(69)90187-6} {\bibfield  {journal} {\bibinfo
   {journal} {Physica}\ }\textbf {\bibinfo {volume} {43}},\ \bibinfo {pages}
  {549} (\bibinfo {year} {1969})}\BibitemShut {NoStop}%
\bibitem [{\citenamefont {Martin}\ \emph {et~al.}(2017)\citenamefont {Martin},
  \citenamefont {Bonville}, \citenamefont {Lhotel}, \citenamefont {Guitteny},
  \citenamefont {Wildes}, \citenamefont {Decorse}, \citenamefont
  {Ciomaga~Hatnean}, \citenamefont {Balakrishnan}, \citenamefont {Mirebeau},\
  and\ \citenamefont {Petit}}]{martin17:7}%
  \BibitemOpen
  \bibfield  {author} {\bibinfo {author} {\bibfnamefont {N.}~\bibnamefont
  {Martin}}, \bibinfo {author} {\bibfnamefont {P.}~\bibnamefont {Bonville}},
  \bibinfo {author} {\bibfnamefont {E.}~\bibnamefont {Lhotel}}, \bibinfo
  {author} {\bibfnamefont {S.}~\bibnamefont {Guitteny}}, \bibinfo {author}
  {\bibfnamefont {A.}~\bibnamefont {Wildes}}, \bibinfo {author} {\bibfnamefont
  {C.}~\bibnamefont {Decorse}}, \bibinfo {author} {\bibfnamefont
  {M.}~\bibnamefont {Ciomaga~Hatnean}}, \bibinfo {author} {\bibfnamefont
  {G.}~\bibnamefont {Balakrishnan}}, \bibinfo {author} {\bibfnamefont
  {I.}~\bibnamefont {Mirebeau}}, \ and\ \bibinfo {author} {\bibfnamefont
  {S.}~\bibnamefont {Petit}},\ }\href {\doibase 10.1103/PhysRevX.7.041028}
  {\bibfield  {journal} {\bibinfo  {journal} {Phys. Rev. X}\ }\textbf {\bibinfo
  {volume} {7}},\ \bibinfo {pages} {041028} (\bibinfo {year}
  {2017})}\BibitemShut {NoStop}%
\bibitem [{\citenamefont {Gomez}\ \emph {et~al.}(2021)\citenamefont {Gomez},
  \citenamefont {Sarte}, \citenamefont {Zelensky}, \citenamefont {Hallas},
  \citenamefont {Gonzalez}, \citenamefont {Hong}, \citenamefont {Pace},
  \citenamefont {Calder}, \citenamefont {Stone}, \citenamefont {Su},
  \citenamefont {Feng}, \citenamefont {Le}, \citenamefont {Stock},
  \citenamefont {Attfield}, \citenamefont {Wilson}, \citenamefont {Wiebe},\
  and\ \citenamefont {Aczel}}]{gomez2021absence}%
  \BibitemOpen
  \bibfield  {author} {\bibinfo {author} {\bibfnamefont {S.~J.}\ \bibnamefont
  {Gomez}}, \bibinfo {author} {\bibfnamefont {P.~M.}\ \bibnamefont {Sarte}},
  \bibinfo {author} {\bibfnamefont {M.}~\bibnamefont {Zelensky}}, \bibinfo
  {author} {\bibfnamefont {A.~M.}\ \bibnamefont {Hallas}}, \bibinfo {author}
  {\bibfnamefont {B.~A.}\ \bibnamefont {Gonzalez}}, \bibinfo {author}
  {\bibfnamefont {K.~H.}\ \bibnamefont {Hong}}, \bibinfo {author}
  {\bibfnamefont {E.~J.}\ \bibnamefont {Pace}}, \bibinfo {author}
  {\bibfnamefont {S.}~\bibnamefont {Calder}}, \bibinfo {author} {\bibfnamefont
  {M.~B.}\ \bibnamefont {Stone}}, \bibinfo {author} {\bibfnamefont
  {Y.}~\bibnamefont {Su}}, \bibinfo {author} {\bibfnamefont {E.}~\bibnamefont
  {Feng}}, \bibinfo {author} {\bibfnamefont {M.~D.}\ \bibnamefont {Le}},
  \bibinfo {author} {\bibfnamefont {C.}~\bibnamefont {Stock}}, \bibinfo
  {author} {\bibfnamefont {J.~P.}\ \bibnamefont {Attfield}}, \bibinfo {author}
  {\bibfnamefont {S.~D.}\ \bibnamefont {Wilson}}, \bibinfo {author}
  {\bibfnamefont {C.~R.}\ \bibnamefont {Wiebe}}, \ and\ \bibinfo {author}
  {\bibfnamefont {A.~A.}\ \bibnamefont {Aczel}},\ }\href {\doibase
  10.1103/PhysRevB.103.214419} {\bibfield  {journal} {\bibinfo  {journal}
  {Phys. Rev. B}\ }\textbf {\bibinfo {volume} {103}},\ \bibinfo {pages}
  {214419} (\bibinfo {year} {2021})}\BibitemShut {NoStop}%
\bibitem [{\citenamefont {Sarte}\ \emph {et~al.}(2021)\citenamefont {Sarte},
  \citenamefont {Cruz-Kan}, \citenamefont {Ortiz}, \citenamefont {Hong},
  \citenamefont {Bordelon}, \citenamefont {Reig-i Plessis}, \citenamefont
  {Lee}, \citenamefont {Choi}, \citenamefont {Stone}, \citenamefont {Calder},
  \citenamefont {Pajerowski}, \citenamefont {Mangin-Thro}, \citenamefont {Qiu},
  \citenamefont {Attfield}, \citenamefont {Wilson}, \citenamefont {Stock},
  \citenamefont {Zhou}, \citenamefont {Hallas}, \citenamefont {Paddison},
  \citenamefont {Aczel},\ and\ \citenamefont
  {Wiebe}}]{YbGaSb_sarte2021dynamical}%
  \BibitemOpen
  \bibfield  {author} {\bibinfo {author} {\bibfnamefont {P.~M.}\ \bibnamefont
  {Sarte}}, \bibinfo {author} {\bibfnamefont {K.}~\bibnamefont {Cruz-Kan}},
  \bibinfo {author} {\bibfnamefont {B.~R.}\ \bibnamefont {Ortiz}}, \bibinfo
  {author} {\bibfnamefont {K.~H.}\ \bibnamefont {Hong}}, \bibinfo {author}
  {\bibfnamefont {M.~M.}\ \bibnamefont {Bordelon}}, \bibinfo {author}
  {\bibfnamefont {D.}~\bibnamefont {Reig-i Plessis}}, \bibinfo {author}
  {\bibfnamefont {M.}~\bibnamefont {Lee}}, \bibinfo {author} {\bibfnamefont
  {E.~S.}\ \bibnamefont {Choi}}, \bibinfo {author} {\bibfnamefont {M.~B.}\
  \bibnamefont {Stone}}, \bibinfo {author} {\bibfnamefont {S.}~\bibnamefont
  {Calder}}, \bibinfo {author} {\bibfnamefont {D.~M.}\ \bibnamefont
  {Pajerowski}}, \bibinfo {author} {\bibfnamefont {L.}~\bibnamefont
  {Mangin-Thro}}, \bibinfo {author} {\bibfnamefont {Y.}~\bibnamefont {Qiu}},
  \bibinfo {author} {\bibfnamefont {J.~P.}\ \bibnamefont {Attfield}}, \bibinfo
  {author} {\bibfnamefont {S.~D.}\ \bibnamefont {Wilson}}, \bibinfo {author}
  {\bibfnamefont {C.}~\bibnamefont {Stock}}, \bibinfo {author} {\bibfnamefont
  {H.~D.}\ \bibnamefont {Zhou}}, \bibinfo {author} {\bibfnamefont {A.~M.}\
  \bibnamefont {Hallas}}, \bibinfo {author} {\bibfnamefont {J.~A.~M.}\
  \bibnamefont {Paddison}}, \bibinfo {author} {\bibfnamefont {A.~A.}\
  \bibnamefont {Aczel}}, \ and\ \bibinfo {author} {\bibfnamefont {C.~R.}\
  \bibnamefont {Wiebe}},\ }\href {\doibase
  https://doi.org/10.1038/s41535-021-00343-4} {\bibfield  {journal} {\bibinfo
  {journal} {npj Quantum Mater.}\ }\textbf {\bibinfo {volume} {6}},\ \bibinfo
  {pages} {1} (\bibinfo {year} {2021})}\BibitemShut {NoStop}%
\bibitem [{\citenamefont {Nandi}\ \emph {et~al.}(2021)\citenamefont {Nandi},
  \citenamefont {Jana}, \citenamefont {Alam}, \citenamefont {Bag},
  \citenamefont {Islam},\ and\ \citenamefont {Nath}}]{dy2gasbo7}%
  \BibitemOpen
  \bibfield  {author} {\bibinfo {author} {\bibfnamefont {S.}~\bibnamefont
  {Nandi}}, \bibinfo {author} {\bibfnamefont {Y.}~\bibnamefont {Jana}},
  \bibinfo {author} {\bibfnamefont {J.}~\bibnamefont {Alam}}, \bibinfo {author}
  {\bibfnamefont {P.}~\bibnamefont {Bag}}, \bibinfo {author} {\bibfnamefont
  {S.~S.}\ \bibnamefont {Islam}}, \ and\ \bibinfo {author} {\bibfnamefont
  {R.}~\bibnamefont {Nath}},\ }\href@noop {} {} (\bibinfo {year} {2021}),\
  \Eprint {http://arxiv.org/abs/arXiv:2104.07890v1} {arXiv:2104.07890v1}
  \BibitemShut {NoStop}%
\bibitem [{\citenamefont {Rutherford}(2021)}]{megan_thesis}%
  \BibitemOpen
  \bibfield  {author} {\bibinfo {author} {\bibfnamefont {M.~R.}\ \bibnamefont
  {Rutherford}},\ }\emph {\bibinfo {title} {{Dy$_{2}$ScNbO$_{7}$}: a study of
  the effect of a disordered B-site on the spin ice magnetism typically seen in
  dysprosium pyrochlores}},\ \href@noop {} {Master's thesis},\ \bibinfo
  {school} {McMaster University}, \bibinfo {address} {Hamilton, ON} (\bibinfo
  {year} {2021})\BibitemShut {NoStop}%
\bibitem [{\citenamefont {Beare}(2021)}]{james_thesis}%
  \BibitemOpen
  \bibfield  {author} {\bibinfo {author} {\bibfnamefont {J.}~\bibnamefont
  {Beare}},\ }\emph {\bibinfo {title} {Muon Spin Rotation, Relaxation, and
  Resonance, and AC Susceptibility as a probe of Frustrated Pyrochlore Magnets
  and Type-I Superconductivity}},\ \href@noop {} {Ph.D. thesis},\ \bibinfo
  {school} {McMaster University}, \bibinfo {address} {Hamilton, ON} (\bibinfo
  {year} {2021})\BibitemShut {NoStop}%
\bibitem [{\citenamefont {Mauws}\ \emph {et~al.}(2021)\citenamefont {Mauws},
  \citenamefont {Hiebert}, \citenamefont {Rutherford}, \citenamefont {Zhou},
  \citenamefont {Huang}, \citenamefont {Stone}, \citenamefont {Butch},
  \citenamefont {Su}, \citenamefont {Choi}, \citenamefont {Yamani},\ and\
  \citenamefont {Wiebe}}]{Mauws21:33}%
  \BibitemOpen
  \bibfield  {author} {\bibinfo {author} {\bibfnamefont {C.}~\bibnamefont
  {Mauws}}, \bibinfo {author} {\bibfnamefont {N.}~\bibnamefont {Hiebert}},
  \bibinfo {author} {\bibfnamefont {M.~L.}\ \bibnamefont {Rutherford}},
  \bibinfo {author} {\bibfnamefont {H.~D.}\ \bibnamefont {Zhou}}, \bibinfo
  {author} {\bibfnamefont {Q.}~\bibnamefont {Huang}}, \bibinfo {author}
  {\bibfnamefont {M.~B.}\ \bibnamefont {Stone}}, \bibinfo {author}
  {\bibfnamefont {N.~P.}\ \bibnamefont {Butch}}, \bibinfo {author}
  {\bibfnamefont {Y.}~\bibnamefont {Su}}, \bibinfo {author} {\bibfnamefont
  {E.~S.}\ \bibnamefont {Choi}}, \bibinfo {author} {\bibfnamefont
  {Z.}~\bibnamefont {Yamani}}, \ and\ \bibinfo {author} {\bibfnamefont {C.~R.}\
  \bibnamefont {Wiebe}},\ }\href {\doibase 10.1088/1361-648x/abf594} {\bibfield
   {journal} {\bibinfo  {journal} {J. Phys.: Condens. Matter}\ }\textbf
  {\bibinfo {volume} {33}},\ \bibinfo {pages} {245802} (\bibinfo {year}
  {2021})}\BibitemShut {NoStop}%
\bibitem [{\citenamefont {Strobel}\ \emph {et~al.}(2010)\citenamefont
  {Strobel}, \citenamefont {Zouari}, \citenamefont {Ballou}, \citenamefont
  {Cheikh-Rouhou}, \citenamefont {Jumas},\ and\ \citenamefont
  {Olivier-Fourcade}}]{strobel2010structural}%
  \BibitemOpen
  \bibfield  {author} {\bibinfo {author} {\bibfnamefont {P.}~\bibnamefont
  {Strobel}}, \bibinfo {author} {\bibfnamefont {S.}~\bibnamefont {Zouari}},
  \bibinfo {author} {\bibfnamefont {R.}~\bibnamefont {Ballou}}, \bibinfo
  {author} {\bibfnamefont {A.}~\bibnamefont {Cheikh-Rouhou}}, \bibinfo {author}
  {\bibfnamefont {J.-C.}\ \bibnamefont {Jumas}}, \ and\ \bibinfo {author}
  {\bibfnamefont {J.}~\bibnamefont {Olivier-Fourcade}},\ }\href {\doibase
  https://doi.org/10.1016/j.solidstatesciences.2010.01.007} {\bibfield
  {journal} {\bibinfo  {journal} {Solid State Sci.}\ }\textbf {\bibinfo
  {volume} {12}},\ \bibinfo {pages} {570} (\bibinfo {year} {2010})}\BibitemShut
  {NoStop}%
\bibitem [{\citenamefont {Coelho}(2018)}]{Coelho}%
  \BibitemOpen
  \bibfield  {author} {\bibinfo {author} {\bibfnamefont {A.~A.}\ \bibnamefont
  {Coelho}},\ }\href {\doibase 10.1107/S1600576718000183} {\bibfield  {journal}
  {\bibinfo  {journal} {J. Appl. Crystallogr.}\ }\textbf {\bibinfo {volume}
  {51}},\ \bibinfo {pages} {210} (\bibinfo {year} {2018})}\BibitemShut
  {NoStop}%
\bibitem [{\citenamefont {Momma}\ and\ \citenamefont
  {Izumi}(2011)}]{Momma2011}%
  \BibitemOpen
  \bibfield  {author} {\bibinfo {author} {\bibfnamefont {K.}~\bibnamefont
  {Momma}}\ and\ \bibinfo {author} {\bibfnamefont {F.}~\bibnamefont {Izumi}},\
  }\href {\doibase 10.1107/S0021889811038970} {\bibfield  {journal} {\bibinfo
  {journal} {J. Appl. Crystallogr.}\ }\textbf {\bibinfo {volume} {44}},\
  \bibinfo {pages} {1272} (\bibinfo {year} {2011})}\BibitemShut {NoStop}%
\bibitem [{\citenamefont {Talanov}\ and\ \citenamefont
  {Talanov}(2021)}]{talanov2021structural}%
  \BibitemOpen
  \bibfield  {author} {\bibinfo {author} {\bibfnamefont {M.~V.}\ \bibnamefont
  {Talanov}}\ and\ \bibinfo {author} {\bibfnamefont {V.~M.}\ \bibnamefont
  {Talanov}},\ }\href {\doibase https://doi.org/10.1021/acs.chemmater.0c04864}
  {\bibfield  {journal} {\bibinfo  {journal} {Chem. Mater.}\ }\textbf {\bibinfo
  {volume} {33}},\ \bibinfo {pages} {2706} (\bibinfo {year}
  {2021})}\BibitemShut {NoStop}%
\bibitem [{\citenamefont {Trump}\ \emph {et~al.}(2018)\citenamefont {Trump},
  \citenamefont {Koohpayeh}, \citenamefont {Livi}, \citenamefont {Wen},
  \citenamefont {Arpino}, \citenamefont {Ramasse}, \citenamefont {Brydson},
  \citenamefont {Feygenson}, \citenamefont {Takeda}, \citenamefont {Takigawa},
  \citenamefont {Kimura}, \citenamefont {Nakatsuji}, \citenamefont {Broholm},\
  and\ \citenamefont {McQueen}}]{trump18:9}%
  \BibitemOpen
  \bibfield  {author} {\bibinfo {author} {\bibfnamefont {B.~A.}\ \bibnamefont
  {Trump}}, \bibinfo {author} {\bibfnamefont {S.~M.}\ \bibnamefont
  {Koohpayeh}}, \bibinfo {author} {\bibfnamefont {K.~J.}\ \bibnamefont {Livi}},
  \bibinfo {author} {\bibfnamefont {J.-J.}\ \bibnamefont {Wen}}, \bibinfo
  {author} {\bibfnamefont {K.}~\bibnamefont {Arpino}}, \bibinfo {author}
  {\bibfnamefont {Q.~M.}\ \bibnamefont {Ramasse}}, \bibinfo {author}
  {\bibfnamefont {R.}~\bibnamefont {Brydson}}, \bibinfo {author} {\bibfnamefont
  {M.}~\bibnamefont {Feygenson}}, \bibinfo {author} {\bibfnamefont
  {H.}~\bibnamefont {Takeda}}, \bibinfo {author} {\bibfnamefont
  {M.}~\bibnamefont {Takigawa}}, \bibinfo {author} {\bibfnamefont
  {K.}~\bibnamefont {Kimura}}, \bibinfo {author} {\bibfnamefont
  {S.}~\bibnamefont {Nakatsuji}}, \bibinfo {author} {\bibfnamefont {C.~L.}\
  \bibnamefont {Broholm}}, \ and\ \bibinfo {author} {\bibfnamefont {T.~M.}\
  \bibnamefont {McQueen}},\ }\href {\doibase
  https://doi.org/10.1038/s41467-018-05033-7} {\bibfield  {journal} {\bibinfo
  {journal} {Nat. Commun.}\ }\textbf {\bibinfo {volume} {9}},\ \bibinfo {pages}
  {2619} (\bibinfo {year} {2018})}\BibitemShut {NoStop}%
\bibitem [{\citenamefont {Sleight}(1969)}]{sleight69:8}%
  \BibitemOpen
  \bibfield  {author} {\bibinfo {author} {\bibfnamefont {A.~W.}\ \bibnamefont
  {Sleight}},\ }\href {\doibase https://doi.org/10.1021/ic50078a060} {\bibfield
   {journal} {\bibinfo  {journal} {Inorg. Chem.}\ }\textbf {\bibinfo {volume}
  {8}},\ \bibinfo {pages} {1807} (\bibinfo {year} {1969})}\BibitemShut
  {NoStop}%
\bibitem [{\citenamefont {Klee}\ and\ \citenamefont {Weitz}(1969)}]{KLEE69:31}%
  \BibitemOpen
  \bibfield  {author} {\bibinfo {author} {\bibfnamefont {W.~E.}\ \bibnamefont
  {Klee}}\ and\ \bibinfo {author} {\bibfnamefont {G.}~\bibnamefont {Weitz}},\
  }\href {\doibase https://doi.org/10.1016/0022-1902(69)80566-X} {\bibfield
  {journal} {\bibinfo  {journal} {J. Inorg. Nucl.}\ }\textbf {\bibinfo {volume}
  {31}},\ \bibinfo {pages} {2367} (\bibinfo {year} {1969})}\BibitemShut
  {NoStop}%
\bibitem [{\citenamefont {Michel}\ \emph {et~al.}(1974)\citenamefont {Michel},
  \citenamefont {{Perez y Jorba}},\ and\ \citenamefont
  {Collongues}}]{MICHEL74:9}%
  \BibitemOpen
  \bibfield  {author} {\bibinfo {author} {\bibfnamefont {D.}~\bibnamefont
  {Michel}}, \bibinfo {author} {\bibfnamefont {M.}~\bibnamefont {{Perez y
  Jorba}}}, \ and\ \bibinfo {author} {\bibfnamefont {R.}~\bibnamefont
  {Collongues}},\ }\href {\doibase
  https://doi.org/10.1016/0025-5408(74)90092-0} {\bibfield  {journal} {\bibinfo
   {journal} {Mater. Res. Bull.}\ }\textbf {\bibinfo {volume} {9}},\ \bibinfo
  {pages} {1457} (\bibinfo {year} {1974})}\BibitemShut {NoStop}%
\bibitem [{\citenamefont {{Perez y Jorba}}(1962)}]{perez62:7}%
  \BibitemOpen
  \bibfield  {author} {\bibinfo {author} {\bibfnamefont {M.}~\bibnamefont
  {{Perez y Jorba}}},\ }\href@noop {} {\bibfield  {journal} {\bibinfo
  {journal} {Ann. Chim.}\ }\textbf {\bibinfo {volume} {7}},\ \bibinfo {pages}
  {479} (\bibinfo {year} {1962})}\BibitemShut {NoStop}%
\bibitem [{\citenamefont {Collongues}\ \emph {et~al.}(1961)\citenamefont
  {Collongues}, \citenamefont {Perez Y~Jorba},\ and\ \citenamefont
  {Lefevre}}]{Collongues61:70}%
  \BibitemOpen
  \bibfield  {author} {\bibinfo {author} {\bibfnamefont {R.}~\bibnamefont
  {Collongues}}, \bibinfo {author} {\bibfnamefont {M.}~\bibnamefont {Perez
  Y~Jorba}}, \ and\ \bibinfo {author} {\bibfnamefont {J.}~\bibnamefont
  {Lefevre}},\ }\href@noop {} {\bibfield  {journal} {\bibinfo  {journal} {Bull.
  Soc. Chim. France}\ }\textbf {\bibinfo {volume} {28}},\ \bibinfo {pages} {70}
  (\bibinfo {year} {1961})}\BibitemShut {NoStop}%
\bibitem [{\citenamefont {Roth}(1956)}]{Roth56:56}%
  \BibitemOpen
  \bibfield  {author} {\bibinfo {author} {\bibfnamefont {R.~S.}\ \bibnamefont
  {Roth}},\ }\href@noop {} {\bibfield  {journal} {\bibinfo  {journal} {J. Rs.
  Natl. Bur. Stds. (USA)}\ }\textbf {\bibinfo {volume} {56}},\ \bibinfo {pages}
  {17} (\bibinfo {year} {1956})}\BibitemShut {NoStop}%
\bibitem [{\citenamefont {Fomina}\ and\ \citenamefont
  {Pal'guev}(1977)}]{SmZr_fomina1977formation}%
  \BibitemOpen
  \bibfield  {author} {\bibinfo {author} {\bibfnamefont {L.~N.}\ \bibnamefont
  {Fomina}}\ and\ \bibinfo {author} {\bibfnamefont {S.~F.}\ \bibnamefont
  {Pal'guev}},\ }\href@noop {} {\bibfield  {journal} {\bibinfo  {journal}
  {Russ. J. Phys. Chem.}\ }\textbf {\bibinfo {volume} {22}},\ \bibinfo {pages}
  {326} (\bibinfo {year} {1977})}\BibitemShut {NoStop}%
\bibitem [{\citenamefont {Gundovin}\ \emph {et~al.}(1975)\citenamefont
  {Gundovin}, \citenamefont {Spiridonov}, \citenamefont {Komissarova},\ and\
  \citenamefont {Petrov}}]{Gundovin75:20}%
  \BibitemOpen
  \bibfield  {author} {\bibinfo {author} {\bibfnamefont {N.}~\bibnamefont
  {Gundovin}}, \bibinfo {author} {\bibfnamefont {F.~M.}\ \bibnamefont
  {Spiridonov}}, \bibinfo {author} {\bibfnamefont {L.~N.}\ \bibnamefont
  {Komissarova}}, \ and\ \bibinfo {author} {\bibfnamefont {K.~I.}\ \bibnamefont
  {Petrov}},\ }\href@noop {} {\bibfield  {journal} {\bibinfo  {journal} {Russ.
  J. Inorg. Chem.}\ }\textbf {\bibinfo {volume} {20}},\ \bibinfo {pages} {325}
  (\bibinfo {year} {1975})}\BibitemShut {NoStop}%
\bibitem [{\citenamefont {Besson}\ \emph {et~al.}(1966)\citenamefont {Besson},
  \citenamefont {Deportes},\ and\ \citenamefont {Robert}}]{Besson66:262C}%
  \BibitemOpen
  \bibfield  {author} {\bibinfo {author} {\bibfnamefont {J.}~\bibnamefont
  {Besson}}, \bibinfo {author} {\bibfnamefont {C.}~\bibnamefont {Deportes}}, \
  and\ \bibinfo {author} {\bibfnamefont {G.}~\bibnamefont {Robert}},\
  }\href@noop {} {\bibfield  {journal} {\bibinfo  {journal} {Compt. Rend.
  (Paris)}\ }\textbf {\bibinfo {volume} {262C}},\ \bibinfo {pages} {527}
  (\bibinfo {year} {1966})}\BibitemShut {NoStop}%
\bibitem [{\citenamefont {Spiridinov}\ \emph {et~al.}(1968)\citenamefont
  {Spiridinov}, \citenamefont {Stepanov}, \citenamefont {Komissarova},\ and\
  \citenamefont {Spitsyn}}]{SPIRIDINOV68:14}%
  \BibitemOpen
  \bibfield  {author} {\bibinfo {author} {\bibfnamefont {F.~M.}\ \bibnamefont
  {Spiridinov}}, \bibinfo {author} {\bibfnamefont {V.}~\bibnamefont
  {Stepanov}}, \bibinfo {author} {\bibfnamefont {L.~N.}\ \bibnamefont
  {Komissarova}}, \ and\ \bibinfo {author} {\bibfnamefont {V.~I.}\ \bibnamefont
  {Spitsyn}},\ }\href {\doibase https://doi.org/10.1016/0022-5088(68)90167-7}
  {\bibfield  {journal} {\bibinfo  {journal} {J. Less. Comm. Metals}\ }\textbf
  {\bibinfo {volume} {14}},\ \bibinfo {pages} {435} (\bibinfo {year}
  {1968})}\BibitemShut {NoStop}%
\bibitem [{\citenamefont {Sibille}\ \emph
  {et~al.}(2016{\natexlab{a}})\citenamefont {Sibille}, \citenamefont {Fennell},
  \citenamefont {Hatnean}, \citenamefont {Lhotel}, \citenamefont {Keen},
  \citenamefont {Balakrishnan},\ and\ \citenamefont
  {Kenzelmann}}]{TbHf_sibille2016ms21}%
  \BibitemOpen
  \bibfield  {author} {\bibinfo {author} {\bibfnamefont {R.}~\bibnamefont
  {Sibille}}, \bibinfo {author} {\bibfnamefont {T.}~\bibnamefont {Fennell}},
  \bibinfo {author} {\bibfnamefont {M.~C.}\ \bibnamefont {Hatnean}}, \bibinfo
  {author} {\bibfnamefont {E.}~\bibnamefont {Lhotel}}, \bibinfo {author}
  {\bibfnamefont {D.}~\bibnamefont {Keen}}, \bibinfo {author} {\bibfnamefont
  {G.}~\bibnamefont {Balakrishnan}}, \ and\ \bibinfo {author} {\bibfnamefont
  {M.}~\bibnamefont {Kenzelmann}},\ }\href {\doibase 10.1107/S2053273316095401}
  {\bibfield  {journal} {\bibinfo  {journal} {Acta. Cryst. A}\ }\textbf
  {\bibinfo {volume} {72}},\ \bibinfo {pages} {s306} (\bibinfo {year}
  {2016}{\natexlab{a}})}\BibitemShut {NoStop}%
\bibitem [{\citenamefont {Shlyakhtina}\ \emph {et~al.}(2006)\citenamefont
  {Shlyakhtina}, \citenamefont {Boguslavskii}, \citenamefont {Stefanovich},
  \citenamefont {Kolbanev}, \citenamefont {Knotko}, \citenamefont {Karyagina},
  \citenamefont {Borisov},\ and\ \citenamefont
  {Shcherbakova}}]{TbHf_shlyakhtina2006structure}%
  \BibitemOpen
  \bibfield  {author} {\bibinfo {author} {\bibfnamefont {A.~V.}\ \bibnamefont
  {Shlyakhtina}}, \bibinfo {author} {\bibfnamefont {M.~V.}\ \bibnamefont
  {Boguslavskii}}, \bibinfo {author} {\bibfnamefont {S.~Y.}\ \bibnamefont
  {Stefanovich}}, \bibinfo {author} {\bibfnamefont {I.~V.}\ \bibnamefont
  {Kolbanev}}, \bibinfo {author} {\bibfnamefont {A.~V.}\ \bibnamefont
  {Knotko}}, \bibinfo {author} {\bibfnamefont {O.~K.}\ \bibnamefont
  {Karyagina}}, \bibinfo {author} {\bibfnamefont {S.~A.}\ \bibnamefont
  {Borisov}}, \ and\ \bibinfo {author} {\bibfnamefont {L.~G.}\ \bibnamefont
  {Shcherbakova}},\ }\href {\doibase https://doi.org/10.1134/S002016850605013X}
  {\bibfield  {journal} {\bibinfo  {journal} {Inorg. Mater.}\ }\textbf
  {\bibinfo {volume} {42}},\ \bibinfo {pages} {519} (\bibinfo {year}
  {2006})}\BibitemShut {NoStop}%
\bibitem [{\citenamefont {Shlyakhtina}\ \emph {et~al.}(2014)\citenamefont
  {Shlyakhtina}, \citenamefont {Belov}, \citenamefont {Pigalskiy},
  \citenamefont {Shchegolikhin}, \citenamefont {Kolbanev},\ and\ \citenamefont
  {Karyagina}}]{shlyakhtina2014synthesis}%
  \BibitemOpen
  \bibfield  {author} {\bibinfo {author} {\bibfnamefont {A.~V.}\ \bibnamefont
  {Shlyakhtina}}, \bibinfo {author} {\bibfnamefont {D.~A.}\ \bibnamefont
  {Belov}}, \bibinfo {author} {\bibfnamefont {K.~S.}\ \bibnamefont
  {Pigalskiy}}, \bibinfo {author} {\bibfnamefont {A.~N.}\ \bibnamefont
  {Shchegolikhin}}, \bibinfo {author} {\bibfnamefont {I.~V.}\ \bibnamefont
  {Kolbanev}}, \ and\ \bibinfo {author} {\bibfnamefont {O.~K.}\ \bibnamefont
  {Karyagina}},\ }\href {\doibase
  https://doi.org/10.1016/j.materresbull.2013.10.004} {\bibfield  {journal}
  {\bibinfo  {journal} {Mater. Res. Bull.}\ }\textbf {\bibinfo {volume} {49}},\
  \bibinfo {pages} {625} (\bibinfo {year} {2014})}\BibitemShut {NoStop}%
\bibitem [{\citenamefont {Shlyakhtina}\ \emph {et~al.}(2018)\citenamefont
  {Shlyakhtina}, \citenamefont {Pigalskiy}, \citenamefont {Belov},
  \citenamefont {Lyskov}, \citenamefont {Kharitonova}, \citenamefont
  {Kolbanev}, \citenamefont {Borunova}, \citenamefont {Karyagina},
  \citenamefont {Sadovskaya}, \citenamefont {Sadykov},\ and\ \citenamefont
  {Eremeev}}]{shlyakhtina2018proton}%
  \BibitemOpen
  \bibfield  {author} {\bibinfo {author} {\bibfnamefont {A.~V.}\ \bibnamefont
  {Shlyakhtina}}, \bibinfo {author} {\bibfnamefont {K.~S.}\ \bibnamefont
  {Pigalskiy}}, \bibinfo {author} {\bibfnamefont {D.~A.}\ \bibnamefont
  {Belov}}, \bibinfo {author} {\bibfnamefont {N.~V.}\ \bibnamefont {Lyskov}},
  \bibinfo {author} {\bibfnamefont {E.~P.}\ \bibnamefont {Kharitonova}},
  \bibinfo {author} {\bibfnamefont {I.~V.}\ \bibnamefont {Kolbanev}}, \bibinfo
  {author} {\bibfnamefont {A.~B.}\ \bibnamefont {Borunova}}, \bibinfo {author}
  {\bibfnamefont {O.~K.}\ \bibnamefont {Karyagina}}, \bibinfo {author}
  {\bibfnamefont {E.~M.}\ \bibnamefont {Sadovskaya}}, \bibinfo {author}
  {\bibfnamefont {V.~A.}\ \bibnamefont {Sadykov}}, \ and\ \bibinfo {author}
  {\bibfnamefont {N.~F.}\ \bibnamefont {Eremeev}},\ }\href {\doibase
  10.1039/C7DT03912C} {\bibfield  {journal} {\bibinfo  {journal} {Dalton
  Trans.}\ }\textbf {\bibinfo {volume} {47}},\ \bibinfo {pages} {2376}
  (\bibinfo {year} {2018})}\BibitemShut {NoStop}%
\bibitem [{\citenamefont {Song}\ \emph {et~al.}(2019)\citenamefont {Song},
  \citenamefont {Pecanha-Antonio},\ and\ \citenamefont {Su}}]{MLZNd}%
  \BibitemOpen
  \bibinfo {editor} {\bibfnamefont {J.}~\bibnamefont {Song}}, \bibinfo {editor}
  {\bibfnamefont {V.}~\bibnamefont {Pecanha-Antonio}}, \ and\ \bibinfo {editor}
  {\bibfnamefont {Y.}~\bibnamefont {Su}},\ eds.,\ \href@noop {} {\emph
  {\bibinfo {title} {MLZ Conference: Neutrons for information and quantum
  technologies}}}\ (\bibinfo  {publisher} {ACM},\ \bibinfo {year}
  {2019})\BibitemShut {NoStop}%
\bibitem [{\citenamefont {Sato}\ \emph {et~al.}(1976)\citenamefont {Sato},
  \citenamefont {Adachi},\ and\ \citenamefont {Shiokawa}}]{SATO76:38}%
  \BibitemOpen
  \bibfield  {author} {\bibinfo {author} {\bibfnamefont {K.}~\bibnamefont
  {Sato}}, \bibinfo {author} {\bibfnamefont {G.-Y.}\ \bibnamefont {Adachi}}, \
  and\ \bibinfo {author} {\bibfnamefont {J.}~\bibnamefont {Shiokawa}},\ }\href
  {\doibase https://doi.org/10.1016/0022-1902(76)80137-6} {\bibfield  {journal}
  {\bibinfo  {journal} {J. Inorg. Nucl.}\ }\textbf {\bibinfo {volume} {38}},\
  \bibinfo {pages} {1287} (\bibinfo {year} {1976})}\BibitemShut {NoStop}%
\bibitem [{\citenamefont {McQueen}\ \emph {et~al.}(2008)\citenamefont
  {McQueen}, \citenamefont {West}, \citenamefont {Muegge}, \citenamefont
  {Huang}, \citenamefont {Noble}, \citenamefont {Zandbergen},\ and\
  \citenamefont {Cava}}]{mcqueen2008frustrated}%
  \BibitemOpen
  \bibfield  {author} {\bibinfo {author} {\bibfnamefont {T.~M.}\ \bibnamefont
  {McQueen}}, \bibinfo {author} {\bibfnamefont {D.~V.}\ \bibnamefont {West}},
  \bibinfo {author} {\bibfnamefont {B.}~\bibnamefont {Muegge}}, \bibinfo
  {author} {\bibfnamefont {Q.}~\bibnamefont {Huang}}, \bibinfo {author}
  {\bibfnamefont {K.}~\bibnamefont {Noble}}, \bibinfo {author} {\bibfnamefont
  {H.~W.}\ \bibnamefont {Zandbergen}}, \ and\ \bibinfo {author} {\bibfnamefont
  {R.~J.}\ \bibnamefont {Cava}},\ }\href {\doibase
  https://doi.org/10.1088/0953-8984/20/23/235210} {\bibfield  {journal}
  {\bibinfo  {journal} {J. Phys.: Condens. Matter}\ }\textbf {\bibinfo {volume}
  {20}},\ \bibinfo {pages} {235210} (\bibinfo {year} {2008})}\BibitemShut
  {NoStop}%
\bibitem [{\citenamefont {Zouari}\ \emph {et~al.}(2008)\citenamefont {Zouari},
  \citenamefont {Ballou}, \citenamefont {Cheikh-Rouhou},\ and\ \citenamefont
  {Strobel}}]{zouari2008synthesis}%
  \BibitemOpen
  \bibfield  {author} {\bibinfo {author} {\bibfnamefont {S.}~\bibnamefont
  {Zouari}}, \bibinfo {author} {\bibfnamefont {R.}~\bibnamefont {Ballou}},
  \bibinfo {author} {\bibfnamefont {A.}~\bibnamefont {Cheikh-Rouhou}}, \ and\
  \bibinfo {author} {\bibfnamefont {P.}~\bibnamefont {Strobel}},\ }\href
  {\doibase https://doi.org/10.1016/j.matlet.2008.04.060} {\bibfield  {journal}
  {\bibinfo  {journal} {Mater. Lett.}\ }\textbf {\bibinfo {volume} {62}},\
  \bibinfo {pages} {3767} (\bibinfo {year} {2008})}\BibitemShut {NoStop}%
\bibitem [{\citenamefont {Kennedy}\ \emph {et~al.}(1997)\citenamefont
  {Kennedy}, \citenamefont {Hunter},\ and\ \citenamefont
  {Howard}}]{Kennedy97:58}%
  \BibitemOpen
  \bibfield  {author} {\bibinfo {author} {\bibfnamefont {B.~J.}\ \bibnamefont
  {Kennedy}}, \bibinfo {author} {\bibfnamefont {B.~A.}\ \bibnamefont {Hunter}},
  \ and\ \bibinfo {author} {\bibfnamefont {C.~J.}\ \bibnamefont {Howard}},\
  }\href {\doibase https://doi.org/10.1006/jssc.1997.7277} {\bibfield
  {journal} {\bibinfo  {journal} {J. Solid State Chem.}\ }\textbf {\bibinfo
  {volume} {130}},\ \bibinfo {pages} {58} (\bibinfo {year} {1997})}\BibitemShut
  {NoStop}%
\bibitem [{\citenamefont {Ortiz}\ \emph {et~al.}()\citenamefont {Ortiz},
  \citenamefont {Sarte}, \citenamefont {Pokharel},\ and\ \citenamefont
  {Wilson}}]{inprepOrtiz}%
  \BibitemOpen
  \bibfield  {author} {\bibinfo {author} {\bibfnamefont {B.~R.}\ \bibnamefont
  {Ortiz}}, \bibinfo {author} {\bibfnamefont {P.~M.}\ \bibnamefont {Sarte}},
  \bibinfo {author} {\bibfnamefont {G.}~\bibnamefont {Pokharel}}, \ and\
  \bibinfo {author} {\bibfnamefont {S.~D.}\ \bibnamefont {Wilson}},\
  }\href@noop {} {\bibinfo  {journal} {In Preparation}\ }\BibitemShut {NoStop}%
\bibitem [{\citenamefont {Clark}\ \emph {et~al.}(2013)\citenamefont {Clark},
  \citenamefont {Ritter}, \citenamefont {Harrison},\ and\ \citenamefont
  {Attfield}}]{CLARK13:203}%
  \BibitemOpen
\bibfield  {journal} {  }\bibfield  {author} {\bibinfo {author} {\bibfnamefont
  {L.}~\bibnamefont {Clark}}, \bibinfo {author} {\bibfnamefont
  {C.}~\bibnamefont {Ritter}}, \bibinfo {author} {\bibfnamefont
  {A.}~\bibnamefont {Harrison}}, \ and\ \bibinfo {author} {\bibfnamefont
  {J.~P.}\ \bibnamefont {Attfield}},\ }\href {\doibase
  https://doi.org/10.1016/j.jssc.2013.04.012} {\bibfield  {journal} {\bibinfo
  {journal} {J. Solid State Chem.}\ }\textbf {\bibinfo {volume} {203}},\
  \bibinfo {pages} {199} (\bibinfo {year} {2013})}\BibitemShut {NoStop}%
\bibitem [{\citenamefont {Greedan}\ \emph {et~al.}(1987)\citenamefont
  {Greedan}, \citenamefont {Sato}, \citenamefont {Ali},\ and\ \citenamefont
  {Datars}}]{GREEDAN87:68}%
  \BibitemOpen
  \bibfield  {author} {\bibinfo {author} {\bibfnamefont {J.~E.}\ \bibnamefont
  {Greedan}}, \bibinfo {author} {\bibfnamefont {M.}~\bibnamefont {Sato}},
  \bibinfo {author} {\bibfnamefont {N.}~\bibnamefont {Ali}}, \ and\ \bibinfo
  {author} {\bibfnamefont {W.~R.}\ \bibnamefont {Datars}},\ }\href {\doibase
  https://doi.org/10.1016/0022-4596(87)90316-1} {\bibfield  {journal} {\bibinfo
   {journal} {J. Solid State Chem.}\ }\textbf {\bibinfo {volume} {68}},\
  \bibinfo {pages} {300} (\bibinfo {year} {1987})}\BibitemShut {NoStop}%
\bibitem [{\citenamefont {Subramanian}\ \emph {et~al.}(1980)\citenamefont
  {Subramanian}, \citenamefont {Aravamudan},\ and\ \citenamefont
  {Subba~Rao}}]{subramanian80:15}%
  \BibitemOpen
  \bibfield  {author} {\bibinfo {author} {\bibfnamefont {M.~A.}\ \bibnamefont
  {Subramanian}}, \bibinfo {author} {\bibfnamefont {G.}~\bibnamefont
  {Aravamudan}}, \ and\ \bibinfo {author} {\bibfnamefont {G.~V.}\ \bibnamefont
  {Subba~Rao}},\ }\href {\doibase https://doi.org/10.1016/0025-5408(80)90094-X}
  {\bibfield  {journal} {\bibinfo  {journal} {Mater. Res. Bull.}\ }\textbf
  {\bibinfo {volume} {15}},\ \bibinfo {pages} {1401} (\bibinfo {year}
  {1980})}\BibitemShut {NoStop}%
\bibitem [{\citenamefont {Tkachenko}\ and\ \citenamefont
  {Fedorov}(2003)}]{tkachenko03:39}%
  \BibitemOpen
  \bibfield  {author} {\bibinfo {author} {\bibfnamefont {E.~A.}\ \bibnamefont
  {Tkachenko}}\ and\ \bibinfo {author} {\bibfnamefont {P.~P.}\ \bibnamefont
  {Fedorov}},\ }\href {\doibase https://doi.org/10.1023/A:1024132818445}
  {\bibfield  {journal} {\bibinfo  {journal} {Inorg. Mater.}\ }\textbf
  {\bibinfo {volume} {39}},\ \bibinfo {pages} {S25} (\bibinfo {year}
  {2003})}\BibitemShut {NoStop}%
\bibitem [{\citenamefont {McCarthy}(1971)}]{MCCARTHY71:6}%
  \BibitemOpen
  \bibfield  {author} {\bibinfo {author} {\bibfnamefont {G.~J.}\ \bibnamefont
  {McCarthy}},\ }\href {\doibase https://doi.org/10.1016/0025-5408(71)90156-5}
  {\bibfield  {journal} {\bibinfo  {journal} {Mater. Res. Bull.}\ }\textbf
  {\bibinfo {volume} {6}},\ \bibinfo {pages} {31} (\bibinfo {year}
  {1971})}\BibitemShut {NoStop}%
\bibitem [{\citenamefont {Ranganathan}\ \emph {et~al.}(1983)\citenamefont
  {Ranganathan}, \citenamefont {Rangarajan}, \citenamefont {Srinivasan},
  \citenamefont {Subramanian},\ and\ \citenamefont {Rao}}]{ranganathan83:52}%
  \BibitemOpen
  \bibfield  {author} {\bibinfo {author} {\bibfnamefont {R.}~\bibnamefont
  {Ranganathan}}, \bibinfo {author} {\bibfnamefont {G.}~\bibnamefont
  {Rangarajan}}, \bibinfo {author} {\bibfnamefont {R.}~\bibnamefont
  {Srinivasan}}, \bibinfo {author} {\bibfnamefont {M.~A.}\ \bibnamefont
  {Subramanian}}, \ and\ \bibinfo {author} {\bibfnamefont {G.~V.}\ \bibnamefont
  {Rao}},\ }\href {\doibase https://doi.org/10.1007/BF00682127} {\bibfield
  {journal} {\bibinfo  {journal} {J. Low Temp. Phys.}\ }\textbf {\bibinfo
  {volume} {52}},\ \bibinfo {pages} {481} (\bibinfo {year} {1983})}\BibitemShut
  {NoStop}%
\bibitem [{\citenamefont {Shaplygin}\ and\ \citenamefont
  {Lazarev}(1973)}]{SHAPLYGIN73:8}%
  \BibitemOpen
  \bibfield  {author} {\bibinfo {author} {\bibfnamefont {I.}~\bibnamefont
  {Shaplygin}}\ and\ \bibinfo {author} {\bibfnamefont {V.}~\bibnamefont
  {Lazarev}},\ }\href {\doibase https://doi.org/10.1016/0025-5408(73)90181-5}
  {\bibfield  {journal} {\bibinfo  {journal} {Mater. Res. Bull.}\ }\textbf
  {\bibinfo {volume} {8}},\ \bibinfo {pages} {761} (\bibinfo {year}
  {1973})}\BibitemShut {NoStop}%
\bibitem [{\citenamefont {Sleight}(1968)}]{sleight1968new}%
  \BibitemOpen
  \bibfield  {author} {\bibinfo {author} {\bibfnamefont {A.~W.}\ \bibnamefont
  {Sleight}},\ }\href {\doibase https://doi.org/10.1016/0025-5408(68)90120-7}
  {\bibfield  {journal} {\bibinfo  {journal} {Mater. Res. Bull.}\ }\textbf
  {\bibinfo {volume} {3}},\ \bibinfo {pages} {699} (\bibinfo {year}
  {1968})}\BibitemShut {NoStop}%
\bibitem [{\citenamefont {Chien}\ and\ \citenamefont
  {Sleight}(1978{\natexlab{a}})}]{chien1978mossbauer}%
  \BibitemOpen
  \bibfield  {author} {\bibinfo {author} {\bibfnamefont {C.~L.}\ \bibnamefont
  {Chien}}\ and\ \bibinfo {author} {\bibfnamefont {A.~W.}\ \bibnamefont
  {Sleight}},\ }\href {\doibase 10.1103/PhysRevB.18.2031} {\bibfield  {journal}
  {\bibinfo  {journal} {Phys. Rev. B}\ }\textbf {\bibinfo {volume} {18}},\
  \bibinfo {pages} {2031} (\bibinfo {year} {1978}{\natexlab{a}})}\BibitemShut
  {NoStop}%
\bibitem [{\citenamefont {Matsuhira}\ \emph
  {et~al.}(2011{\natexlab{a}})\citenamefont {Matsuhira}, \citenamefont
  {Wakeshima}, \citenamefont {Hinatsu},\ and\ \citenamefont
  {Takagi}}]{Matsuhira11:80}%
  \BibitemOpen
  \bibfield  {author} {\bibinfo {author} {\bibfnamefont {K.}~\bibnamefont
  {Matsuhira}}, \bibinfo {author} {\bibfnamefont {M.}~\bibnamefont
  {Wakeshima}}, \bibinfo {author} {\bibfnamefont {Y.}~\bibnamefont {Hinatsu}},
  \ and\ \bibinfo {author} {\bibfnamefont {S.}~\bibnamefont {Takagi}},\ }\href
  {\doibase 10.1143/JPSJ.80.094701} {\bibfield  {journal} {\bibinfo  {journal}
  {J. Phys. Soc. Jpn.}\ }\textbf {\bibinfo {volume} {80}},\ \bibinfo {pages}
  {094701} (\bibinfo {year} {2011}{\natexlab{a}})}\BibitemShut {NoStop}%
\bibitem [{\citenamefont {Klicpera}\ \emph {et~al.}(2020)\citenamefont
  {Klicpera}, \citenamefont {Vla\u{s}kov\'{a}},\ and\ \citenamefont
  {Divi\u{s}}}]{klicpera2020characterization}%
  \BibitemOpen
  \bibfield  {author} {\bibinfo {author} {\bibfnamefont {M.}~\bibnamefont
  {Klicpera}}, \bibinfo {author} {\bibfnamefont {K.}~\bibnamefont
  {Vla\u{s}kov\'{a}}}, \ and\ \bibinfo {author} {\bibfnamefont
  {M.}~\bibnamefont {Divi\u{s}}},\ }\href {\doibase
  https://doi.org/10.1021/acs.jpcc.0c04847} {\bibfield  {journal} {\bibinfo
  {journal} {J. Phys. Chem. C}\ }\textbf {\bibinfo {volume} {124}},\ \bibinfo
  {pages} {20367} (\bibinfo {year} {2020})}\BibitemShut {NoStop}%
\bibitem [{\citenamefont {Blacklock}\ and\ \citenamefont
  {White}(1980)}]{Blackrock80:72}%
  \BibitemOpen
  \bibfield  {author} {\bibinfo {author} {\bibfnamefont {K.}~\bibnamefont
  {Blacklock}}\ and\ \bibinfo {author} {\bibfnamefont {H.~W.}\ \bibnamefont
  {White}},\ }\href {\doibase 10.1063/1.439315} {\bibfield  {journal} {\bibinfo
   {journal} {J. Chem. Phys.}\ }\textbf {\bibinfo {volume} {72}},\ \bibinfo
  {pages} {2191} (\bibinfo {year} {1980})}\BibitemShut {NoStop}%
\bibitem [{\citenamefont {Yanagishima}\ and\ \citenamefont
  {Maeno}(2001)}]{yanagishima2001metal}%
  \BibitemOpen
  \bibfield  {author} {\bibinfo {author} {\bibfnamefont {D.}~\bibnamefont
  {Yanagishima}}\ and\ \bibinfo {author} {\bibfnamefont {Y.}~\bibnamefont
  {Maeno}},\ }\href {\doibase 10.1143/JPSJ.70.2880} {\bibfield  {journal}
  {\bibinfo  {journal} {J. Phys. Soc. Jpn.}\ }\textbf {\bibinfo {volume}
  {70}},\ \bibinfo {pages} {2880} (\bibinfo {year} {2001})}\BibitemShut
  {NoStop}%
\bibitem [{\citenamefont {Lazarev}\ and\ \citenamefont
  {Shaplygin}(1978)}]{LAZAREV78:13}%
  \BibitemOpen
  \bibfield  {author} {\bibinfo {author} {\bibfnamefont {V.~B.}\ \bibnamefont
  {Lazarev}}\ and\ \bibinfo {author} {\bibfnamefont {I.~S.}\ \bibnamefont
  {Shaplygin}},\ }\href {\doibase https://doi.org/10.1016/0025-5408(78)90227-1}
  {\bibfield  {journal} {\bibinfo  {journal} {Mater. Res. Bull.}\ }\textbf
  {\bibinfo {volume} {13}},\ \bibinfo {pages} {229} (\bibinfo {year}
  {1978})}\BibitemShut {NoStop}%
\bibitem [{\citenamefont {Ismunandar}\ \emph {et~al.}(2000)\citenamefont
  {Ismunandar}, \citenamefont {Kennedy},\ and\ \citenamefont
  {Hunter}}]{ISMUNANDAR00:302}%
  \BibitemOpen
  \bibfield  {author} {\bibinfo {author} {\bibfnamefont {D.}~\bibnamefont
  {Ismunandar}}, \bibinfo {author} {\bibfnamefont {B.~J.}\ \bibnamefont
  {Kennedy}}, \ and\ \bibinfo {author} {\bibfnamefont {B.~A.}\ \bibnamefont
  {Hunter}},\ }\href {\doibase https://doi.org/10.1016/S0925-8388(00)00677-0}
  {\bibfield  {journal} {\bibinfo  {journal} {J. Alloys Compd.}\ }\textbf
  {\bibinfo {volume} {302}},\ \bibinfo {pages} {94} (\bibinfo {year}
  {2000})}\BibitemShut {NoStop}%
\bibitem [{\citenamefont {Berndt}\ \emph {et~al.}(2015)\citenamefont {Berndt},
  \citenamefont {Silva}, \citenamefont {Ivashita}, \citenamefont {Paesano},
  \citenamefont {Blanco}, \citenamefont {Miner}, \citenamefont {Carbonio},
  \citenamefont {Dantas}, \citenamefont {Ayala},\ and\ \citenamefont
  {Isnard}}]{BERNDT15:618}%
  \BibitemOpen
  \bibfield  {author} {\bibinfo {author} {\bibfnamefont {G.}~\bibnamefont
  {Berndt}}, \bibinfo {author} {\bibfnamefont {K.}~\bibnamefont {Silva}},
  \bibinfo {author} {\bibfnamefont {F.}~\bibnamefont {Ivashita}}, \bibinfo
  {author} {\bibfnamefont {A.}~\bibnamefont {Paesano}}, \bibinfo {author}
  {\bibfnamefont {M.}~\bibnamefont {Blanco}}, \bibinfo {author} {\bibfnamefont
  {E.}~\bibnamefont {Miner}}, \bibinfo {author} {\bibfnamefont
  {R.}~\bibnamefont {Carbonio}}, \bibinfo {author} {\bibfnamefont
  {S.}~\bibnamefont {Dantas}}, \bibinfo {author} {\bibfnamefont
  {A.}~\bibnamefont {Ayala}}, \ and\ \bibinfo {author} {\bibfnamefont
  {O.}~\bibnamefont {Isnard}},\ }\href {\doibase
  https://doi.org/10.1016/j.jallcom.2014.08.068} {\bibfield  {journal}
  {\bibinfo  {journal} {J. Alloys Compd.}\ }\textbf {\bibinfo {volume} {618}},\
  \bibinfo {pages} {635} (\bibinfo {year} {2015})}\BibitemShut {NoStop}%
\bibitem [{\citenamefont {Knop}\ \emph {et~al.}(1968)\citenamefont {Knop},
  \citenamefont {Brisse}, \citenamefont {Meads},\ and\ \citenamefont
  {Bainbridge}}]{Knop68:46}%
  \BibitemOpen
  \bibfield  {author} {\bibinfo {author} {\bibfnamefont {O.}~\bibnamefont
  {Knop}}, \bibinfo {author} {\bibfnamefont {F.}~\bibnamefont {Brisse}},
  \bibinfo {author} {\bibfnamefont {R.~E.}\ \bibnamefont {Meads}}, \ and\
  \bibinfo {author} {\bibfnamefont {J.}~\bibnamefont {Bainbridge}},\ }\href
  {\doibase 10.1139/v68-635} {\bibfield  {journal} {\bibinfo  {journal} {Can.
  J. Chem.}\ }\textbf {\bibinfo {volume} {46}},\ \bibinfo {pages} {3829}
  (\bibinfo {year} {1968})}\BibitemShut {NoStop}%
\bibitem [{\citenamefont {Montmory}\ and\ \citenamefont
  {Bertaut}(1961)}]{Montmory61:252}%
  \BibitemOpen
  \bibfield  {author} {\bibinfo {author} {\bibfnamefont {M.-C.}\ \bibnamefont
  {Montmory}}\ and\ \bibinfo {author} {\bibfnamefont {F.}~\bibnamefont
  {Bertaut}},\ }\href@noop {} {\bibfield  {journal} {\bibinfo  {journal}
  {Compt. Rend.}\ }\textbf {\bibinfo {volume} {252}},\ \bibinfo {pages} {4171}
  (\bibinfo {year} {1961})}\BibitemShut {NoStop}%
\bibitem [{\citenamefont {Taira}\ \emph {et~al.}(1999)\citenamefont {Taira},
  \citenamefont {Wakeshima},\ and\ \citenamefont {Hinatsu}}]{Taira99:11}%
  \BibitemOpen
  \bibfield  {author} {\bibinfo {author} {\bibfnamefont {N.}~\bibnamefont
  {Taira}}, \bibinfo {author} {\bibfnamefont {M.}~\bibnamefont {Wakeshima}}, \
  and\ \bibinfo {author} {\bibfnamefont {Y.}~\bibnamefont {Hinatsu}},\ }\href
  {\doibase 10.1088/0953-8984/11/36/314} {\bibfield  {journal} {\bibinfo
  {journal} {J. Phys.: Condens. Matter}\ }\textbf {\bibinfo {volume} {11}},\
  \bibinfo {pages} {6983} (\bibinfo {year} {1999})}\BibitemShut {NoStop}%
\bibitem [{\citenamefont {Gaultois}(2015)}]{gaultois2015design}%
  \BibitemOpen
  \bibfield  {author} {\bibinfo {author} {\bibfnamefont {M.~W.}\ \bibnamefont
  {Gaultois}},\ }\href@noop {} {\emph {\bibinfo {title} {Design principles for
  oxide thermoelectric materials}}}\ (\bibinfo  {publisher} {University of
  California, Santa Barbara},\ \bibinfo {year} {2015})\BibitemShut {NoStop}%
\bibitem [{\citenamefont {Kennedy}(1995)}]{Kennedy95:51}%
  \BibitemOpen
  \bibfield  {author} {\bibinfo {author} {\bibfnamefont {B.~J.}\ \bibnamefont
  {Kennedy}},\ }\href {\doibase 10.1107/S0108270194013168} {\bibfield
  {journal} {\bibinfo  {journal} {Acta Crystallogr. C}\ }\textbf {\bibinfo
  {volume} {51}},\ \bibinfo {pages} {790} (\bibinfo {year} {1995})}\BibitemShut
  {NoStop}%
\bibitem [{\citenamefont {Bertaut}\ \emph {et~al.}(1959)\citenamefont
  {Bertaut}, \citenamefont {Forrat},\ and\ \citenamefont
  {Montmory}}]{Bertaut59:249}%
  \BibitemOpen
  \bibfield  {author} {\bibinfo {author} {\bibfnamefont {E.~F.}\ \bibnamefont
  {Bertaut}}, \bibinfo {author} {\bibfnamefont {F.}~\bibnamefont {Forrat}}, \
  and\ \bibinfo {author} {\bibfnamefont {M.~C.}\ \bibnamefont {Montmory}},\
  }\href@noop {} {\bibfield  {journal} {\bibinfo  {journal} {C.R. H. Seances
  Acad. Sci., Paris C}\ }\textbf {\bibinfo {volume} {249}},\ \bibinfo {pages}
  {829} (\bibinfo {year} {1959})}\BibitemShut {NoStop}%
\bibitem [{\citenamefont {Kobayashi}\ \emph {et~al.}(2001)\citenamefont
  {Kobayashi}, \citenamefont {Miyashita}, \citenamefont {Fukamachi},\ and\
  \citenamefont {Sato}}]{kobayashi2001nmr}%
  \BibitemOpen
  \bibfield  {author} {\bibinfo {author} {\bibfnamefont {Y.}~\bibnamefont
  {Kobayashi}}, \bibinfo {author} {\bibfnamefont {T.}~\bibnamefont
  {Miyashita}}, \bibinfo {author} {\bibfnamefont {T.}~\bibnamefont
  {Fukamachi}}, \ and\ \bibinfo {author} {\bibfnamefont {M.}~\bibnamefont
  {Sato}},\ }\href {\doibase https://doi.org/10.1016/S0022-3697(00)00161-X}
  {\bibfield  {journal} {\bibinfo  {journal} {J. Phys. Chem. Solids}\ }\textbf
  {\bibinfo {volume} {62}},\ \bibinfo {pages} {347} (\bibinfo {year}
  {2001})}\BibitemShut {NoStop}%
\bibitem [{\citenamefont {Whitaker}\ and\ \citenamefont
  {Greaves}(2014)}]{WHITAKER14:215}%
  \BibitemOpen
  \bibfield  {author} {\bibinfo {author} {\bibfnamefont {M.~J.}\ \bibnamefont
  {Whitaker}}\ and\ \bibinfo {author} {\bibfnamefont {C.}~\bibnamefont
  {Greaves}},\ }\href {\doibase https://doi.org/10.1016/j.jssc.2014.03.039}
  {\bibfield  {journal} {\bibinfo  {journal} {J. Solid State Chem.}\ }\textbf
  {\bibinfo {volume} {215}},\ \bibinfo {pages} {171} (\bibinfo {year}
  {2014})}\BibitemShut {NoStop}%
\bibitem [{\citenamefont {Bongers}\ and\ \citenamefont
  {Van~Meurs}(1967)}]{Bongers67:38}%
  \BibitemOpen
  \bibfield  {author} {\bibinfo {author} {\bibfnamefont {P.~F.}\ \bibnamefont
  {Bongers}}\ and\ \bibinfo {author} {\bibfnamefont {E.~R.}\ \bibnamefont
  {Van~Meurs}},\ }\href {\doibase 10.1063/1.1709693} {\bibfield  {journal}
  {\bibinfo  {journal} {J. Appl. Phys.}\ }\textbf {\bibinfo {volume} {38}},\
  \bibinfo {pages} {944} (\bibinfo {year} {1967})}\BibitemShut {NoStop}%
\bibitem [{\citenamefont {Casado}\ \emph {et~al.}(1985)\citenamefont {Casado},
  \citenamefont {Mendiola},\ and\ \citenamefont {Rasines}}]{CASADO85:46}%
  \BibitemOpen
  \bibfield  {author} {\bibinfo {author} {\bibfnamefont {P.}~\bibnamefont
  {Casado}}, \bibinfo {author} {\bibfnamefont {A.}~\bibnamefont {Mendiola}}, \
  and\ \bibinfo {author} {\bibfnamefont {I.}~\bibnamefont {Rasines}},\ }\href
  {\doibase https://doi.org/10.1016/0022-3697(85)90094-0} {\bibfield  {journal}
  {\bibinfo  {journal} {J. Phys. Chem. Solids}\ }\textbf {\bibinfo {volume}
  {46}},\ \bibinfo {pages} {921} (\bibinfo {year} {1985})}\BibitemShut
  {NoStop}%
\bibitem [{\citenamefont {Villars}\ \emph {et~al.}(2015)\citenamefont
  {Villars}, \citenamefont {Cenzual},\ and\ \citenamefont
  {Gladyshevskii}}]{villars2017handbook}%
  \BibitemOpen
  \bibfield  {author} {\bibinfo {author} {\bibfnamefont {P.}~\bibnamefont
  {Villars}}, \bibinfo {author} {\bibfnamefont {K.}~\bibnamefont {Cenzual}}, \
  and\ \bibinfo {author} {\bibfnamefont {R.}~\bibnamefont {Gladyshevskii}},\
  }\href@noop {} {\emph {\bibinfo {title} {Handbook of Inorganic Substances}}}\
  (\bibinfo  {publisher} {Walter de Gruyter GmbH},\ \bibinfo {address}
  {Berlin},\ \bibinfo {year} {2015})\BibitemShut {NoStop}%
\bibitem [{\citenamefont {T.~F.}\ and\ \citenamefont
  {Copenhaver}(1972)}]{woolfson1973solid}%
  \BibitemOpen
  \bibfield  {author} {\bibinfo {author} {\bibfnamefont {C.}~\bibnamefont
  {T.~F.}}\ and\ \bibinfo {author} {\bibfnamefont {E.~D.}\ \bibnamefont
  {Copenhaver}},\ }\href {\doibase https://doi.org/10.1007/978-1-4684-1396-0}
  {\emph {\bibinfo {title} {Bibliography of Magnetic Materials and Tabulation
  of Magnetic Transition Temperatures}}}\ (\bibinfo  {publisher} {Springer New
  York},\ \bibinfo {address} {New York},\ \bibinfo {year} {1972})\BibitemShut
  {NoStop}%
\bibitem [{\citenamefont {Knop}\ \emph {et~al.}(1969)\citenamefont {Knop},
  \citenamefont {Brisse},\ and\ \citenamefont {Castelliz}}]{Knop69:47}%
  \BibitemOpen
  \bibfield  {author} {\bibinfo {author} {\bibfnamefont {O.}~\bibnamefont
  {Knop}}, \bibinfo {author} {\bibfnamefont {F.}~\bibnamefont {Brisse}}, \ and\
  \bibinfo {author} {\bibfnamefont {L.}~\bibnamefont {Castelliz}},\ }\href
  {\doibase 10.1139/v69-155} {\bibfield  {journal} {\bibinfo  {journal} {Can.
  J. Chem.}\ }\textbf {\bibinfo {volume} {47}},\ \bibinfo {pages} {971}
  (\bibinfo {year} {1969})}\BibitemShut {NoStop}%
\bibitem [{\citenamefont {Knop}\ \emph {et~al.}(1965)\citenamefont {Knop},
  \citenamefont {Brisse},\ and\ \citenamefont {Castelliz}}]{Knop65:43}%
  \BibitemOpen
  \bibfield  {author} {\bibinfo {author} {\bibfnamefont {O.}~\bibnamefont
  {Knop}}, \bibinfo {author} {\bibfnamefont {F.}~\bibnamefont {Brisse}}, \ and\
  \bibinfo {author} {\bibfnamefont {L.}~\bibnamefont {Castelliz}},\ }\href
  {\doibase 10.1139/v65-392} {\bibfield  {journal} {\bibinfo  {journal} {Can.
  J. Chem.}\ }\textbf {\bibinfo {volume} {43}},\ \bibinfo {pages} {2812}
  (\bibinfo {year} {1965})}\BibitemShut {NoStop}%
\bibitem [{\citenamefont {Haipeng}\ \emph {et~al.}(2016)\citenamefont
  {Haipeng}, \citenamefont {Xiaoge}, \citenamefont {Hongsong}, \citenamefont
  {Haoming}, \citenamefont {Yongde}, \citenamefont {Yanxu},\ and\ \citenamefont
  {An}}]{haipeng2016preparation}%
  \BibitemOpen
  \bibfield  {author} {\bibinfo {author} {\bibfnamefont {Y.}~\bibnamefont
  {Haipeng}}, \bibinfo {author} {\bibfnamefont {C.}~\bibnamefont {Xiaoge}},
  \bibinfo {author} {\bibfnamefont {Z.}~\bibnamefont {Hongsong}}, \bibinfo
  {author} {\bibfnamefont {Z.}~\bibnamefont {Haoming}}, \bibinfo {author}
  {\bibfnamefont {Z.}~\bibnamefont {Yongde}}, \bibinfo {author} {\bibfnamefont
  {L.}~\bibnamefont {Yanxu}}, \ and\ \bibinfo {author} {\bibfnamefont
  {T.}~\bibnamefont {An}},\ }\href {\doibase 10.1080/23311940.2016.1244244}
  {\bibfield  {journal} {\bibinfo  {journal} {Cogent Physics}\ }\textbf
  {\bibinfo {volume} {3}},\ \bibinfo {pages} {1244244} (\bibinfo {year}
  {2016})}\BibitemShut {NoStop}%
\bibitem [{\citenamefont {Xiaoge}\ \emph {et~al.}(2016)\citenamefont {Xiaoge},
  \citenamefont {An}, \citenamefont {Hongsong}, \citenamefont {Yanxu},
  \citenamefont {Haoming},\ and\ \citenamefont {Yongde}}]{xiaoge2016thermal}%
  \BibitemOpen
  \bibfield  {author} {\bibinfo {author} {\bibfnamefont {C.}~\bibnamefont
  {Xiaoge}}, \bibinfo {author} {\bibfnamefont {T.}~\bibnamefont {An}}, \bibinfo
  {author} {\bibfnamefont {Z.}~\bibnamefont {Hongsong}}, \bibinfo {author}
  {\bibfnamefont {L.}~\bibnamefont {Yanxu}}, \bibinfo {author} {\bibfnamefont
  {Z.}~\bibnamefont {Haoming}}, \ and\ \bibinfo {author} {\bibfnamefont
  {Z.}~\bibnamefont {Yongde}},\ }\href {\doibase
  10.1016/j.ceramint.2016.05.141} {\bibfield  {journal} {\bibinfo  {journal}
  {Ceram. Int.}\ }\textbf {\bibinfo {volume} {42}},\ \bibinfo {pages} {13491}
  (\bibinfo {year} {2016})}\BibitemShut {NoStop}%
\bibitem [{\citenamefont {Li}\ \emph {et~al.}(2009)\citenamefont {Li},
  \citenamefont {Chen}, \citenamefont {Zhang},\ and\ \citenamefont
  {Li}}]{li2009photocatalytic}%
  \BibitemOpen
  \bibfield  {author} {\bibinfo {author} {\bibfnamefont {Y.}~\bibnamefont
  {Li}}, \bibinfo {author} {\bibfnamefont {G.}~\bibnamefont {Chen}}, \bibinfo
  {author} {\bibfnamefont {H.}~\bibnamefont {Zhang}}, \ and\ \bibinfo {author}
  {\bibfnamefont {Z.}~\bibnamefont {Li}},\ }\href {\doibase
  https://doi.org/10.1016/j.jpcs.2008.12.005} {\bibfield  {journal} {\bibinfo
  {journal} {J. Phys. Chem. Solids}\ }\textbf {\bibinfo {volume} {70}},\
  \bibinfo {pages} {536} (\bibinfo {year} {2009})}\BibitemShut {NoStop}%
\bibitem [{\citenamefont {Kitayama}\ and\ \citenamefont
  {Katsura}(1976)}]{kitayama76:5}%
  \BibitemOpen
  \bibfield  {author} {\bibinfo {author} {\bibfnamefont {K.}~\bibnamefont
  {Kitayama}}\ and\ \bibinfo {author} {\bibfnamefont {T.}~\bibnamefont
  {Katsura}},\ }\href {\doibase https://doi.org/10.1246/cl.1976.815} {\bibfield
   {journal} {\bibinfo  {journal} {Chem. Lett.}\ }\textbf {\bibinfo {volume}
  {5}},\ \bibinfo {pages} {815} (\bibinfo {year} {1976})}\BibitemShut {NoStop}%
\bibitem [{\citenamefont {Soderholm}\ \emph {et~al.}(1982)\citenamefont
  {Soderholm}, \citenamefont {Stager},\ and\ \citenamefont
  {Greedan}}]{SODERHOLM82:43}%
  \BibitemOpen
  \bibfield  {author} {\bibinfo {author} {\bibfnamefont {L.}~\bibnamefont
  {Soderholm}}, \bibinfo {author} {\bibfnamefont {C.~V.}\ \bibnamefont
  {Stager}}, \ and\ \bibinfo {author} {\bibfnamefont {J.~E.}\ \bibnamefont
  {Greedan}},\ }\href {\doibase https://doi.org/10.1016/0022-4596(82)90226-2}
  {\bibfield  {journal} {\bibinfo  {journal} {J. Solid State Chem.}\ }\textbf
  {\bibinfo {volume} {43}},\ \bibinfo {pages} {175} (\bibinfo {year}
  {1982})}\BibitemShut {NoStop}%
\bibitem [{\citenamefont {Troyanchuk}(1990)}]{troyanchuk1990preparation}%
  \BibitemOpen
  \bibfield  {author} {\bibinfo {author} {\bibfnamefont {I.~O.}\ \bibnamefont
  {Troyanchuk}},\ }\href@noop {} {\bibfield  {journal} {\bibinfo  {journal}
  {Inorg. Mater}\ }\textbf {\bibinfo {volume} {26}},\ \bibinfo {pages} {182}
  (\bibinfo {year} {1990})}\BibitemShut {NoStop}%
\bibitem [{\citenamefont {Yokokawa}\ \emph {et~al.}(1992)\citenamefont
  {Yokokawa}, \citenamefont {Sakai}, \citenamefont {Kawada},\ and\
  \citenamefont {Dokiya}}]{YOKOKAWA92:52}%
  \BibitemOpen
  \bibfield  {author} {\bibinfo {author} {\bibfnamefont {H.}~\bibnamefont
  {Yokokawa}}, \bibinfo {author} {\bibfnamefont {N.}~\bibnamefont {Sakai}},
  \bibinfo {author} {\bibfnamefont {T.}~\bibnamefont {Kawada}}, \ and\ \bibinfo
  {author} {\bibfnamefont {M.}~\bibnamefont {Dokiya}},\ }\href {\doibase
  https://doi.org/10.1016/0167-2738(92)90090-C} {\bibfield  {journal} {\bibinfo
   {journal} {Solid State Ion.}\ }\textbf {\bibinfo {volume} {52}},\ \bibinfo
  {pages} {43} (\bibinfo {year} {1992})}\BibitemShut {NoStop}%
\bibitem [{\citenamefont {Shin-ike}\ \emph {et~al.}(1977)\citenamefont
  {Shin-ike}, \citenamefont {Adachi},\ and\ \citenamefont
  {Shiokawa}}]{SHINIKE77:12}%
  \BibitemOpen
  \bibfield  {author} {\bibinfo {author} {\bibfnamefont {T.}~\bibnamefont
  {Shin-ike}}, \bibinfo {author} {\bibfnamefont {G.}~\bibnamefont {Adachi}}, \
  and\ \bibinfo {author} {\bibfnamefont {J.}~\bibnamefont {Shiokawa}},\ }\href
  {\doibase https://doi.org/10.1016/0025-5408(77)90168-4} {\bibfield  {journal}
  {\bibinfo  {journal} {Mater. Res. Bull.}\ }\textbf {\bibinfo {volume} {12}},\
  \bibinfo {pages} {1149} (\bibinfo {year} {1977})}\BibitemShut {NoStop}%
\bibitem [{\citenamefont {Shimakawa}\ \emph {et~al.}(1999)\citenamefont
  {Shimakawa}, \citenamefont {Kubo}, \citenamefont {Hamada}, \citenamefont
  {Jorgensen}, \citenamefont {Hu}, \citenamefont {Short}, \citenamefont
  {Nohara},\ and\ \citenamefont {Takagi}}]{Shimakawa99:59}%
  \BibitemOpen
  \bibfield  {author} {\bibinfo {author} {\bibfnamefont {Y.}~\bibnamefont
  {Shimakawa}}, \bibinfo {author} {\bibfnamefont {Y.}~\bibnamefont {Kubo}},
  \bibinfo {author} {\bibfnamefont {N.}~\bibnamefont {Hamada}}, \bibinfo
  {author} {\bibfnamefont {J.~D.}\ \bibnamefont {Jorgensen}}, \bibinfo {author}
  {\bibfnamefont {Z.}~\bibnamefont {Hu}}, \bibinfo {author} {\bibfnamefont
  {S.}~\bibnamefont {Short}}, \bibinfo {author} {\bibfnamefont
  {M.}~\bibnamefont {Nohara}}, \ and\ \bibinfo {author} {\bibfnamefont
  {H.}~\bibnamefont {Takagi}},\ }\href {\doibase 10.1103/PhysRevB.59.1249}
  {\bibfield  {journal} {\bibinfo  {journal} {Phys. Rev. B}\ }\textbf {\bibinfo
  {volume} {59}},\ \bibinfo {pages} {1249} (\bibinfo {year}
  {1999})}\BibitemShut {NoStop}%
\bibitem [{\citenamefont {Subramanian}\ \emph {et~al.}(1988)\citenamefont
  {Subramanian}, \citenamefont {Torardi}, \citenamefont {Johnson},
  \citenamefont {Pannetier},\ and\ \citenamefont {Sleight}}]{Subramanian88:72}%
  \BibitemOpen
  \bibfield  {author} {\bibinfo {author} {\bibfnamefont {M.~A.}\ \bibnamefont
  {Subramanian}}, \bibinfo {author} {\bibfnamefont {C.~C.}\ \bibnamefont
  {Torardi}}, \bibinfo {author} {\bibfnamefont {D.~C.}\ \bibnamefont
  {Johnson}}, \bibinfo {author} {\bibfnamefont {J.}~\bibnamefont {Pannetier}},
  \ and\ \bibinfo {author} {\bibfnamefont {A.~W.}\ \bibnamefont {Sleight}},\
  }\href {\doibase https://doi.org/10.1016/0022-4596(88)90004-7} {\bibfield
  {journal} {\bibinfo  {journal} {J. Solid State Chem.}\ }\textbf {\bibinfo
  {volume} {72}},\ \bibinfo {pages} {24} (\bibinfo {year} {1988})}\BibitemShut
  {NoStop}%
\bibitem [{\citenamefont {Greedan}\ \emph {et~al.}(1996)\citenamefont
  {Greedan}, \citenamefont {Raju}, \citenamefont {Maignan}, \citenamefont
  {Simon}, \citenamefont {Pedersen}, \citenamefont {Niraimathi}, \citenamefont
  {Gmelin},\ and\ \citenamefont {Subramanian}}]{Greedan96:54}%
  \BibitemOpen
  \bibfield  {author} {\bibinfo {author} {\bibfnamefont {J.~E.}\ \bibnamefont
  {Greedan}}, \bibinfo {author} {\bibfnamefont {N.~P.}\ \bibnamefont {Raju}},
  \bibinfo {author} {\bibfnamefont {A.}~\bibnamefont {Maignan}}, \bibinfo
  {author} {\bibfnamefont {C.}~\bibnamefont {Simon}}, \bibinfo {author}
  {\bibfnamefont {J.~S.}\ \bibnamefont {Pedersen}}, \bibinfo {author}
  {\bibfnamefont {A.~M.}\ \bibnamefont {Niraimathi}}, \bibinfo {author}
  {\bibfnamefont {E.}~\bibnamefont {Gmelin}}, \ and\ \bibinfo {author}
  {\bibfnamefont {M.~A.}\ \bibnamefont {Subramanian}},\ }\href {\doibase
  10.1103/PhysRevB.54.7189} {\bibfield  {journal} {\bibinfo  {journal} {Phys.
  Rev. B}\ }\textbf {\bibinfo {volume} {54}},\ \bibinfo {pages} {7189}
  (\bibinfo {year} {1996})}\BibitemShut {NoStop}%
\bibitem [{\citenamefont {Shimazaki}\ \emph {et~al.}(2000)\citenamefont
  {Shimazaki}, \citenamefont {Yamazaki}, \citenamefont {Terayama},\ and\
  \citenamefont {Yoshimura}}]{shimazaki00:19}%
  \BibitemOpen
  \bibfield  {author} {\bibinfo {author} {\bibfnamefont {T.}~\bibnamefont
  {Shimazaki}}, \bibinfo {author} {\bibfnamefont {T.}~\bibnamefont {Yamazaki}},
  \bibinfo {author} {\bibfnamefont {K.}~\bibnamefont {Terayama}}, \ and\
  \bibinfo {author} {\bibfnamefont {M.}~\bibnamefont {Yoshimura}},\ }\href
  {\doibase https://doi.org/10.1023/A:1026775321982} {\bibfield  {journal}
  {\bibinfo  {journal} {J. Mater. Sci. Lett.}\ }\textbf {\bibinfo {volume}
  {19}},\ \bibinfo {pages} {2029} (\bibinfo {year} {2000})}\BibitemShut
  {NoStop}%
\bibitem [{\citenamefont {Shannon}\ and\ \citenamefont
  {Sleight}(1968)}]{shannon1968synthesis}%
  \BibitemOpen
  \bibfield  {author} {\bibinfo {author} {\bibfnamefont {R.~D.}\ \bibnamefont
  {Shannon}}\ and\ \bibinfo {author} {\bibfnamefont {A.~W.}\ \bibnamefont
  {Sleight}},\ }\href {\doibase https://doi.org/10.1021/ic50066a038} {\bibfield
   {journal} {\bibinfo  {journal} {Inor. Chem.}\ }\textbf {\bibinfo {volume}
  {7}},\ \bibinfo {pages} {1649} (\bibinfo {year} {1968})}\BibitemShut
  {NoStop}%
\bibitem [{\citenamefont {Becker}\ and\ \citenamefont
  {Felsche}(1987)}]{BECKER1987269}%
  \BibitemOpen
  \bibfield  {author} {\bibinfo {author} {\bibfnamefont {U.~W.}\ \bibnamefont
  {Becker}}\ and\ \bibinfo {author} {\bibfnamefont {J.}~\bibnamefont
  {Felsche}},\ }\href {\doibase https://doi.org/10.1016/0022-5088(87)90215-3}
  {\bibfield  {journal} {\bibinfo  {journal} {J. Less Common Met.}\ }\textbf
  {\bibinfo {volume} {128}},\ \bibinfo {pages} {269} (\bibinfo {year}
  {1987})}\BibitemShut {NoStop}%
\bibitem [{\citenamefont {Clark}\ and\ \citenamefont
  {Morley}(1976)}]{Clark76:5}%
  \BibitemOpen
  \bibfield  {author} {\bibinfo {author} {\bibfnamefont {G.~M.}\ \bibnamefont
  {Clark}}\ and\ \bibinfo {author} {\bibfnamefont {R.}~\bibnamefont {Morley}},\
  }\href {\doibase 10.1039/CS9760500269} {\bibfield  {journal} {\bibinfo
  {journal} {Chem. Soc. Rev.}\ }\textbf {\bibinfo {volume} {5}},\ \bibinfo
  {pages} {269} (\bibinfo {year} {1976})}\BibitemShut {NoStop}%
\bibitem [{\citenamefont {Wells}(1987)}]{OF}%
  \BibitemOpen
  \bibfield  {author} {\bibinfo {author} {\bibfnamefont {F.}~\bibnamefont
  {Wells}},\ }\href@noop {} {\emph {\bibinfo {title} {Structural Inorganic
  Chemistry}}}\ (\bibinfo  {publisher} {Clarendon Press},\ \bibinfo {address}
  {Oxford},\ \bibinfo {year} {1987})\BibitemShut {NoStop}%
\bibitem [{\citenamefont {Phraewphiphat}\ \emph {et~al.}(2018)\citenamefont
  {Phraewphiphat}, \citenamefont {Iqbal}, \citenamefont {Suzuki}, \citenamefont
  {Hirayama},\ and\ \citenamefont {Kanno}}]{PHRAEWPHIPHAT2018}%
  \BibitemOpen
  \bibfield  {author} {\bibinfo {author} {\bibfnamefont {T.}~\bibnamefont
  {Phraewphiphat}}, \bibinfo {author} {\bibfnamefont {M.}~\bibnamefont
  {Iqbal}}, \bibinfo {author} {\bibfnamefont {K.}~\bibnamefont {Suzuki}},
  \bibinfo {author} {\bibfnamefont {M.}~\bibnamefont {Hirayama}}, \ and\
  \bibinfo {author} {\bibfnamefont {R.}~\bibnamefont {Kanno}},\ }\href
  {\doibase 10.2497/jjspm.65.26} {\bibfield  {journal} {\bibinfo  {journal} {J.
  Jpn. Soc. Powder Powder Metall.}\ }\textbf {\bibinfo {volume} {65}},\
  \bibinfo {pages} {26} (\bibinfo {year} {2018})}\BibitemShut {NoStop}%
\bibitem [{\citenamefont {Ross}\ \emph {et~al.}(2017)\citenamefont {Ross},
  \citenamefont {Brown}, \citenamefont {Cava}, \citenamefont {Krizan},
  \citenamefont {Nagler}, \citenamefont {Rodriguez-Rivera},\ and\ \citenamefont
  {Stone}}]{PhysRevB.95.144414}%
  \BibitemOpen
  \bibfield  {author} {\bibinfo {author} {\bibfnamefont {K.~A.}\ \bibnamefont
  {Ross}}, \bibinfo {author} {\bibfnamefont {J.~M.}\ \bibnamefont {Brown}},
  \bibinfo {author} {\bibfnamefont {R.~J.}\ \bibnamefont {Cava}}, \bibinfo
  {author} {\bibfnamefont {J.~W.}\ \bibnamefont {Krizan}}, \bibinfo {author}
  {\bibfnamefont {S.~E.}\ \bibnamefont {Nagler}}, \bibinfo {author}
  {\bibfnamefont {J.~A.}\ \bibnamefont {Rodriguez-Rivera}}, \ and\ \bibinfo
  {author} {\bibfnamefont {M.~B.}\ \bibnamefont {Stone}},\ }\href {\doibase
  10.1103/PhysRevB.95.144414} {\bibfield  {journal} {\bibinfo  {journal} {Phys.
  Rev. B}\ }\textbf {\bibinfo {volume} {95}},\ \bibinfo {pages} {144414}
  (\bibinfo {year} {2017})}\BibitemShut {NoStop}%
\bibitem [{\citenamefont {Che}\ \emph {et~al.}(2020)\citenamefont {Che},
  \citenamefont {Zhao}, \citenamefont {Rao}, \citenamefont {Chu}, \citenamefont
  {Li}, \citenamefont {Chu}, \citenamefont {Gao}, \citenamefont {Yue},
  \citenamefont {Zhou}, \citenamefont {Li}, \citenamefont {Huang},
  \citenamefont {Choi}, \citenamefont {Han}, \citenamefont {He}, \citenamefont
  {Zhou}, \citenamefont {Zhao},\ and\ \citenamefont
  {Sun}}]{AlSb_che2020absence}%
  \BibitemOpen
  \bibfield  {author} {\bibinfo {author} {\bibfnamefont {H.~L.}\ \bibnamefont
  {Che}}, \bibinfo {author} {\bibfnamefont {Z.~Y.}\ \bibnamefont {Zhao}},
  \bibinfo {author} {\bibfnamefont {X.}~\bibnamefont {Rao}}, \bibinfo {author}
  {\bibfnamefont {L.~G.}\ \bibnamefont {Chu}}, \bibinfo {author} {\bibfnamefont
  {N.}~\bibnamefont {Li}}, \bibinfo {author} {\bibfnamefont {W.~J.}\
  \bibnamefont {Chu}}, \bibinfo {author} {\bibfnamefont {P.}~\bibnamefont
  {Gao}}, \bibinfo {author} {\bibfnamefont {X.~Y.}\ \bibnamefont {Yue}},
  \bibinfo {author} {\bibfnamefont {Y.}~\bibnamefont {Zhou}}, \bibinfo {author}
  {\bibfnamefont {Q.~J.}\ \bibnamefont {Li}}, \bibinfo {author} {\bibfnamefont
  {Q.}~\bibnamefont {Huang}}, \bibinfo {author} {\bibfnamefont {E.~S.}\
  \bibnamefont {Choi}}, \bibinfo {author} {\bibfnamefont {Y.~Y.}\ \bibnamefont
  {Han}}, \bibinfo {author} {\bibfnamefont {Z.~Z.}\ \bibnamefont {He}},
  \bibinfo {author} {\bibfnamefont {H.~D.}\ \bibnamefont {Zhou}}, \bibinfo
  {author} {\bibfnamefont {X.}~\bibnamefont {Zhao}}, \ and\ \bibinfo {author}
  {\bibfnamefont {X.~F.}\ \bibnamefont {Sun}},\ }\href {\doibase
  10.1103/PhysRevMaterials.4.054406} {\bibfield  {journal} {\bibinfo  {journal}
  {Phys. Rev. Materials}\ }\textbf {\bibinfo {volume} {4}},\ \bibinfo {pages}
  {054406} (\bibinfo {year} {2020})}\BibitemShut {NoStop}%
\bibitem [{\citenamefont {Simeone}\ \emph {et~al.}(2017)\citenamefont
  {Simeone}, \citenamefont {Thorogood}, \citenamefont {Huo}, \citenamefont
  {Luneville}, \citenamefont {Baldinozzi}, \citenamefont {Petricek},
  \citenamefont {Porcher}, \citenamefont {Ribis}, \citenamefont {Mazerolles},
  \citenamefont {Largeau}, \citenamefont {Berar},\ and\ \citenamefont
  {Surble}}]{simeone2017intricate}%
  \BibitemOpen
  \bibfield  {author} {\bibinfo {author} {\bibfnamefont {D.}~\bibnamefont
  {Simeone}}, \bibinfo {author} {\bibfnamefont {G.~J.}\ \bibnamefont
  {Thorogood}}, \bibinfo {author} {\bibfnamefont {D.}~\bibnamefont {Huo}},
  \bibinfo {author} {\bibfnamefont {L.}~\bibnamefont {Luneville}}, \bibinfo
  {author} {\bibfnamefont {G.}~\bibnamefont {Baldinozzi}}, \bibinfo {author}
  {\bibfnamefont {V.}~\bibnamefont {Petricek}}, \bibinfo {author}
  {\bibfnamefont {F.}~\bibnamefont {Porcher}}, \bibinfo {author} {\bibfnamefont
  {J.}~\bibnamefont {Ribis}}, \bibinfo {author} {\bibfnamefont
  {L.}~\bibnamefont {Mazerolles}}, \bibinfo {author} {\bibfnamefont
  {L.}~\bibnamefont {Largeau}}, \bibinfo {author} {\bibfnamefont {J.~F.}\
  \bibnamefont {Berar}}, \ and\ \bibinfo {author} {\bibfnamefont
  {S.}~\bibnamefont {Surble}},\ }\href {\doibase
  https://doi.org/10.1038/s41598-017-02787-w} {\bibfield  {journal} {\bibinfo
  {journal} {Sci. Rep.}\ }\textbf {\bibinfo {volume} {7}},\ \bibinfo {pages}
  {1} (\bibinfo {year} {2017})}\BibitemShut {NoStop}%
\bibitem [{\citenamefont {Sibille}\ \emph {et~al.}(2017)\citenamefont
  {Sibille}, \citenamefont {Lhotel}, \citenamefont {Ciomaga~Hatnean},
  \citenamefont {Nilsen}, \citenamefont {Ehlers}, \citenamefont {Cervellino},
  \citenamefont {Ressouche}, \citenamefont {Frontzek}, \citenamefont {Zaharko},
  \citenamefont {Pomjakushin}, \citenamefont {Stuhr}, \citenamefont {Walker},
  \citenamefont {Adroja}, \citenamefont {Luetkens}, \citenamefont {Baines},
  \citenamefont {Amato}, \citenamefont {Fennell},\ and\ \citenamefont
  {Kenzelmann}}]{TbHf_sibille2017coulomb}%
  \BibitemOpen
  \bibfield  {author} {\bibinfo {author} {\bibfnamefont {R.}~\bibnamefont
  {Sibille}}, \bibinfo {author} {\bibfnamefont {E.}~\bibnamefont {Lhotel}},
  \bibinfo {author} {\bibfnamefont {M.}~\bibnamefont {Ciomaga~Hatnean}},
  \bibinfo {author} {\bibfnamefont {G.~J.}\ \bibnamefont {Nilsen}}, \bibinfo
  {author} {\bibfnamefont {G.}~\bibnamefont {Ehlers}}, \bibinfo {author}
  {\bibfnamefont {A.}~\bibnamefont {Cervellino}}, \bibinfo {author}
  {\bibfnamefont {E.}~\bibnamefont {Ressouche}}, \bibinfo {author}
  {\bibfnamefont {M.}~\bibnamefont {Frontzek}}, \bibinfo {author}
  {\bibfnamefont {O.}~\bibnamefont {Zaharko}}, \bibinfo {author} {\bibfnamefont
  {V.}~\bibnamefont {Pomjakushin}}, \bibinfo {author} {\bibfnamefont
  {U.}~\bibnamefont {Stuhr}}, \bibinfo {author} {\bibfnamefont {H.~C.}\
  \bibnamefont {Walker}}, \bibinfo {author} {\bibfnamefont {D.~T.}\
  \bibnamefont {Adroja}}, \bibinfo {author} {\bibfnamefont {H.}~\bibnamefont
  {Luetkens}}, \bibinfo {author} {\bibfnamefont {C.}~\bibnamefont {Baines}},
  \bibinfo {author} {\bibfnamefont {G.}~\bibnamefont {Amato}, \bibfnamefont
  {A~Balakrishnan}}, \bibinfo {author} {\bibfnamefont {T.}~\bibnamefont
  {Fennell}}, \ and\ \bibinfo {author} {\bibfnamefont {M.}~\bibnamefont
  {Kenzelmann}},\ }\href {\doibase https://doi.org/10.1038/s41467-017-00905-w}
  {\bibfield  {journal} {\bibinfo  {journal} {Nat. Commun.}\ }\textbf {\bibinfo
  {volume} {8}},\ \bibinfo {pages} {892} (\bibinfo {year} {2017})}\BibitemShut
  {NoStop}%
\bibitem [{\citenamefont {Blanchard}\ \emph {et~al.}(2012)\citenamefont
  {Blanchard}, \citenamefont {Clements}, \citenamefont {Kennedy}, \citenamefont
  {Ling}, \citenamefont {Reynolds}, \citenamefont {Avdeev}, \citenamefont
  {Stampfl}, \citenamefont {Zhang},\ and\ \citenamefont
  {Jang}}]{blanchard2012does}%
  \BibitemOpen
  \bibfield  {author} {\bibinfo {author} {\bibfnamefont {P.~E.~R.}\
  \bibnamefont {Blanchard}}, \bibinfo {author} {\bibfnamefont {R.}~\bibnamefont
  {Clements}}, \bibinfo {author} {\bibfnamefont {B.~J.}\ \bibnamefont
  {Kennedy}}, \bibinfo {author} {\bibfnamefont {C.~D.}\ \bibnamefont {Ling}},
  \bibinfo {author} {\bibfnamefont {E.}~\bibnamefont {Reynolds}}, \bibinfo
  {author} {\bibfnamefont {M.}~\bibnamefont {Avdeev}}, \bibinfo {author}
  {\bibfnamefont {A.~P.~J.}\ \bibnamefont {Stampfl}}, \bibinfo {author}
  {\bibfnamefont {Z.}~\bibnamefont {Zhang}}, \ and\ \bibinfo {author}
  {\bibfnamefont {L.-Y.}\ \bibnamefont {Jang}},\ }\href {\doibase
  10.1021/ic301677b} {\bibfield  {journal} {\bibinfo  {journal} {Inorg. Chem.}\
  }\textbf {\bibinfo {volume} {51}},\ \bibinfo {pages} {13237} (\bibinfo {year}
  {2012})}\BibitemShut {NoStop}%
\bibitem [{\citenamefont {Chang}\ \emph {et~al.}(2012)\citenamefont {Chang},
  \citenamefont {Onoda}, \citenamefont {Su}, \citenamefont {Kao}, \citenamefont
  {Tsuei}, \citenamefont {Yasui}, \citenamefont {Kakurai},\ and\ \citenamefont
  {Lees}}]{YbTi_chang2012higgs}%
  \BibitemOpen
  \bibfield  {author} {\bibinfo {author} {\bibfnamefont {L.-J.}\ \bibnamefont
  {Chang}}, \bibinfo {author} {\bibfnamefont {S.}~\bibnamefont {Onoda}},
  \bibinfo {author} {\bibfnamefont {Y.}~\bibnamefont {Su}}, \bibinfo {author}
  {\bibfnamefont {Y.-J.}\ \bibnamefont {Kao}}, \bibinfo {author} {\bibfnamefont
  {K.-D.}\ \bibnamefont {Tsuei}}, \bibinfo {author} {\bibfnamefont
  {Y.}~\bibnamefont {Yasui}}, \bibinfo {author} {\bibfnamefont
  {K.}~\bibnamefont {Kakurai}}, \ and\ \bibinfo {author} {\bibfnamefont
  {M.~R.}\ \bibnamefont {Lees}},\ }\href {\doibase 10.1038/ncomms1989}
  {\bibfield  {journal} {\bibinfo  {journal} {Nat. Commun.}\ }\textbf {\bibinfo
  {volume} {3}},\ \bibinfo {pages} {992} (\bibinfo {year} {2012})}\BibitemShut
  {NoStop}%
\bibitem [{\citenamefont {Gaudet}\ \emph
  {et~al.}(2016{\natexlab{a}})\citenamefont {Gaudet}, \citenamefont {Ross},
  \citenamefont {Kermarrec}, \citenamefont {Butch}, \citenamefont {Ehlers},
  \citenamefont {Dabkowska},\ and\ \citenamefont
  {Gaulin}}]{YbTi_gaudet2016gapless}%
  \BibitemOpen
  \bibfield  {author} {\bibinfo {author} {\bibfnamefont {J.}~\bibnamefont
  {Gaudet}}, \bibinfo {author} {\bibfnamefont {K.~A.}\ \bibnamefont {Ross}},
  \bibinfo {author} {\bibfnamefont {E.}~\bibnamefont {Kermarrec}}, \bibinfo
  {author} {\bibfnamefont {N.}~\bibnamefont {Butch}}, \bibinfo {author}
  {\bibfnamefont {G.}~\bibnamefont {Ehlers}}, \bibinfo {author} {\bibfnamefont
  {H.~A.}\ \bibnamefont {Dabkowska}}, \ and\ \bibinfo {author} {\bibfnamefont
  {B.~D.}\ \bibnamefont {Gaulin}},\ }\href {\doibase
  10.1103/PhysRevB.93.064406} {\bibfield  {journal} {\bibinfo  {journal} {Phys.
  Rev. B}\ }\textbf {\bibinfo {volume} {93}},\ \bibinfo {pages} {064406}
  (\bibinfo {year} {2016}{\natexlab{a}})}\BibitemShut {NoStop}%
\bibitem [{\citenamefont {Ross}\ \emph {et~al.}(2012)\citenamefont {Ross},
  \citenamefont {Proffen}, \citenamefont {Dabkowska}, \citenamefont {Quilliam},
  \citenamefont {Yaraskavitch}, \citenamefont {Kycia},\ and\ \citenamefont
  {Gaulin}}]{YbTi_ross2012lightly}%
  \BibitemOpen
  \bibfield  {author} {\bibinfo {author} {\bibfnamefont {K.~A.}\ \bibnamefont
  {Ross}}, \bibinfo {author} {\bibfnamefont {T.}~\bibnamefont {Proffen}},
  \bibinfo {author} {\bibfnamefont {H.~A.}\ \bibnamefont {Dabkowska}}, \bibinfo
  {author} {\bibfnamefont {J.~A.}\ \bibnamefont {Quilliam}}, \bibinfo {author}
  {\bibfnamefont {L.~R.}\ \bibnamefont {Yaraskavitch}}, \bibinfo {author}
  {\bibfnamefont {J.~B.}\ \bibnamefont {Kycia}}, \ and\ \bibinfo {author}
  {\bibfnamefont {B.~D.}\ \bibnamefont {Gaulin}},\ }\href {\doibase
  10.1103/PhysRevB.86.174424} {\bibfield  {journal} {\bibinfo  {journal} {Phys.
  Rev. B}\ }\textbf {\bibinfo {volume} {86}},\ \bibinfo {pages} {174424}
  (\bibinfo {year} {2012})}\BibitemShut {NoStop}%
\bibitem [{\citenamefont {Yasui}\ \emph {et~al.}(2003)\citenamefont {Yasui},
  \citenamefont {Soda}, \citenamefont {Iikubo}, \citenamefont {Ito},
  \citenamefont {Sato}, \citenamefont {Hamaguchi}, \citenamefont {Matsushita},
  \citenamefont {Wada}, \citenamefont {Takeuchi}, \citenamefont {Aso},\ and\
  \citenamefont {Kakurai}}]{YbTi_yasui2003ferromagnetic}%
  \BibitemOpen
  \bibfield  {author} {\bibinfo {author} {\bibfnamefont {Y.}~\bibnamefont
  {Yasui}}, \bibinfo {author} {\bibfnamefont {M.}~\bibnamefont {Soda}},
  \bibinfo {author} {\bibfnamefont {S.}~\bibnamefont {Iikubo}}, \bibinfo
  {author} {\bibfnamefont {M.}~\bibnamefont {Ito}}, \bibinfo {author}
  {\bibfnamefont {M.}~\bibnamefont {Sato}}, \bibinfo {author} {\bibfnamefont
  {N.}~\bibnamefont {Hamaguchi}}, \bibinfo {author} {\bibfnamefont
  {T.}~\bibnamefont {Matsushita}}, \bibinfo {author} {\bibfnamefont
  {N.}~\bibnamefont {Wada}}, \bibinfo {author} {\bibfnamefont {T.}~\bibnamefont
  {Takeuchi}}, \bibinfo {author} {\bibfnamefont {N.}~\bibnamefont {Aso}}, \
  and\ \bibinfo {author} {\bibfnamefont {K.}~\bibnamefont {Kakurai}},\ }\href
  {\doibase 10.1143/JPSJ.72.3014} {\bibfield  {journal} {\bibinfo  {journal}
  {J. Phys. Soc. Jpn.}\ }\textbf {\bibinfo {volume} {72}},\ \bibinfo {pages}
  {3014} (\bibinfo {year} {2003})}\BibitemShut {NoStop}%
\bibitem [{\citenamefont {Yaouanc}\ \emph
  {et~al.}(2011{\natexlab{a}})\citenamefont {Yaouanc}, \citenamefont {Dalmas~de
  R\'eotier}, \citenamefont {Marin},\ and\ \citenamefont
  {Glazkov}}]{PhysRevB.84.172408}%
  \BibitemOpen
  \bibfield  {author} {\bibinfo {author} {\bibfnamefont {A.}~\bibnamefont
  {Yaouanc}}, \bibinfo {author} {\bibfnamefont {P.}~\bibnamefont {Dalmas~de
  R\'eotier}}, \bibinfo {author} {\bibfnamefont {C.}~\bibnamefont {Marin}}, \
  and\ \bibinfo {author} {\bibfnamefont {V.}~\bibnamefont {Glazkov}},\ }\href
  {\doibase 10.1103/PhysRevB.84.172408} {\bibfield  {journal} {\bibinfo
  {journal} {Phys. Rev. B}\ }\textbf {\bibinfo {volume} {84}},\ \bibinfo
  {pages} {172408} (\bibinfo {year} {2011}{\natexlab{a}})}\BibitemShut
  {NoStop}%
\bibitem [{\citenamefont {Ross}\ \emph
  {et~al.}(2011{\natexlab{b}})\citenamefont {Ross}, \citenamefont
  {Yaraskavitch}, \citenamefont {Laver}, \citenamefont {Gardner}, \citenamefont
  {Quilliam}, \citenamefont {Meng}, \citenamefont {Kycia}, \citenamefont
  {Singh}, \citenamefont {Proffen}, \citenamefont {Dabkowska},\ and\
  \citenamefont {Gaulin}}]{PhysRevB.84.174442}%
  \BibitemOpen
  \bibfield  {author} {\bibinfo {author} {\bibfnamefont {K.~A.}\ \bibnamefont
  {Ross}}, \bibinfo {author} {\bibfnamefont {L.~R.}\ \bibnamefont
  {Yaraskavitch}}, \bibinfo {author} {\bibfnamefont {M.}~\bibnamefont {Laver}},
  \bibinfo {author} {\bibfnamefont {J.~S.}\ \bibnamefont {Gardner}}, \bibinfo
  {author} {\bibfnamefont {J.~A.}\ \bibnamefont {Quilliam}}, \bibinfo {author}
  {\bibfnamefont {S.}~\bibnamefont {Meng}}, \bibinfo {author} {\bibfnamefont
  {J.~B.}\ \bibnamefont {Kycia}}, \bibinfo {author} {\bibfnamefont {D.~K.}\
  \bibnamefont {Singh}}, \bibinfo {author} {\bibfnamefont {T.}~\bibnamefont
  {Proffen}}, \bibinfo {author} {\bibfnamefont {H.~A.}\ \bibnamefont
  {Dabkowska}}, \ and\ \bibinfo {author} {\bibfnamefont {B.~D.}\ \bibnamefont
  {Gaulin}},\ }\href {\doibase 10.1103/PhysRevB.84.174442} {\bibfield
  {journal} {\bibinfo  {journal} {Phys. Rev. B}\ }\textbf {\bibinfo {volume}
  {84}},\ \bibinfo {pages} {174442} (\bibinfo {year}
  {2011}{\natexlab{b}})}\BibitemShut {NoStop}%
\bibitem [{\citenamefont {D'Ortenzio}\ \emph {et~al.}(2013)\citenamefont
  {D'Ortenzio}, \citenamefont {Dabkowska}, \citenamefont {Dunsiger},
  \citenamefont {Gaulin}, \citenamefont {Gingras}, \citenamefont {Goko},
  \citenamefont {Kycia}, \citenamefont {Liu}, \citenamefont {Medina},
  \citenamefont {Munsie}, \citenamefont {Pomaranski}, \citenamefont {Ross},
  \citenamefont {Uemura}, \citenamefont {Williams},\ and\ \citenamefont
  {Luke}}]{PhysRevB.88.134428}%
  \BibitemOpen
  \bibfield  {author} {\bibinfo {author} {\bibfnamefont {R.~M.}\ \bibnamefont
  {D'Ortenzio}}, \bibinfo {author} {\bibfnamefont {H.~A.}\ \bibnamefont
  {Dabkowska}}, \bibinfo {author} {\bibfnamefont {S.~R.}\ \bibnamefont
  {Dunsiger}}, \bibinfo {author} {\bibfnamefont {B.~D.}\ \bibnamefont
  {Gaulin}}, \bibinfo {author} {\bibfnamefont {M.~J.~P.}\ \bibnamefont
  {Gingras}}, \bibinfo {author} {\bibfnamefont {T.}~\bibnamefont {Goko}},
  \bibinfo {author} {\bibfnamefont {J.~B.}\ \bibnamefont {Kycia}}, \bibinfo
  {author} {\bibfnamefont {L.}~\bibnamefont {Liu}}, \bibinfo {author}
  {\bibfnamefont {T.}~\bibnamefont {Medina}}, \bibinfo {author} {\bibfnamefont
  {T.~J.}\ \bibnamefont {Munsie}}, \bibinfo {author} {\bibfnamefont
  {D.}~\bibnamefont {Pomaranski}}, \bibinfo {author} {\bibfnamefont {K.~A.}\
  \bibnamefont {Ross}}, \bibinfo {author} {\bibfnamefont {Y.~J.}\ \bibnamefont
  {Uemura}}, \bibinfo {author} {\bibfnamefont {T.~J.}\ \bibnamefont
  {Williams}}, \ and\ \bibinfo {author} {\bibfnamefont {G.~M.}\ \bibnamefont
  {Luke}},\ }\href {\doibase 10.1103/PhysRevB.88.134428} {\bibfield  {journal}
  {\bibinfo  {journal} {Phys. Rev. B}\ }\textbf {\bibinfo {volume} {88}},\
  \bibinfo {pages} {134428} (\bibinfo {year} {2013})}\BibitemShut {NoStop}%
\bibitem [{\citenamefont {Chang}\ \emph {et~al.}(2014)\citenamefont {Chang},
  \citenamefont {Lees}, \citenamefont {Watanabe}, \citenamefont {Hillier},
  \citenamefont {Yasui},\ and\ \citenamefont {Onoda}}]{PhysRevB.89.184416}%
  \BibitemOpen
  \bibfield  {author} {\bibinfo {author} {\bibfnamefont {L.-J.}\ \bibnamefont
  {Chang}}, \bibinfo {author} {\bibfnamefont {M.~R.}\ \bibnamefont {Lees}},
  \bibinfo {author} {\bibfnamefont {I.}~\bibnamefont {Watanabe}}, \bibinfo
  {author} {\bibfnamefont {A.~D.}\ \bibnamefont {Hillier}}, \bibinfo {author}
  {\bibfnamefont {Y.}~\bibnamefont {Yasui}}, \ and\ \bibinfo {author}
  {\bibfnamefont {S.}~\bibnamefont {Onoda}},\ }\href {\doibase
  10.1103/PhysRevB.89.184416} {\bibfield  {journal} {\bibinfo  {journal} {Phys.
  Rev. B}\ }\textbf {\bibinfo {volume} {89}},\ \bibinfo {pages} {184416}
  (\bibinfo {year} {2014})}\BibitemShut {NoStop}%
\bibitem [{\citenamefont {Arpino}\ \emph {et~al.}(2017)\citenamefont {Arpino},
  \citenamefont {Trump}, \citenamefont {Scheie}, \citenamefont {McQueen},\ and\
  \citenamefont {Koohpayeh}}]{PhysRevB.95.094407}%
  \BibitemOpen
  \bibfield  {author} {\bibinfo {author} {\bibfnamefont {K.~E.}\ \bibnamefont
  {Arpino}}, \bibinfo {author} {\bibfnamefont {B.~A.}\ \bibnamefont {Trump}},
  \bibinfo {author} {\bibfnamefont {A.~O.}\ \bibnamefont {Scheie}}, \bibinfo
  {author} {\bibfnamefont {T.~M.}\ \bibnamefont {McQueen}}, \ and\ \bibinfo
  {author} {\bibfnamefont {S.~M.}\ \bibnamefont {Koohpayeh}},\ }\href {\doibase
  10.1103/PhysRevB.95.094407} {\bibfield  {journal} {\bibinfo  {journal} {Phys.
  Rev. B}\ }\textbf {\bibinfo {volume} {95}},\ \bibinfo {pages} {094407}
  (\bibinfo {year} {2017})}\BibitemShut {NoStop}%
\bibitem [{\citenamefont {Ruminy}\ \emph {et~al.}(2016)\citenamefont {Ruminy},
  \citenamefont {Bovo}, \citenamefont {Pomjakushina}, \citenamefont {Haas},
  \citenamefont {Stuhr}, \citenamefont {Cervellino}, \citenamefont {Cava},
  \citenamefont {Kenzelmann},\ and\ \citenamefont
  {Fennell}}]{PhysRevB.93.144407}%
  \BibitemOpen
  \bibfield  {author} {\bibinfo {author} {\bibfnamefont {M.}~\bibnamefont
  {Ruminy}}, \bibinfo {author} {\bibfnamefont {L.}~\bibnamefont {Bovo}},
  \bibinfo {author} {\bibfnamefont {E.}~\bibnamefont {Pomjakushina}}, \bibinfo
  {author} {\bibfnamefont {M.~K.}\ \bibnamefont {Haas}}, \bibinfo {author}
  {\bibfnamefont {U.}~\bibnamefont {Stuhr}}, \bibinfo {author} {\bibfnamefont
  {A.}~\bibnamefont {Cervellino}}, \bibinfo {author} {\bibfnamefont {R.~J.}\
  \bibnamefont {Cava}}, \bibinfo {author} {\bibfnamefont {M.}~\bibnamefont
  {Kenzelmann}}, \ and\ \bibinfo {author} {\bibfnamefont {T.}~\bibnamefont
  {Fennell}},\ }\href {\doibase 10.1103/PhysRevB.93.144407} {\bibfield
  {journal} {\bibinfo  {journal} {Phys. Rev. B}\ }\textbf {\bibinfo {volume}
  {93}},\ \bibinfo {pages} {144407} (\bibinfo {year} {2016})}\BibitemShut
  {NoStop}%
\bibitem [{\citenamefont {Gardner}\ \emph {et~al.}(2003)\citenamefont
  {Gardner}, \citenamefont {Keren}, \citenamefont {Ehlers}, \citenamefont
  {Stock}, \citenamefont {Segal}, \citenamefont {Roper}, \citenamefont
  {F\aa{}k}, \citenamefont {Stone}, \citenamefont {Hammar}, \citenamefont
  {Reich},\ and\ \citenamefont {Gaulin}}]{PhysRevB.68.180401}%
  \BibitemOpen
  \bibfield  {author} {\bibinfo {author} {\bibfnamefont {J.~S.}\ \bibnamefont
  {Gardner}}, \bibinfo {author} {\bibfnamefont {A.}~\bibnamefont {Keren}},
  \bibinfo {author} {\bibfnamefont {G.}~\bibnamefont {Ehlers}}, \bibinfo
  {author} {\bibfnamefont {C.}~\bibnamefont {Stock}}, \bibinfo {author}
  {\bibfnamefont {E.}~\bibnamefont {Segal}}, \bibinfo {author} {\bibfnamefont
  {J.~M.}\ \bibnamefont {Roper}}, \bibinfo {author} {\bibfnamefont
  {B.}~\bibnamefont {F\aa{}k}}, \bibinfo {author} {\bibfnamefont {M.~B.}\
  \bibnamefont {Stone}}, \bibinfo {author} {\bibfnamefont {P.~R.}\ \bibnamefont
  {Hammar}}, \bibinfo {author} {\bibfnamefont {D.~H.}\ \bibnamefont {Reich}}, \
  and\ \bibinfo {author} {\bibfnamefont {B.~D.}\ \bibnamefont {Gaulin}},\
  }\href {\doibase 10.1103/PhysRevB.68.180401} {\bibfield  {journal} {\bibinfo
  {journal} {Phys. Rev. B}\ }\textbf {\bibinfo {volume} {68}},\ \bibinfo
  {pages} {180401} (\bibinfo {year} {2003})}\BibitemShut {NoStop}%
\bibitem [{\citenamefont {Gingras}\ \emph {et~al.}(2000)\citenamefont
  {Gingras}, \citenamefont {den Hertog}, \citenamefont {Faucher}, \citenamefont
  {Gardner}, \citenamefont {Dunsiger}, \citenamefont {Chang}, \citenamefont
  {Gaulin}, \citenamefont {Raju},\ and\ \citenamefont
  {Greedan}}]{PhysRevB.62.6496}%
  \BibitemOpen
  \bibfield  {author} {\bibinfo {author} {\bibfnamefont {M.~J.~P.}\
  \bibnamefont {Gingras}}, \bibinfo {author} {\bibfnamefont {B.~C.}\
  \bibnamefont {den Hertog}}, \bibinfo {author} {\bibfnamefont
  {M.}~\bibnamefont {Faucher}}, \bibinfo {author} {\bibfnamefont {J.~S.}\
  \bibnamefont {Gardner}}, \bibinfo {author} {\bibfnamefont {S.~R.}\
  \bibnamefont {Dunsiger}}, \bibinfo {author} {\bibfnamefont {L.~J.}\
  \bibnamefont {Chang}}, \bibinfo {author} {\bibfnamefont {B.~D.}\ \bibnamefont
  {Gaulin}}, \bibinfo {author} {\bibfnamefont {N.~P.}\ \bibnamefont {Raju}}, \
  and\ \bibinfo {author} {\bibfnamefont {J.~E.}\ \bibnamefont {Greedan}},\
  }\href {\doibase 10.1103/PhysRevB.62.6496} {\bibfield  {journal} {\bibinfo
  {journal} {Phys. Rev. B}\ }\textbf {\bibinfo {volume} {62}},\ \bibinfo
  {pages} {6496} (\bibinfo {year} {2000})}\BibitemShut {NoStop}%
\bibitem [{\citenamefont {Chapuis}\ \emph {et~al.}(2010)\citenamefont
  {Chapuis}, \citenamefont {Yaouanc}, \citenamefont {Dalmas~de R\'eotier},
  \citenamefont {Marin}, \citenamefont {Vanishri}, \citenamefont {Curnoe},
  \citenamefont {V\^aju},\ and\ \citenamefont {Forget}}]{PhysRevB.82.100402}%
  \BibitemOpen
  \bibfield  {author} {\bibinfo {author} {\bibfnamefont {Y.}~\bibnamefont
  {Chapuis}}, \bibinfo {author} {\bibfnamefont {A.}~\bibnamefont {Yaouanc}},
  \bibinfo {author} {\bibfnamefont {P.}~\bibnamefont {Dalmas~de R\'eotier}},
  \bibinfo {author} {\bibfnamefont {C.}~\bibnamefont {Marin}}, \bibinfo
  {author} {\bibfnamefont {S.}~\bibnamefont {Vanishri}}, \bibinfo {author}
  {\bibfnamefont {S.~H.}\ \bibnamefont {Curnoe}}, \bibinfo {author}
  {\bibfnamefont {C.}~\bibnamefont {V\^aju}}, \ and\ \bibinfo {author}
  {\bibfnamefont {A.}~\bibnamefont {Forget}},\ }\href {\doibase
  10.1103/PhysRevB.82.100402} {\bibfield  {journal} {\bibinfo  {journal} {Phys.
  Rev. B}\ }\textbf {\bibinfo {volume} {82}},\ \bibinfo {pages} {100402}
  (\bibinfo {year} {2010})}\BibitemShut {NoStop}%
\bibitem [{\citenamefont {Takatsu}\ \emph {et~al.}(2011)\citenamefont
  {Takatsu}, \citenamefont {Kadowaki}, \citenamefont {Sato}, \citenamefont
  {Lynn}, \citenamefont {Tabata}, \citenamefont {Yamazaki},\ and\ \citenamefont
  {Matsuhira}}]{Takatsu_2011}%
  \BibitemOpen
  \bibfield  {author} {\bibinfo {author} {\bibfnamefont {H.}~\bibnamefont
  {Takatsu}}, \bibinfo {author} {\bibfnamefont {H.}~\bibnamefont {Kadowaki}},
  \bibinfo {author} {\bibfnamefont {T.~J.}\ \bibnamefont {Sato}}, \bibinfo
  {author} {\bibfnamefont {J.~W.}\ \bibnamefont {Lynn}}, \bibinfo {author}
  {\bibfnamefont {Y.}~\bibnamefont {Tabata}}, \bibinfo {author} {\bibfnamefont
  {T.}~\bibnamefont {Yamazaki}}, \ and\ \bibinfo {author} {\bibfnamefont
  {K.}~\bibnamefont {Matsuhira}},\ }\href {\doibase
  10.1088/0953-8984/24/5/052201} {\bibfield  {journal} {\bibinfo  {journal} {J.
  Condens. Matter Phys.}\ }\textbf {\bibinfo {volume} {24}},\ \bibinfo {pages}
  {052201} (\bibinfo {year} {2011})}\BibitemShut {NoStop}%
\bibitem [{\citenamefont {Yaouanc}\ \emph
  {et~al.}(2011{\natexlab{b}})\citenamefont {Yaouanc}, \citenamefont {Dalmas~de
  R\'eotier}, \citenamefont {Chapuis}, \citenamefont {Marin}, \citenamefont
  {Vanishri}, \citenamefont {Aoki}, \citenamefont {F\aa{}k}, \citenamefont
  {Regnault}, \citenamefont {Buisson}, \citenamefont {Amato}, \citenamefont
  {Baines},\ and\ \citenamefont {Hillier}}]{PhysRevB.84.184403}%
  \BibitemOpen
  \bibfield  {author} {\bibinfo {author} {\bibfnamefont {A.}~\bibnamefont
  {Yaouanc}}, \bibinfo {author} {\bibfnamefont {P.}~\bibnamefont {Dalmas~de
  R\'eotier}}, \bibinfo {author} {\bibfnamefont {Y.}~\bibnamefont {Chapuis}},
  \bibinfo {author} {\bibfnamefont {C.}~\bibnamefont {Marin}}, \bibinfo
  {author} {\bibfnamefont {S.}~\bibnamefont {Vanishri}}, \bibinfo {author}
  {\bibfnamefont {D.}~\bibnamefont {Aoki}}, \bibinfo {author} {\bibfnamefont
  {B.}~\bibnamefont {F\aa{}k}}, \bibinfo {author} {\bibfnamefont {L.-P.}\
  \bibnamefont {Regnault}}, \bibinfo {author} {\bibfnamefont {C.}~\bibnamefont
  {Buisson}}, \bibinfo {author} {\bibfnamefont {A.}~\bibnamefont {Amato}},
  \bibinfo {author} {\bibfnamefont {C.}~\bibnamefont {Baines}}, \ and\ \bibinfo
  {author} {\bibfnamefont {A.~D.}\ \bibnamefont {Hillier}},\ }\href {\doibase
  10.1103/PhysRevB.84.184403} {\bibfield  {journal} {\bibinfo  {journal} {Phys.
  Rev. B}\ }\textbf {\bibinfo {volume} {84}},\ \bibinfo {pages} {184403}
  (\bibinfo {year} {2011}{\natexlab{b}})}\BibitemShut {NoStop}%
\bibitem [{\citenamefont {Hamaguchi}\ \emph {et~al.}(2004)\citenamefont
  {Hamaguchi}, \citenamefont {Matsushita}, \citenamefont {Wada}, \citenamefont
  {Yasui},\ and\ \citenamefont {Sato}}]{PhysRevB.69.132413}%
  \BibitemOpen
  \bibfield  {author} {\bibinfo {author} {\bibfnamefont {N.}~\bibnamefont
  {Hamaguchi}}, \bibinfo {author} {\bibfnamefont {T.}~\bibnamefont
  {Matsushita}}, \bibinfo {author} {\bibfnamefont {N.}~\bibnamefont {Wada}},
  \bibinfo {author} {\bibfnamefont {Y.}~\bibnamefont {Yasui}}, \ and\ \bibinfo
  {author} {\bibfnamefont {M.}~\bibnamefont {Sato}},\ }\href {\doibase
  10.1103/PhysRevB.69.132413} {\bibfield  {journal} {\bibinfo  {journal} {Phys.
  Rev. B}\ }\textbf {\bibinfo {volume} {69}},\ \bibinfo {pages} {132413}
  (\bibinfo {year} {2004})}\BibitemShut {NoStop}%
\bibitem [{\citenamefont {Fennell}\ \emph {et~al.}(2014)\citenamefont
  {Fennell}, \citenamefont {Kenzelmann}, \citenamefont {Roessli}, \citenamefont
  {Mutka}, \citenamefont {Ollivier}, \citenamefont {Ruminy}, \citenamefont
  {Stuhr}, \citenamefont {Zaharko}, \citenamefont {Bovo}, \citenamefont
  {Cervellino}, \citenamefont {Haas},\ and\ \citenamefont
  {Cava}}]{PhysRevLett.112.017203}%
  \BibitemOpen
  \bibfield  {author} {\bibinfo {author} {\bibfnamefont {T.}~\bibnamefont
  {Fennell}}, \bibinfo {author} {\bibfnamefont {M.}~\bibnamefont {Kenzelmann}},
  \bibinfo {author} {\bibfnamefont {B.}~\bibnamefont {Roessli}}, \bibinfo
  {author} {\bibfnamefont {H.}~\bibnamefont {Mutka}}, \bibinfo {author}
  {\bibfnamefont {J.}~\bibnamefont {Ollivier}}, \bibinfo {author}
  {\bibfnamefont {M.}~\bibnamefont {Ruminy}}, \bibinfo {author} {\bibfnamefont
  {U.}~\bibnamefont {Stuhr}}, \bibinfo {author} {\bibfnamefont
  {O.}~\bibnamefont {Zaharko}}, \bibinfo {author} {\bibfnamefont
  {L.}~\bibnamefont {Bovo}}, \bibinfo {author} {\bibfnamefont {A.}~\bibnamefont
  {Cervellino}}, \bibinfo {author} {\bibfnamefont {M.~K.}\ \bibnamefont
  {Haas}}, \ and\ \bibinfo {author} {\bibfnamefont {R.~J.}\ \bibnamefont
  {Cava}},\ }\href {\doibase 10.1103/PhysRevLett.112.017203} {\bibfield
  {journal} {\bibinfo  {journal} {Phys. Rev. Lett.}\ }\textbf {\bibinfo
  {volume} {112}},\ \bibinfo {pages} {017203} (\bibinfo {year}
  {2014})}\BibitemShut {NoStop}%
\bibitem [{\citenamefont {Siddharthan}\ \emph {et~al.}(1999)\citenamefont
  {Siddharthan}, \citenamefont {Shastry}, \citenamefont {Ramirez},
  \citenamefont {Hayashi}, \citenamefont {Cava},\ and\ \citenamefont
  {Rosenkranz}}]{PhysRevLett.83.1854}%
  \BibitemOpen
  \bibfield  {author} {\bibinfo {author} {\bibfnamefont {R.}~\bibnamefont
  {Siddharthan}}, \bibinfo {author} {\bibfnamefont {B.~S.}\ \bibnamefont
  {Shastry}}, \bibinfo {author} {\bibfnamefont {A.~P.}\ \bibnamefont
  {Ramirez}}, \bibinfo {author} {\bibfnamefont {A.}~\bibnamefont {Hayashi}},
  \bibinfo {author} {\bibfnamefont {R.~J.}\ \bibnamefont {Cava}}, \ and\
  \bibinfo {author} {\bibfnamefont {S.}~\bibnamefont {Rosenkranz}},\ }\href
  {\doibase 10.1103/PhysRevLett.83.1854} {\bibfield  {journal} {\bibinfo
  {journal} {Phys. Rev. Lett.}\ }\textbf {\bibinfo {volume} {83}},\ \bibinfo
  {pages} {1854} (\bibinfo {year} {1999})}\BibitemShut {NoStop}%
\bibitem [{\citenamefont {Shannon}(1976)}]{Shannon:a12967}%
  \BibitemOpen
  \bibfield  {author} {\bibinfo {author} {\bibfnamefont {R.~D.}\ \bibnamefont
  {Shannon}},\ }\href {\doibase 10.1107/S0567739476001551} {\bibfield
  {journal} {\bibinfo  {journal} {Acta Cryst.}\ }\textbf {\bibinfo {volume}
  {32}},\ \bibinfo {pages} {751} (\bibinfo {year} {1976})}\BibitemShut
  {NoStop}%
\bibitem [{\citenamefont {Levin}\ \emph {et~al.}(2019)\citenamefont {Levin},
  \citenamefont {Grebenkemper}, \citenamefont {Pollock},\ and\ \citenamefont
  {Seshadri}}]{levin2019protocols}%
  \BibitemOpen
  \bibfield  {author} {\bibinfo {author} {\bibfnamefont {E.~E.}\ \bibnamefont
  {Levin}}, \bibinfo {author} {\bibfnamefont {J.~H.}\ \bibnamefont
  {Grebenkemper}}, \bibinfo {author} {\bibfnamefont {T.~M.}\ \bibnamefont
  {Pollock}}, \ and\ \bibinfo {author} {\bibfnamefont {R.}~\bibnamefont
  {Seshadri}},\ }\href {\doibase https://doi.org/10.1021/acs.chemmater.9b02594}
  {\bibfield  {journal} {\bibinfo  {journal} {Chem. Mater.}\ }\textbf {\bibinfo
  {volume} {31}},\ \bibinfo {pages} {7151} (\bibinfo {year}
  {2019})}\BibitemShut {NoStop}%
\bibitem [{\citenamefont {Michael P~{\'a}Mingos}\ and\ \citenamefont
  {Baghurst}(1991)}]{michael1991tilden}%
  \BibitemOpen
  \bibfield  {author} {\bibinfo {author} {\bibfnamefont {D.}~\bibnamefont
  {Michael P~{\'a}Mingos}}\ and\ \bibinfo {author} {\bibfnamefont {D.~R.}\
  \bibnamefont {Baghurst}},\ }\href {\doibase 10.1039/cs9912000001} {\bibfield
  {journal} {\bibinfo  {journal} {Chem. Soc. Rev.}\ }\textbf {\bibinfo {volume}
  {20}},\ \bibinfo {pages} {1} (\bibinfo {year} {1991})}\BibitemShut {NoStop}%
\bibitem [{\citenamefont {Rao}\ \emph {et~al.}(1999)\citenamefont {Rao},
  \citenamefont {Vaidhyanathan}, \citenamefont {Ganguli},\ and\ \citenamefont
  {Ramakrishnan}}]{rao1999synthesis}%
  \BibitemOpen
  \bibfield  {author} {\bibinfo {author} {\bibfnamefont {K.~J.}\ \bibnamefont
  {Rao}}, \bibinfo {author} {\bibfnamefont {B.}~\bibnamefont {Vaidhyanathan}},
  \bibinfo {author} {\bibfnamefont {M.}~\bibnamefont {Ganguli}}, \ and\
  \bibinfo {author} {\bibfnamefont {P.~A.}\ \bibnamefont {Ramakrishnan}},\
  }\href {\doibase 10.1021/cm9803859} {\bibfield  {journal} {\bibinfo
  {journal} {Chem. Mater.}\ }\textbf {\bibinfo {volume} {11}},\ \bibinfo
  {pages} {882} (\bibinfo {year} {1999})}\BibitemShut {NoStop}%
\bibitem [{\citenamefont {Mingos}(1993)}]{mingos1993microwave}%
  \BibitemOpen
  \bibfield  {author} {\bibinfo {author} {\bibfnamefont {D.~M.~P.}\
  \bibnamefont {Mingos}},\ }\href {\doibase 10.1002/adma.19930051115}
  {\bibfield  {journal} {\bibinfo  {journal} {Adv. Mater.}\ }\textbf {\bibinfo
  {volume} {5}},\ \bibinfo {pages} {857} (\bibinfo {year} {1993})}\BibitemShut
  {NoStop}%
\bibitem [{\citenamefont {Whittaker}\ and\ \citenamefont
  {Mingos}(1994)}]{whittaker1994application}%
  \BibitemOpen
  \bibfield  {author} {\bibinfo {author} {\bibfnamefont {A.~G.}\ \bibnamefont
  {Whittaker}}\ and\ \bibinfo {author} {\bibfnamefont {D.~M.}\ \bibnamefont
  {Mingos}},\ }\href {\doibase 10.1080/08327823.1994.11688249} {\bibfield
  {journal} {\bibinfo  {journal} {J. Microwave Power EE}\ }\textbf {\bibinfo
  {volume} {29}},\ \bibinfo {pages} {195} (\bibinfo {year} {1994})}\BibitemShut
  {NoStop}%
\bibitem [{\citenamefont {Maram}\ \emph {et~al.}(2018)\citenamefont {Maram},
  \citenamefont {Ushakov}, \citenamefont {Weber}, \citenamefont {Benmore},\
  and\ \citenamefont {Navrotsky}}]{maram2018probing}%
  \BibitemOpen
  \bibfield  {author} {\bibinfo {author} {\bibfnamefont {P.~S.}\ \bibnamefont
  {Maram}}, \bibinfo {author} {\bibfnamefont {S.~V.}\ \bibnamefont {Ushakov}},
  \bibinfo {author} {\bibfnamefont {R.~J.~K.}\ \bibnamefont {Weber}}, \bibinfo
  {author} {\bibfnamefont {C.~J.}\ \bibnamefont {Benmore}}, \ and\ \bibinfo
  {author} {\bibfnamefont {A.}~\bibnamefont {Navrotsky}},\ }\href {\doibase
  https://doi.org/10.1038/s41598-018-28877-x} {\bibfield  {journal} {\bibinfo
  {journal} {Sci. Rep.}\ }\textbf {\bibinfo {volume} {8}},\ \bibinfo {pages}
  {1} (\bibinfo {year} {2018})}\BibitemShut {NoStop}%
\bibitem [{\citenamefont {Ciomaga~Hatnean}\ \emph {et~al.}(2016)\citenamefont
  {Ciomaga~Hatnean}, \citenamefont {Decorse}, \citenamefont {Lees},
  \citenamefont {Petrenko},\ and\ \citenamefont {Balakrishnan}}]{cryst6070079}%
  \BibitemOpen
  \bibfield  {author} {\bibinfo {author} {\bibfnamefont {M.}~\bibnamefont
  {Ciomaga~Hatnean}}, \bibinfo {author} {\bibfnamefont {C.}~\bibnamefont
  {Decorse}}, \bibinfo {author} {\bibfnamefont {M.~R.}\ \bibnamefont {Lees}},
  \bibinfo {author} {\bibfnamefont {O.~A.}\ \bibnamefont {Petrenko}}, \ and\
  \bibinfo {author} {\bibfnamefont {G.}~\bibnamefont {Balakrishnan}},\ }\href
  {\doibase 10.3390/cryst6070079} {\bibfield  {journal} {\bibinfo  {journal}
  {Crystals}\ }\textbf {\bibinfo {volume} {6}} (\bibinfo {year} {2016}),\
  10.3390/cryst6070079}\BibitemShut {NoStop}%
\bibitem [{\citenamefont {Popov}\ \emph {et~al.}(2016)\citenamefont {Popov},
  \citenamefont {Menushenkov}, \citenamefont {Yaroslavtsev}, \citenamefont
  {Zubavichus}, \citenamefont {Gaynanov}, \citenamefont {Yastrebtsev},
  \citenamefont {Leshchev},\ and\ \citenamefont {Chernikov}}]{POPOV2016669}%
  \BibitemOpen
  \bibfield  {author} {\bibinfo {author} {\bibfnamefont {V.}~\bibnamefont
  {Popov}}, \bibinfo {author} {\bibfnamefont {A.}~\bibnamefont {Menushenkov}},
  \bibinfo {author} {\bibfnamefont {A.}~\bibnamefont {Yaroslavtsev}}, \bibinfo
  {author} {\bibfnamefont {Y.}~\bibnamefont {Zubavichus}}, \bibinfo {author}
  {\bibfnamefont {B.}~\bibnamefont {Gaynanov}}, \bibinfo {author}
  {\bibfnamefont {A.}~\bibnamefont {Yastrebtsev}}, \bibinfo {author}
  {\bibfnamefont {D.}~\bibnamefont {Leshchev}}, \ and\ \bibinfo {author}
  {\bibfnamefont {R.}~\bibnamefont {Chernikov}},\ }\href {\doibase
  https://doi.org/10.1016/j.jallcom.2016.08.019} {\bibfield  {journal}
  {\bibinfo  {journal} {J. Alloys Compd.}\ }\textbf {\bibinfo {volume} {689}},\
  \bibinfo {pages} {669} (\bibinfo {year} {2016})}\BibitemShut {NoStop}%
\bibitem [{\citenamefont {Jiang}\ \emph {et~al.}(2020)\citenamefont {Jiang},
  \citenamefont {Bridges}, \citenamefont {Unocic}, \citenamefont {Pitike},
  \citenamefont {Cooper}, \citenamefont {Zhang}, \citenamefont {Lin},\ and\
  \citenamefont {Page}}]{jiang2020probing}%
  \BibitemOpen
  \bibfield  {author} {\bibinfo {author} {\bibfnamefont {B.}~\bibnamefont
  {Jiang}}, \bibinfo {author} {\bibfnamefont {C.~A.}\ \bibnamefont {Bridges}},
  \bibinfo {author} {\bibfnamefont {R.~R.}\ \bibnamefont {Unocic}}, \bibinfo
  {author} {\bibfnamefont {K.~C.}\ \bibnamefont {Pitike}}, \bibinfo {author}
  {\bibfnamefont {V.~R.}\ \bibnamefont {Cooper}}, \bibinfo {author}
  {\bibfnamefont {Y.}~\bibnamefont {Zhang}}, \bibinfo {author} {\bibfnamefont
  {D.-Y.}\ \bibnamefont {Lin}}, \ and\ \bibinfo {author} {\bibfnamefont
  {K.}~\bibnamefont {Page}},\ }\href {\doibase
  https://doi.org/10.1021/jacs.0c10739} {\bibfield  {journal} {\bibinfo
  {journal} {Journal of the American Chemical Society}\ }\textbf {\bibinfo
  {volume} {143}},\ \bibinfo {pages} {4193} (\bibinfo {year}
  {2020})}\BibitemShut {NoStop}%
\bibitem [{\citenamefont {Bramwell}\ \emph {et~al.}(1999)\citenamefont
  {Bramwell}, \citenamefont {Field}, \citenamefont {Harris},\ and\
  \citenamefont {Parkin}}]{Bramwell99:12}%
  \BibitemOpen
  \bibfield  {author} {\bibinfo {author} {\bibfnamefont {S.~T.}\ \bibnamefont
  {Bramwell}}, \bibinfo {author} {\bibfnamefont {M.~N.}\ \bibnamefont {Field}},
  \bibinfo {author} {\bibfnamefont {M.~J.}\ \bibnamefont {Harris}}, \ and\
  \bibinfo {author} {\bibfnamefont {I.~P.}\ \bibnamefont {Parkin}},\ }\href
  {\doibase 10.1088/0953-8984/12/4/308} {\bibfield  {journal} {\bibinfo
  {journal} {J. Phys. Condens. Matter}\ }\textbf {\bibinfo {volume} {12}},\
  \bibinfo {pages} {483} (\bibinfo {year} {1999})}\BibitemShut {NoStop}%
\bibitem [{\citenamefont {Scheie}\ \emph {et~al.}(2021)\citenamefont {Scheie},
  \citenamefont {Sanders}, \citenamefont {Gui}, \citenamefont {Qiu},
  \citenamefont {Prisk}, \citenamefont {Cava},\ and\ \citenamefont
  {Broholm}}]{Scheie21:104}%
  \BibitemOpen
  \bibfield  {author} {\bibinfo {author} {\bibfnamefont {A.}~\bibnamefont
  {Scheie}}, \bibinfo {author} {\bibfnamefont {M.}~\bibnamefont {Sanders}},
  \bibinfo {author} {\bibfnamefont {X.}~\bibnamefont {Gui}}, \bibinfo {author}
  {\bibfnamefont {Y.}~\bibnamefont {Qiu}}, \bibinfo {author} {\bibfnamefont
  {T.~R.}\ \bibnamefont {Prisk}}, \bibinfo {author} {\bibfnamefont {R.~J.}\
  \bibnamefont {Cava}}, \ and\ \bibinfo {author} {\bibfnamefont
  {C.}~\bibnamefont {Broholm}},\ }\href {\doibase 10.1103/PhysRevB.104.134418}
  {\bibfield  {journal} {\bibinfo  {journal} {Phys. Rev. B}\ }\textbf {\bibinfo
  {volume} {104}},\ \bibinfo {pages} {134418} (\bibinfo {year}
  {2021})}\BibitemShut {NoStop}%
\bibitem [{\citenamefont {Bertin}\ \emph {et~al.}(2015)\citenamefont {Bertin},
  \citenamefont {Dalmas~de R\'eotier}, \citenamefont {F\aa{}k}, \citenamefont
  {Marin}, \citenamefont {Yaouanc}, \citenamefont {Forget}, \citenamefont
  {Sheptyakov}, \citenamefont {Frick}, \citenamefont {Ritter}, \citenamefont
  {Amato}, \citenamefont {Baines},\ and\ \citenamefont {King}}]{Bertin15:92}%
  \BibitemOpen
  \bibfield  {author} {\bibinfo {author} {\bibfnamefont {A.}~\bibnamefont
  {Bertin}}, \bibinfo {author} {\bibfnamefont {P.}~\bibnamefont {Dalmas~de
  R\'eotier}}, \bibinfo {author} {\bibfnamefont {B.}~\bibnamefont {F\aa{}k}},
  \bibinfo {author} {\bibfnamefont {C.}~\bibnamefont {Marin}}, \bibinfo
  {author} {\bibfnamefont {A.}~\bibnamefont {Yaouanc}}, \bibinfo {author}
  {\bibfnamefont {A.}~\bibnamefont {Forget}}, \bibinfo {author} {\bibfnamefont
  {D.}~\bibnamefont {Sheptyakov}}, \bibinfo {author} {\bibfnamefont
  {B.}~\bibnamefont {Frick}}, \bibinfo {author} {\bibfnamefont
  {C.}~\bibnamefont {Ritter}}, \bibinfo {author} {\bibfnamefont
  {A.}~\bibnamefont {Amato}}, \bibinfo {author} {\bibfnamefont
  {C.}~\bibnamefont {Baines}}, \ and\ \bibinfo {author} {\bibfnamefont
  {P.~J.~C.}\ \bibnamefont {King}},\ }\href {\doibase
  10.1103/PhysRevB.92.144423} {\bibfield  {journal} {\bibinfo  {journal} {Phys.
  Rev. B}\ }\textbf {\bibinfo {volume} {92}},\ \bibinfo {pages} {144423}
  (\bibinfo {year} {2015})}\BibitemShut {NoStop}%
\bibitem [{\citenamefont {Xu}\ \emph {et~al.}(2015)\citenamefont {Xu},
  \citenamefont {Anand}, \citenamefont {Bera}, \citenamefont {Frontzek},
  \citenamefont {Abernathy}, \citenamefont {Casati}, \citenamefont
  {Siemensmeyer},\ and\ \citenamefont {Lake}}]{xu15:92}%
  \BibitemOpen
  \bibfield  {author} {\bibinfo {author} {\bibfnamefont {J.}~\bibnamefont
  {Xu}}, \bibinfo {author} {\bibfnamefont {V.~K.}\ \bibnamefont {Anand}},
  \bibinfo {author} {\bibfnamefont {A.~K.}\ \bibnamefont {Bera}}, \bibinfo
  {author} {\bibfnamefont {M.}~\bibnamefont {Frontzek}}, \bibinfo {author}
  {\bibfnamefont {D.~L.}\ \bibnamefont {Abernathy}}, \bibinfo {author}
  {\bibfnamefont {N.}~\bibnamefont {Casati}}, \bibinfo {author} {\bibfnamefont
  {K.}~\bibnamefont {Siemensmeyer}}, \ and\ \bibinfo {author} {\bibfnamefont
  {B.}~\bibnamefont {Lake}},\ }\href {\doibase 10.1103/PhysRevB.92.224430}
  {\bibfield  {journal} {\bibinfo  {journal} {Phys. Rev. B}\ }\textbf {\bibinfo
  {volume} {92}},\ \bibinfo {pages} {224430} (\bibinfo {year}
  {2015})}\BibitemShut {NoStop}%
\bibitem [{\citenamefont {Lhotel}\ \emph {et~al.}(2015)\citenamefont {Lhotel},
  \citenamefont {Petit}, \citenamefont {Guitteny}, \citenamefont {Florea},
  \citenamefont {Ciomaga~Hatnean}, \citenamefont {Colin}, \citenamefont
  {Ressouche}, \citenamefont {Lees},\ and\ \citenamefont
  {Balakrishnan}}]{Lhotel15:115}%
  \BibitemOpen
  \bibfield  {author} {\bibinfo {author} {\bibfnamefont {E.}~\bibnamefont
  {Lhotel}}, \bibinfo {author} {\bibfnamefont {S.}~\bibnamefont {Petit}},
  \bibinfo {author} {\bibfnamefont {S.}~\bibnamefont {Guitteny}}, \bibinfo
  {author} {\bibfnamefont {O.}~\bibnamefont {Florea}}, \bibinfo {author}
  {\bibfnamefont {M.}~\bibnamefont {Ciomaga~Hatnean}}, \bibinfo {author}
  {\bibfnamefont {C.}~\bibnamefont {Colin}}, \bibinfo {author} {\bibfnamefont
  {E.}~\bibnamefont {Ressouche}}, \bibinfo {author} {\bibfnamefont {M.~R.}\
  \bibnamefont {Lees}}, \ and\ \bibinfo {author} {\bibfnamefont
  {G.}~\bibnamefont {Balakrishnan}},\ }\href {\doibase
  10.1103/PhysRevLett.115.197202} {\bibfield  {journal} {\bibinfo  {journal}
  {Phys. Rev. Lett.}\ }\textbf {\bibinfo {volume} {115}},\ \bibinfo {pages}
  {197202} (\bibinfo {year} {2015})}\BibitemShut {NoStop}%
\bibitem [{\citenamefont {Anand}\ \emph {et~al.}(2017)\citenamefont {Anand},
  \citenamefont {Abernathy}, \citenamefont {Adroja}, \citenamefont {Hillier},
  \citenamefont {Biswas},\ and\ \citenamefont {Lake}}]{Anand17:95}%
  \BibitemOpen
  \bibfield  {author} {\bibinfo {author} {\bibfnamefont {V.~K.}\ \bibnamefont
  {Anand}}, \bibinfo {author} {\bibfnamefont {D.~L.}\ \bibnamefont
  {Abernathy}}, \bibinfo {author} {\bibfnamefont {D.~T.}\ \bibnamefont
  {Adroja}}, \bibinfo {author} {\bibfnamefont {A.~D.}\ \bibnamefont {Hillier}},
  \bibinfo {author} {\bibfnamefont {P.~K.}\ \bibnamefont {Biswas}}, \ and\
  \bibinfo {author} {\bibfnamefont {B.}~\bibnamefont {Lake}},\ }\href {\doibase
  10.1103/PhysRevB.95.224420} {\bibfield  {journal} {\bibinfo  {journal} {Phys.
  Rev. B}\ }\textbf {\bibinfo {volume} {95}},\ \bibinfo {pages} {224420}
  (\bibinfo {year} {2017})}\BibitemShut {NoStop}%
\bibitem [{\citenamefont {Anand}\ \emph {et~al.}(2015)\citenamefont {Anand},
  \citenamefont {Bera}, \citenamefont {Xu}, \citenamefont {Herrmannsd\"orfer},
  \citenamefont {Ritter},\ and\ \citenamefont {Lake}}]{Anand15:92}%
  \BibitemOpen
  \bibfield  {author} {\bibinfo {author} {\bibfnamefont {V.~K.}\ \bibnamefont
  {Anand}}, \bibinfo {author} {\bibfnamefont {A.~K.}\ \bibnamefont {Bera}},
  \bibinfo {author} {\bibfnamefont {J.}~\bibnamefont {Xu}}, \bibinfo {author}
  {\bibfnamefont {T.}~\bibnamefont {Herrmannsd\"orfer}}, \bibinfo {author}
  {\bibfnamefont {C.}~\bibnamefont {Ritter}}, \ and\ \bibinfo {author}
  {\bibfnamefont {B.}~\bibnamefont {Lake}},\ }\href {\doibase
  10.1103/PhysRevB.92.184418} {\bibfield  {journal} {\bibinfo  {journal} {Phys.
  Rev. B}\ }\textbf {\bibinfo {volume} {92}},\ \bibinfo {pages} {184418}
  (\bibinfo {year} {2015})}\BibitemShut {NoStop}%
\bibitem [{\citenamefont {Matsuhira}\ \emph {et~al.}(2013)\citenamefont
  {Matsuhira}, \citenamefont {Tokunaga}, \citenamefont {Wakeshima},
  \citenamefont {Hinatsu},\ and\ \citenamefont {Takagi}}]{matsuhira13:82}%
  \BibitemOpen
  \bibfield  {author} {\bibinfo {author} {\bibfnamefont {K.}~\bibnamefont
  {Matsuhira}}, \bibinfo {author} {\bibfnamefont {M.}~\bibnamefont {Tokunaga}},
  \bibinfo {author} {\bibfnamefont {M.}~\bibnamefont {Wakeshima}}, \bibinfo
  {author} {\bibfnamefont {Y.}~\bibnamefont {Hinatsu}}, \ and\ \bibinfo
  {author} {\bibfnamefont {S.}~\bibnamefont {Takagi}},\ }\href {\doibase
  10.7566/JPSJ.82.023706} {\bibfield  {journal} {\bibinfo  {journal} {J. Phys.
  Soc. Jpn.}\ }\textbf {\bibinfo {volume} {82}},\ \bibinfo {pages} {023706}
  (\bibinfo {year} {2013})}\BibitemShut {NoStop}%
\bibitem [{\citenamefont {Tomiyasu}\ \emph {et~al.}(2012)\citenamefont
  {Tomiyasu}, \citenamefont {Matsuhira}, \citenamefont {Iwasa}, \citenamefont
  {Watahiki}, \citenamefont {Takagi}, \citenamefont {Wakeshima}, \citenamefont
  {Hinatsu}, \citenamefont {Yokoyama}, \citenamefont {Ohoyama},\ and\
  \citenamefont {Yamada}}]{tomiyasu12:81}%
  \BibitemOpen
  \bibfield  {author} {\bibinfo {author} {\bibfnamefont {K.}~\bibnamefont
  {Tomiyasu}}, \bibinfo {author} {\bibfnamefont {K.}~\bibnamefont {Matsuhira}},
  \bibinfo {author} {\bibfnamefont {K.}~\bibnamefont {Iwasa}}, \bibinfo
  {author} {\bibfnamefont {M.}~\bibnamefont {Watahiki}}, \bibinfo {author}
  {\bibfnamefont {S.}~\bibnamefont {Takagi}}, \bibinfo {author} {\bibfnamefont
  {M.}~\bibnamefont {Wakeshima}}, \bibinfo {author} {\bibfnamefont
  {Y.}~\bibnamefont {Hinatsu}}, \bibinfo {author} {\bibfnamefont
  {M.}~\bibnamefont {Yokoyama}}, \bibinfo {author} {\bibfnamefont
  {K.}~\bibnamefont {Ohoyama}}, \ and\ \bibinfo {author} {\bibfnamefont
  {K.}~\bibnamefont {Yamada}},\ }\href {\doibase 10.1143/JPSJ.81.034709}
  {\bibfield  {journal} {\bibinfo  {journal} {J. Phys. Soc. Jpn.}\ }\textbf
  {\bibinfo {volume} {81}},\ \bibinfo {pages} {034709} (\bibinfo {year}
  {2012})}\BibitemShut {NoStop}%
\bibitem [{\citenamefont {Yasui}\ \emph {et~al.}(2001)\citenamefont {Yasui},
  \citenamefont {Kondo}, \citenamefont {Kanada}, \citenamefont {Ito},
  \citenamefont {Harashina}, \citenamefont {Sato},\ and\ \citenamefont
  {Kakurai}}]{yasui01:70}%
  \BibitemOpen
  \bibfield  {author} {\bibinfo {author} {\bibfnamefont {Y.}~\bibnamefont
  {Yasui}}, \bibinfo {author} {\bibfnamefont {Y.}~\bibnamefont {Kondo}},
  \bibinfo {author} {\bibfnamefont {M.}~\bibnamefont {Kanada}}, \bibinfo
  {author} {\bibfnamefont {M.}~\bibnamefont {Ito}}, \bibinfo {author}
  {\bibfnamefont {H.}~\bibnamefont {Harashina}}, \bibinfo {author}
  {\bibfnamefont {M.}~\bibnamefont {Sato}}, \ and\ \bibinfo {author}
  {\bibfnamefont {K.}~\bibnamefont {Kakurai}},\ }\href {\doibase
  10.1143/JPSJ.70.284} {\bibfield  {journal} {\bibinfo  {journal} {J. Phys.
  Soc. Jpn.}\ }\textbf {\bibinfo {volume} {70}},\ \bibinfo {pages} {284}
  (\bibinfo {year} {2001})}\BibitemShut {NoStop}%
\bibitem [{\citenamefont {Ku}\ \emph {et~al.}(2018)\citenamefont {Ku},
  \citenamefont {Kumar}, \citenamefont {Lees}, \citenamefont {Lee},
  \citenamefont {Aldus}, \citenamefont {Studer}, \citenamefont {Imperia},
  \citenamefont {Asai}, \citenamefont {Masuda}, \citenamefont {Chen},
  \citenamefont {Chen},\ and\ \citenamefont {Chang}}]{Ku18:30}%
  \BibitemOpen
  \bibfield  {author} {\bibinfo {author} {\bibfnamefont {S.~T.}\ \bibnamefont
  {Ku}}, \bibinfo {author} {\bibfnamefont {D.}~\bibnamefont {Kumar}}, \bibinfo
  {author} {\bibfnamefont {M.~R.}\ \bibnamefont {Lees}}, \bibinfo {author}
  {\bibfnamefont {W.-T.}\ \bibnamefont {Lee}}, \bibinfo {author} {\bibfnamefont
  {R.}~\bibnamefont {Aldus}}, \bibinfo {author} {\bibfnamefont
  {A.}~\bibnamefont {Studer}}, \bibinfo {author} {\bibfnamefont
  {P.}~\bibnamefont {Imperia}}, \bibinfo {author} {\bibfnamefont
  {S.}~\bibnamefont {Asai}}, \bibinfo {author} {\bibfnamefont {T.}~\bibnamefont
  {Masuda}}, \bibinfo {author} {\bibfnamefont {S.~W.}\ \bibnamefont {Chen}},
  \bibinfo {author} {\bibfnamefont {J.~M.}\ \bibnamefont {Chen}}, \ and\
  \bibinfo {author} {\bibfnamefont {L.~J.}\ \bibnamefont {Chang}},\ }\href
  {\doibase 10.1088/1361-648x/aab398} {\bibfield  {journal} {\bibinfo
  {journal} {J. Phys.: Condens. Matter}\ }\textbf {\bibinfo {volume} {30}},\
  \bibinfo {pages} {155601} (\bibinfo {year} {2018})}\BibitemShut {NoStop}%
\bibitem [{\citenamefont {Brooks-Bartlett}\ \emph {et~al.}(2014)\citenamefont
  {Brooks-Bartlett}, \citenamefont {Banks}, \citenamefont {Jaubert},
  \citenamefont {Harman-Clarke},\ and\ \citenamefont
  {Holdsworth}}]{brooks14:4}%
  \BibitemOpen
  \bibfield  {author} {\bibinfo {author} {\bibfnamefont {M.~E.}\ \bibnamefont
  {Brooks-Bartlett}}, \bibinfo {author} {\bibfnamefont {S.~T.}\ \bibnamefont
  {Banks}}, \bibinfo {author} {\bibfnamefont {L.~D.~C.}\ \bibnamefont
  {Jaubert}}, \bibinfo {author} {\bibfnamefont {A.}~\bibnamefont
  {Harman-Clarke}}, \ and\ \bibinfo {author} {\bibfnamefont {P.~C.~W.}\
  \bibnamefont {Holdsworth}},\ }\href {\doibase 10.1103/PhysRevX.4.011007}
  {\bibfield  {journal} {\bibinfo  {journal} {Phys. Rev. X}\ }\textbf {\bibinfo
  {volume} {4}},\ \bibinfo {pages} {011007} (\bibinfo {year}
  {2014})}\BibitemShut {NoStop}%
\bibitem [{\citenamefont {Benton}(2016)}]{Benton16:94}%
  \BibitemOpen
  \bibfield  {author} {\bibinfo {author} {\bibfnamefont {O.}~\bibnamefont
  {Benton}},\ }\href {\doibase 10.1103/PhysRevB.94.104430} {\bibfield
  {journal} {\bibinfo  {journal} {Phys. Rev. B}\ }\textbf {\bibinfo {volume}
  {94}},\ \bibinfo {pages} {104430} (\bibinfo {year} {2016})}\BibitemShut
  {NoStop}%
\bibitem [{\citenamefont {Sibille}\ \emph {et~al.}(2018)\citenamefont
  {Sibille}, \citenamefont {Gauthier}, \citenamefont {Yan}, \citenamefont
  {Ciomaga~Hatnean}, \citenamefont {Ollivier}, \citenamefont {Winn},
  \citenamefont {Filges}, \citenamefont {Balakrishnan}, \citenamefont
  {Kenzelmann}, \citenamefont {Shannon},\ and\ \citenamefont
  {Fennell}}]{PrHf_sibille2018experimental}%
  \BibitemOpen
  \bibfield  {author} {\bibinfo {author} {\bibfnamefont {R.}~\bibnamefont
  {Sibille}}, \bibinfo {author} {\bibfnamefont {N.}~\bibnamefont {Gauthier}},
  \bibinfo {author} {\bibfnamefont {H.}~\bibnamefont {Yan}}, \bibinfo {author}
  {\bibfnamefont {M.}~\bibnamefont {Ciomaga~Hatnean}}, \bibinfo {author}
  {\bibfnamefont {J.}~\bibnamefont {Ollivier}}, \bibinfo {author}
  {\bibfnamefont {B.}~\bibnamefont {Winn}}, \bibinfo {author} {\bibfnamefont
  {U.}~\bibnamefont {Filges}}, \bibinfo {author} {\bibfnamefont
  {G.}~\bibnamefont {Balakrishnan}}, \bibinfo {author} {\bibfnamefont
  {M.}~\bibnamefont {Kenzelmann}}, \bibinfo {author} {\bibfnamefont
  {N.}~\bibnamefont {Shannon}}, \ and\ \bibinfo {author} {\bibfnamefont
  {T.}~\bibnamefont {Fennell}},\ }\href {\doibase 10.1038/s41567-018-0116-x}
  {\bibfield  {journal} {\bibinfo  {journal} {Nat. Phys.}\ }\textbf {\bibinfo
  {volume} {14}},\ \bibinfo {pages} {711} (\bibinfo {year} {2018})}\BibitemShut
  {NoStop}%
\bibitem [{\citenamefont {Petit}\ \emph
  {et~al.}(2016{\natexlab{a}})\citenamefont {Petit}, \citenamefont {Lhotel},
  \citenamefont {Canals}, \citenamefont {Ciomaga~Hatnean}, \citenamefont
  {Ollivier}, \citenamefont {Mutka}, \citenamefont {Ressouche}, \citenamefont
  {Wildes}, \citenamefont {Lees},\ and\ \citenamefont
  {Balakrishnan}}]{petit16:12}%
  \BibitemOpen
  \bibfield  {author} {\bibinfo {author} {\bibfnamefont {S.}~\bibnamefont
  {Petit}}, \bibinfo {author} {\bibfnamefont {E.}~\bibnamefont {Lhotel}},
  \bibinfo {author} {\bibfnamefont {B.}~\bibnamefont {Canals}}, \bibinfo
  {author} {\bibfnamefont {M.}~\bibnamefont {Ciomaga~Hatnean}}, \bibinfo
  {author} {\bibfnamefont {J.}~\bibnamefont {Ollivier}}, \bibinfo {author}
  {\bibfnamefont {H.}~\bibnamefont {Mutka}}, \bibinfo {author} {\bibfnamefont
  {E.}~\bibnamefont {Ressouche}}, \bibinfo {author} {\bibfnamefont {A.~R.}\
  \bibnamefont {Wildes}}, \bibinfo {author} {\bibfnamefont {M.~R.}\
  \bibnamefont {Lees}}, \ and\ \bibinfo {author} {\bibfnamefont
  {G.}~\bibnamefont {Balakrishnan}},\ }\href {\doibase
  https://doi.org/10.1038/nphys3710} {\bibfield  {journal} {\bibinfo  {journal}
  {Nat. Phys.}\ }\textbf {\bibinfo {volume} {12}},\ \bibinfo {pages} {746}
  (\bibinfo {year} {2016}{\natexlab{a}})}\BibitemShut {NoStop}%
\bibitem [{\citenamefont {Zoghlin}\ \emph {et~al.}(2021)\citenamefont
  {Zoghlin}, \citenamefont {Schmehr}, \citenamefont {Holgate}, \citenamefont
  {Dally}, \citenamefont {Liu}, \citenamefont {Laurita},\ and\ \citenamefont
  {Wilson}}]{zoghlin21:5}%
  \BibitemOpen
  \bibfield  {author} {\bibinfo {author} {\bibfnamefont {E.}~\bibnamefont
  {Zoghlin}}, \bibinfo {author} {\bibfnamefont {J.}~\bibnamefont {Schmehr}},
  \bibinfo {author} {\bibfnamefont {C.}~\bibnamefont {Holgate}}, \bibinfo
  {author} {\bibfnamefont {R.}~\bibnamefont {Dally}}, \bibinfo {author}
  {\bibfnamefont {Y.}~\bibnamefont {Liu}}, \bibinfo {author} {\bibfnamefont
  {G.}~\bibnamefont {Laurita}}, \ and\ \bibinfo {author} {\bibfnamefont
  {S.~D.}\ \bibnamefont {Wilson}},\ }\href {\doibase
  10.1103/PhysRevMaterials.5.084403} {\bibfield  {journal} {\bibinfo  {journal}
  {Phys. Rev. Materials}\ }\textbf {\bibinfo {volume} {5}},\ \bibinfo {pages}
  {084403} (\bibinfo {year} {2021})}\BibitemShut {NoStop}%
\bibitem [{\citenamefont {Xu}\ \emph {et~al.}(2020)\citenamefont {Xu},
  \citenamefont {Benton}, \citenamefont {Islam}, \citenamefont {Guidi},
  \citenamefont {Ehlers},\ and\ \citenamefont {Lake}}]{Xu20:124}%
  \BibitemOpen
  \bibfield  {author} {\bibinfo {author} {\bibfnamefont {J.}~\bibnamefont
  {Xu}}, \bibinfo {author} {\bibfnamefont {O.}~\bibnamefont {Benton}}, \bibinfo
  {author} {\bibfnamefont {A.~T. M.~N.}\ \bibnamefont {Islam}}, \bibinfo
  {author} {\bibfnamefont {T.}~\bibnamefont {Guidi}}, \bibinfo {author}
  {\bibfnamefont {G.}~\bibnamefont {Ehlers}}, \ and\ \bibinfo {author}
  {\bibfnamefont {B.}~\bibnamefont {Lake}},\ }\href {\doibase
  10.1103/PhysRevLett.124.097203} {\bibfield  {journal} {\bibinfo  {journal}
  {Phys. Rev. Lett.}\ }\textbf {\bibinfo {volume} {124}},\ \bibinfo {pages}
  {097203} (\bibinfo {year} {2020})}\BibitemShut {NoStop}%
\bibitem [{\citenamefont {Kimura}\ \emph {et~al.}(2013)\citenamefont {Kimura},
  \citenamefont {Nakatsuji}, \citenamefont {Wen}, \citenamefont {Broholm},
  \citenamefont {Stone}, \citenamefont {Nishibori},\ and\ \citenamefont
  {Sawa}}]{kimura13:4}%
  \BibitemOpen
  \bibfield  {author} {\bibinfo {author} {\bibfnamefont {K.}~\bibnamefont
  {Kimura}}, \bibinfo {author} {\bibfnamefont {S.}~\bibnamefont {Nakatsuji}},
  \bibinfo {author} {\bibfnamefont {J.}~\bibnamefont {Wen}}, \bibinfo {author}
  {\bibfnamefont {C.}~\bibnamefont {Broholm}}, \bibinfo {author} {\bibfnamefont
  {M.~B.}\ \bibnamefont {Stone}}, \bibinfo {author} {\bibfnamefont
  {E.}~\bibnamefont {Nishibori}}, \ and\ \bibinfo {author} {\bibfnamefont
  {H.}~\bibnamefont {Sawa}},\ }\href {\doibase
  https://doi.org/10.1038/ncomms2914} {\bibfield  {journal} {\bibinfo
  {journal} {Nat. Commun.}\ }\textbf {\bibinfo {volume} {4}},\ \bibinfo {pages}
  {1934} (\bibinfo {year} {2013})}\BibitemShut {NoStop}%
\bibitem [{\citenamefont {Jensen}\ and\ \citenamefont
  {Mackintosh}(1991)}]{jensen1991}%
  \BibitemOpen
  \bibfield  {author} {\bibinfo {author} {\bibfnamefont {J.}~\bibnamefont
  {Jensen}}\ and\ \bibinfo {author} {\bibfnamefont {A.~R.}\ \bibnamefont
  {Mackintosh}},\ }\href@noop {} {\emph {\bibinfo {title} {Rare earth
  magnetism}}}\ (\bibinfo  {publisher} {Clarendon Press Oxford},\ \bibinfo
  {address} {Oxford},\ \bibinfo {year} {1991})\BibitemShut {NoStop}%
\bibitem [{\citenamefont {Cashion}\ \emph {et~al.}(1968)\citenamefont
  {Cashion}, \citenamefont {Cooke}, \citenamefont {Leask}, \citenamefont
  {Thorp},\ and\ \citenamefont {Wells}}]{cashion1968crystal}%
  \BibitemOpen
  \bibfield  {author} {\bibinfo {author} {\bibfnamefont {J.~D.}\ \bibnamefont
  {Cashion}}, \bibinfo {author} {\bibfnamefont {A.~H.}\ \bibnamefont {Cooke}},
  \bibinfo {author} {\bibfnamefont {M.~J.~M.}\ \bibnamefont {Leask}}, \bibinfo
  {author} {\bibfnamefont {T.~L.}\ \bibnamefont {Thorp}}, \ and\ \bibinfo
  {author} {\bibfnamefont {M.~R.}\ \bibnamefont {Wells}},\ }\href {\doibase
  https://doi.org/10.1007/BF00550984} {\bibfield  {journal} {\bibinfo
  {journal} {J. Mater. Sci.}\ }\textbf {\bibinfo {volume} {3}},\ \bibinfo
  {pages} {402} (\bibinfo {year} {1968})}\BibitemShut {NoStop}%
\bibitem [{\citenamefont {Glazkov}\ \emph {et~al.}(2005)\citenamefont
  {Glazkov}, \citenamefont {Zhitomirsky}, \citenamefont {Smirnov},
  \citenamefont {Krug~von Nidda}, \citenamefont {Loidl}, \citenamefont
  {Marin},\ and\ \citenamefont {Sanchez}}]{PhysRevB.72.020409}%
  \BibitemOpen
  \bibfield  {author} {\bibinfo {author} {\bibfnamefont {V.~N.}\ \bibnamefont
  {Glazkov}}, \bibinfo {author} {\bibfnamefont {M.~E.}\ \bibnamefont
  {Zhitomirsky}}, \bibinfo {author} {\bibfnamefont {A.~I.}\ \bibnamefont
  {Smirnov}}, \bibinfo {author} {\bibfnamefont {H.-A.}\ \bibnamefont {Krug~von
  Nidda}}, \bibinfo {author} {\bibfnamefont {A.}~\bibnamefont {Loidl}},
  \bibinfo {author} {\bibfnamefont {C.}~\bibnamefont {Marin}}, \ and\ \bibinfo
  {author} {\bibfnamefont {J.-P.}\ \bibnamefont {Sanchez}},\ }\href {\doibase
  10.1103/PhysRevB.72.020409} {\bibfield  {journal} {\bibinfo  {journal} {Phys.
  Rev. B}\ }\textbf {\bibinfo {volume} {72}},\ \bibinfo {pages} {020409}
  (\bibinfo {year} {2005})}\BibitemShut {NoStop}%
\bibitem [{\citenamefont {Canals}\ and\ \citenamefont
  {Lacroix}(1998)}]{canals98:80}%
  \BibitemOpen
  \bibfield  {author} {\bibinfo {author} {\bibfnamefont {B.}~\bibnamefont
  {Canals}}\ and\ \bibinfo {author} {\bibfnamefont {C.}~\bibnamefont
  {Lacroix}},\ }\href {\doibase 10.1103/PhysRevLett.80.2933} {\bibfield
  {journal} {\bibinfo  {journal} {Phys. Rev. Lett.}\ }\textbf {\bibinfo
  {volume} {80}},\ \bibinfo {pages} {2933} (\bibinfo {year}
  {1998})}\BibitemShut {NoStop}%
\bibitem [{\citenamefont {Canals}\ and\ \citenamefont
  {Lacroix}(2000)}]{canals00:61}%
  \BibitemOpen
  \bibfield  {author} {\bibinfo {author} {\bibfnamefont {B.}~\bibnamefont
  {Canals}}\ and\ \bibinfo {author} {\bibfnamefont {C.}~\bibnamefont
  {Lacroix}},\ }\href {\doibase 10.1103/PhysRevB.61.1149} {\bibfield  {journal}
  {\bibinfo  {journal} {Phys. Rev. B}\ }\textbf {\bibinfo {volume} {61}},\
  \bibinfo {pages} {1149} (\bibinfo {year} {2000})}\BibitemShut {NoStop}%
\bibitem [{\citenamefont {Moessner}\ and\ \citenamefont
  {Chalker}(1998{\natexlab{a}})}]{Moessner98:80}%
  \BibitemOpen
  \bibfield  {author} {\bibinfo {author} {\bibfnamefont {R.}~\bibnamefont
  {Moessner}}\ and\ \bibinfo {author} {\bibfnamefont {J.~T.}\ \bibnamefont
  {Chalker}},\ }\href {\doibase 10.1103/PhysRevLett.80.2929} {\bibfield
  {journal} {\bibinfo  {journal} {Phys. Rev. Lett.}\ }\textbf {\bibinfo
  {volume} {80}},\ \bibinfo {pages} {2929} (\bibinfo {year}
  {1998}{\natexlab{a}})}\BibitemShut {NoStop}%
\bibitem [{\citenamefont {Moessner}\ and\ \citenamefont
  {Chalker}(1998{\natexlab{b}})}]{Moessner98:58}%
  \BibitemOpen
  \bibfield  {author} {\bibinfo {author} {\bibfnamefont {R.}~\bibnamefont
  {Moessner}}\ and\ \bibinfo {author} {\bibfnamefont {J.~T.}\ \bibnamefont
  {Chalker}},\ }\href {\doibase 10.1103/PhysRevB.58.12049} {\bibfield
  {journal} {\bibinfo  {journal} {Phys. Rev. B}\ }\textbf {\bibinfo {volume}
  {58}},\ \bibinfo {pages} {12049} (\bibinfo {year}
  {1998}{\natexlab{b}})}\BibitemShut {NoStop}%
\bibitem [{\citenamefont {Nawa}\ \emph {et~al.}(2018)\citenamefont {Nawa},
  \citenamefont {Okuyama}, \citenamefont {Avdeev}, \citenamefont {Nojiri},
  \citenamefont {Yoshida}, \citenamefont {Ueta}, \citenamefont {Yoshizawa},\
  and\ \citenamefont {Sato}}]{nawa18:98}%
  \BibitemOpen
  \bibfield  {author} {\bibinfo {author} {\bibfnamefont {K.}~\bibnamefont
  {Nawa}}, \bibinfo {author} {\bibfnamefont {D.}~\bibnamefont {Okuyama}},
  \bibinfo {author} {\bibfnamefont {M.}~\bibnamefont {Avdeev}}, \bibinfo
  {author} {\bibfnamefont {H.}~\bibnamefont {Nojiri}}, \bibinfo {author}
  {\bibfnamefont {M.}~\bibnamefont {Yoshida}}, \bibinfo {author} {\bibfnamefont
  {D.}~\bibnamefont {Ueta}}, \bibinfo {author} {\bibfnamefont {H.}~\bibnamefont
  {Yoshizawa}}, \ and\ \bibinfo {author} {\bibfnamefont {T.~J.}\ \bibnamefont
  {Sato}},\ }\href {\doibase 10.1103/PhysRevB.98.144426} {\bibfield  {journal}
  {\bibinfo  {journal} {Phys. Rev. B}\ }\textbf {\bibinfo {volume} {98}},\
  \bibinfo {pages} {144426} (\bibinfo {year} {2018})}\BibitemShut {NoStop}%
\bibitem [{\citenamefont {Reimers}(1992)}]{Reimers92:45}%
  \BibitemOpen
  \bibfield  {author} {\bibinfo {author} {\bibfnamefont {J.~N.}\ \bibnamefont
  {Reimers}},\ }\href {\doibase 10.1103/PhysRevB.45.7287} {\bibfield  {journal}
  {\bibinfo  {journal} {Phys. Rev. B}\ }\textbf {\bibinfo {volume} {45}},\
  \bibinfo {pages} {7287} (\bibinfo {year} {1992})}\BibitemShut {NoStop}%
\bibitem [{\citenamefont {Reimers}\ \emph
  {et~al.}(1991{\natexlab{b}})\citenamefont {Reimers}, \citenamefont
  {Berlinsky},\ and\ \citenamefont {Shi}}]{reimers91:43}%
  \BibitemOpen
  \bibfield  {author} {\bibinfo {author} {\bibfnamefont {J.~N.}\ \bibnamefont
  {Reimers}}, \bibinfo {author} {\bibfnamefont {A.~J.}\ \bibnamefont
  {Berlinsky}}, \ and\ \bibinfo {author} {\bibfnamefont {A.-C.}\ \bibnamefont
  {Shi}},\ }\href {\doibase 10.1103/PhysRevB.43.865} {\bibfield  {journal}
  {\bibinfo  {journal} {Phys. Rev. B}\ }\textbf {\bibinfo {volume} {43}},\
  \bibinfo {pages} {865} (\bibinfo {year} {1991}{\natexlab{b}})}\BibitemShut
  {NoStop}%
\bibitem [{\citenamefont {Villain}(1979)}]{villain1979insulating}%
  \BibitemOpen
  \bibfield  {author} {\bibinfo {author} {\bibfnamefont {J.}~\bibnamefont
  {Villain}},\ }\href {\doibase https://doi.org/10.1007/BF01325811} {\bibfield
  {journal} {\bibinfo  {journal} {Z. Physik B}\ }\textbf {\bibinfo {volume}
  {33}},\ \bibinfo {pages} {31} (\bibinfo {year} {1979})}\BibitemShut {NoStop}%
\bibitem [{\citenamefont {Bonville}\ \emph {et~al.}(2003)\citenamefont
  {Bonville}, \citenamefont {Hodges}, \citenamefont {Ocio}, \citenamefont
  {Sanchez}, \citenamefont {Vulliet}, \citenamefont {Sosin},\ and\
  \citenamefont {Braithwaite}}]{Bonville_2003}%
  \BibitemOpen
  \bibfield  {author} {\bibinfo {author} {\bibfnamefont {P.}~\bibnamefont
  {Bonville}}, \bibinfo {author} {\bibfnamefont {J.~A.}\ \bibnamefont
  {Hodges}}, \bibinfo {author} {\bibfnamefont {M.}~\bibnamefont {Ocio}},
  \bibinfo {author} {\bibfnamefont {J.~P.}\ \bibnamefont {Sanchez}}, \bibinfo
  {author} {\bibfnamefont {P.}~\bibnamefont {Vulliet}}, \bibinfo {author}
  {\bibfnamefont {S.}~\bibnamefont {Sosin}}, \ and\ \bibinfo {author}
  {\bibfnamefont {D.}~\bibnamefont {Braithwaite}},\ }\href {\doibase
  10.1088/0953-8984/15/45/016} {\bibfield  {journal} {\bibinfo  {journal} {J.
  Phys.: Condens. Matter}\ }\textbf {\bibinfo {volume} {15}},\ \bibinfo {pages}
  {7777} (\bibinfo {year} {2003})}\BibitemShut {NoStop}%
\bibitem [{\citenamefont {Luo}\ \emph {et~al.}(2001)\citenamefont {Luo},
  \citenamefont {Hess},\ and\ \citenamefont {Corruccini}}]{LUO2001306}%
  \BibitemOpen
  \bibfield  {author} {\bibinfo {author} {\bibfnamefont {G.}~\bibnamefont
  {Luo}}, \bibinfo {author} {\bibfnamefont {S.~T.}\ \bibnamefont {Hess}}, \
  and\ \bibinfo {author} {\bibfnamefont {L.}~\bibnamefont {Corruccini}},\
  }\href {\doibase https://doi.org/10.1016/S0375-9601(01)00734-4} {\bibfield
  {journal} {\bibinfo  {journal} {Phys. Lett. A}\ }\textbf {\bibinfo {volume}
  {291}},\ \bibinfo {pages} {306} (\bibinfo {year} {2001})}\BibitemShut
  {NoStop}%
\bibitem [{\citenamefont {Matsuhira}\ \emph {et~al.}(2002)\citenamefont
  {Matsuhira}, \citenamefont {Hinatsu}, \citenamefont {Tenya}, \citenamefont
  {Amitsuka},\ and\ \citenamefont {Sakakibara}}]{Sn_Matsuhira02:71}%
  \BibitemOpen
  \bibfield  {author} {\bibinfo {author} {\bibfnamefont {K.}~\bibnamefont
  {Matsuhira}}, \bibinfo {author} {\bibfnamefont {Y.}~\bibnamefont {Hinatsu}},
  \bibinfo {author} {\bibfnamefont {K.}~\bibnamefont {Tenya}}, \bibinfo
  {author} {\bibfnamefont {H.}~\bibnamefont {Amitsuka}}, \ and\ \bibinfo
  {author} {\bibfnamefont {T.}~\bibnamefont {Sakakibara}},\ }\href {\doibase
  10.1143/JPSJ.71.1576} {\bibfield  {journal} {\bibinfo  {journal} {J. Phys.
  Soc. Jpn.}\ }\textbf {\bibinfo {volume} {71}},\ \bibinfo {pages} {1576}
  (\bibinfo {year} {2002})}\BibitemShut {NoStop}%
\bibitem [{\citenamefont {Durand}\ \emph {et~al.}(2008)\citenamefont {Durand},
  \citenamefont {Klavins},\ and\ \citenamefont {Corruccini}}]{Durand08:20}%
  \BibitemOpen
  \bibfield  {author} {\bibinfo {author} {\bibfnamefont {A.~M.}\ \bibnamefont
  {Durand}}, \bibinfo {author} {\bibfnamefont {P.}~\bibnamefont {Klavins}}, \
  and\ \bibinfo {author} {\bibfnamefont {L.~R.}\ \bibnamefont {Corruccini}},\
  }\href {\doibase 10.1088/0953-8984/20/23/235208} {\bibfield  {journal}
  {\bibinfo  {journal} {J. Phys.: Condens. Matter}\ }\textbf {\bibinfo {volume}
  {20}},\ \bibinfo {pages} {235208} (\bibinfo {year} {2008})}\BibitemShut
  {NoStop}%
\bibitem [{\citenamefont {Freitas}\ and\ \citenamefont
  {Gardner}(2011)}]{Freitas11:23}%
  \BibitemOpen
  \bibfield  {author} {\bibinfo {author} {\bibfnamefont {R.~S.}\ \bibnamefont
  {Freitas}}\ and\ \bibinfo {author} {\bibfnamefont {J.~S.}\ \bibnamefont
  {Gardner}},\ }\href {\doibase https://doi.org/10.1088/0953-8984/23/16/164215}
  {\bibfield  {journal} {\bibinfo  {journal} {J. Phys. Condens. Matter}\
  }\textbf {\bibinfo {volume} {23}},\ \bibinfo {pages} {164215} (\bibinfo
  {year} {2011})}\BibitemShut {NoStop}%
\bibitem [{\citenamefont {Mydosh}(1993)}]{mydosh1993spin}%
  \BibitemOpen
  \bibfield  {author} {\bibinfo {author} {\bibfnamefont {J.~A.}\ \bibnamefont
  {Mydosh}},\ }\href {\doibase https://doi.org/10.1201/9781482295191} {\emph
  {\bibinfo {title} {Spin glasses: an experimental introduction}}}\ (\bibinfo
  {publisher} {Taylor \& Francis},\ \bibinfo {address} {London},\ \bibinfo
  {year} {1993})\BibitemShut {NoStop}%
\bibitem [{\citenamefont {Wills}\ \emph {et~al.}(2006)\citenamefont {Wills},
  \citenamefont {Zhitomirsky}, \citenamefont {Canals}, \citenamefont {Sanchez},
  \citenamefont {Bonville}, \citenamefont {de~R{\'{e}}otier},\ and\
  \citenamefont {Yaouanc}}]{Wills06:18}%
  \BibitemOpen
  \bibfield  {author} {\bibinfo {author} {\bibfnamefont {A.~S.}\ \bibnamefont
  {Wills}}, \bibinfo {author} {\bibfnamefont {M.~E.}\ \bibnamefont
  {Zhitomirsky}}, \bibinfo {author} {\bibfnamefont {B.}~\bibnamefont {Canals}},
  \bibinfo {author} {\bibfnamefont {J.~P.}\ \bibnamefont {Sanchez}}, \bibinfo
  {author} {\bibfnamefont {P.}~\bibnamefont {Bonville}}, \bibinfo {author}
  {\bibfnamefont {P.~D.}\ \bibnamefont {de~R{\'{e}}otier}}, \ and\ \bibinfo
  {author} {\bibfnamefont {A.}~\bibnamefont {Yaouanc}},\ }\href {\doibase
  10.1088/0953-8984/18/3/l02} {\bibfield  {journal} {\bibinfo  {journal} {J.
  Phys. Condens. Matter}\ }\textbf {\bibinfo {volume} {18}},\ \bibinfo {pages}
  {L37} (\bibinfo {year} {2006})}\BibitemShut {NoStop}%
\bibitem [{\citenamefont {Champion}\ \emph {et~al.}(2001)\citenamefont
  {Champion}, \citenamefont {Wills}, \citenamefont {Fennell}, \citenamefont
  {Bramwell}, \citenamefont {Gardner},\ and\ \citenamefont
  {Green}}]{champion01:64}%
  \BibitemOpen
  \bibfield  {author} {\bibinfo {author} {\bibfnamefont {J.~D.~M.}\
  \bibnamefont {Champion}}, \bibinfo {author} {\bibfnamefont {A.~S.}\
  \bibnamefont {Wills}}, \bibinfo {author} {\bibfnamefont {T.}~\bibnamefont
  {Fennell}}, \bibinfo {author} {\bibfnamefont {S.~T.}\ \bibnamefont
  {Bramwell}}, \bibinfo {author} {\bibfnamefont {J.~S.}\ \bibnamefont
  {Gardner}}, \ and\ \bibinfo {author} {\bibfnamefont {M.~A.}\ \bibnamefont
  {Green}},\ }\href {\doibase 10.1103/PhysRevB.64.140407} {\bibfield  {journal}
  {\bibinfo  {journal} {Phys. Rev. B}\ }\textbf {\bibinfo {volume} {64}},\
  \bibinfo {pages} {140407} (\bibinfo {year} {2001})}\BibitemShut {NoStop}%
\bibitem [{\citenamefont {Stewart}\ \emph {et~al.}(2004)\citenamefont
  {Stewart}, \citenamefont {Ehlers}, \citenamefont {Wills}, \citenamefont
  {Bramwell},\ and\ \citenamefont {Gardner}}]{Stewart_2004}%
  \BibitemOpen
  \bibfield  {author} {\bibinfo {author} {\bibfnamefont {J.~R.}\ \bibnamefont
  {Stewart}}, \bibinfo {author} {\bibfnamefont {G.}~\bibnamefont {Ehlers}},
  \bibinfo {author} {\bibfnamefont {A.~S.}\ \bibnamefont {Wills}}, \bibinfo
  {author} {\bibfnamefont {S.~T.}\ \bibnamefont {Bramwell}}, \ and\ \bibinfo
  {author} {\bibfnamefont {J.~S.}\ \bibnamefont {Gardner}},\ }\href {\doibase
  10.1088/0953-8984/16/28/l01} {\bibfield  {journal} {\bibinfo  {journal} {J.
  Phys. Condens. Matter}\ }\textbf {\bibinfo {volume} {16}},\ \bibinfo {pages}
  {L321} (\bibinfo {year} {2004})}\BibitemShut {NoStop}%
\bibitem [{\citenamefont {Stewart}\ \emph {et~al.}(2008)\citenamefont
  {Stewart}, \citenamefont {Gardner}, \citenamefont {Qiu},\ and\ \citenamefont
  {Ehlers}}]{Stewart08:78}%
  \BibitemOpen
  \bibfield  {author} {\bibinfo {author} {\bibfnamefont {J.~R.}\ \bibnamefont
  {Stewart}}, \bibinfo {author} {\bibfnamefont {J.~S.}\ \bibnamefont
  {Gardner}}, \bibinfo {author} {\bibfnamefont {Y.}~\bibnamefont {Qiu}}, \ and\
  \bibinfo {author} {\bibfnamefont {G.}~\bibnamefont {Ehlers}},\ }\href
  {\doibase 10.1103/PhysRevB.78.132410} {\bibfield  {journal} {\bibinfo
  {journal} {Phys. Rev. B}\ }\textbf {\bibinfo {volume} {78}},\ \bibinfo
  {pages} {132410} (\bibinfo {year} {2008})}\BibitemShut {NoStop}%
\bibitem [{\citenamefont {Palmer}\ and\ \citenamefont {Chalker}(2000)}]{PC}%
  \BibitemOpen
  \bibfield  {author} {\bibinfo {author} {\bibfnamefont {S.~E.}\ \bibnamefont
  {Palmer}}\ and\ \bibinfo {author} {\bibfnamefont {J.~T.}\ \bibnamefont
  {Chalker}},\ }\href {\doibase 10.1103/PhysRevB.62.488} {\bibfield  {journal}
  {\bibinfo  {journal} {Phys. Rev. B}\ }\textbf {\bibinfo {volume} {62}},\
  \bibinfo {pages} {488} (\bibinfo {year} {2000})}\BibitemShut {NoStop}%
\bibitem [{\citenamefont {Hallas}\ \emph
  {et~al.}(2016{\natexlab{b}})\citenamefont {Hallas}, \citenamefont {Sharma},
  \citenamefont {Cai}, \citenamefont {Munsie}, \citenamefont {Wilson},
  \citenamefont {Tachibana}, \citenamefont {Wiebe},\ and\ \citenamefont
  {Luke}}]{hallas16:94}%
  \BibitemOpen
  \bibfield  {author} {\bibinfo {author} {\bibfnamefont {A.~M.}\ \bibnamefont
  {Hallas}}, \bibinfo {author} {\bibfnamefont {A.~Z.}\ \bibnamefont {Sharma}},
  \bibinfo {author} {\bibfnamefont {Y.}~\bibnamefont {Cai}}, \bibinfo {author}
  {\bibfnamefont {T.~J.}\ \bibnamefont {Munsie}}, \bibinfo {author}
  {\bibfnamefont {M.~N.}\ \bibnamefont {Wilson}}, \bibinfo {author}
  {\bibfnamefont {M.}~\bibnamefont {Tachibana}}, \bibinfo {author}
  {\bibfnamefont {C.~R.}\ \bibnamefont {Wiebe}}, \ and\ \bibinfo {author}
  {\bibfnamefont {G.~M.}\ \bibnamefont {Luke}},\ }\href {\doibase
  10.1103/PhysRevB.94.134417} {\bibfield  {journal} {\bibinfo  {journal} {Phys.
  Rev. B}\ }\textbf {\bibinfo {volume} {94}},\ \bibinfo {pages} {134417}
  (\bibinfo {year} {2016}{\natexlab{b}})}\BibitemShut {NoStop}%
\bibitem [{\citenamefont {Paddison}\ \emph {et~al.}(2021)\citenamefont
  {Paddison}, \citenamefont {Ehlers}, \citenamefont {Cairns}, \citenamefont
  {Gardner}, \citenamefont {Petrenko}, \citenamefont {Butch}, \citenamefont
  {Khalyavin}, \citenamefont {Manuel}, \citenamefont {Fischer}, \citenamefont
  {Zhou}, \citenamefont {Goodwin},\ and\ \citenamefont
  {Stewart}}]{paddison2021suppressed}%
  \BibitemOpen
  \bibfield  {author} {\bibinfo {author} {\bibfnamefont {J.~A.~M.}\
  \bibnamefont {Paddison}}, \bibinfo {author} {\bibfnamefont {G.}~\bibnamefont
  {Ehlers}}, \bibinfo {author} {\bibfnamefont {A.~B.}\ \bibnamefont {Cairns}},
  \bibinfo {author} {\bibfnamefont {J.~S.}\ \bibnamefont {Gardner}}, \bibinfo
  {author} {\bibfnamefont {O.~A.}\ \bibnamefont {Petrenko}}, \bibinfo {author}
  {\bibfnamefont {N.~P.}\ \bibnamefont {Butch}}, \bibinfo {author}
  {\bibfnamefont {D.~D.}\ \bibnamefont {Khalyavin}}, \bibinfo {author}
  {\bibfnamefont {P.}~\bibnamefont {Manuel}}, \bibinfo {author} {\bibfnamefont
  {H.~E.}\ \bibnamefont {Fischer}}, \bibinfo {author} {\bibfnamefont
  {H.}~\bibnamefont {Zhou}}, \bibinfo {author} {\bibfnamefont {A.~L.}\
  \bibnamefont {Goodwin}}, \ and\ \bibinfo {author} {\bibfnamefont {J.~R.}\
  \bibnamefont {Stewart}},\ }\href {\doibase
  https://doi.org/10.1038/s41535-021-00391-w} {\bibfield  {journal} {\bibinfo
  {journal} {npj Quantum Mater.}\ }\textbf {\bibinfo {volume} {6}},\ \bibinfo
  {pages} {1} (\bibinfo {year} {2021})}\BibitemShut {NoStop}%
\bibitem [{\citenamefont {Petrenko}\ \emph {et~al.}(2004)\citenamefont
  {Petrenko}, \citenamefont {Lees}, \citenamefont {Balakrishnan},\ and\
  \citenamefont {Paul}}]{Petrenko04:70}%
  \BibitemOpen
  \bibfield  {author} {\bibinfo {author} {\bibfnamefont {O.~A.}\ \bibnamefont
  {Petrenko}}, \bibinfo {author} {\bibfnamefont {M.~R.}\ \bibnamefont {Lees}},
  \bibinfo {author} {\bibfnamefont {G.}~\bibnamefont {Balakrishnan}}, \ and\
  \bibinfo {author} {\bibfnamefont {D.~M.}\ \bibnamefont {Paul}},\ }\href
  {\doibase 10.1103/PhysRevB.70.012402} {\bibfield  {journal} {\bibinfo
  {journal} {Phys. Rev. B}\ }\textbf {\bibinfo {volume} {70}},\ \bibinfo
  {pages} {012402} (\bibinfo {year} {2004})}\BibitemShut {NoStop}%
\bibitem [{\citenamefont {Brammall}\ \emph {et~al.}(2011)\citenamefont
  {Brammall}, \citenamefont {Briffa},\ and\ \citenamefont
  {Long}}]{brammall11:83}%
  \BibitemOpen
  \bibfield  {author} {\bibinfo {author} {\bibfnamefont {M.~I.}\ \bibnamefont
  {Brammall}}, \bibinfo {author} {\bibfnamefont {A.~K.~R.}\ \bibnamefont
  {Briffa}}, \ and\ \bibinfo {author} {\bibfnamefont {M.~W.}\ \bibnamefont
  {Long}},\ }\href {\doibase 10.1103/PhysRevB.83.054422} {\bibfield  {journal}
  {\bibinfo  {journal} {Phys. Rev. B}\ }\textbf {\bibinfo {volume} {83}},\
  \bibinfo {pages} {054422} (\bibinfo {year} {2011})}\BibitemShut {NoStop}%
\bibitem [{\citenamefont {Javanparast}\ \emph {et~al.}(2015)\citenamefont
  {Javanparast}, \citenamefont {Hao}, \citenamefont {Enjalran},\ and\
  \citenamefont {Gingras}}]{PhysRevLett.114.130601}%
  \BibitemOpen
  \bibfield  {author} {\bibinfo {author} {\bibfnamefont {B.}~\bibnamefont
  {Javanparast}}, \bibinfo {author} {\bibfnamefont {Z.}~\bibnamefont {Hao}},
  \bibinfo {author} {\bibfnamefont {M.}~\bibnamefont {Enjalran}}, \ and\
  \bibinfo {author} {\bibfnamefont {M.~J.~P.}\ \bibnamefont {Gingras}},\ }\href
  {\doibase 10.1103/PhysRevLett.114.130601} {\bibfield  {journal} {\bibinfo
  {journal} {Phys. Rev. Lett.}\ }\textbf {\bibinfo {volume} {114}},\ \bibinfo
  {pages} {130601} (\bibinfo {year} {2015})}\BibitemShut {NoStop}%
\bibitem [{\citenamefont {Xu}(2017)}]{xu_thesis}%
  \BibitemOpen
  \bibfield  {author} {\bibinfo {author} {\bibfnamefont {J.}~\bibnamefont
  {Xu}},\ }\emph {\bibinfo {title} {Magnetic properties of rare earth zirconate
  pyrochlores}},\ \href@noop {} {Ph.D. thesis},\ \bibinfo  {school} {Technische
  Universit\"{a}t Berlin}, \bibinfo {address} {Berlin} (\bibinfo {year}
  {2017})\BibitemShut {NoStop}%
\bibitem [{\citenamefont {C\'epas}\ and\ \citenamefont
  {Shastry}(2004)}]{cepas4:69}%
  \BibitemOpen
  \bibfield  {author} {\bibinfo {author} {\bibfnamefont {O.}~\bibnamefont
  {C\'epas}}\ and\ \bibinfo {author} {\bibfnamefont {B.~S.}\ \bibnamefont
  {Shastry}},\ }\href {\doibase 10.1103/PhysRevB.69.184402} {\bibfield
  {journal} {\bibinfo  {journal} {Phys. Rev. B}\ }\textbf {\bibinfo {volume}
  {69}},\ \bibinfo {pages} {184402} (\bibinfo {year} {2004})}\BibitemShut
  {NoStop}%
\bibitem [{\citenamefont {Guitteny}\ \emph
  {et~al.}(2013{\natexlab{a}})\citenamefont {Guitteny}, \citenamefont {Petit},
  \citenamefont {Lhotel}, \citenamefont {Robert}, \citenamefont {Bonville},
  \citenamefont {Forget},\ and\ \citenamefont {Mirebeau}}]{guitteny2013palmer}%
  \BibitemOpen
  \bibfield  {author} {\bibinfo {author} {\bibfnamefont {S.}~\bibnamefont
  {Guitteny}}, \bibinfo {author} {\bibfnamefont {S.}~\bibnamefont {Petit}},
  \bibinfo {author} {\bibfnamefont {E.}~\bibnamefont {Lhotel}}, \bibinfo
  {author} {\bibfnamefont {J.}~\bibnamefont {Robert}}, \bibinfo {author}
  {\bibfnamefont {P.}~\bibnamefont {Bonville}}, \bibinfo {author}
  {\bibfnamefont {A.}~\bibnamefont {Forget}}, \ and\ \bibinfo {author}
  {\bibfnamefont {I.}~\bibnamefont {Mirebeau}},\ }\href {\doibase
  10.1103/PhysRevB.88.134408} {\bibfield  {journal} {\bibinfo  {journal}
  {Physical Review B}\ }\textbf {\bibinfo {volume} {88}},\ \bibinfo {pages}
  {134408} (\bibinfo {year} {2013}{\natexlab{a}})}\BibitemShut {NoStop}%
\bibitem [{\citenamefont {Dun}\ \emph {et~al.}(2015)\citenamefont {Dun},
  \citenamefont {Li}, \citenamefont {Freitas}, \citenamefont {Arrighi},
  \citenamefont {Dela~Cruz}, \citenamefont {Lee}, \citenamefont {Choi},
  \citenamefont {Cao}, \citenamefont {Silverstein}, \citenamefont {Wiebe},
  \citenamefont {Cheng},\ and\ \citenamefont {Zhou}}]{dun15:92}%
  \BibitemOpen
  \bibfield  {author} {\bibinfo {author} {\bibfnamefont {Z.~L.}\ \bibnamefont
  {Dun}}, \bibinfo {author} {\bibfnamefont {X.}~\bibnamefont {Li}}, \bibinfo
  {author} {\bibfnamefont {R.~S.}\ \bibnamefont {Freitas}}, \bibinfo {author}
  {\bibfnamefont {E.}~\bibnamefont {Arrighi}}, \bibinfo {author} {\bibfnamefont
  {C.~R.}\ \bibnamefont {Dela~Cruz}}, \bibinfo {author} {\bibfnamefont
  {M.}~\bibnamefont {Lee}}, \bibinfo {author} {\bibfnamefont {E.~S.}\
  \bibnamefont {Choi}}, \bibinfo {author} {\bibfnamefont {H.~B.}\ \bibnamefont
  {Cao}}, \bibinfo {author} {\bibfnamefont {H.~J.}\ \bibnamefont
  {Silverstein}}, \bibinfo {author} {\bibfnamefont {C.~R.}\ \bibnamefont
  {Wiebe}}, \bibinfo {author} {\bibfnamefont {J.~G.}\ \bibnamefont {Cheng}}, \
  and\ \bibinfo {author} {\bibfnamefont {H.~D.}\ \bibnamefont {Zhou}},\ }\href
  {\doibase 10.1103/PhysRevB.92.140407} {\bibfield  {journal} {\bibinfo
  {journal} {Phys. Rev. B}\ }\textbf {\bibinfo {volume} {92}},\ \bibinfo
  {pages} {140407} (\bibinfo {year} {2015})}\BibitemShut {NoStop}%
\bibitem [{\citenamefont {Poole}\ \emph {et~al.}(2007)\citenamefont {Poole},
  \citenamefont {Wills},\ and\ \citenamefont
  {Leli{\`{e}}vre-Berna}}]{Poole_2007}%
  \BibitemOpen
  \bibfield  {author} {\bibinfo {author} {\bibfnamefont {A.}~\bibnamefont
  {Poole}}, \bibinfo {author} {\bibfnamefont {A.~S.}\ \bibnamefont {Wills}}, \
  and\ \bibinfo {author} {\bibfnamefont {E.}~\bibnamefont
  {Leli{\`{e}}vre-Berna}},\ }\href {\doibase 10.1088/0953-8984/19/45/452201}
  {\bibfield  {journal} {\bibinfo  {journal} {J. Phys.: Condens. Matter}\
  }\textbf {\bibinfo {volume} {19}},\ \bibinfo {pages} {452201} (\bibinfo
  {year} {2007})}\BibitemShut {NoStop}%
\bibitem [{\citenamefont {Lago}\ \emph {et~al.}(2014)\citenamefont {Lago},
  \citenamefont {{\v{Z}}ivkovi{\'c}}, \citenamefont {Piatek}, \citenamefont
  {{\'A}lvarez}, \citenamefont {H{\"u}vonen}, \citenamefont {Pratt},
  \citenamefont {D{\'\i}az},\ and\ \citenamefont {Rojo}}]{YbSn_lago2014glassy}%
  \BibitemOpen
  \bibfield  {author} {\bibinfo {author} {\bibfnamefont {J.}~\bibnamefont
  {Lago}}, \bibinfo {author} {\bibfnamefont {I.}~\bibnamefont
  {{\v{Z}}ivkovi{\'c}}}, \bibinfo {author} {\bibfnamefont {J.}~\bibnamefont
  {Piatek}}, \bibinfo {author} {\bibfnamefont {P.}~\bibnamefont {{\'A}lvarez}},
  \bibinfo {author} {\bibfnamefont {D.}~\bibnamefont {H{\"u}vonen}}, \bibinfo
  {author} {\bibfnamefont {F.}~\bibnamefont {Pratt}}, \bibinfo {author}
  {\bibfnamefont {M.}~\bibnamefont {D{\'\i}az}}, \ and\ \bibinfo {author}
  {\bibfnamefont {T.}~\bibnamefont {Rojo}},\ }\href {\doibase
  10.1103/PhysRevB.89.024421} {\bibfield  {journal} {\bibinfo  {journal} {Phys.
  Rev. B}\ }\textbf {\bibinfo {volume} {89}},\ \bibinfo {pages} {024421}
  (\bibinfo {year} {2014})}\BibitemShut {NoStop}%
\bibitem [{\citenamefont {Yaouanc}\ \emph
  {et~al.}(2013{\natexlab{a}})\citenamefont {Yaouanc}, \citenamefont
  {De~R{\'e}otier}, \citenamefont {Bonville}, \citenamefont {Hodges},
  \citenamefont {Glazkov}, \citenamefont {Keller}, \citenamefont {Sikolenko},
  \citenamefont {Bartkowiak}, \citenamefont {Amato}, \citenamefont {Baines},
  \citenamefont {King}, \citenamefont {Gubbens},\ and\ \citenamefont
  {Forget}}]{YbSn_yaouanc2013dynamical}%
  \BibitemOpen
  \bibfield  {author} {\bibinfo {author} {\bibfnamefont {A.}~\bibnamefont
  {Yaouanc}}, \bibinfo {author} {\bibfnamefont {P.~D.}\ \bibnamefont
  {De~R{\'e}otier}}, \bibinfo {author} {\bibfnamefont {P.}~\bibnamefont
  {Bonville}}, \bibinfo {author} {\bibfnamefont {J.}~\bibnamefont {Hodges}},
  \bibinfo {author} {\bibfnamefont {V.}~\bibnamefont {Glazkov}}, \bibinfo
  {author} {\bibfnamefont {L.}~\bibnamefont {Keller}}, \bibinfo {author}
  {\bibfnamefont {V.}~\bibnamefont {Sikolenko}}, \bibinfo {author}
  {\bibfnamefont {M.}~\bibnamefont {Bartkowiak}}, \bibinfo {author}
  {\bibfnamefont {A.}~\bibnamefont {Amato}}, \bibinfo {author} {\bibfnamefont
  {C.}~\bibnamefont {Baines}}, \bibinfo {author} {\bibfnamefont {P.~J.~C.}\
  \bibnamefont {King}}, \bibinfo {author} {\bibfnamefont {P.~C.~M.}\
  \bibnamefont {Gubbens}}, \ and\ \bibinfo {author} {\bibfnamefont
  {A.}~\bibnamefont {Forget}},\ }\href {\doibase
  10.1103/PhysRevLett.110.127207} {\bibfield  {journal} {\bibinfo  {journal}
  {Phys. Rev. Lett.}\ }\textbf {\bibinfo {volume} {110}},\ \bibinfo {pages}
  {127207} (\bibinfo {year} {2013}{\natexlab{a}})}\BibitemShut {NoStop}%
\bibitem [{\citenamefont {Cai}\ \emph {et~al.}(2016)\citenamefont {Cai},
  \citenamefont {Cui}, \citenamefont {Li}, \citenamefont {Dun}, \citenamefont
  {Ma}, \citenamefont {dela Cruz}, \citenamefont {Jiao}, \citenamefont {Liao},
  \citenamefont {Sun}, \citenamefont {Li}, \citenamefont {Zhou}, \citenamefont
  {Goodenough}, \citenamefont {Zhou},\ and\ \citenamefont {Cheng}}]{cai16:93}%
  \BibitemOpen
  \bibfield  {author} {\bibinfo {author} {\bibfnamefont {Y.~Q.}\ \bibnamefont
  {Cai}}, \bibinfo {author} {\bibfnamefont {Q.}~\bibnamefont {Cui}}, \bibinfo
  {author} {\bibfnamefont {X.}~\bibnamefont {Li}}, \bibinfo {author}
  {\bibfnamefont {Z.~L.}\ \bibnamefont {Dun}}, \bibinfo {author} {\bibfnamefont
  {J.}~\bibnamefont {Ma}}, \bibinfo {author} {\bibfnamefont {C.}~\bibnamefont
  {dela Cruz}}, \bibinfo {author} {\bibfnamefont {Y.~Y.}\ \bibnamefont {Jiao}},
  \bibinfo {author} {\bibfnamefont {J.}~\bibnamefont {Liao}}, \bibinfo {author}
  {\bibfnamefont {P.~J.}\ \bibnamefont {Sun}}, \bibinfo {author} {\bibfnamefont
  {Y.~Q.}\ \bibnamefont {Li}}, \bibinfo {author} {\bibfnamefont {J.~S.}\
  \bibnamefont {Zhou}}, \bibinfo {author} {\bibfnamefont {J.~B.}\ \bibnamefont
  {Goodenough}}, \bibinfo {author} {\bibfnamefont {H.~D.}\ \bibnamefont
  {Zhou}}, \ and\ \bibinfo {author} {\bibfnamefont {J.-G.}\ \bibnamefont
  {Cheng}},\ }\href {\doibase 10.1103/PhysRevB.93.014443} {\bibfield  {journal}
  {\bibinfo  {journal} {Phys. Rev. B}\ }\textbf {\bibinfo {volume} {93}},\
  \bibinfo {pages} {014443} (\bibinfo {year} {2016})}\BibitemShut {NoStop}%
\bibitem [{\citenamefont {Yan}\ \emph {et~al.}(2017)\citenamefont {Yan},
  \citenamefont {Benton}, \citenamefont {Jaubert},\ and\ \citenamefont
  {Shannon}}]{YbCompetition_yan2017theory}%
  \BibitemOpen
  \bibfield  {author} {\bibinfo {author} {\bibfnamefont {H.}~\bibnamefont
  {Yan}}, \bibinfo {author} {\bibfnamefont {O.}~\bibnamefont {Benton}},
  \bibinfo {author} {\bibfnamefont {L.}~\bibnamefont {Jaubert}}, \ and\
  \bibinfo {author} {\bibfnamefont {N.}~\bibnamefont {Shannon}},\ }\href
  {\doibase 10.1103/PhysRevB.95.094422} {\bibfield  {journal} {\bibinfo
  {journal} {Phys. Rev. B}\ }\textbf {\bibinfo {volume} {95}},\ \bibinfo
  {pages} {094422} (\bibinfo {year} {2017})}\BibitemShut {NoStop}%
\bibitem [{\citenamefont {Jaubert}\ \emph {et~al.}(2015)\citenamefont
  {Jaubert}, \citenamefont {Benton}, \citenamefont {Rau}, \citenamefont
  {Oitmaa}, \citenamefont {Singh}, \citenamefont {Shannon},\ and\ \citenamefont
  {Gingras}}]{jaubert15:115}%
  \BibitemOpen
  \bibfield  {author} {\bibinfo {author} {\bibfnamefont {L.~D.~C.}\
  \bibnamefont {Jaubert}}, \bibinfo {author} {\bibfnamefont {O.}~\bibnamefont
  {Benton}}, \bibinfo {author} {\bibfnamefont {J.~G.}\ \bibnamefont {Rau}},
  \bibinfo {author} {\bibfnamefont {J.}~\bibnamefont {Oitmaa}}, \bibinfo
  {author} {\bibfnamefont {R.~R.~P.}\ \bibnamefont {Singh}}, \bibinfo {author}
  {\bibfnamefont {N.}~\bibnamefont {Shannon}}, \ and\ \bibinfo {author}
  {\bibfnamefont {M.~J.~P.}\ \bibnamefont {Gingras}},\ }\href {\doibase
  10.1103/PhysRevLett.115.267208} {\bibfield  {journal} {\bibinfo  {journal}
  {Phys. Rev. Lett.}\ }\textbf {\bibinfo {volume} {115}},\ \bibinfo {pages}
  {267208} (\bibinfo {year} {2015})}\BibitemShut {NoStop}%
\bibitem [{\citenamefont {Guratinder}\ \emph {et~al.}(2019)\citenamefont
  {Guratinder}, \citenamefont {Rau}, \citenamefont {Tsurkan}, \citenamefont
  {Ritter}, \citenamefont {Embs}, \citenamefont {Fennell}, \citenamefont
  {Walker}, \citenamefont {Medarde}, \citenamefont {Shang}, \citenamefont
  {Cervellino}, \citenamefont {R\"uegg},\ and\ \citenamefont
  {Zaharko}}]{PhysRevB.100.094420}%
  \BibitemOpen
  \bibfield  {author} {\bibinfo {author} {\bibfnamefont {K.}~\bibnamefont
  {Guratinder}}, \bibinfo {author} {\bibfnamefont {J.~G.}\ \bibnamefont {Rau}},
  \bibinfo {author} {\bibfnamefont {V.}~\bibnamefont {Tsurkan}}, \bibinfo
  {author} {\bibfnamefont {C.}~\bibnamefont {Ritter}}, \bibinfo {author}
  {\bibfnamefont {J.}~\bibnamefont {Embs}}, \bibinfo {author} {\bibfnamefont
  {T.}~\bibnamefont {Fennell}}, \bibinfo {author} {\bibfnamefont {H.~C.}\
  \bibnamefont {Walker}}, \bibinfo {author} {\bibfnamefont {M.}~\bibnamefont
  {Medarde}}, \bibinfo {author} {\bibfnamefont {T.}~\bibnamefont {Shang}},
  \bibinfo {author} {\bibfnamefont {A.}~\bibnamefont {Cervellino}}, \bibinfo
  {author} {\bibfnamefont {C.}~\bibnamefont {R\"uegg}}, \ and\ \bibinfo
  {author} {\bibfnamefont {O.}~\bibnamefont {Zaharko}},\ }\href {\doibase
  10.1103/PhysRevB.100.094420} {\bibfield  {journal} {\bibinfo  {journal}
  {Phys. Rev. B}\ }\textbf {\bibinfo {volume} {100}},\ \bibinfo {pages}
  {094420} (\bibinfo {year} {2019})}\BibitemShut {NoStop}%
\bibitem [{\citenamefont {Scheie}\ \emph {et~al.}(2020)\citenamefont {Scheie},
  \citenamefont {Kindervater}, \citenamefont {Zhang}, \citenamefont
  {Changlani}, \citenamefont {Sala}, \citenamefont {Ehlers}, \citenamefont
  {Heinemann}, \citenamefont {Tucker}, \citenamefont {Koohpayeh},\ and\
  \citenamefont {Broholm}}]{Scheie20:117}%
  \BibitemOpen
  \bibfield  {author} {\bibinfo {author} {\bibfnamefont {A.}~\bibnamefont
  {Scheie}}, \bibinfo {author} {\bibfnamefont {J.}~\bibnamefont {Kindervater}},
  \bibinfo {author} {\bibfnamefont {S.}~\bibnamefont {Zhang}}, \bibinfo
  {author} {\bibfnamefont {H.~J.}\ \bibnamefont {Changlani}}, \bibinfo {author}
  {\bibfnamefont {G.}~\bibnamefont {Sala}}, \bibinfo {author} {\bibfnamefont
  {G.}~\bibnamefont {Ehlers}}, \bibinfo {author} {\bibfnamefont
  {A.}~\bibnamefont {Heinemann}}, \bibinfo {author} {\bibfnamefont {G.~S.}\
  \bibnamefont {Tucker}}, \bibinfo {author} {\bibfnamefont {S.~M.}\
  \bibnamefont {Koohpayeh}}, \ and\ \bibinfo {author} {\bibfnamefont
  {C.}~\bibnamefont {Broholm}},\ }\href {\doibase 10.1073/pnas.2008791117}
  {\bibfield  {journal} {\bibinfo  {journal} {Proc. Natl. Acad. Sci.}\ }\textbf
  {\bibinfo {volume} {117}},\ \bibinfo {pages} {27245} (\bibinfo {year}
  {2020})}\BibitemShut {NoStop}%
\bibitem [{\citenamefont {Hallas}\ \emph {et~al.}(2017)\citenamefont {Hallas},
  \citenamefont {Gaudet}, \citenamefont {Butch}, \citenamefont {Xu},
  \citenamefont {Tachibana}, \citenamefont {Wiebe}, \citenamefont {Luke},\ and\
  \citenamefont {Gaulin}}]{Hallas17:119}%
  \BibitemOpen
  \bibfield  {author} {\bibinfo {author} {\bibfnamefont {A.~M.}\ \bibnamefont
  {Hallas}}, \bibinfo {author} {\bibfnamefont {J.}~\bibnamefont {Gaudet}},
  \bibinfo {author} {\bibfnamefont {N.~P.}\ \bibnamefont {Butch}}, \bibinfo
  {author} {\bibfnamefont {G.}~\bibnamefont {Xu}}, \bibinfo {author}
  {\bibfnamefont {M.}~\bibnamefont {Tachibana}}, \bibinfo {author}
  {\bibfnamefont {C.~R.}\ \bibnamefont {Wiebe}}, \bibinfo {author}
  {\bibfnamefont {G.~M.}\ \bibnamefont {Luke}}, \ and\ \bibinfo {author}
  {\bibfnamefont {B.~D.}\ \bibnamefont {Gaulin}},\ }\href {\doibase
  10.1103/PhysRevLett.119.187201} {\bibfield  {journal} {\bibinfo  {journal}
  {Phys. Rev. Lett.}\ }\textbf {\bibinfo {volume} {119}},\ \bibinfo {pages}
  {187201} (\bibinfo {year} {2017})}\BibitemShut {NoStop}%
\bibitem [{\citenamefont {Petit}\ \emph {et~al.}(2017)\citenamefont {Petit},
  \citenamefont {Lhotel}, \citenamefont {Damay}, \citenamefont {Boutrouille},
  \citenamefont {Forget},\ and\ \citenamefont {Colson}}]{petit2017long}%
  \BibitemOpen
  \bibfield  {author} {\bibinfo {author} {\bibfnamefont {S.}~\bibnamefont
  {Petit}}, \bibinfo {author} {\bibfnamefont {E.}~\bibnamefont {Lhotel}},
  \bibinfo {author} {\bibfnamefont {F.}~\bibnamefont {Damay}}, \bibinfo
  {author} {\bibfnamefont {P.}~\bibnamefont {Boutrouille}}, \bibinfo {author}
  {\bibfnamefont {A.}~\bibnamefont {Forget}}, \ and\ \bibinfo {author}
  {\bibfnamefont {D.}~\bibnamefont {Colson}},\ }\href {\doibase
  10.1103/PhysRevLett.119.187202} {\bibfield  {journal} {\bibinfo  {journal}
  {Phys. Rev. Lett.}\ }\textbf {\bibinfo {volume} {119}},\ \bibinfo {pages}
  {187202} (\bibinfo {year} {2017})}\BibitemShut {NoStop}%
\bibitem [{\citenamefont {Rau}\ \emph {et~al.}(2016)\citenamefont {Rau},
  \citenamefont {Petit},\ and\ \citenamefont {Gingras}}]{Rau16:93}%
  \BibitemOpen
  \bibfield  {author} {\bibinfo {author} {\bibfnamefont {J.~G.}\ \bibnamefont
  {Rau}}, \bibinfo {author} {\bibfnamefont {S.}~\bibnamefont {Petit}}, \ and\
  \bibinfo {author} {\bibfnamefont {M.~J.~P.}\ \bibnamefont {Gingras}},\ }\href
  {\doibase 10.1103/PhysRevB.93.184408} {\bibfield  {journal} {\bibinfo
  {journal} {Phys. Rev. B}\ }\textbf {\bibinfo {volume} {93}},\ \bibinfo
  {pages} {184408} (\bibinfo {year} {2016})}\BibitemShut {NoStop}%
\bibitem [{\citenamefont {Yahne}\ \emph {et~al.}(2021)\citenamefont {Yahne},
  \citenamefont {Pereira}, \citenamefont {Jaubert}, \citenamefont {Sanjeewa},
  \citenamefont {Powell}, \citenamefont {Kolis}, \citenamefont {Xu},
  \citenamefont {Enjalran}, \citenamefont {Gingras},\ and\ \citenamefont
  {Ross}}]{Yahne21:127}%
  \BibitemOpen
  \bibfield  {author} {\bibinfo {author} {\bibfnamefont {D.~R.}\ \bibnamefont
  {Yahne}}, \bibinfo {author} {\bibfnamefont {D.}~\bibnamefont {Pereira}},
  \bibinfo {author} {\bibfnamefont {L.~D.~C.}\ \bibnamefont {Jaubert}},
  \bibinfo {author} {\bibfnamefont {L.~D.}\ \bibnamefont {Sanjeewa}}, \bibinfo
  {author} {\bibfnamefont {M.}~\bibnamefont {Powell}}, \bibinfo {author}
  {\bibfnamefont {J.~W.}\ \bibnamefont {Kolis}}, \bibinfo {author}
  {\bibfnamefont {G.}~\bibnamefont {Xu}}, \bibinfo {author} {\bibfnamefont
  {M.}~\bibnamefont {Enjalran}}, \bibinfo {author} {\bibfnamefont {M.~J.~P.}\
  \bibnamefont {Gingras}}, \ and\ \bibinfo {author} {\bibfnamefont {K.~A.}\
  \bibnamefont {Ross}},\ }\href {\doibase 10.1103/PhysRevLett.127.277206}
  {\bibfield  {journal} {\bibinfo  {journal} {Phys. Rev. Lett.}\ }\textbf
  {\bibinfo {volume} {127}},\ \bibinfo {pages} {277206} (\bibinfo {year}
  {2021})}\BibitemShut {NoStop}%
\bibitem [{\citenamefont {Gaudet}\ \emph
  {et~al.}(2016{\natexlab{b}})\citenamefont {Gaudet}, \citenamefont {Hallas},
  \citenamefont {Maharaj}, \citenamefont {Buhariwalla}, \citenamefont
  {Kermarrec}, \citenamefont {Butch}, \citenamefont {Munsie}, \citenamefont
  {Dabkowska}, \citenamefont {Luke},\ and\ \citenamefont
  {Gaulin}}]{gaudet16:94}%
  \BibitemOpen
  \bibfield  {author} {\bibinfo {author} {\bibfnamefont {J.}~\bibnamefont
  {Gaudet}}, \bibinfo {author} {\bibfnamefont {A.~M.}\ \bibnamefont {Hallas}},
  \bibinfo {author} {\bibfnamefont {D.~D.}\ \bibnamefont {Maharaj}}, \bibinfo
  {author} {\bibfnamefont {C.~R.~C.}\ \bibnamefont {Buhariwalla}}, \bibinfo
  {author} {\bibfnamefont {E.}~\bibnamefont {Kermarrec}}, \bibinfo {author}
  {\bibfnamefont {N.~P.}\ \bibnamefont {Butch}}, \bibinfo {author}
  {\bibfnamefont {T.~J.~S.}\ \bibnamefont {Munsie}}, \bibinfo {author}
  {\bibfnamefont {H.~A.}\ \bibnamefont {Dabkowska}}, \bibinfo {author}
  {\bibfnamefont {G.~M.}\ \bibnamefont {Luke}}, \ and\ \bibinfo {author}
  {\bibfnamefont {B.~D.}\ \bibnamefont {Gaulin}},\ }\href {\doibase
  10.1103/PhysRevB.94.060407} {\bibfield  {journal} {\bibinfo  {journal} {Phys.
  Rev. B}\ }\textbf {\bibinfo {volume} {94}},\ \bibinfo {pages} {060407}
  (\bibinfo {year} {2016}{\natexlab{b}})}\BibitemShut {NoStop}%
\bibitem [{\citenamefont {Cao}\ \emph {et~al.}(2009)\citenamefont {Cao},
  \citenamefont {Gukasov}, \citenamefont {Mirebeau}, \citenamefont {Bonville},
  \citenamefont {Decorse},\ and\ \citenamefont {Dhalenne}}]{cao09:103}%
  \BibitemOpen
  \bibfield  {author} {\bibinfo {author} {\bibfnamefont {H.}~\bibnamefont
  {Cao}}, \bibinfo {author} {\bibfnamefont {A.}~\bibnamefont {Gukasov}},
  \bibinfo {author} {\bibfnamefont {I.}~\bibnamefont {Mirebeau}}, \bibinfo
  {author} {\bibfnamefont {P.}~\bibnamefont {Bonville}}, \bibinfo {author}
  {\bibfnamefont {C.}~\bibnamefont {Decorse}}, \ and\ \bibinfo {author}
  {\bibfnamefont {G.}~\bibnamefont {Dhalenne}},\ }\href {\doibase
  10.1103/PhysRevLett.103.056402} {\bibfield  {journal} {\bibinfo  {journal}
  {Phys. Rev. Lett.}\ }\textbf {\bibinfo {volume} {103}},\ \bibinfo {pages}
  {056402} (\bibinfo {year} {2009})}\BibitemShut {NoStop}%
\bibitem [{\citenamefont {Bonville}\ \emph {et~al.}(2013)\citenamefont
  {Bonville}, \citenamefont {Petit}, \citenamefont {Mirebeau}, \citenamefont
  {Robert}, \citenamefont {Lhotel},\ and\ \citenamefont
  {Paulsen}}]{Bonville_2013}%
  \BibitemOpen
  \bibfield  {author} {\bibinfo {author} {\bibfnamefont {P.}~\bibnamefont
  {Bonville}}, \bibinfo {author} {\bibfnamefont {S.}~\bibnamefont {Petit}},
  \bibinfo {author} {\bibfnamefont {I.}~\bibnamefont {Mirebeau}}, \bibinfo
  {author} {\bibfnamefont {J.}~\bibnamefont {Robert}}, \bibinfo {author}
  {\bibfnamefont {E.}~\bibnamefont {Lhotel}}, \ and\ \bibinfo {author}
  {\bibfnamefont {C.}~\bibnamefont {Paulsen}},\ }\href {\doibase
  10.1088/0953-8984/25/27/275601} {\bibfield  {journal} {\bibinfo  {journal}
  {J. Phys.: Condens. Matter}\ }\textbf {\bibinfo {volume} {25}},\ \bibinfo
  {pages} {275601} (\bibinfo {year} {2013})}\BibitemShut {NoStop}%
\bibitem [{\citenamefont {Petrenko}\ \emph {et~al.}(2011)\citenamefont
  {Petrenko}, \citenamefont {Lees},\ and\ \citenamefont
  {Balakrishnan}}]{Petrenko_2011}%
  \BibitemOpen
  \bibfield  {author} {\bibinfo {author} {\bibfnamefont {O.~A.}\ \bibnamefont
  {Petrenko}}, \bibinfo {author} {\bibfnamefont {M.~R.}\ \bibnamefont {Lees}},
  \ and\ \bibinfo {author} {\bibfnamefont {G.}~\bibnamefont {Balakrishnan}},\
  }\href {\doibase 10.1088/0953-8984/23/16/164218} {\bibfield  {journal}
  {\bibinfo  {journal} {J. Phys.: Condens. Matter}\ }\textbf {\bibinfo {volume}
  {23}},\ \bibinfo {pages} {164218} (\bibinfo {year} {2011})}\BibitemShut
  {NoStop}%
\bibitem [{\citenamefont {Sarte}\ \emph {et~al.}(2011)\citenamefont {Sarte},
  \citenamefont {Silverstein}, \citenamefont {Wyk}, \citenamefont {Gardner},
  \citenamefont {Qiu}, \citenamefont {Zhou},\ and\ \citenamefont
  {Wiebe}}]{Sarte11:23}%
  \BibitemOpen
  \bibfield  {author} {\bibinfo {author} {\bibfnamefont {P.~M.}\ \bibnamefont
  {Sarte}}, \bibinfo {author} {\bibfnamefont {H.~J.}\ \bibnamefont
  {Silverstein}}, \bibinfo {author} {\bibfnamefont {B.~T. K.~V.}\ \bibnamefont
  {Wyk}}, \bibinfo {author} {\bibfnamefont {J.~S.}\ \bibnamefont {Gardner}},
  \bibinfo {author} {\bibfnamefont {Y.}~\bibnamefont {Qiu}}, \bibinfo {author}
  {\bibfnamefont {H.~D.}\ \bibnamefont {Zhou}}, \ and\ \bibinfo {author}
  {\bibfnamefont {C.~R.}\ \bibnamefont {Wiebe}},\ }\href {\doibase
  https://doi.org/10.1088/0953-8984/23/38/382201} {\bibfield  {journal}
  {\bibinfo  {journal} {J. Phys. Condens. Matter}\ }\textbf {\bibinfo {volume}
  {23}},\ \bibinfo {pages} {382201} (\bibinfo {year} {2011})}\BibitemShut
  {NoStop}%
\bibitem [{\citenamefont {Bondah-Jagalu}\ and\ \citenamefont
  {Bramwell}(2001)}]{EuTmSn_bondah2001magnetic}%
  \BibitemOpen
  \bibfield  {author} {\bibinfo {author} {\bibfnamefont {V.}~\bibnamefont
  {Bondah-Jagalu}}\ and\ \bibinfo {author} {\bibfnamefont {S.~T.}\ \bibnamefont
  {Bramwell}},\ }\href {\doibase 10.1139/cjp-79-11/12-1381} {\bibfield
  {journal} {\bibinfo  {journal} {Can. J. Phys.}\ }\textbf {\bibinfo {volume}
  {79}},\ \bibinfo {pages} {1381} (\bibinfo {year} {2001})}\BibitemShut
  {NoStop}%
\bibitem [{\citenamefont {{Al Ghamdi}}\ \emph {et~al.}(2014)\citenamefont {{Al
  Ghamdi}}, \citenamefont {Orend{\'a}{\v{c}}ov{\'a}}, \citenamefont
  {Pavl{\'\i}k},\ and\ \citenamefont
  {Orend{\'a}{\v{c}}}}]{al2014thermodynamic}%
  \BibitemOpen
  \bibfield  {author} {\bibinfo {author} {\bibfnamefont {N.}~\bibnamefont {{Al
  Ghamdi}}}, \bibinfo {author} {\bibfnamefont {A.}~\bibnamefont
  {Orend{\'a}{\v{c}}ov{\'a}}}, \bibinfo {author} {\bibfnamefont
  {V.}~\bibnamefont {Pavl{\'\i}k}}, \ and\ \bibinfo {author} {\bibfnamefont
  {M.}~\bibnamefont {Orend{\'a}{\v{c}}}},\ }\href {\doibase
  10.12693/aphyspola.126.264} {\bibfield  {journal} {\bibinfo  {journal} {Acta
  Phys. Pol.}\ }\textbf {\bibinfo {volume} {126}} (\bibinfo {year} {2014}),\
  10.12693/aphyspola.126.264}\BibitemShut {NoStop}%
\bibitem [{\citenamefont {Shirai}\ \emph {et~al.}(2017)\citenamefont {Shirai},
  \citenamefont {Freitas}, \citenamefont {Lago}, \citenamefont {Bramwell},
  \citenamefont {Ritter},\ and\ \citenamefont {\ifmmode \check{Z}\else
  \v{Z}\fi{}ivkovi\ifmmode~\acute{c}\else \'{c}\fi{}}}]{Shirai17:96}%
  \BibitemOpen
  \bibfield  {author} {\bibinfo {author} {\bibfnamefont {M.}~\bibnamefont
  {Shirai}}, \bibinfo {author} {\bibfnamefont {R.~S.}\ \bibnamefont {Freitas}},
  \bibinfo {author} {\bibfnamefont {J.}~\bibnamefont {Lago}}, \bibinfo {author}
  {\bibfnamefont {S.~T.}\ \bibnamefont {Bramwell}}, \bibinfo {author}
  {\bibfnamefont {C.}~\bibnamefont {Ritter}}, \ and\ \bibinfo {author}
  {\bibfnamefont {I.}~\bibnamefont {\ifmmode \check{Z}\else
  \v{Z}\fi{}ivkovi\ifmmode~\acute{c}\else \'{c}\fi{}}},\ }\href {\doibase
  10.1103/PhysRevB.96.180411} {\bibfield  {journal} {\bibinfo  {journal} {Phys.
  Rev. B}\ }\textbf {\bibinfo {volume} {96}},\ \bibinfo {pages} {180411}
  (\bibinfo {year} {2017})}\BibitemShut {NoStop}%
\bibitem [{\citenamefont {Tardif}\ \emph {et~al.}(2015)\citenamefont {Tardif},
  \citenamefont {Takeshita}, \citenamefont {Ohsumi}, \citenamefont {Yamaura},
  \citenamefont {Okuyama}, \citenamefont {Hiroi}, \citenamefont {Takata},\ and\
  \citenamefont {Arima}}]{Tardif15:114}%
  \BibitemOpen
  \bibfield  {author} {\bibinfo {author} {\bibfnamefont {S.}~\bibnamefont
  {Tardif}}, \bibinfo {author} {\bibfnamefont {S.}~\bibnamefont {Takeshita}},
  \bibinfo {author} {\bibfnamefont {H.}~\bibnamefont {Ohsumi}}, \bibinfo
  {author} {\bibfnamefont {J.-i.}\ \bibnamefont {Yamaura}}, \bibinfo {author}
  {\bibfnamefont {D.}~\bibnamefont {Okuyama}}, \bibinfo {author} {\bibfnamefont
  {Z.}~\bibnamefont {Hiroi}}, \bibinfo {author} {\bibfnamefont
  {M.}~\bibnamefont {Takata}}, \ and\ \bibinfo {author} {\bibfnamefont {T.-h.}\
  \bibnamefont {Arima}},\ }\href {\doibase 10.1103/PhysRevLett.114.147205}
  {\bibfield  {journal} {\bibinfo  {journal} {Phys. Rev. Lett.}\ }\textbf
  {\bibinfo {volume} {114}},\ \bibinfo {pages} {147205} (\bibinfo {year}
  {2015})}\BibitemShut {NoStop}%
\bibitem [{\citenamefont {Lhotel}\ \emph {et~al.}(2011)\citenamefont {Lhotel},
  \citenamefont {Simonet}, \citenamefont {Ortloff}, \citenamefont {Canals},
  \citenamefont {Paulsen}, \citenamefont {Suard}, \citenamefont {Hansen},
  \citenamefont {Price}, \citenamefont {Wood}, \citenamefont {Powell},\ and\
  \citenamefont {Ballou}}]{Lhotel11:107}%
  \BibitemOpen
  \bibfield  {author} {\bibinfo {author} {\bibfnamefont {E.}~\bibnamefont
  {Lhotel}}, \bibinfo {author} {\bibfnamefont {V.}~\bibnamefont {Simonet}},
  \bibinfo {author} {\bibfnamefont {J.}~\bibnamefont {Ortloff}}, \bibinfo
  {author} {\bibfnamefont {B.}~\bibnamefont {Canals}}, \bibinfo {author}
  {\bibfnamefont {C.}~\bibnamefont {Paulsen}}, \bibinfo {author} {\bibfnamefont
  {E.}~\bibnamefont {Suard}}, \bibinfo {author} {\bibfnamefont
  {T.}~\bibnamefont {Hansen}}, \bibinfo {author} {\bibfnamefont {D.~J.}\
  \bibnamefont {Price}}, \bibinfo {author} {\bibfnamefont {P.~T.}\ \bibnamefont
  {Wood}}, \bibinfo {author} {\bibfnamefont {A.~K.}\ \bibnamefont {Powell}}, \
  and\ \bibinfo {author} {\bibfnamefont {R.}~\bibnamefont {Ballou}},\ }\href
  {\doibase 10.1103/PhysRevLett.107.257205} {\bibfield  {journal} {\bibinfo
  {journal} {Phys. Rev. Lett.}\ }\textbf {\bibinfo {volume} {107}},\ \bibinfo
  {pages} {257205} (\bibinfo {year} {2011})}\BibitemShut {NoStop}%
\bibitem [{\citenamefont {Snyder}\ \emph
  {et~al.}(2004{\natexlab{a}})\citenamefont {Snyder}, \citenamefont {Ueland},
  \citenamefont {Mizel}, \citenamefont {Slusky}, \citenamefont {Karunadasa},
  \citenamefont {Cava},\ and\ \citenamefont {Schiffer}}]{snyder04:70}%
  \BibitemOpen
  \bibfield  {author} {\bibinfo {author} {\bibfnamefont {J.}~\bibnamefont
  {Snyder}}, \bibinfo {author} {\bibfnamefont {B.~G.}\ \bibnamefont {Ueland}},
  \bibinfo {author} {\bibfnamefont {A.}~\bibnamefont {Mizel}}, \bibinfo
  {author} {\bibfnamefont {J.~S.}\ \bibnamefont {Slusky}}, \bibinfo {author}
  {\bibfnamefont {H.}~\bibnamefont {Karunadasa}}, \bibinfo {author}
  {\bibfnamefont {R.~J.}\ \bibnamefont {Cava}}, \ and\ \bibinfo {author}
  {\bibfnamefont {P.}~\bibnamefont {Schiffer}},\ }\href {\doibase
  10.1103/PhysRevB.70.184431} {\bibfield  {journal} {\bibinfo  {journal} {Phys.
  Rev. B}\ }\textbf {\bibinfo {volume} {70}},\ \bibinfo {pages} {184431}
  (\bibinfo {year} {2004}{\natexlab{a}})}\BibitemShut {NoStop}%
\bibitem [{\citenamefont {Snyder}\ \emph
  {et~al.}(2001{\natexlab{a}})\citenamefont {Snyder}, \citenamefont {Slusky},
  \citenamefont {Cava},\ and\ \citenamefont {Schiffer}}]{synder01:413}%
  \BibitemOpen
  \bibfield  {author} {\bibinfo {author} {\bibfnamefont {J.}~\bibnamefont
  {Snyder}}, \bibinfo {author} {\bibfnamefont {J.~S.}\ \bibnamefont {Slusky}},
  \bibinfo {author} {\bibfnamefont {R.~J.}\ \bibnamefont {Cava}}, \ and\
  \bibinfo {author} {\bibfnamefont {P.}~\bibnamefont {Schiffer}},\ }\href
  {\doibase https://doi.org/10.1038/35092516} {\bibfield  {journal} {\bibinfo
  {journal} {Nature}\ }\textbf {\bibinfo {volume} {413}},\ \bibinfo {pages}
  {48} (\bibinfo {year} {2001}{\natexlab{a}})}\BibitemShut {NoStop}%
\bibitem [{\citenamefont {Ehlers}\ \emph {et~al.}(2006)\citenamefont {Ehlers},
  \citenamefont {Gardner}, \citenamefont {Booth}, \citenamefont {Daniel},
  \citenamefont {Kam}, \citenamefont {Cheetham}, \citenamefont {Antonio},
  \citenamefont {Brooks}, \citenamefont {Cornelius}, \citenamefont {Bramwell},
  \citenamefont {Lago}, \citenamefont {H\"aussler},\ and\ \citenamefont
  {Rosov}}]{Ehlers06:73}%
  \BibitemOpen
  \bibfield  {author} {\bibinfo {author} {\bibfnamefont {G.}~\bibnamefont
  {Ehlers}}, \bibinfo {author} {\bibfnamefont {J.~S.}\ \bibnamefont {Gardner}},
  \bibinfo {author} {\bibfnamefont {C.~H.}\ \bibnamefont {Booth}}, \bibinfo
  {author} {\bibfnamefont {M.}~\bibnamefont {Daniel}}, \bibinfo {author}
  {\bibfnamefont {K.~C.}\ \bibnamefont {Kam}}, \bibinfo {author} {\bibfnamefont
  {A.~K.}\ \bibnamefont {Cheetham}}, \bibinfo {author} {\bibfnamefont
  {D.}~\bibnamefont {Antonio}}, \bibinfo {author} {\bibfnamefont {H.~E.}\
  \bibnamefont {Brooks}}, \bibinfo {author} {\bibfnamefont {A.~L.}\
  \bibnamefont {Cornelius}}, \bibinfo {author} {\bibfnamefont {S.~T.}\
  \bibnamefont {Bramwell}}, \bibinfo {author} {\bibfnamefont {J.}~\bibnamefont
  {Lago}}, \bibinfo {author} {\bibfnamefont {W.}~\bibnamefont {H\"aussler}}, \
  and\ \bibinfo {author} {\bibfnamefont {N.}~\bibnamefont {Rosov}},\ }\href
  {\doibase 10.1103/PhysRevB.73.174429} {\bibfield  {journal} {\bibinfo
  {journal} {Phys. Rev. B}\ }\textbf {\bibinfo {volume} {73}},\ \bibinfo
  {pages} {174429} (\bibinfo {year} {2006})}\BibitemShut {NoStop}%
\bibitem [{\citenamefont {Ke}\ \emph {et~al.}(2007)\citenamefont {Ke},
  \citenamefont {Ueland}, \citenamefont {West}, \citenamefont {Dahlberg},
  \citenamefont {Cava},\ and\ \citenamefont {Schiffer}}]{Ke07:76}%
  \BibitemOpen
  \bibfield  {author} {\bibinfo {author} {\bibfnamefont {X.}~\bibnamefont
  {Ke}}, \bibinfo {author} {\bibfnamefont {B.~G.}\ \bibnamefont {Ueland}},
  \bibinfo {author} {\bibfnamefont {D.~V.}\ \bibnamefont {West}}, \bibinfo
  {author} {\bibfnamefont {M.~L.}\ \bibnamefont {Dahlberg}}, \bibinfo {author}
  {\bibfnamefont {R.~J.}\ \bibnamefont {Cava}}, \ and\ \bibinfo {author}
  {\bibfnamefont {P.}~\bibnamefont {Schiffer}},\ }\href {\doibase
  10.1103/PhysRevB.76.214413} {\bibfield  {journal} {\bibinfo  {journal} {Phys.
  Rev. B}\ }\textbf {\bibinfo {volume} {76}},\ \bibinfo {pages} {214413}
  (\bibinfo {year} {2007})}\BibitemShut {NoStop}%
\bibitem [{\citenamefont {Gingras}\ and\ \citenamefont
  {McClarty}(2014)}]{Yb_gingras2014quantum}%
  \BibitemOpen
  \bibfield  {author} {\bibinfo {author} {\bibfnamefont {M.~J.}\ \bibnamefont
  {Gingras}}\ and\ \bibinfo {author} {\bibfnamefont {P.~A.}\ \bibnamefont
  {McClarty}},\ }\href {\doibase 10.1088/0034-4885/77/5/056501} {\bibfield
  {journal} {\bibinfo  {journal} {Rep. Prog. Phys.}\ }\textbf {\bibinfo
  {volume} {77}},\ \bibinfo {pages} {056501} (\bibinfo {year}
  {2014})}\BibitemShut {NoStop}%
\bibitem [{\citenamefont {Applegate}\ \emph {et~al.}(2012)\citenamefont
  {Applegate}, \citenamefont {Hayre}, \citenamefont {Singh}, \citenamefont
  {Lin}, \citenamefont {Day},\ and\ \citenamefont {Gingras}}]{applegate12:109}%
  \BibitemOpen
  \bibfield  {author} {\bibinfo {author} {\bibfnamefont {R.}~\bibnamefont
  {Applegate}}, \bibinfo {author} {\bibfnamefont {N.~R.}\ \bibnamefont
  {Hayre}}, \bibinfo {author} {\bibfnamefont {R.~R.~P.}\ \bibnamefont {Singh}},
  \bibinfo {author} {\bibfnamefont {T.}~\bibnamefont {Lin}}, \bibinfo {author}
  {\bibfnamefont {A.~G.~R.}\ \bibnamefont {Day}}, \ and\ \bibinfo {author}
  {\bibfnamefont {M.~J.~P.}\ \bibnamefont {Gingras}},\ }\href {\doibase
  10.1103/PhysRevLett.109.097205} {\bibfield  {journal} {\bibinfo  {journal}
  {Phys. Rev. Lett.}\ }\textbf {\bibinfo {volume} {109}},\ \bibinfo {pages}
  {097205} (\bibinfo {year} {2012})}\BibitemShut {NoStop}%
\bibitem [{\citenamefont {Petit}(2020)}]{petit20:117}%
  \BibitemOpen
  \bibfield  {author} {\bibinfo {author} {\bibfnamefont {S.}~\bibnamefont
  {Petit}},\ }\href {\doibase 10.1073/pnas.2020105117} {\bibfield  {journal}
  {\bibinfo  {journal} {Proc. Natl. Acad. Sci.}\ }\textbf {\bibinfo {volume}
  {117}},\ \bibinfo {pages} {29263} (\bibinfo {year} {2020})}\BibitemShut
  {NoStop}%
\bibitem [{\citenamefont {Pan}\ \emph {et~al.}(2016)\citenamefont {Pan},
  \citenamefont {Laurita}, \citenamefont {Ross}, \citenamefont {Gaulin},\ and\
  \citenamefont {Armitage}}]{pan2016measure}%
  \BibitemOpen
  \bibfield  {author} {\bibinfo {author} {\bibfnamefont {L.}~\bibnamefont
  {Pan}}, \bibinfo {author} {\bibfnamefont {N.~J.}\ \bibnamefont {Laurita}},
  \bibinfo {author} {\bibfnamefont {K.~A.}\ \bibnamefont {Ross}}, \bibinfo
  {author} {\bibfnamefont {B.~D.}\ \bibnamefont {Gaulin}}, \ and\ \bibinfo
  {author} {\bibfnamefont {N.~P.}\ \bibnamefont {Armitage}},\ }\href {\doibase
  https://doi.org/10.1038/nphys3608} {\bibfield  {journal} {\bibinfo  {journal}
  {Nat. Phys.}\ }\textbf {\bibinfo {volume} {12}},\ \bibinfo {pages} {361}
  (\bibinfo {year} {2016})}\BibitemShut {NoStop}%
\bibitem [{\citenamefont {Baenitz}\ \emph {et~al.}(2018)\citenamefont
  {Baenitz}, \citenamefont {Schlender}, \citenamefont {Sichelschmidt},
  \citenamefont {Onykiienko}, \citenamefont {Zangeneh}, \citenamefont
  {Ranjith}, \citenamefont {Sarkar}, \citenamefont {Hozoi}, \citenamefont
  {Walker}, \citenamefont {Orain}, \citenamefont {Yasuoka}, \citenamefont
  {van~den Brink}, \citenamefont {Klauss}, \citenamefont {Inosov},\ and\
  \citenamefont {Doert}}]{PhysRevB.98.220409}%
  \BibitemOpen
  \bibfield  {author} {\bibinfo {author} {\bibfnamefont {M.}~\bibnamefont
  {Baenitz}}, \bibinfo {author} {\bibfnamefont {P.}~\bibnamefont {Schlender}},
  \bibinfo {author} {\bibfnamefont {J.}~\bibnamefont {Sichelschmidt}}, \bibinfo
  {author} {\bibfnamefont {Y.~A.}\ \bibnamefont {Onykiienko}}, \bibinfo
  {author} {\bibfnamefont {Z.}~\bibnamefont {Zangeneh}}, \bibinfo {author}
  {\bibfnamefont {K.~M.}\ \bibnamefont {Ranjith}}, \bibinfo {author}
  {\bibfnamefont {R.}~\bibnamefont {Sarkar}}, \bibinfo {author} {\bibfnamefont
  {L.}~\bibnamefont {Hozoi}}, \bibinfo {author} {\bibfnamefont {H.~C.}\
  \bibnamefont {Walker}}, \bibinfo {author} {\bibfnamefont {J.-C.}\
  \bibnamefont {Orain}}, \bibinfo {author} {\bibfnamefont {H.}~\bibnamefont
  {Yasuoka}}, \bibinfo {author} {\bibfnamefont {J.}~\bibnamefont {van~den
  Brink}}, \bibinfo {author} {\bibfnamefont {H.~H.}\ \bibnamefont {Klauss}},
  \bibinfo {author} {\bibfnamefont {D.~S.}\ \bibnamefont {Inosov}}, \ and\
  \bibinfo {author} {\bibfnamefont {T.}~\bibnamefont {Doert}},\ }\href
  {\doibase 10.1103/PhysRevB.98.220409} {\bibfield  {journal} {\bibinfo
  {journal} {Phys. Rev. B}\ }\textbf {\bibinfo {volume} {98}},\ \bibinfo
  {pages} {220409} (\bibinfo {year} {2018})}\BibitemShut {NoStop}%
\bibitem [{\citenamefont {Liu}\ \emph {et~al.}(2018)\citenamefont {Liu},
  \citenamefont {Zhang}, \citenamefont {Ji}, \citenamefont {Liu}, \citenamefont
  {Li}, \citenamefont {Wang}, \citenamefont {Lei}, \citenamefont {Chen},\ and\
  \citenamefont {Zhang}}]{liu2018rare}%
  \BibitemOpen
  \bibfield  {author} {\bibinfo {author} {\bibfnamefont {W.}~\bibnamefont
  {Liu}}, \bibinfo {author} {\bibfnamefont {Z.}~\bibnamefont {Zhang}}, \bibinfo
  {author} {\bibfnamefont {J.}~\bibnamefont {Ji}}, \bibinfo {author}
  {\bibfnamefont {Y.}~\bibnamefont {Liu}}, \bibinfo {author} {\bibfnamefont
  {J.}~\bibnamefont {Li}}, \bibinfo {author} {\bibfnamefont {X.}~\bibnamefont
  {Wang}}, \bibinfo {author} {\bibfnamefont {H.}~\bibnamefont {Lei}}, \bibinfo
  {author} {\bibfnamefont {G.}~\bibnamefont {Chen}}, \ and\ \bibinfo {author}
  {\bibfnamefont {Q.}~\bibnamefont {Zhang}},\ }\href {\doibase
  10.1088/0256-307X/35/11/117501} {\bibfield  {journal} {\bibinfo  {journal}
  {Chin. Phys. Lett.}\ }\textbf {\bibinfo {volume} {35}},\ \bibinfo {pages}
  {117501} (\bibinfo {year} {2018})}\BibitemShut {NoStop}%
\bibitem [{\citenamefont {Ranjith}\ \emph
  {et~al.}(2019{\natexlab{a}})\citenamefont {Ranjith}, \citenamefont
  {Dmytriieva}, \citenamefont {Khim}, \citenamefont {Sichelschmidt},
  \citenamefont {Luther}, \citenamefont {Ehlers}, \citenamefont {Yasuoka},
  \citenamefont {Wosnitza}, \citenamefont {Tsirlin}, \citenamefont {K\"uhne},\
  and\ \citenamefont {Baenitz}}]{PhysRevB.99.180401}%
  \BibitemOpen
  \bibfield  {author} {\bibinfo {author} {\bibfnamefont {K.~M.}\ \bibnamefont
  {Ranjith}}, \bibinfo {author} {\bibfnamefont {D.}~\bibnamefont {Dmytriieva}},
  \bibinfo {author} {\bibfnamefont {S.}~\bibnamefont {Khim}}, \bibinfo {author}
  {\bibfnamefont {J.}~\bibnamefont {Sichelschmidt}}, \bibinfo {author}
  {\bibfnamefont {S.}~\bibnamefont {Luther}}, \bibinfo {author} {\bibfnamefont
  {D.}~\bibnamefont {Ehlers}}, \bibinfo {author} {\bibfnamefont
  {H.}~\bibnamefont {Yasuoka}}, \bibinfo {author} {\bibfnamefont
  {J.}~\bibnamefont {Wosnitza}}, \bibinfo {author} {\bibfnamefont {A.~A.}\
  \bibnamefont {Tsirlin}}, \bibinfo {author} {\bibfnamefont {H.}~\bibnamefont
  {K\"uhne}}, \ and\ \bibinfo {author} {\bibfnamefont {M.}~\bibnamefont
  {Baenitz}},\ }\href {\doibase 10.1103/PhysRevB.99.180401} {\bibfield
  {journal} {\bibinfo  {journal} {Phys. Rev. B}\ }\textbf {\bibinfo {volume}
  {99}},\ \bibinfo {pages} {180401} (\bibinfo {year}
  {2019}{\natexlab{a}})}\BibitemShut {NoStop}%
\bibitem [{\citenamefont {Bordelon}\ \emph {et~al.}(2019)\citenamefont
  {Bordelon}, \citenamefont {Kenney}, \citenamefont {Liu}, \citenamefont
  {Hogan}, \citenamefont {Posthuma}, \citenamefont {Kavand}, \citenamefont
  {Lyu}, \citenamefont {Sherwin}, \citenamefont {Butch}, \citenamefont {Brown},
  \citenamefont {Graf}, \citenamefont {Balents},\ and\ \citenamefont
  {Wilson}}]{bordelon2019field}%
  \BibitemOpen
  \bibfield  {author} {\bibinfo {author} {\bibfnamefont {M.~M.}\ \bibnamefont
  {Bordelon}}, \bibinfo {author} {\bibfnamefont {E.}~\bibnamefont {Kenney}},
  \bibinfo {author} {\bibfnamefont {C.}~\bibnamefont {Liu}}, \bibinfo {author}
  {\bibfnamefont {T.}~\bibnamefont {Hogan}}, \bibinfo {author} {\bibfnamefont
  {L.}~\bibnamefont {Posthuma}}, \bibinfo {author} {\bibfnamefont
  {M.}~\bibnamefont {Kavand}}, \bibinfo {author} {\bibfnamefont
  {Y.}~\bibnamefont {Lyu}}, \bibinfo {author} {\bibfnamefont {M.}~\bibnamefont
  {Sherwin}}, \bibinfo {author} {\bibfnamefont {N.~P.}\ \bibnamefont {Butch}},
  \bibinfo {author} {\bibfnamefont {C.}~\bibnamefont {Brown}}, \bibinfo
  {author} {\bibfnamefont {M.~J.}\ \bibnamefont {Graf}}, \bibinfo {author}
  {\bibfnamefont {L.}~\bibnamefont {Balents}}, \ and\ \bibinfo {author}
  {\bibfnamefont {S.~D.}\ \bibnamefont {Wilson}},\ }\href {\doibase
  https://doi.org/10.1038/s41567-019-0594-5} {\bibfield  {journal} {\bibinfo
  {journal} {Nat. Phys.}\ }\textbf {\bibinfo {volume} {15}},\ \bibinfo {pages}
  {1058} (\bibinfo {year} {2019})}\BibitemShut {NoStop}%
\bibitem [{\citenamefont {Ranjith}\ \emph
  {et~al.}(2019{\natexlab{b}})\citenamefont {Ranjith}, \citenamefont {Luther},
  \citenamefont {Reimann}, \citenamefont {Schmidt}, \citenamefont {Schlender},
  \citenamefont {Sichelschmidt}, \citenamefont {Yasuoka}, \citenamefont
  {Strydom}, \citenamefont {Skourski}, \citenamefont {Wosnitza}, \citenamefont
  {K\"uhne}, \citenamefont {Doert},\ and\ \citenamefont
  {Baenitz}}]{PhysRevB.100.224417}%
  \BibitemOpen
  \bibfield  {author} {\bibinfo {author} {\bibfnamefont {K.~M.}\ \bibnamefont
  {Ranjith}}, \bibinfo {author} {\bibfnamefont {S.}~\bibnamefont {Luther}},
  \bibinfo {author} {\bibfnamefont {T.}~\bibnamefont {Reimann}}, \bibinfo
  {author} {\bibfnamefont {B.}~\bibnamefont {Schmidt}}, \bibinfo {author}
  {\bibfnamefont {P.}~\bibnamefont {Schlender}}, \bibinfo {author}
  {\bibfnamefont {J.}~\bibnamefont {Sichelschmidt}}, \bibinfo {author}
  {\bibfnamefont {H.}~\bibnamefont {Yasuoka}}, \bibinfo {author} {\bibfnamefont
  {A.~M.}\ \bibnamefont {Strydom}}, \bibinfo {author} {\bibfnamefont
  {Y.}~\bibnamefont {Skourski}}, \bibinfo {author} {\bibfnamefont
  {J.}~\bibnamefont {Wosnitza}}, \bibinfo {author} {\bibfnamefont
  {H.}~\bibnamefont {K\"uhne}}, \bibinfo {author} {\bibfnamefont
  {T.}~\bibnamefont {Doert}}, \ and\ \bibinfo {author} {\bibfnamefont
  {M.}~\bibnamefont {Baenitz}},\ }\href {\doibase 10.1103/PhysRevB.100.224417}
  {\bibfield  {journal} {\bibinfo  {journal} {Phys. Rev. B}\ }\textbf {\bibinfo
  {volume} {100}},\ \bibinfo {pages} {224417} (\bibinfo {year}
  {2019}{\natexlab{b}})}\BibitemShut {NoStop}%
\bibitem [{\citenamefont {Schmidt}\ \emph {et~al.}(2021)\citenamefont
  {Schmidt}, \citenamefont {Sichelschmidt}, \citenamefont {Ranjith},
  \citenamefont {Doert},\ and\ \citenamefont {Baenitz}}]{PhysRevB.103.214445}%
  \BibitemOpen
  \bibfield  {author} {\bibinfo {author} {\bibfnamefont {B.}~\bibnamefont
  {Schmidt}}, \bibinfo {author} {\bibfnamefont {J.}~\bibnamefont
  {Sichelschmidt}}, \bibinfo {author} {\bibfnamefont {K.~M.}\ \bibnamefont
  {Ranjith}}, \bibinfo {author} {\bibfnamefont {T.}~\bibnamefont {Doert}}, \
  and\ \bibinfo {author} {\bibfnamefont {M.}~\bibnamefont {Baenitz}},\ }\href
  {\doibase 10.1103/PhysRevB.103.214445} {\bibfield  {journal} {\bibinfo
  {journal} {Phys. Rev. B}\ }\textbf {\bibinfo {volume} {103}},\ \bibinfo
  {pages} {214445} (\bibinfo {year} {2021})}\BibitemShut {NoStop}%
\bibitem [{\citenamefont {Li}\ \emph {et~al.}(2015)\citenamefont {Li},
  \citenamefont {Chen}, \citenamefont {Tong}, \citenamefont {Pi}, \citenamefont
  {Liu}, \citenamefont {Yang}, \citenamefont {Wang},\ and\ \citenamefont
  {Zhang}}]{PhysRevLett.115.167203}%
  \BibitemOpen
  \bibfield  {author} {\bibinfo {author} {\bibfnamefont {Y.}~\bibnamefont
  {Li}}, \bibinfo {author} {\bibfnamefont {G.}~\bibnamefont {Chen}}, \bibinfo
  {author} {\bibfnamefont {W.}~\bibnamefont {Tong}}, \bibinfo {author}
  {\bibfnamefont {L.}~\bibnamefont {Pi}}, \bibinfo {author} {\bibfnamefont
  {J.}~\bibnamefont {Liu}}, \bibinfo {author} {\bibfnamefont {Z.}~\bibnamefont
  {Yang}}, \bibinfo {author} {\bibfnamefont {X.}~\bibnamefont {Wang}}, \ and\
  \bibinfo {author} {\bibfnamefont {Q.}~\bibnamefont {Zhang}},\ }\href
  {\doibase 10.1103/PhysRevLett.115.167203} {\bibfield  {journal} {\bibinfo
  {journal} {Phys. Rev. Lett.}\ }\textbf {\bibinfo {volume} {115}},\ \bibinfo
  {pages} {167203} (\bibinfo {year} {2015})}\BibitemShut {NoStop}%
\bibitem [{\citenamefont {Paddison}\ \emph {et~al.}(2017)\citenamefont
  {Paddison}, \citenamefont {Daum}, \citenamefont {Dun}, \citenamefont
  {Ehlers}, \citenamefont {Liu}, \citenamefont {Stone}, \citenamefont {Zhou},\
  and\ \citenamefont {Mourigal}}]{paddison2017continuous}%
  \BibitemOpen
  \bibfield  {author} {\bibinfo {author} {\bibfnamefont {J.~A.~M.}\
  \bibnamefont {Paddison}}, \bibinfo {author} {\bibfnamefont {M.}~\bibnamefont
  {Daum}}, \bibinfo {author} {\bibfnamefont {Z.}~\bibnamefont {Dun}}, \bibinfo
  {author} {\bibfnamefont {G.}~\bibnamefont {Ehlers}}, \bibinfo {author}
  {\bibfnamefont {Y.}~\bibnamefont {Liu}}, \bibinfo {author} {\bibfnamefont
  {M.~B.}\ \bibnamefont {Stone}}, \bibinfo {author} {\bibfnamefont
  {H.}~\bibnamefont {Zhou}}, \ and\ \bibinfo {author} {\bibfnamefont
  {M.}~\bibnamefont {Mourigal}},\ }\href {\doibase
  https://doi.org/10.1038/nphys3971} {\bibfield  {journal} {\bibinfo  {journal}
  {Nat. Phys.}\ }\textbf {\bibinfo {volume} {13}},\ \bibinfo {pages} {117}
  (\bibinfo {year} {2017})}\BibitemShut {NoStop}%
\bibitem [{\citenamefont {{Sriram Shastry}}\ and\ \citenamefont
  {Sutherland}(1981)}]{SRIRAMSHASTRY19811069}%
  \BibitemOpen
  \bibfield  {author} {\bibinfo {author} {\bibfnamefont {B.}~\bibnamefont
  {{Sriram Shastry}}}\ and\ \bibinfo {author} {\bibfnamefont {B.}~\bibnamefont
  {Sutherland}},\ }\href {\doibase
  https://doi.org/10.1016/0378-4363(81)90838-X} {\bibfield  {journal} {\bibinfo
   {journal} {Physica}\ }\textbf {\bibinfo {volume} {108}},\ \bibinfo {pages}
  {1069} (\bibinfo {year} {1981})}\BibitemShut {NoStop}%
\bibitem [{\citenamefont {Kim}\ and\ \citenamefont
  {Aronson}(2013)}]{PhysRevLett.110.017201}%
  \BibitemOpen
  \bibfield  {author} {\bibinfo {author} {\bibfnamefont {M.~S.}\ \bibnamefont
  {Kim}}\ and\ \bibinfo {author} {\bibfnamefont {M.~C.}\ \bibnamefont
  {Aronson}},\ }\href {\doibase 10.1103/PhysRevLett.110.017201} {\bibfield
  {journal} {\bibinfo  {journal} {Phys. Rev. Lett.}\ }\textbf {\bibinfo
  {volume} {110}},\ \bibinfo {pages} {017201} (\bibinfo {year}
  {2013})}\BibitemShut {NoStop}%
\bibitem [{\citenamefont {Bernhard}\ \emph {et~al.}(2011)\citenamefont
  {Bernhard}, \citenamefont {Coqblin},\ and\ \citenamefont
  {Lacroix}}]{PhysRevB.83.214427}%
  \BibitemOpen
  \bibfield  {author} {\bibinfo {author} {\bibfnamefont {B.~H.}\ \bibnamefont
  {Bernhard}}, \bibinfo {author} {\bibfnamefont {B.}~\bibnamefont {Coqblin}}, \
  and\ \bibinfo {author} {\bibfnamefont {C.}~\bibnamefont {Lacroix}},\ }\href
  {\doibase 10.1103/PhysRevB.83.214427} {\bibfield  {journal} {\bibinfo
  {journal} {Phys. Rev. B}\ }\textbf {\bibinfo {volume} {83}},\ \bibinfo
  {pages} {214427} (\bibinfo {year} {2011})}\BibitemShut {NoStop}%
\bibitem [{\citenamefont {Kikuchi}\ \emph {et~al.}(2009)\citenamefont
  {Kikuchi}, \citenamefont {Hara}, \citenamefont {Matsuoka}, \citenamefont
  {Onodera}, \citenamefont {Nakamura}, \citenamefont {Nojima}, \citenamefont
  {Katoh},\ and\ \citenamefont {Ochiai}}]{doi:10.1143/JPSJ.78.083708}%
  \BibitemOpen
  \bibfield  {author} {\bibinfo {author} {\bibfnamefont {F.}~\bibnamefont
  {Kikuchi}}, \bibinfo {author} {\bibfnamefont {K.}~\bibnamefont {Hara}},
  \bibinfo {author} {\bibfnamefont {E.}~\bibnamefont {Matsuoka}}, \bibinfo
  {author} {\bibfnamefont {H.}~\bibnamefont {Onodera}}, \bibinfo {author}
  {\bibfnamefont {S.}~\bibnamefont {Nakamura}}, \bibinfo {author}
  {\bibfnamefont {T.}~\bibnamefont {Nojima}}, \bibinfo {author} {\bibfnamefont
  {K.}~\bibnamefont {Katoh}}, \ and\ \bibinfo {author} {\bibfnamefont
  {A.}~\bibnamefont {Ochiai}},\ }\href {\doibase 10.1143/JPSJ.78.083708}
  {\bibfield  {journal} {\bibinfo  {journal} {J. Phys. Soc. Jpn.}\ }\textbf
  {\bibinfo {volume} {78}},\ \bibinfo {pages} {083708} (\bibinfo {year}
  {2009})}\BibitemShut {NoStop}%
\bibitem [{\citenamefont {Liu}\ \emph {et~al.}(2014{\natexlab{a}})\citenamefont
  {Liu}, \citenamefont {Chen}, \citenamefont {Lin}, \citenamefont {Tao},\ and\
  \citenamefont {Liu}}]{liu2014antiferromagnetic}%
  \BibitemOpen
  \bibfield  {author} {\bibinfo {author} {\bibfnamefont {H.-D.}\ \bibnamefont
  {Liu}}, \bibinfo {author} {\bibfnamefont {Y.-H.}\ \bibnamefont {Chen}},
  \bibinfo {author} {\bibfnamefont {H.-F.}\ \bibnamefont {Lin}}, \bibinfo
  {author} {\bibfnamefont {H.-S.}\ \bibnamefont {Tao}}, \ and\ \bibinfo
  {author} {\bibfnamefont {W.-M.}\ \bibnamefont {Liu}},\ }\href {\doibase
  https://doi.org/10.1038/srep04829} {\bibfield  {journal} {\bibinfo  {journal}
  {Sci. Rep.}\ }\textbf {\bibinfo {volume} {4}},\ \bibinfo {pages} {1}
  (\bibinfo {year} {2014}{\natexlab{a}})}\BibitemShut {NoStop}%
\bibitem [{\citenamefont {Dublenych}(2012)}]{PhysRevLett.109.167202}%
  \BibitemOpen
  \bibfield  {author} {\bibinfo {author} {\bibfnamefont {Y.~I.}\ \bibnamefont
  {Dublenych}},\ }\href {\doibase 10.1103/PhysRevLett.109.167202} {\bibfield
  {journal} {\bibinfo  {journal} {Phys. Rev. Lett.}\ }\textbf {\bibinfo
  {volume} {109}},\ \bibinfo {pages} {167202} (\bibinfo {year}
  {2012})}\BibitemShut {NoStop}%
\bibitem [{\citenamefont {Hodges}\ \emph {et~al.}(2001)\citenamefont {Hodges},
  \citenamefont {Bonville}, \citenamefont {Forget}, \citenamefont {Rams},
  \citenamefont {Kr{\'{o}}las},\ and\ \citenamefont {Dhalenne}}]{Hodges_2001}%
  \BibitemOpen
  \bibfield  {author} {\bibinfo {author} {\bibfnamefont {J.~A.}\ \bibnamefont
  {Hodges}}, \bibinfo {author} {\bibfnamefont {P.}~\bibnamefont {Bonville}},
  \bibinfo {author} {\bibfnamefont {A.}~\bibnamefont {Forget}}, \bibinfo
  {author} {\bibfnamefont {M.}~\bibnamefont {Rams}}, \bibinfo {author}
  {\bibfnamefont {K.}~\bibnamefont {Kr{\'{o}}las}}, \ and\ \bibinfo {author}
  {\bibfnamefont {G.}~\bibnamefont {Dhalenne}},\ }\href {\doibase
  10.1088/0953-8984/13/41/318} {\bibfield  {journal} {\bibinfo  {journal} {J.
  Phys.: Condens. Matter}\ }\textbf {\bibinfo {volume} {13}},\ \bibinfo {pages}
  {9301} (\bibinfo {year} {2001})}\BibitemShut {NoStop}%
\bibitem [{\citenamefont {Gaudet}\ \emph {et~al.}(2015)\citenamefont {Gaudet},
  \citenamefont {Maharaj}, \citenamefont {Sala}, \citenamefont {Kermarrec},
  \citenamefont {Ross}, \citenamefont {Dabkowska}, \citenamefont {Kolesnikov},
  \citenamefont {Granroth},\ and\ \citenamefont {Gaulin}}]{PhysRevB.92.134420}%
  \BibitemOpen
  \bibfield  {author} {\bibinfo {author} {\bibfnamefont {J.}~\bibnamefont
  {Gaudet}}, \bibinfo {author} {\bibfnamefont {D.~D.}\ \bibnamefont {Maharaj}},
  \bibinfo {author} {\bibfnamefont {G.}~\bibnamefont {Sala}}, \bibinfo {author}
  {\bibfnamefont {E.}~\bibnamefont {Kermarrec}}, \bibinfo {author}
  {\bibfnamefont {K.~A.}\ \bibnamefont {Ross}}, \bibinfo {author}
  {\bibfnamefont {H.~A.}\ \bibnamefont {Dabkowska}}, \bibinfo {author}
  {\bibfnamefont {A.~I.}\ \bibnamefont {Kolesnikov}}, \bibinfo {author}
  {\bibfnamefont {G.~E.}\ \bibnamefont {Granroth}}, \ and\ \bibinfo {author}
  {\bibfnamefont {B.~D.}\ \bibnamefont {Gaulin}},\ }\href {\doibase
  10.1103/PhysRevB.92.134420} {\bibfield  {journal} {\bibinfo  {journal} {Phys.
  Rev. B}\ }\textbf {\bibinfo {volume} {92}},\ \bibinfo {pages} {134420}
  (\bibinfo {year} {2015})}\BibitemShut {NoStop}%
\bibitem [{\citenamefont {Malkin}\ \emph {et~al.}(2004)\citenamefont {Malkin},
  \citenamefont {Zakirov}, \citenamefont {Popova}, \citenamefont {Klimin},
  \citenamefont {Chukalina}, \citenamefont {Antic-Fidancev}, \citenamefont
  {Goldner}, \citenamefont {Aschehoug},\ and\ \citenamefont
  {Dhalenne}}]{PhysRevB.70.075112}%
  \BibitemOpen
  \bibfield  {author} {\bibinfo {author} {\bibfnamefont {B.~Z.}\ \bibnamefont
  {Malkin}}, \bibinfo {author} {\bibfnamefont {A.~R.}\ \bibnamefont {Zakirov}},
  \bibinfo {author} {\bibfnamefont {M.~N.}\ \bibnamefont {Popova}}, \bibinfo
  {author} {\bibfnamefont {S.~A.}\ \bibnamefont {Klimin}}, \bibinfo {author}
  {\bibfnamefont {E.~P.}\ \bibnamefont {Chukalina}}, \bibinfo {author}
  {\bibfnamefont {E.}~\bibnamefont {Antic-Fidancev}}, \bibinfo {author}
  {\bibfnamefont {P.}~\bibnamefont {Goldner}}, \bibinfo {author} {\bibfnamefont
  {P.}~\bibnamefont {Aschehoug}}, \ and\ \bibinfo {author} {\bibfnamefont
  {G.}~\bibnamefont {Dhalenne}},\ }\href {\doibase 10.1103/PhysRevB.70.075112}
  {\bibfield  {journal} {\bibinfo  {journal} {Phys. Rev. B}\ }\textbf {\bibinfo
  {volume} {70}},\ \bibinfo {pages} {075112} (\bibinfo {year}
  {2004})}\BibitemShut {NoStop}%
\bibitem [{\citenamefont {Hermele}\ \emph {et~al.}(2004)\citenamefont
  {Hermele}, \citenamefont {Fisher},\ and\ \citenamefont
  {Balents}}]{PhysRevB.69.064404}%
  \BibitemOpen
  \bibfield  {author} {\bibinfo {author} {\bibfnamefont {M.}~\bibnamefont
  {Hermele}}, \bibinfo {author} {\bibfnamefont {M.~P.~A.}\ \bibnamefont
  {Fisher}}, \ and\ \bibinfo {author} {\bibfnamefont {L.}~\bibnamefont
  {Balents}},\ }\href {\doibase 10.1103/PhysRevB.69.064404} {\bibfield
  {journal} {\bibinfo  {journal} {Phys. Rev. B}\ }\textbf {\bibinfo {volume}
  {69}},\ \bibinfo {pages} {064404} (\bibinfo {year} {2004})}\BibitemShut
  {NoStop}%
\bibitem [{\citenamefont {Banerjee}\ \emph {et~al.}(2008)\citenamefont
  {Banerjee}, \citenamefont {Isakov}, \citenamefont {Damle},\ and\
  \citenamefont {Kim}}]{PhysRevLett.100.047208}%
  \BibitemOpen
  \bibfield  {author} {\bibinfo {author} {\bibfnamefont {A.}~\bibnamefont
  {Banerjee}}, \bibinfo {author} {\bibfnamefont {S.~V.}\ \bibnamefont
  {Isakov}}, \bibinfo {author} {\bibfnamefont {K.}~\bibnamefont {Damle}}, \
  and\ \bibinfo {author} {\bibfnamefont {Y.~B.}\ \bibnamefont {Kim}},\ }\href
  {\doibase 10.1103/PhysRevLett.100.047208} {\bibfield  {journal} {\bibinfo
  {journal} {Phys. Rev. Lett.}\ }\textbf {\bibinfo {volume} {100}},\ \bibinfo
  {pages} {047208} (\bibinfo {year} {2008})}\BibitemShut {NoStop}%
\bibitem [{\citenamefont {Savary}\ and\ \citenamefont
  {Balents}(2012)}]{PhysRevLett.108.037202}%
  \BibitemOpen
  \bibfield  {author} {\bibinfo {author} {\bibfnamefont {L.}~\bibnamefont
  {Savary}}\ and\ \bibinfo {author} {\bibfnamefont {L.}~\bibnamefont
  {Balents}},\ }\href {\doibase 10.1103/PhysRevLett.108.037202} {\bibfield
  {journal} {\bibinfo  {journal} {Phys. Rev. Lett.}\ }\textbf {\bibinfo
  {volume} {108}},\ \bibinfo {pages} {037202} (\bibinfo {year}
  {2012})}\BibitemShut {NoStop}%
\bibitem [{\citenamefont {Benton}\ \emph {et~al.}(2012)\citenamefont {Benton},
  \citenamefont {Sikora},\ and\ \citenamefont {Shannon}}]{PhysRevB.86.075154}%
  \BibitemOpen
  \bibfield  {author} {\bibinfo {author} {\bibfnamefont {O.}~\bibnamefont
  {Benton}}, \bibinfo {author} {\bibfnamefont {O.}~\bibnamefont {Sikora}}, \
  and\ \bibinfo {author} {\bibfnamefont {N.}~\bibnamefont {Shannon}},\ }\href
  {\doibase 10.1103/PhysRevB.86.075154} {\bibfield  {journal} {\bibinfo
  {journal} {Phys. Rev. B}\ }\textbf {\bibinfo {volume} {86}},\ \bibinfo
  {pages} {075154} (\bibinfo {year} {2012})}\BibitemShut {NoStop}%
\bibitem [{\citenamefont {Dun}\ \emph {et~al.}(2014)\citenamefont {Dun},
  \citenamefont {Lee}, \citenamefont {Choi}, \citenamefont {Hallas},
  \citenamefont {Wiebe}, \citenamefont {Gardner}, \citenamefont {Arrighi},
  \citenamefont {Freitas}, \citenamefont {Arevalo-Lopez}, \citenamefont
  {Attfield}, \citenamefont {Zhou},\ and\ \citenamefont
  {Cheng}}]{YbSn_dun2014chemical}%
  \BibitemOpen
  \bibfield  {author} {\bibinfo {author} {\bibfnamefont {Z.}~\bibnamefont
  {Dun}}, \bibinfo {author} {\bibfnamefont {M.}~\bibnamefont {Lee}}, \bibinfo
  {author} {\bibfnamefont {E.}~\bibnamefont {Choi}}, \bibinfo {author}
  {\bibfnamefont {A.}~\bibnamefont {Hallas}}, \bibinfo {author} {\bibfnamefont
  {C.}~\bibnamefont {Wiebe}}, \bibinfo {author} {\bibfnamefont
  {J.}~\bibnamefont {Gardner}}, \bibinfo {author} {\bibfnamefont
  {E.}~\bibnamefont {Arrighi}}, \bibinfo {author} {\bibfnamefont
  {R.}~\bibnamefont {Freitas}}, \bibinfo {author} {\bibfnamefont
  {A.}~\bibnamefont {Arevalo-Lopez}}, \bibinfo {author} {\bibfnamefont
  {J.}~\bibnamefont {Attfield}}, \bibinfo {author} {\bibfnamefont {H.~D.}\
  \bibnamefont {Zhou}}, \ and\ \bibinfo {author} {\bibfnamefont {J.~G.}\
  \bibnamefont {Cheng}},\ }\href {\doibase 10.1103/PhysRevB.89.064401}
  {\bibfield  {journal} {\bibinfo  {journal} {Phys. Rev. B}\ }\textbf {\bibinfo
  {volume} {89}},\ \bibinfo {pages} {064401} (\bibinfo {year}
  {2014})}\BibitemShut {NoStop}%
\bibitem [{\citenamefont {Hallas}\ \emph
  {et~al.}(2016{\natexlab{c}})\citenamefont {Hallas}, \citenamefont {Gaudet},
  \citenamefont {Butch}, \citenamefont {Tachibana}, \citenamefont {Freitas},
  \citenamefont {Luke}, \citenamefont {Wiebe},\ and\ \citenamefont
  {Gaulin}}]{YbSn_hallas2016universal}%
  \BibitemOpen
  \bibfield  {author} {\bibinfo {author} {\bibfnamefont {A.~M.}\ \bibnamefont
  {Hallas}}, \bibinfo {author} {\bibfnamefont {J.}~\bibnamefont {Gaudet}},
  \bibinfo {author} {\bibfnamefont {N.~P.}\ \bibnamefont {Butch}}, \bibinfo
  {author} {\bibfnamefont {M.}~\bibnamefont {Tachibana}}, \bibinfo {author}
  {\bibfnamefont {R.~S.}\ \bibnamefont {Freitas}}, \bibinfo {author}
  {\bibfnamefont {G.}~\bibnamefont {Luke}}, \bibinfo {author} {\bibfnamefont
  {C.~R.}\ \bibnamefont {Wiebe}}, \ and\ \bibinfo {author} {\bibfnamefont
  {B.~D.}\ \bibnamefont {Gaulin}},\ }\href {\doibase
  10.1103/PhysRevB.93.100403} {\bibfield  {journal} {\bibinfo  {journal} {Phys.
  Rev. B}\ }\textbf {\bibinfo {volume} {93}},\ \bibinfo {pages} {100403}
  (\bibinfo {year} {2016}{\natexlab{c}})}\BibitemShut {NoStop}%
\bibitem [{\citenamefont {Lhotel}\ \emph {et~al.}(2014)\citenamefont {Lhotel},
  \citenamefont {Giblin}, \citenamefont {Lees}, \citenamefont {Balakrishnan},
  \citenamefont {Chang},\ and\ \citenamefont {Yasui}}]{PhysRevB.89.224419}%
  \BibitemOpen
  \bibfield  {author} {\bibinfo {author} {\bibfnamefont {E.}~\bibnamefont
  {Lhotel}}, \bibinfo {author} {\bibfnamefont {S.~R.}\ \bibnamefont {Giblin}},
  \bibinfo {author} {\bibfnamefont {M.~R.}\ \bibnamefont {Lees}}, \bibinfo
  {author} {\bibfnamefont {G.}~\bibnamefont {Balakrishnan}}, \bibinfo {author}
  {\bibfnamefont {L.~J.}\ \bibnamefont {Chang}}, \ and\ \bibinfo {author}
  {\bibfnamefont {Y.}~\bibnamefont {Yasui}},\ }\href {\doibase
  10.1103/PhysRevB.89.224419} {\bibfield  {journal} {\bibinfo  {journal} {Phys.
  Rev. B}\ }\textbf {\bibinfo {volume} {89}},\ \bibinfo {pages} {224419}
  (\bibinfo {year} {2014})}\BibitemShut {NoStop}%
\bibitem [{\citenamefont {Yaouanc}\ \emph
  {et~al.}(2013{\natexlab{b}})\citenamefont {Yaouanc}, \citenamefont
  {Maisuradze},\ and\ \citenamefont {Dalmas~de
  R\'eotier}}]{PhysRevB.87.134405}%
  \BibitemOpen
  \bibfield  {author} {\bibinfo {author} {\bibfnamefont {A.}~\bibnamefont
  {Yaouanc}}, \bibinfo {author} {\bibfnamefont {A.}~\bibnamefont {Maisuradze}},
  \ and\ \bibinfo {author} {\bibfnamefont {P.}~\bibnamefont {Dalmas~de
  R\'eotier}},\ }\href {\doibase 10.1103/PhysRevB.87.134405} {\bibfield
  {journal} {\bibinfo  {journal} {Phys. Rev. B}\ }\textbf {\bibinfo {volume}
  {87}},\ \bibinfo {pages} {134405} (\bibinfo {year}
  {2013}{\natexlab{b}})}\BibitemShut {NoStop}%
\bibitem [{\citenamefont {Bowman}\ \emph {et~al.}(2019)\citenamefont {Bowman},
  \citenamefont {Cemal}, \citenamefont {Lehner}, \citenamefont {Wildes},
  \citenamefont {Mangin-Thro}, \citenamefont {Nilsen}, \citenamefont {Gutmann},
  \citenamefont {Voneshen}, \citenamefont {Prabhakaran}, \citenamefont
  {Boothroyd}, \citenamefont {Porter}, \citenamefont {castelnovo},
  \citenamefont {Refson},\ and\ \citenamefont {Goff}}]{bowman2019role}%
  \BibitemOpen
  \bibfield  {author} {\bibinfo {author} {\bibfnamefont {D.~F.}\ \bibnamefont
  {Bowman}}, \bibinfo {author} {\bibfnamefont {E.}~\bibnamefont {Cemal}},
  \bibinfo {author} {\bibfnamefont {T.}~\bibnamefont {Lehner}}, \bibinfo
  {author} {\bibfnamefont {A.~R.}\ \bibnamefont {Wildes}}, \bibinfo {author}
  {\bibfnamefont {L.}~\bibnamefont {Mangin-Thro}}, \bibinfo {author}
  {\bibfnamefont {G.~J.}\ \bibnamefont {Nilsen}}, \bibinfo {author}
  {\bibfnamefont {M.~J.}\ \bibnamefont {Gutmann}}, \bibinfo {author}
  {\bibfnamefont {D.}~\bibnamefont {Voneshen}}, \bibinfo {author}
  {\bibfnamefont {D.}~\bibnamefont {Prabhakaran}}, \bibinfo {author}
  {\bibfnamefont {A.~T.}\ \bibnamefont {Boothroyd}}, \bibinfo {author}
  {\bibfnamefont {D.~G.}\ \bibnamefont {Porter}}, \bibinfo {author}
  {\bibfnamefont {C.}~\bibnamefont {castelnovo}}, \bibinfo {author}
  {\bibfnamefont {K.}~\bibnamefont {Refson}}, \ and\ \bibinfo {author}
  {\bibfnamefont {J.~P.}\ \bibnamefont {Goff}},\ }\href {\doibase
  https://doi.org/10.1038/s41467-019-08598-z} {\bibfield  {journal} {\bibinfo
  {journal} {Nat. Commun.}\ }\textbf {\bibinfo {volume} {10}},\ \bibinfo
  {pages} {637} (\bibinfo {year} {2019})}\BibitemShut {NoStop}%
\bibitem [{\citenamefont {Sala}\ \emph {et~al.}(2014)\citenamefont {Sala},
  \citenamefont {Gutmann}, \citenamefont {Prabhakaran}, \citenamefont
  {Pomaranski}, \citenamefont {Mitchelitis}, \citenamefont {Kycia},
  \citenamefont {Porter}, \citenamefont {Castelnovo},\ and\ \citenamefont
  {Goff}}]{sala2014vacancy}%
  \BibitemOpen
  \bibfield  {author} {\bibinfo {author} {\bibfnamefont {G.}~\bibnamefont
  {Sala}}, \bibinfo {author} {\bibfnamefont {M.~J.}\ \bibnamefont {Gutmann}},
  \bibinfo {author} {\bibfnamefont {D.}~\bibnamefont {Prabhakaran}}, \bibinfo
  {author} {\bibfnamefont {D.}~\bibnamefont {Pomaranski}}, \bibinfo {author}
  {\bibfnamefont {C.}~\bibnamefont {Mitchelitis}}, \bibinfo {author}
  {\bibfnamefont {J.~B.}\ \bibnamefont {Kycia}}, \bibinfo {author}
  {\bibfnamefont {D.~G.}\ \bibnamefont {Porter}}, \bibinfo {author}
  {\bibfnamefont {C.}~\bibnamefont {Castelnovo}}, \ and\ \bibinfo {author}
  {\bibfnamefont {J.~P.}\ \bibnamefont {Goff}},\ }\href {\doibase
  https://doi.org/10.1038/nmat3924} {\bibfield  {journal} {\bibinfo  {journal}
  {Nature Mater.}\ }\textbf {\bibinfo {volume} {13}},\ \bibinfo {pages} {488}
  (\bibinfo {year} {2014})}\BibitemShut {NoStop}%
\bibitem [{\citenamefont {Mostaed}\ \emph {et~al.}(2017)\citenamefont
  {Mostaed}, \citenamefont {Balakrishnan}, \citenamefont {Lees}, \citenamefont
  {Yasui}, \citenamefont {Chang},\ and\ \citenamefont
  {Beanland}}]{PhysRevB.95.094431}%
  \BibitemOpen
  \bibfield  {author} {\bibinfo {author} {\bibfnamefont {A.}~\bibnamefont
  {Mostaed}}, \bibinfo {author} {\bibfnamefont {G.}~\bibnamefont
  {Balakrishnan}}, \bibinfo {author} {\bibfnamefont {M.~R.}\ \bibnamefont
  {Lees}}, \bibinfo {author} {\bibfnamefont {Y.}~\bibnamefont {Yasui}},
  \bibinfo {author} {\bibfnamefont {L.-J.}\ \bibnamefont {Chang}}, \ and\
  \bibinfo {author} {\bibfnamefont {R.}~\bibnamefont {Beanland}},\ }\href
  {\doibase 10.1103/PhysRevB.95.094431} {\bibfield  {journal} {\bibinfo
  {journal} {Phys. Rev. B}\ }\textbf {\bibinfo {volume} {95}},\ \bibinfo
  {pages} {094431} (\bibinfo {year} {2017})}\BibitemShut {NoStop}%
\bibitem [{\citenamefont {Kermarrec}\ \emph {et~al.}(2017)\citenamefont
  {Kermarrec}, \citenamefont {Gaudet}, \citenamefont {Fritsch}, \citenamefont
  {Khasanov}, \citenamefont {Guguchia}, \citenamefont {Ritter}, \citenamefont
  {Ross}, \citenamefont {Dabkowska},\ and\ \citenamefont
  {Gaulin}}]{kermarrec2017ground}%
  \BibitemOpen
  \bibfield  {author} {\bibinfo {author} {\bibfnamefont {E.}~\bibnamefont
  {Kermarrec}}, \bibinfo {author} {\bibfnamefont {J.}~\bibnamefont {Gaudet}},
  \bibinfo {author} {\bibfnamefont {K.}~\bibnamefont {Fritsch}}, \bibinfo
  {author} {\bibfnamefont {R.}~\bibnamefont {Khasanov}}, \bibinfo {author}
  {\bibfnamefont {Z.}~\bibnamefont {Guguchia}}, \bibinfo {author}
  {\bibfnamefont {C.}~\bibnamefont {Ritter}}, \bibinfo {author} {\bibfnamefont
  {K.~A.}\ \bibnamefont {Ross}}, \bibinfo {author} {\bibfnamefont {H.~A.}\
  \bibnamefont {Dabkowska}}, \ and\ \bibinfo {author} {\bibfnamefont {B.~D.}\
  \bibnamefont {Gaulin}},\ }\href {\doibase
  https://doi.org/10.1038/ncomms14810} {\bibfield  {journal} {\bibinfo
  {journal} {Nat. Commun.}\ }\textbf {\bibinfo {volume} {8}},\ \bibinfo {pages}
  {14810} (\bibinfo {year} {2017})}\BibitemShut {NoStop}%
\bibitem [{\citenamefont {Dun}\ \emph {et~al.}(2013)\citenamefont {Dun},
  \citenamefont {Choi}, \citenamefont {Zhou}, \citenamefont {Hallas},
  \citenamefont {Silverstein}, \citenamefont {Qiu}, \citenamefont {Copley},
  \citenamefont {Gardner},\ and\ \citenamefont {Wiebe}}]{YbSn_dun2013yb}%
  \BibitemOpen
  \bibfield  {author} {\bibinfo {author} {\bibfnamefont {Z.~L.}\ \bibnamefont
  {Dun}}, \bibinfo {author} {\bibfnamefont {E.~S.}\ \bibnamefont {Choi}},
  \bibinfo {author} {\bibfnamefont {H.~D.}\ \bibnamefont {Zhou}}, \bibinfo
  {author} {\bibfnamefont {A.~M.}\ \bibnamefont {Hallas}}, \bibinfo {author}
  {\bibfnamefont {H.~J.}\ \bibnamefont {Silverstein}}, \bibinfo {author}
  {\bibfnamefont {Y.}~\bibnamefont {Qiu}}, \bibinfo {author} {\bibfnamefont
  {J.~R.~D.}\ \bibnamefont {Copley}}, \bibinfo {author} {\bibfnamefont {J.~S.}\
  \bibnamefont {Gardner}}, \ and\ \bibinfo {author} {\bibfnamefont {C.~R.}\
  \bibnamefont {Wiebe}},\ }\href {\doibase 10.1103/PhysRevB.87.134408}
  {\bibfield  {journal} {\bibinfo  {journal} {Phys. Rev. B}\ }\textbf {\bibinfo
  {volume} {87}},\ \bibinfo {pages} {134408} (\bibinfo {year}
  {2013})}\BibitemShut {NoStop}%
\bibitem [{\citenamefont {Hodges}\ \emph {et~al.}(2002)\citenamefont {Hodges},
  \citenamefont {Bonville}, \citenamefont {Forget}, \citenamefont {Yaouanc},
  \citenamefont {Dalmas~de R\'eotier}, \citenamefont {Andr\'e}, \citenamefont
  {Rams}, \citenamefont {Kr\'olas}, \citenamefont {Ritter}, \citenamefont
  {Gubbens}, \citenamefont {Kaiser}, \citenamefont {King},\ and\ \citenamefont
  {Baines}}]{PhysRevLett.88.077204}%
  \BibitemOpen
  \bibfield  {author} {\bibinfo {author} {\bibfnamefont {J.~A.}\ \bibnamefont
  {Hodges}}, \bibinfo {author} {\bibfnamefont {P.}~\bibnamefont {Bonville}},
  \bibinfo {author} {\bibfnamefont {A.}~\bibnamefont {Forget}}, \bibinfo
  {author} {\bibfnamefont {A.}~\bibnamefont {Yaouanc}}, \bibinfo {author}
  {\bibfnamefont {P.}~\bibnamefont {Dalmas~de R\'eotier}}, \bibinfo {author}
  {\bibfnamefont {G.}~\bibnamefont {Andr\'e}}, \bibinfo {author} {\bibfnamefont
  {M.}~\bibnamefont {Rams}}, \bibinfo {author} {\bibfnamefont {K.}~\bibnamefont
  {Kr\'olas}}, \bibinfo {author} {\bibfnamefont {C.}~\bibnamefont {Ritter}},
  \bibinfo {author} {\bibfnamefont {P.~C.~M.}\ \bibnamefont {Gubbens}},
  \bibinfo {author} {\bibfnamefont {C.~T.}\ \bibnamefont {Kaiser}}, \bibinfo
  {author} {\bibfnamefont {P.~J.~C.}\ \bibnamefont {King}}, \ and\ \bibinfo
  {author} {\bibfnamefont {C.}~\bibnamefont {Baines}},\ }\href {\doibase
  10.1103/PhysRevLett.88.077204} {\bibfield  {journal} {\bibinfo  {journal}
  {Phys. Rev. Lett.}\ }\textbf {\bibinfo {volume} {88}},\ \bibinfo {pages}
  {077204} (\bibinfo {year} {2002})}\BibitemShut {NoStop}%
\bibitem [{\citenamefont {Hodges}\ \emph {et~al.}(2011)\citenamefont {Hodges},
  \citenamefont {De~Reotier}, \citenamefont {Yaouanc}, \citenamefont {Gubbens},
  \citenamefont {King},\ and\ \citenamefont {Baines}}]{hodges2011magnetic}%
  \BibitemOpen
  \bibfield  {author} {\bibinfo {author} {\bibfnamefont {J.~A.}\ \bibnamefont
  {Hodges}}, \bibinfo {author} {\bibfnamefont {P.~D.}\ \bibnamefont
  {De~Reotier}}, \bibinfo {author} {\bibfnamefont {A.}~\bibnamefont {Yaouanc}},
  \bibinfo {author} {\bibfnamefont {P.~C.~M.}\ \bibnamefont {Gubbens}},
  \bibinfo {author} {\bibfnamefont {P.~J.~C.}\ \bibnamefont {King}}, \ and\
  \bibinfo {author} {\bibfnamefont {C.}~\bibnamefont {Baines}},\ }\href
  {\doibase 10.1088/0953-8984/23/16/164217} {\bibfield  {journal} {\bibinfo
  {journal} {J. Condens. Matter Phys.}\ }\textbf {\bibinfo {volume} {23}},\
  \bibinfo {pages} {164217} (\bibinfo {year} {2011})}\BibitemShut {NoStop}%
\bibitem [{\citenamefont {Kermarrec}\ \emph {et~al.}(2015)\citenamefont
  {Kermarrec}, \citenamefont {Maharaj}, \citenamefont {Gaudet}, \citenamefont
  {Fritsch}, \citenamefont {Pomaranski}, \citenamefont {Kycia}, \citenamefont
  {Qiu}, \citenamefont {Copley}, \citenamefont {Couchman}, \citenamefont
  {Morningstar}, \citenamefont {Dabkowska},\ and\ \citenamefont
  {Gaulin}}]{PhysRevB.92.245114}%
  \BibitemOpen
  \bibfield  {author} {\bibinfo {author} {\bibfnamefont {E.}~\bibnamefont
  {Kermarrec}}, \bibinfo {author} {\bibfnamefont {D.~D.}\ \bibnamefont
  {Maharaj}}, \bibinfo {author} {\bibfnamefont {J.}~\bibnamefont {Gaudet}},
  \bibinfo {author} {\bibfnamefont {K.}~\bibnamefont {Fritsch}}, \bibinfo
  {author} {\bibfnamefont {D.}~\bibnamefont {Pomaranski}}, \bibinfo {author}
  {\bibfnamefont {J.~B.}\ \bibnamefont {Kycia}}, \bibinfo {author}
  {\bibfnamefont {Y.}~\bibnamefont {Qiu}}, \bibinfo {author} {\bibfnamefont
  {J.~R.~D.}\ \bibnamefont {Copley}}, \bibinfo {author} {\bibfnamefont
  {M.~M.~P.}\ \bibnamefont {Couchman}}, \bibinfo {author} {\bibfnamefont
  {A.~O.~R.}\ \bibnamefont {Morningstar}}, \bibinfo {author} {\bibfnamefont
  {H.~A.}\ \bibnamefont {Dabkowska}}, \ and\ \bibinfo {author} {\bibfnamefont
  {B.~D.}\ \bibnamefont {Gaulin}},\ }\href {\doibase
  10.1103/PhysRevB.92.245114} {\bibfield  {journal} {\bibinfo  {journal} {Phys.
  Rev. B}\ }\textbf {\bibinfo {volume} {92}},\ \bibinfo {pages} {245114}
  (\bibinfo {year} {2015})}\BibitemShut {NoStop}%
\bibitem [{\citenamefont {Taniguchi}\ \emph {et~al.}(2013)\citenamefont
  {Taniguchi}, \citenamefont {Kadowaki}, \citenamefont {Takatsu}, \citenamefont
  {F\aa{}k}, \citenamefont {Ollivier}, \citenamefont {Yamazaki}, \citenamefont
  {Sato}, \citenamefont {Yoshizawa}, \citenamefont {Shimura}, \citenamefont
  {Sakakibara}, \citenamefont {Hong}, \citenamefont {Goto}, \citenamefont
  {Yaraskavitch},\ and\ \citenamefont {Kycia}}]{PhysRevB.87.060408}%
  \BibitemOpen
  \bibfield  {author} {\bibinfo {author} {\bibfnamefont {T.}~\bibnamefont
  {Taniguchi}}, \bibinfo {author} {\bibfnamefont {H.}~\bibnamefont {Kadowaki}},
  \bibinfo {author} {\bibfnamefont {H.}~\bibnamefont {Takatsu}}, \bibinfo
  {author} {\bibfnamefont {B.}~\bibnamefont {F\aa{}k}}, \bibinfo {author}
  {\bibfnamefont {J.}~\bibnamefont {Ollivier}}, \bibinfo {author}
  {\bibfnamefont {T.}~\bibnamefont {Yamazaki}}, \bibinfo {author}
  {\bibfnamefont {T.~J.}\ \bibnamefont {Sato}}, \bibinfo {author}
  {\bibfnamefont {H.}~\bibnamefont {Yoshizawa}}, \bibinfo {author}
  {\bibfnamefont {Y.}~\bibnamefont {Shimura}}, \bibinfo {author} {\bibfnamefont
  {T.}~\bibnamefont {Sakakibara}}, \bibinfo {author} {\bibfnamefont
  {T.}~\bibnamefont {Hong}}, \bibinfo {author} {\bibfnamefont {K.}~\bibnamefont
  {Goto}}, \bibinfo {author} {\bibfnamefont {L.~R.}\ \bibnamefont
  {Yaraskavitch}}, \ and\ \bibinfo {author} {\bibfnamefont {J.~B.}\
  \bibnamefont {Kycia}},\ }\href {\doibase 10.1103/PhysRevB.87.060408}
  {\bibfield  {journal} {\bibinfo  {journal} {Phys. Rev. B}\ }\textbf {\bibinfo
  {volume} {87}},\ \bibinfo {pages} {060408} (\bibinfo {year}
  {2013})}\BibitemShut {NoStop}%
\bibitem [{\citenamefont {Kadowaki}\ \emph {et~al.}(2018)\citenamefont
  {Kadowaki}, \citenamefont {Wakita}, \citenamefont {F\r{a}k}, \citenamefont
  {Ollivier}, \citenamefont {Ohira-Kawamura}, \citenamefont {Nakajima},
  \citenamefont {Takatsu},\ and\ \citenamefont
  {Tamai}}]{doi:10.7566/JPSJ.87.064704}%
  \BibitemOpen
  \bibfield  {author} {\bibinfo {author} {\bibfnamefont {H.}~\bibnamefont
  {Kadowaki}}, \bibinfo {author} {\bibfnamefont {M.}~\bibnamefont {Wakita}},
  \bibinfo {author} {\bibfnamefont {B.}~\bibnamefont {F\r{a}k}}, \bibinfo
  {author} {\bibfnamefont {J.}~\bibnamefont {Ollivier}}, \bibinfo {author}
  {\bibfnamefont {S.}~\bibnamefont {Ohira-Kawamura}}, \bibinfo {author}
  {\bibfnamefont {K.}~\bibnamefont {Nakajima}}, \bibinfo {author}
  {\bibfnamefont {H.}~\bibnamefont {Takatsu}}, \ and\ \bibinfo {author}
  {\bibfnamefont {M.}~\bibnamefont {Tamai}},\ }\href {\doibase
  10.7566/JPSJ.87.064704} {\bibfield  {journal} {\bibinfo  {journal} {J. Phys.
  Soc. Jpn.}\ }\textbf {\bibinfo {volume} {87}},\ \bibinfo {pages} {064704}
  (\bibinfo {year} {2018})}\BibitemShut {NoStop}%
\bibitem [{\citenamefont {Kanada}\ \emph {et~al.}(1999)\citenamefont {Kanada},
  \citenamefont {Yasui}, \citenamefont {Ito}, \citenamefont {Harashina},
  \citenamefont {Sato}, \citenamefont {Okumura},\ and\ \citenamefont
  {Kakurai}}]{doi:10.1143/JPSJ.68.3802}%
  \BibitemOpen
  \bibfield  {author} {\bibinfo {author} {\bibfnamefont {M.}~\bibnamefont
  {Kanada}}, \bibinfo {author} {\bibfnamefont {Y.}~\bibnamefont {Yasui}},
  \bibinfo {author} {\bibfnamefont {M.}~\bibnamefont {Ito}}, \bibinfo {author}
  {\bibfnamefont {H.}~\bibnamefont {Harashina}}, \bibinfo {author}
  {\bibfnamefont {M.}~\bibnamefont {Sato}}, \bibinfo {author} {\bibfnamefont
  {H.}~\bibnamefont {Okumura}}, \ and\ \bibinfo {author} {\bibfnamefont
  {K.}~\bibnamefont {Kakurai}},\ }\href {\doibase 10.1143/JPSJ.68.3802}
  {\bibfield  {journal} {\bibinfo  {journal} {J. Phys. Soc. Jpn.}\ }\textbf
  {\bibinfo {volume} {68}},\ \bibinfo {pages} {3802} (\bibinfo {year}
  {1999})}\BibitemShut {NoStop}%
\bibitem [{\citenamefont {Anand}\ \emph {et~al.}(2018)\citenamefont {Anand},
  \citenamefont {Opherden}, \citenamefont {Xu}, \citenamefont {Adroja},
  \citenamefont {Hillier}, \citenamefont {Biswas}, \citenamefont
  {Herrmannsd\"orfer}, \citenamefont {Uhlarz}, \citenamefont {Hornung},
  \citenamefont {Wosnitza}, \citenamefont {Can\'evet},\ and\ \citenamefont
  {Lake}}]{TbHf_anand2018evidence}%
  \BibitemOpen
  \bibfield  {author} {\bibinfo {author} {\bibfnamefont {V.~K.}\ \bibnamefont
  {Anand}}, \bibinfo {author} {\bibfnamefont {L.}~\bibnamefont {Opherden}},
  \bibinfo {author} {\bibfnamefont {J.}~\bibnamefont {Xu}}, \bibinfo {author}
  {\bibfnamefont {D.~T.}\ \bibnamefont {Adroja}}, \bibinfo {author}
  {\bibfnamefont {A.~D.}\ \bibnamefont {Hillier}}, \bibinfo {author}
  {\bibfnamefont {P.~K.}\ \bibnamefont {Biswas}}, \bibinfo {author}
  {\bibfnamefont {T.}~\bibnamefont {Herrmannsd\"orfer}}, \bibinfo {author}
  {\bibfnamefont {M.}~\bibnamefont {Uhlarz}}, \bibinfo {author} {\bibfnamefont
  {J.}~\bibnamefont {Hornung}}, \bibinfo {author} {\bibfnamefont
  {J.}~\bibnamefont {Wosnitza}}, \bibinfo {author} {\bibfnamefont
  {E.}~\bibnamefont {Can\'evet}}, \ and\ \bibinfo {author} {\bibfnamefont
  {B.}~\bibnamefont {Lake}},\ }\href {\doibase 10.1103/PhysRevB.97.094402}
  {\bibfield  {journal} {\bibinfo  {journal} {Phys. Rev. B}\ }\textbf {\bibinfo
  {volume} {97}},\ \bibinfo {pages} {094402} (\bibinfo {year}
  {2018})}\BibitemShut {NoStop}%
\bibitem [{\citenamefont {Gaulin}\ \emph {et~al.}(2015)\citenamefont {Gaulin},
  \citenamefont {Kermarrec}, \citenamefont {Dahlberg}, \citenamefont
  {Matthews}, \citenamefont {Bert}, \citenamefont {Zhang}, \citenamefont
  {Mendels}, \citenamefont {Fritsch}, \citenamefont {Granroth}, \citenamefont
  {Jiramongkolchai}, \citenamefont {Amato}, \citenamefont {Baines},
  \citenamefont {Cava},\ and\ \citenamefont {Schiffer}}]{Gaulin15:91}%
  \BibitemOpen
  \bibfield  {author} {\bibinfo {author} {\bibfnamefont {B.~D.}\ \bibnamefont
  {Gaulin}}, \bibinfo {author} {\bibfnamefont {E.}~\bibnamefont {Kermarrec}},
  \bibinfo {author} {\bibfnamefont {M.~L.}\ \bibnamefont {Dahlberg}}, \bibinfo
  {author} {\bibfnamefont {M.~J.}\ \bibnamefont {Matthews}}, \bibinfo {author}
  {\bibfnamefont {F.}~\bibnamefont {Bert}}, \bibinfo {author} {\bibfnamefont
  {J.}~\bibnamefont {Zhang}}, \bibinfo {author} {\bibfnamefont
  {P.}~\bibnamefont {Mendels}}, \bibinfo {author} {\bibfnamefont
  {K.}~\bibnamefont {Fritsch}}, \bibinfo {author} {\bibfnamefont {G.~E.}\
  \bibnamefont {Granroth}}, \bibinfo {author} {\bibfnamefont {P.}~\bibnamefont
  {Jiramongkolchai}}, \bibinfo {author} {\bibfnamefont {A.}~\bibnamefont
  {Amato}}, \bibinfo {author} {\bibfnamefont {C.}~\bibnamefont {Baines}},
  \bibinfo {author} {\bibfnamefont {R.~J.}\ \bibnamefont {Cava}}, \ and\
  \bibinfo {author} {\bibfnamefont {P.}~\bibnamefont {Schiffer}},\ }\href
  {\doibase 10.1103/PhysRevB.91.245141} {\bibfield  {journal} {\bibinfo
  {journal} {Phys. Rev. B}\ }\textbf {\bibinfo {volume} {91}},\ \bibinfo
  {pages} {245141} (\bibinfo {year} {2015})}\BibitemShut {NoStop}%
\bibitem [{\citenamefont {Scheie}\ \emph {et~al.}(2017)\citenamefont {Scheie},
  \citenamefont {Kindervater}, \citenamefont {S\"aubert}, \citenamefont
  {Duvinage}, \citenamefont {Pfleiderer}, \citenamefont {Changlani},
  \citenamefont {Zhang}, \citenamefont {Harriger}, \citenamefont {Arpino},
  \citenamefont {Koohpayeh}, \citenamefont {Tchernyshyov},\ and\ \citenamefont
  {Broholm}}]{PhysRevLett.119.127201}%
  \BibitemOpen
  \bibfield  {author} {\bibinfo {author} {\bibfnamefont {A.}~\bibnamefont
  {Scheie}}, \bibinfo {author} {\bibfnamefont {J.}~\bibnamefont {Kindervater}},
  \bibinfo {author} {\bibfnamefont {S.}~\bibnamefont {S\"aubert}}, \bibinfo
  {author} {\bibfnamefont {C.}~\bibnamefont {Duvinage}}, \bibinfo {author}
  {\bibfnamefont {C.}~\bibnamefont {Pfleiderer}}, \bibinfo {author}
  {\bibfnamefont {H.~J.}\ \bibnamefont {Changlani}}, \bibinfo {author}
  {\bibfnamefont {S.}~\bibnamefont {Zhang}}, \bibinfo {author} {\bibfnamefont
  {L.}~\bibnamefont {Harriger}}, \bibinfo {author} {\bibfnamefont
  {K.}~\bibnamefont {Arpino}}, \bibinfo {author} {\bibfnamefont {S.~M.}\
  \bibnamefont {Koohpayeh}}, \bibinfo {author} {\bibfnamefont {O.}~\bibnamefont
  {Tchernyshyov}}, \ and\ \bibinfo {author} {\bibfnamefont {C.}~\bibnamefont
  {Broholm}},\ }\href {\doibase 10.1103/PhysRevLett.119.127201} {\bibfield
  {journal} {\bibinfo  {journal} {Phys. Rev. Lett.}\ }\textbf {\bibinfo
  {volume} {119}},\ \bibinfo {pages} {127201} (\bibinfo {year}
  {2017})}\BibitemShut {NoStop}%
\bibitem [{\citenamefont {Yaouanc}\ \emph {et~al.}(2016)\citenamefont
  {Yaouanc}, \citenamefont {De~R{\'e}otier}, \citenamefont {Keller},
  \citenamefont {Roessli},\ and\ \citenamefont
  {Forget}}]{YbTi_yaouanc2016novel}%
  \BibitemOpen
  \bibfield  {author} {\bibinfo {author} {\bibfnamefont {A.}~\bibnamefont
  {Yaouanc}}, \bibinfo {author} {\bibfnamefont {P.~D.}\ \bibnamefont
  {De~R{\'e}otier}}, \bibinfo {author} {\bibfnamefont {L.}~\bibnamefont
  {Keller}}, \bibinfo {author} {\bibfnamefont {B.}~\bibnamefont {Roessli}}, \
  and\ \bibinfo {author} {\bibfnamefont {A.}~\bibnamefont {Forget}},\ }\href
  {\doibase 10.1088/0953-8984/28/42/426002} {\bibfield  {journal} {\bibinfo
  {journal} {J. Condens. Matter Phys.}\ }\textbf {\bibinfo {volume} {28}},\
  \bibinfo {pages} {426002} (\bibinfo {year} {2016})}\BibitemShut {NoStop}%
\bibitem [{\citenamefont {Princep}\ \emph {et~al.}(2015)\citenamefont
  {Princep}, \citenamefont {Walker}, \citenamefont {Adroja}, \citenamefont
  {Prabhakaran},\ and\ \citenamefont {Boothroyd}}]{PhysRevB.91.224430}%
  \BibitemOpen
  \bibfield  {author} {\bibinfo {author} {\bibfnamefont {A.~J.}\ \bibnamefont
  {Princep}}, \bibinfo {author} {\bibfnamefont {H.~C.}\ \bibnamefont {Walker}},
  \bibinfo {author} {\bibfnamefont {D.~T.}\ \bibnamefont {Adroja}}, \bibinfo
  {author} {\bibfnamefont {D.}~\bibnamefont {Prabhakaran}}, \ and\ \bibinfo
  {author} {\bibfnamefont {A.~T.}\ \bibnamefont {Boothroyd}},\ }\href {\doibase
  10.1103/PhysRevB.91.224430} {\bibfield  {journal} {\bibinfo  {journal} {Phys.
  Rev. B}\ }\textbf {\bibinfo {volume} {91}},\ \bibinfo {pages} {224430}
  (\bibinfo {year} {2015})}\BibitemShut {NoStop}%
\bibitem [{\citenamefont {Zhang}\ \emph {et~al.}(2014)\citenamefont {Zhang},
  \citenamefont {Fritsch}, \citenamefont {Hao}, \citenamefont {Bagheri},
  \citenamefont {Gingras}, \citenamefont {Granroth}, \citenamefont
  {Jiramongkolchai}, \citenamefont {Cava},\ and\ \citenamefont
  {Gaulin}}]{PhysRevB.89.134410}%
  \BibitemOpen
  \bibfield  {author} {\bibinfo {author} {\bibfnamefont {J.}~\bibnamefont
  {Zhang}}, \bibinfo {author} {\bibfnamefont {K.}~\bibnamefont {Fritsch}},
  \bibinfo {author} {\bibfnamefont {Z.}~\bibnamefont {Hao}}, \bibinfo {author}
  {\bibfnamefont {B.~V.}\ \bibnamefont {Bagheri}}, \bibinfo {author}
  {\bibfnamefont {M.~J.~P.}\ \bibnamefont {Gingras}}, \bibinfo {author}
  {\bibfnamefont {G.~E.}\ \bibnamefont {Granroth}}, \bibinfo {author}
  {\bibfnamefont {P.}~\bibnamefont {Jiramongkolchai}}, \bibinfo {author}
  {\bibfnamefont {R.~J.}\ \bibnamefont {Cava}}, \ and\ \bibinfo {author}
  {\bibfnamefont {B.~D.}\ \bibnamefont {Gaulin}},\ }\href {\doibase
  10.1103/PhysRevB.89.134410} {\bibfield  {journal} {\bibinfo  {journal} {Phys.
  Rev. B}\ }\textbf {\bibinfo {volume} {89}},\ \bibinfo {pages} {134410}
  (\bibinfo {year} {2014})}\BibitemShut {NoStop}%
\bibitem [{\citenamefont {Mirebeau}\ \emph {et~al.}(2007)\citenamefont
  {Mirebeau}, \citenamefont {Bonville},\ and\ \citenamefont
  {Hennion}}]{PhysRevB.76.184436}%
  \BibitemOpen
  \bibfield  {author} {\bibinfo {author} {\bibfnamefont {I.}~\bibnamefont
  {Mirebeau}}, \bibinfo {author} {\bibfnamefont {P.}~\bibnamefont {Bonville}},
  \ and\ \bibinfo {author} {\bibfnamefont {M.}~\bibnamefont {Hennion}},\ }\href
  {\doibase 10.1103/PhysRevB.76.184436} {\bibfield  {journal} {\bibinfo
  {journal} {Phys. Rev. B}\ }\textbf {\bibinfo {volume} {76}},\ \bibinfo
  {pages} {184436} (\bibinfo {year} {2007})}\BibitemShut {NoStop}%
\bibitem [{\citenamefont {Molavian}\ \emph {et~al.}(2007)\citenamefont
  {Molavian}, \citenamefont {Gingras},\ and\ \citenamefont
  {Canals}}]{PhysRevLett.98.157204}%
  \BibitemOpen
  \bibfield  {author} {\bibinfo {author} {\bibfnamefont {H.~R.}\ \bibnamefont
  {Molavian}}, \bibinfo {author} {\bibfnamefont {M.~J.~P.}\ \bibnamefont
  {Gingras}}, \ and\ \bibinfo {author} {\bibfnamefont {B.}~\bibnamefont
  {Canals}},\ }\href {\doibase 10.1103/PhysRevLett.98.157204} {\bibfield
  {journal} {\bibinfo  {journal} {Phys. Rev. Lett.}\ }\textbf {\bibinfo
  {volume} {98}},\ \bibinfo {pages} {157204} (\bibinfo {year}
  {2007})}\BibitemShut {NoStop}%
\bibitem [{\citenamefont {Molavian}\ \emph {et~al.}(2009)\citenamefont
  {Molavian}, \citenamefont {McClarty},\ and\ \citenamefont
  {Gingras}}]{https://doi.org/10.48550/arxiv.0912.2957}%
  \BibitemOpen
  \bibfield  {author} {\bibinfo {author} {\bibfnamefont {H.~R.}\ \bibnamefont
  {Molavian}}, \bibinfo {author} {\bibfnamefont {P.~A.}\ \bibnamefont
  {McClarty}}, \ and\ \bibinfo {author} {\bibfnamefont {M.~J.~P.}\ \bibnamefont
  {Gingras}},\ }\href@noop {} {} (\bibinfo {year} {2009}),\ \Eprint
  {http://arxiv.org/abs/arXiv.0912.2957v1} {arXiv.0912.2957v1} \BibitemShut
  {NoStop}%
\bibitem [{\citenamefont {Hallas}\ \emph {et~al.}(2020)\citenamefont {Hallas},
  \citenamefont {Jin}, \citenamefont {Gaudet}, \citenamefont {Tonita},
  \citenamefont {Pomaranski}, \citenamefont {Buhariwalla}, \citenamefont
  {Tachibana}, \citenamefont {Butch}, \citenamefont {Calder}, \citenamefont
  {Stone}, \citenamefont {Luke}, \citenamefont {Wiebe}, \citenamefont {Kycia},
  \citenamefont {Gingras},\ and\ \citenamefont
  {Gaulin}}]{https://doi.org/10.48550/arxiv.2009.05036}%
  \BibitemOpen
  \bibfield  {author} {\bibinfo {author} {\bibfnamefont {A.~M.}\ \bibnamefont
  {Hallas}}, \bibinfo {author} {\bibfnamefont {W.}~\bibnamefont {Jin}},
  \bibinfo {author} {\bibfnamefont {J.}~\bibnamefont {Gaudet}}, \bibinfo
  {author} {\bibfnamefont {E.~M.}\ \bibnamefont {Tonita}}, \bibinfo {author}
  {\bibfnamefont {D.}~\bibnamefont {Pomaranski}}, \bibinfo {author}
  {\bibfnamefont {C.~R.~C.}\ \bibnamefont {Buhariwalla}}, \bibinfo {author}
  {\bibfnamefont {M.}~\bibnamefont {Tachibana}}, \bibinfo {author}
  {\bibfnamefont {N.~P.}\ \bibnamefont {Butch}}, \bibinfo {author}
  {\bibfnamefont {S.}~\bibnamefont {Calder}}, \bibinfo {author} {\bibfnamefont
  {M.~B.}\ \bibnamefont {Stone}}, \bibinfo {author} {\bibfnamefont {G.~M.}\
  \bibnamefont {Luke}}, \bibinfo {author} {\bibfnamefont {C.~R.}\ \bibnamefont
  {Wiebe}}, \bibinfo {author} {\bibfnamefont {J.~B.}\ \bibnamefont {Kycia}},
  \bibinfo {author} {\bibfnamefont {M.~J.~P.}\ \bibnamefont {Gingras}}, \ and\
  \bibinfo {author} {\bibfnamefont {B.~D.}\ \bibnamefont {Gaulin}},\
  }\href@noop {} {} (\bibinfo {year} {2020}),\ \Eprint
  {http://arxiv.org/abs/arXiv:2009.05036v2} {arXiv:2009.05036v2} \BibitemShut
  {NoStop}%
\bibitem [{\citenamefont {Liu}\ \emph {et~al.}(2019{\natexlab{b}})\citenamefont
  {Liu}, \citenamefont {Li},\ and\ \citenamefont {Chen}}]{PhysRevB.99.224407}%
  \BibitemOpen
  \bibfield  {author} {\bibinfo {author} {\bibfnamefont {C.}~\bibnamefont
  {Liu}}, \bibinfo {author} {\bibfnamefont {F.-Y.}\ \bibnamefont {Li}}, \ and\
  \bibinfo {author} {\bibfnamefont {G.}~\bibnamefont {Chen}},\ }\href {\doibase
  10.1103/PhysRevB.99.224407} {\bibfield  {journal} {\bibinfo  {journal} {Phys.
  Rev. B}\ }\textbf {\bibinfo {volume} {99}},\ \bibinfo {pages} {224407}
  (\bibinfo {year} {2019}{\natexlab{b}})}\BibitemShut {NoStop}%
\bibitem [{\citenamefont {Kao}\ \emph {et~al.}(2003)\citenamefont {Kao},
  \citenamefont {Enjalran}, \citenamefont {Del~Maestro}, \citenamefont
  {Molavian},\ and\ \citenamefont {Gingras}}]{PhysRevB.68.172407}%
  \BibitemOpen
  \bibfield  {author} {\bibinfo {author} {\bibfnamefont {Y.-J.}\ \bibnamefont
  {Kao}}, \bibinfo {author} {\bibfnamefont {M.}~\bibnamefont {Enjalran}},
  \bibinfo {author} {\bibfnamefont {A.}~\bibnamefont {Del~Maestro}}, \bibinfo
  {author} {\bibfnamefont {H.~R.}\ \bibnamefont {Molavian}}, \ and\ \bibinfo
  {author} {\bibfnamefont {M.~J.~P.}\ \bibnamefont {Gingras}},\ }\href
  {\doibase 10.1103/PhysRevB.68.172407} {\bibfield  {journal} {\bibinfo
  {journal} {Phys. Rev. B}\ }\textbf {\bibinfo {volume} {68}},\ \bibinfo
  {pages} {172407} (\bibinfo {year} {2003})}\BibitemShut {NoStop}%
\bibitem [{\citenamefont {Molavian}\ and\ \citenamefont
  {Gingras}(2009)}]{Molavian_2009}%
  \BibitemOpen
  \bibfield  {author} {\bibinfo {author} {\bibfnamefont {H.~R.}\ \bibnamefont
  {Molavian}}\ and\ \bibinfo {author} {\bibfnamefont {M.~J.~P.}\ \bibnamefont
  {Gingras}},\ }\href {\doibase 10.1088/0953-8984/21/17/172201} {\bibfield
  {journal} {\bibinfo  {journal} {J. Condens. Matter Phys.}\ }\textbf {\bibinfo
  {volume} {21}},\ \bibinfo {pages} {172201} (\bibinfo {year}
  {2009})}\BibitemShut {NoStop}%
\bibitem [{\citenamefont {Takatsu}\ \emph {et~al.}(2016)\citenamefont
  {Takatsu}, \citenamefont {Onoda}, \citenamefont {Kittaka}, \citenamefont
  {Kasahara}, \citenamefont {Kono}, \citenamefont {Sakakibara}, \citenamefont
  {Kato}, \citenamefont {F\aa{}k}, \citenamefont {Ollivier}, \citenamefont
  {Lynn}, \citenamefont {Taniguchi}, \citenamefont {Wakita},\ and\
  \citenamefont {Kadowaki}}]{PhysRevLett.116.217201}%
  \BibitemOpen
  \bibfield  {author} {\bibinfo {author} {\bibfnamefont {H.}~\bibnamefont
  {Takatsu}}, \bibinfo {author} {\bibfnamefont {S.}~\bibnamefont {Onoda}},
  \bibinfo {author} {\bibfnamefont {S.}~\bibnamefont {Kittaka}}, \bibinfo
  {author} {\bibfnamefont {A.}~\bibnamefont {Kasahara}}, \bibinfo {author}
  {\bibfnamefont {Y.}~\bibnamefont {Kono}}, \bibinfo {author} {\bibfnamefont
  {T.}~\bibnamefont {Sakakibara}}, \bibinfo {author} {\bibfnamefont
  {Y.}~\bibnamefont {Kato}}, \bibinfo {author} {\bibfnamefont {B.}~\bibnamefont
  {F\aa{}k}}, \bibinfo {author} {\bibfnamefont {J.}~\bibnamefont {Ollivier}},
  \bibinfo {author} {\bibfnamefont {J.~W.}\ \bibnamefont {Lynn}}, \bibinfo
  {author} {\bibfnamefont {T.}~\bibnamefont {Taniguchi}}, \bibinfo {author}
  {\bibfnamefont {M.}~\bibnamefont {Wakita}}, \ and\ \bibinfo {author}
  {\bibfnamefont {H.}~\bibnamefont {Kadowaki}},\ }\href {\doibase
  10.1103/PhysRevLett.116.217201} {\bibfield  {journal} {\bibinfo  {journal}
  {Phys. Rev. Lett.}\ }\textbf {\bibinfo {volume} {116}},\ \bibinfo {pages}
  {217201} (\bibinfo {year} {2016})}\BibitemShut {NoStop}%
\bibitem [{\citenamefont {Guitteny}\ \emph
  {et~al.}(2013{\natexlab{b}})\citenamefont {Guitteny}, \citenamefont {Robert},
  \citenamefont {Bonville}, \citenamefont {Ollivier}, \citenamefont {Decorse},
  \citenamefont {Steffens}, \citenamefont {Boehm}, \citenamefont {Mutka},
  \citenamefont {Mirebeau},\ and\ \citenamefont
  {Petit}}]{PhysRevLett.111.087201}%
  \BibitemOpen
  \bibfield  {author} {\bibinfo {author} {\bibfnamefont {S.}~\bibnamefont
  {Guitteny}}, \bibinfo {author} {\bibfnamefont {J.}~\bibnamefont {Robert}},
  \bibinfo {author} {\bibfnamefont {P.}~\bibnamefont {Bonville}}, \bibinfo
  {author} {\bibfnamefont {J.}~\bibnamefont {Ollivier}}, \bibinfo {author}
  {\bibfnamefont {C.}~\bibnamefont {Decorse}}, \bibinfo {author} {\bibfnamefont
  {P.}~\bibnamefont {Steffens}}, \bibinfo {author} {\bibfnamefont
  {M.}~\bibnamefont {Boehm}}, \bibinfo {author} {\bibfnamefont
  {H.}~\bibnamefont {Mutka}}, \bibinfo {author} {\bibfnamefont
  {I.}~\bibnamefont {Mirebeau}}, \ and\ \bibinfo {author} {\bibfnamefont
  {S.}~\bibnamefont {Petit}},\ }\href {\doibase 10.1103/PhysRevLett.111.087201}
  {\bibfield  {journal} {\bibinfo  {journal} {Phys. Rev. Lett.}\ }\textbf
  {\bibinfo {volume} {111}},\ \bibinfo {pages} {087201} (\bibinfo {year}
  {2013}{\natexlab{b}})}\BibitemShut {NoStop}%
\bibitem [{\citenamefont {Petit}\ \emph {et~al.}(2012)\citenamefont {Petit},
  \citenamefont {Bonville}, \citenamefont {Robert}, \citenamefont {Decorse},\
  and\ \citenamefont {Mirebeau}}]{PhysRevB.86.174403}%
  \BibitemOpen
  \bibfield  {author} {\bibinfo {author} {\bibfnamefont {S.}~\bibnamefont
  {Petit}}, \bibinfo {author} {\bibfnamefont {P.}~\bibnamefont {Bonville}},
  \bibinfo {author} {\bibfnamefont {J.}~\bibnamefont {Robert}}, \bibinfo
  {author} {\bibfnamefont {C.}~\bibnamefont {Decorse}}, \ and\ \bibinfo
  {author} {\bibfnamefont {I.}~\bibnamefont {Mirebeau}},\ }\href {\doibase
  10.1103/PhysRevB.86.174403} {\bibfield  {journal} {\bibinfo  {journal} {Phys.
  Rev. B}\ }\textbf {\bibinfo {volume} {86}},\ \bibinfo {pages} {174403}
  (\bibinfo {year} {2012})}\BibitemShut {NoStop}%
\bibitem [{\citenamefont {Wakita}\ \emph {et~al.}(2016)\citenamefont {Wakita},
  \citenamefont {Taniguchi}, \citenamefont {Edamoto}, \citenamefont {Takatsu},\
  and\ \citenamefont {Kadowaki}}]{Wakita_2016}%
  \BibitemOpen
  \bibfield  {author} {\bibinfo {author} {\bibfnamefont {M.}~\bibnamefont
  {Wakita}}, \bibinfo {author} {\bibfnamefont {T.}~\bibnamefont {Taniguchi}},
  \bibinfo {author} {\bibfnamefont {H.}~\bibnamefont {Edamoto}}, \bibinfo
  {author} {\bibfnamefont {H.}~\bibnamefont {Takatsu}}, \ and\ \bibinfo
  {author} {\bibfnamefont {H.}~\bibnamefont {Kadowaki}},\ }\href {\doibase
  10.1088/1742-6596/683/1/012023} {\bibfield  {journal} {\bibinfo  {journal}
  {J. Phys.: Conf. Ser.}\ }\textbf {\bibinfo {volume} {683}},\ \bibinfo {pages}
  {012023} (\bibinfo {year} {2016})}\BibitemShut {NoStop}%
\bibitem [{\citenamefont {Kadowaki}\ \emph {et~al.}(2015)\citenamefont
  {Kadowaki}, \citenamefont {Takatsu}, \citenamefont {Taniguchi}, \citenamefont
  {F\r{a}k},\ and\ \citenamefont {Ollivier}}]{doi:10.1142/S2010324715400032}%
  \BibitemOpen
  \bibfield  {author} {\bibinfo {author} {\bibfnamefont {H.}~\bibnamefont
  {Kadowaki}}, \bibinfo {author} {\bibfnamefont {H.}~\bibnamefont {Takatsu}},
  \bibinfo {author} {\bibfnamefont {T.}~\bibnamefont {Taniguchi}}, \bibinfo
  {author} {\bibfnamefont {B.}~\bibnamefont {F\r{a}k}}, \ and\ \bibinfo
  {author} {\bibfnamefont {J.}~\bibnamefont {Ollivier}},\ }\href {\doibase
  10.1142/S2010324715400032} {\bibfield  {journal} {\bibinfo  {journal} {SPIN}\
  }\textbf {\bibinfo {volume} {05}},\ \bibinfo {pages} {1540003} (\bibinfo
  {year} {2015})}\BibitemShut {NoStop}%
\bibitem [{\citenamefont {Mirebeau}\ \emph {et~al.}(2005)\citenamefont
  {Mirebeau}, \citenamefont {Apetrei}, \citenamefont {Rodr{\'\i}guez-Carvajal},
  \citenamefont {Bonville}, \citenamefont {Forget}, \citenamefont {Colson},
  \citenamefont {Glazkov}, \citenamefont {Sanchez}, \citenamefont {Isnard},\
  and\ \citenamefont {Suard}}]{TbSn_mirebeau2005ordered}%
  \BibitemOpen
  \bibfield  {author} {\bibinfo {author} {\bibfnamefont {I.}~\bibnamefont
  {Mirebeau}}, \bibinfo {author} {\bibfnamefont {A.}~\bibnamefont {Apetrei}},
  \bibinfo {author} {\bibfnamefont {J.}~\bibnamefont
  {Rodr{\'\i}guez-Carvajal}}, \bibinfo {author} {\bibfnamefont
  {P.}~\bibnamefont {Bonville}}, \bibinfo {author} {\bibfnamefont
  {A.}~\bibnamefont {Forget}}, \bibinfo {author} {\bibfnamefont
  {D.}~\bibnamefont {Colson}}, \bibinfo {author} {\bibfnamefont
  {V.}~\bibnamefont {Glazkov}}, \bibinfo {author} {\bibfnamefont
  {J.}~\bibnamefont {Sanchez}}, \bibinfo {author} {\bibfnamefont
  {O.}~\bibnamefont {Isnard}}, \ and\ \bibinfo {author} {\bibfnamefont
  {E.}~\bibnamefont {Suard}},\ }\href {\doibase 10.1103/PhysRevLett.94.246402}
  {\bibfield  {journal} {\bibinfo  {journal} {Phys. Rev. Lett.}\ }\textbf
  {\bibinfo {volume} {94}},\ \bibinfo {pages} {246402} (\bibinfo {year}
  {2005})}\BibitemShut {NoStop}%
\bibitem [{\citenamefont {Bert}\ \emph {et~al.}(2006)\citenamefont {Bert},
  \citenamefont {Mendels}, \citenamefont {Olariu}, \citenamefont {Blanchard},
  \citenamefont {Collin}, \citenamefont {Amato}, \citenamefont {Baines},\ and\
  \citenamefont {Hillier}}]{TbSn_bert2006direct}%
  \BibitemOpen
  \bibfield  {author} {\bibinfo {author} {\bibfnamefont {F.}~\bibnamefont
  {Bert}}, \bibinfo {author} {\bibfnamefont {P.}~\bibnamefont {Mendels}},
  \bibinfo {author} {\bibfnamefont {A.}~\bibnamefont {Olariu}}, \bibinfo
  {author} {\bibfnamefont {N.}~\bibnamefont {Blanchard}}, \bibinfo {author}
  {\bibfnamefont {G.}~\bibnamefont {Collin}}, \bibinfo {author} {\bibfnamefont
  {A.}~\bibnamefont {Amato}}, \bibinfo {author} {\bibfnamefont
  {C.}~\bibnamefont {Baines}}, \ and\ \bibinfo {author} {\bibfnamefont {A.~D.}\
  \bibnamefont {Hillier}},\ }\href {\doibase 10.1103/PhysRevLett.97.117203}
  {\bibfield  {journal} {\bibinfo  {journal} {Phys. Rev. Lett.}\ }\textbf
  {\bibinfo {volume} {97}},\ \bibinfo {pages} {117203} (\bibinfo {year}
  {2006})}\BibitemShut {NoStop}%
\bibitem [{\citenamefont {Dahlberg}\ \emph {et~al.}(2011)\citenamefont
  {Dahlberg}, \citenamefont {Matthews}, \citenamefont {Jiramongkolchai},
  \citenamefont {Cava},\ and\ \citenamefont {Schiffer}}]{TbSn_cava2011low}%
  \BibitemOpen
  \bibfield  {author} {\bibinfo {author} {\bibfnamefont {M.~L.}\ \bibnamefont
  {Dahlberg}}, \bibinfo {author} {\bibfnamefont {M.~J.}\ \bibnamefont
  {Matthews}}, \bibinfo {author} {\bibfnamefont {P.}~\bibnamefont
  {Jiramongkolchai}}, \bibinfo {author} {\bibfnamefont {R.~J.}\ \bibnamefont
  {Cava}}, \ and\ \bibinfo {author} {\bibfnamefont {P.}~\bibnamefont
  {Schiffer}},\ }\href {\doibase 10.1103/PhysRevB.83.140410} {\bibfield
  {journal} {\bibinfo  {journal} {Phys. Rev. B}\ }\textbf {\bibinfo {volume}
  {83}},\ \bibinfo {pages} {140410} (\bibinfo {year} {2011})}\BibitemShut
  {NoStop}%
\bibitem [{\citenamefont {Chapuis}\ \emph {et~al.}(2007)\citenamefont
  {Chapuis}, \citenamefont {Yaouanc}, \citenamefont {de~R{\'e}otier},
  \citenamefont {Pouget}, \citenamefont {Fouquet}, \citenamefont {Cervellino},\
  and\ \citenamefont {Forget}}]{TbSn_chapuis2007ground}%
  \BibitemOpen
  \bibfield  {author} {\bibinfo {author} {\bibfnamefont {Y.}~\bibnamefont
  {Chapuis}}, \bibinfo {author} {\bibfnamefont {A.}~\bibnamefont {Yaouanc}},
  \bibinfo {author} {\bibfnamefont {P.~D.}\ \bibnamefont {de~R{\'e}otier}},
  \bibinfo {author} {\bibfnamefont {S.}~\bibnamefont {Pouget}}, \bibinfo
  {author} {\bibfnamefont {P.}~\bibnamefont {Fouquet}}, \bibinfo {author}
  {\bibfnamefont {A.}~\bibnamefont {Cervellino}}, \ and\ \bibinfo {author}
  {\bibfnamefont {A.}~\bibnamefont {Forget}},\ }\href {\doibase
  10.1088/0953-8984/19/44/446206} {\bibfield  {journal} {\bibinfo  {journal}
  {Journal of Physics: Condensed Matter}\ }\textbf {\bibinfo {volume} {19}},\
  \bibinfo {pages} {446206} (\bibinfo {year} {2007})}\BibitemShut {NoStop}%
\bibitem [{\citenamefont {De~Reotier}\ \emph {et~al.}(2006)\citenamefont
  {De~Reotier}, \citenamefont {Yaouanc}, \citenamefont {Keller}, \citenamefont
  {Cervellino}, \citenamefont {Roessli}, \citenamefont {Baines}, \citenamefont
  {Forget}, \citenamefont {Vaju}, \citenamefont {Gubbens}, \citenamefont
  {Amato},\ and\ \citenamefont {King}}]{TbSn_de2006spin}%
  \BibitemOpen
  \bibfield  {author} {\bibinfo {author} {\bibfnamefont {P.~D.}\ \bibnamefont
  {De~Reotier}}, \bibinfo {author} {\bibfnamefont {A.}~\bibnamefont {Yaouanc}},
  \bibinfo {author} {\bibfnamefont {L.}~\bibnamefont {Keller}}, \bibinfo
  {author} {\bibfnamefont {A.}~\bibnamefont {Cervellino}}, \bibinfo {author}
  {\bibfnamefont {B.}~\bibnamefont {Roessli}}, \bibinfo {author} {\bibfnamefont
  {C.}~\bibnamefont {Baines}}, \bibinfo {author} {\bibfnamefont
  {A.}~\bibnamefont {Forget}}, \bibinfo {author} {\bibfnamefont
  {C.}~\bibnamefont {Vaju}}, \bibinfo {author} {\bibfnamefont {P.}~\bibnamefont
  {Gubbens}}, \bibinfo {author} {\bibfnamefont {A.}~\bibnamefont {Amato}}, \
  and\ \bibinfo {author} {\bibfnamefont {P.~J.~C.}\ \bibnamefont {King}},\
  }\href {\doibase 10.1103/PhysRevLett.96.127202} {\bibfield  {journal}
  {\bibinfo  {journal} {Phys. Rev. Lett.}\ }\textbf {\bibinfo {volume} {96}},\
  \bibinfo {pages} {127202} (\bibinfo {year} {2006})}\BibitemShut {NoStop}%
\bibitem [{\citenamefont {Rule}\ \emph {et~al.}(2007)\citenamefont {Rule},
  \citenamefont {Ehlers}, \citenamefont {Stewart}, \citenamefont {Cornelius},
  \citenamefont {Deen}, \citenamefont {Qiu}, \citenamefont {Wiebe},
  \citenamefont {Janik}, \citenamefont {Zhou}, \citenamefont {Antonio},
  \citenamefont {Woytko}, \citenamefont {Ruff}, \citenamefont {Dabkowska},
  \citenamefont {Gaulin},\ and\ \citenamefont
  {Gardner}}]{TbSn_rule2007polarized}%
  \BibitemOpen
  \bibfield  {author} {\bibinfo {author} {\bibfnamefont {K.~C.}\ \bibnamefont
  {Rule}}, \bibinfo {author} {\bibfnamefont {G.}~\bibnamefont {Ehlers}},
  \bibinfo {author} {\bibfnamefont {J.~R.}\ \bibnamefont {Stewart}}, \bibinfo
  {author} {\bibfnamefont {A.~L.}\ \bibnamefont {Cornelius}}, \bibinfo {author}
  {\bibfnamefont {P.~P.}\ \bibnamefont {Deen}}, \bibinfo {author}
  {\bibfnamefont {Y.}~\bibnamefont {Qiu}}, \bibinfo {author} {\bibfnamefont
  {C.~R.}\ \bibnamefont {Wiebe}}, \bibinfo {author} {\bibfnamefont {J.~A.}\
  \bibnamefont {Janik}}, \bibinfo {author} {\bibfnamefont {H.~D.}\ \bibnamefont
  {Zhou}}, \bibinfo {author} {\bibfnamefont {D.}~\bibnamefont {Antonio}},
  \bibinfo {author} {\bibfnamefont {B.~W.}\ \bibnamefont {Woytko}}, \bibinfo
  {author} {\bibfnamefont {J.~P.}\ \bibnamefont {Ruff}}, \bibinfo {author}
  {\bibfnamefont {H.~A.}\ \bibnamefont {Dabkowska}}, \bibinfo {author}
  {\bibfnamefont {B.~D.}\ \bibnamefont {Gaulin}}, \ and\ \bibinfo {author}
  {\bibfnamefont {J.~S.}\ \bibnamefont {Gardner}},\ }\href {\doibase
  10.1103/PhysRevB.76.212405} {\bibfield  {journal} {\bibinfo  {journal} {Phys.
  Rev. B}\ }\textbf {\bibinfo {volume} {76}},\ \bibinfo {pages} {212405}
  (\bibinfo {year} {2007})}\BibitemShut {NoStop}%
\bibitem [{\citenamefont {Rule}\ \emph {et~al.}(2009)\citenamefont {Rule},
  \citenamefont {Ehlers}, \citenamefont {Gardner}, \citenamefont {Qiu},
  \citenamefont {Moskvin}, \citenamefont {Kiefer},\ and\ \citenamefont
  {Gerischer}}]{TbSn_rule2009neutron}%
  \BibitemOpen
  \bibfield  {author} {\bibinfo {author} {\bibfnamefont {K.~C.}\ \bibnamefont
  {Rule}}, \bibinfo {author} {\bibfnamefont {G.}~\bibnamefont {Ehlers}},
  \bibinfo {author} {\bibfnamefont {J.~S.}\ \bibnamefont {Gardner}}, \bibinfo
  {author} {\bibfnamefont {Y.}~\bibnamefont {Qiu}}, \bibinfo {author}
  {\bibfnamefont {E.}~\bibnamefont {Moskvin}}, \bibinfo {author} {\bibfnamefont
  {K.}~\bibnamefont {Kiefer}}, \ and\ \bibinfo {author} {\bibfnamefont
  {S.}~\bibnamefont {Gerischer}},\ }\href {\doibase 10.1103/PhysRevB.76.212405}
  {\bibfield  {journal} {\bibinfo  {journal} {J. Condens. Matter Phys.}\
  }\textbf {\bibinfo {volume} {21}},\ \bibinfo {pages} {486005} (\bibinfo
  {year} {2009})}\BibitemShut {NoStop}%
\bibitem [{\citenamefont {Lhotel}\ \emph {et~al.}(2012)\citenamefont {Lhotel},
  \citenamefont {Paulsen}, \citenamefont {de~R\'eotier}, \citenamefont
  {Yaouanc}, \citenamefont {Marin},\ and\ \citenamefont
  {Vanishri}}]{PhysRevB.86.020410}%
  \BibitemOpen
  \bibfield  {author} {\bibinfo {author} {\bibfnamefont {E.}~\bibnamefont
  {Lhotel}}, \bibinfo {author} {\bibfnamefont {C.}~\bibnamefont {Paulsen}},
  \bibinfo {author} {\bibfnamefont {P.~D.}\ \bibnamefont {de~R\'eotier}},
  \bibinfo {author} {\bibfnamefont {A.}~\bibnamefont {Yaouanc}}, \bibinfo
  {author} {\bibfnamefont {C.}~\bibnamefont {Marin}}, \ and\ \bibinfo {author}
  {\bibfnamefont {S.}~\bibnamefont {Vanishri}},\ }\href {\doibase
  10.1103/PhysRevB.86.020410} {\bibfield  {journal} {\bibinfo  {journal} {Phys.
  Rev. B}\ }\textbf {\bibinfo {volume} {86}},\ \bibinfo {pages} {020410}
  (\bibinfo {year} {2012})}\BibitemShut {NoStop}%
\bibitem [{\citenamefont {Mauws}\ \emph {et~al.}(2018)\citenamefont {Mauws},
  \citenamefont {Hallas}, \citenamefont {Sala}, \citenamefont {Aczel},
  \citenamefont {Sarte}, \citenamefont {Gaudet}, \citenamefont {Ziat},
  \citenamefont {Quilliam}, \citenamefont {Lussier}, \citenamefont {Bieringer},
  \citenamefont {Zhou}, \citenamefont {Wildes}, \citenamefont {Stone},
  \citenamefont {Abernathy}, \citenamefont {Luke}, \citenamefont {Gaulin},\
  and\ \citenamefont {Wiebe}}]{Mauws18:98}%
  \BibitemOpen
  \bibfield  {author} {\bibinfo {author} {\bibfnamefont {C.}~\bibnamefont
  {Mauws}}, \bibinfo {author} {\bibfnamefont {A.~M.}\ \bibnamefont {Hallas}},
  \bibinfo {author} {\bibfnamefont {G.}~\bibnamefont {Sala}}, \bibinfo {author}
  {\bibfnamefont {A.~A.}\ \bibnamefont {Aczel}}, \bibinfo {author}
  {\bibfnamefont {P.~M.}\ \bibnamefont {Sarte}}, \bibinfo {author}
  {\bibfnamefont {J.}~\bibnamefont {Gaudet}}, \bibinfo {author} {\bibfnamefont
  {D.}~\bibnamefont {Ziat}}, \bibinfo {author} {\bibfnamefont {J.~A.}\
  \bibnamefont {Quilliam}}, \bibinfo {author} {\bibfnamefont {J.~A.}\
  \bibnamefont {Lussier}}, \bibinfo {author} {\bibfnamefont {M.}~\bibnamefont
  {Bieringer}}, \bibinfo {author} {\bibfnamefont {H.~D.}\ \bibnamefont {Zhou}},
  \bibinfo {author} {\bibfnamefont {A.}~\bibnamefont {Wildes}}, \bibinfo
  {author} {\bibfnamefont {M.~B.}\ \bibnamefont {Stone}}, \bibinfo {author}
  {\bibfnamefont {D.}~\bibnamefont {Abernathy}}, \bibinfo {author}
  {\bibfnamefont {G.~M.}\ \bibnamefont {Luke}}, \bibinfo {author}
  {\bibfnamefont {B.~D.}\ \bibnamefont {Gaulin}}, \ and\ \bibinfo {author}
  {\bibfnamefont {C.~R.}\ \bibnamefont {Wiebe}},\ }\href {\doibase
  10.1103/PhysRevB.98.100401} {\bibfield  {journal} {\bibinfo  {journal} {Phys.
  Rev. B}\ }\textbf {\bibinfo {volume} {98}},\ \bibinfo {pages} {100401}
  (\bibinfo {year} {2018})}\BibitemShut {NoStop}%
\bibitem [{\citenamefont {Pe\c{c}anha-Antonio}\ \emph
  {et~al.}(2019)\citenamefont {Pe\c{c}anha-Antonio}, \citenamefont {Feng},
  \citenamefont {Sun}, \citenamefont {Adroja}, \citenamefont {Walker},
  \citenamefont {Gibbs}, \citenamefont {Orlandi}, \citenamefont {Su},\ and\
  \citenamefont {Br\"uckel}}]{Antonio19:99}%
  \BibitemOpen
  \bibfield  {author} {\bibinfo {author} {\bibfnamefont {V.}~\bibnamefont
  {Pe\c{c}anha-Antonio}}, \bibinfo {author} {\bibfnamefont {E.}~\bibnamefont
  {Feng}}, \bibinfo {author} {\bibfnamefont {X.}~\bibnamefont {Sun}}, \bibinfo
  {author} {\bibfnamefont {D.}~\bibnamefont {Adroja}}, \bibinfo {author}
  {\bibfnamefont {H.~C.}\ \bibnamefont {Walker}}, \bibinfo {author}
  {\bibfnamefont {A.~S.}\ \bibnamefont {Gibbs}}, \bibinfo {author}
  {\bibfnamefont {F.}~\bibnamefont {Orlandi}}, \bibinfo {author} {\bibfnamefont
  {Y.}~\bibnamefont {Su}}, \ and\ \bibinfo {author} {\bibfnamefont
  {T.}~\bibnamefont {Br\"uckel}},\ }\href {\doibase 10.1103/PhysRevB.99.134415}
  {\bibfield  {journal} {\bibinfo  {journal} {Phys. Rev. B}\ }\textbf {\bibinfo
  {volume} {99}},\ \bibinfo {pages} {134415} (\bibinfo {year}
  {2019})}\BibitemShut {NoStop}%
\bibitem [{\citenamefont {Singh}\ \emph {et~al.}(2008)\citenamefont {Singh},
  \citenamefont {Saha}, \citenamefont {Dhar}, \citenamefont {Suryanarayanan},
  \citenamefont {Sood},\ and\ \citenamefont {Revcolevschi}}]{Singh08:77}%
  \BibitemOpen
  \bibfield  {author} {\bibinfo {author} {\bibfnamefont {S.}~\bibnamefont
  {Singh}}, \bibinfo {author} {\bibfnamefont {S.}~\bibnamefont {Saha}},
  \bibinfo {author} {\bibfnamefont {S.~K.}\ \bibnamefont {Dhar}}, \bibinfo
  {author} {\bibfnamefont {R.}~\bibnamefont {Suryanarayanan}}, \bibinfo
  {author} {\bibfnamefont {A.~K.}\ \bibnamefont {Sood}}, \ and\ \bibinfo
  {author} {\bibfnamefont {A.}~\bibnamefont {Revcolevschi}},\ }\href {\doibase
  10.1103/PhysRevB.77.054408} {\bibfield  {journal} {\bibinfo  {journal} {Phys.
  Rev. B}\ }\textbf {\bibinfo {volume} {77}},\ \bibinfo {pages} {054408}
  (\bibinfo {year} {2008})}\BibitemShut {NoStop}%
\bibitem [{\citenamefont {Malkin}\ \emph {et~al.}(2010)\citenamefont {Malkin},
  \citenamefont {Lummen}, \citenamefont {van Loosdrecht}, \citenamefont
  {Dhalenne},\ and\ \citenamefont {Zakirov}}]{Malkin10:22}%
  \BibitemOpen
  \bibfield  {author} {\bibinfo {author} {\bibfnamefont {B.~Z.}\ \bibnamefont
  {Malkin}}, \bibinfo {author} {\bibfnamefont {T.~T.~A.}\ \bibnamefont
  {Lummen}}, \bibinfo {author} {\bibfnamefont {P.~H.~M.}\ \bibnamefont {van
  Loosdrecht}}, \bibinfo {author} {\bibfnamefont {G.}~\bibnamefont {Dhalenne}},
  \ and\ \bibinfo {author} {\bibfnamefont {A.~R.}\ \bibnamefont {Zakirov}},\
  }\href {\doibase 10.1088/0953-8984/22/27/276003} {\bibfield  {journal}
  {\bibinfo  {journal} {J. Condens. Matter Phys.}\ }\textbf {\bibinfo {volume}
  {22}},\ \bibinfo {pages} {276003} (\bibinfo {year} {2010})}\BibitemShut
  {NoStop}%
\bibitem [{\citenamefont {Donnerer}\ \emph {et~al.}(2016)\citenamefont
  {Donnerer}, \citenamefont {Rahn}, \citenamefont {Sala}, \citenamefont {Vale},
  \citenamefont {Pincini}, \citenamefont {Strempfer}, \citenamefont {Krisch},
  \citenamefont {Prabhakaran}, \citenamefont {Boothroyd},\ and\ \citenamefont
  {McMorrow}}]{Donnerer16:117}%
  \BibitemOpen
  \bibfield  {author} {\bibinfo {author} {\bibfnamefont {C.}~\bibnamefont
  {Donnerer}}, \bibinfo {author} {\bibfnamefont {M.~C.}\ \bibnamefont {Rahn}},
  \bibinfo {author} {\bibfnamefont {M.~M.}\ \bibnamefont {Sala}}, \bibinfo
  {author} {\bibfnamefont {J.~G.}\ \bibnamefont {Vale}}, \bibinfo {author}
  {\bibfnamefont {D.}~\bibnamefont {Pincini}}, \bibinfo {author} {\bibfnamefont
  {J.}~\bibnamefont {Strempfer}}, \bibinfo {author} {\bibfnamefont
  {M.}~\bibnamefont {Krisch}}, \bibinfo {author} {\bibfnamefont
  {D.}~\bibnamefont {Prabhakaran}}, \bibinfo {author} {\bibfnamefont {A.~T.}\
  \bibnamefont {Boothroyd}}, \ and\ \bibinfo {author} {\bibfnamefont {D.~F.}\
  \bibnamefont {McMorrow}},\ }\href {\doibase 10.1103/PhysRevLett.117.037201}
  {\bibfield  {journal} {\bibinfo  {journal} {Phys. Rev. Lett.}\ }\textbf
  {\bibinfo {volume} {117}},\ \bibinfo {pages} {037201} (\bibinfo {year}
  {2016})}\BibitemShut {NoStop}%
\bibitem [{\citenamefont {Guruciaga}\ \emph {et~al.}(2014)\citenamefont
  {Guruciaga}, \citenamefont {Grigera},\ and\ \citenamefont
  {Borzi}}]{guruciaga14:90}%
  \BibitemOpen
  \bibfield  {author} {\bibinfo {author} {\bibfnamefont {P.~C.}\ \bibnamefont
  {Guruciaga}}, \bibinfo {author} {\bibfnamefont {S.~A.}\ \bibnamefont
  {Grigera}}, \ and\ \bibinfo {author} {\bibfnamefont {R.~A.}\ \bibnamefont
  {Borzi}},\ }\href {\doibase 10.1103/PhysRevB.90.184423} {\bibfield  {journal}
  {\bibinfo  {journal} {Phys. Rev. B}\ }\textbf {\bibinfo {volume} {90}},\
  \bibinfo {pages} {184423} (\bibinfo {year} {2014})}\BibitemShut {NoStop}%
\bibitem [{\citenamefont {Ikeda}\ and\ \citenamefont
  {Kawamura}(2008)}]{Ikeda08:77}%
  \BibitemOpen
  \bibfield  {author} {\bibinfo {author} {\bibfnamefont {A.}~\bibnamefont
  {Ikeda}}\ and\ \bibinfo {author} {\bibfnamefont {H.}~\bibnamefont
  {Kawamura}},\ }\href {\doibase 10.1143/JPSJ.77.073707} {\bibfield  {journal}
  {\bibinfo  {journal} {J. Phys. Soc. Jpn.}\ }\textbf {\bibinfo {volume}
  {77}},\ \bibinfo {pages} {073707} (\bibinfo {year} {2008})}\BibitemShut
  {NoStop}%
\bibitem [{\citenamefont {Bramwell}\ and\ \citenamefont
  {Harris}(1998)}]{Bramwell_1998}%
  \BibitemOpen
  \bibfield  {author} {\bibinfo {author} {\bibfnamefont {S.~T.}\ \bibnamefont
  {Bramwell}}\ and\ \bibinfo {author} {\bibfnamefont {M.~J.}\ \bibnamefont
  {Harris}},\ }\href {\doibase 10.1088/0953-8984/10/14/002} {\bibfield
  {journal} {\bibinfo  {journal} {J. Phys. Condens. Matter}\ }\textbf {\bibinfo
  {volume} {10}},\ \bibinfo {pages} {L215} (\bibinfo {year}
  {1998})}\BibitemShut {NoStop}%
\bibitem [{\citenamefont {Opherden}\ \emph {et~al.}(2017)\citenamefont
  {Opherden}, \citenamefont {Hornung}, \citenamefont {Herrmannsd\"orfer},
  \citenamefont {Xu}, \citenamefont {Islam}, \citenamefont {Lake},\ and\
  \citenamefont {Wosnitza}}]{opherden17:95}%
  \BibitemOpen
  \bibfield  {author} {\bibinfo {author} {\bibfnamefont {L.}~\bibnamefont
  {Opherden}}, \bibinfo {author} {\bibfnamefont {J.}~\bibnamefont {Hornung}},
  \bibinfo {author} {\bibfnamefont {T.}~\bibnamefont {Herrmannsd\"orfer}},
  \bibinfo {author} {\bibfnamefont {J.}~\bibnamefont {Xu}}, \bibinfo {author}
  {\bibfnamefont {A.~T. M.~N.}\ \bibnamefont {Islam}}, \bibinfo {author}
  {\bibfnamefont {B.}~\bibnamefont {Lake}}, \ and\ \bibinfo {author}
  {\bibfnamefont {J.}~\bibnamefont {Wosnitza}},\ }\href {\doibase
  10.1103/PhysRevB.95.184418} {\bibfield  {journal} {\bibinfo  {journal} {Phys.
  Rev. B}\ }\textbf {\bibinfo {volume} {95}},\ \bibinfo {pages} {184418}
  (\bibinfo {year} {2017})}\BibitemShut {NoStop}%
\bibitem [{\citenamefont {Giblin}\ \emph {et~al.}(2018)\citenamefont {Giblin},
  \citenamefont {Twengstr\"om}, \citenamefont {Bovo}, \citenamefont {Ruminy},
  \citenamefont {Bartkowiak}, \citenamefont {Manuel}, \citenamefont {Andresen},
  \citenamefont {Prabhakaran}, \citenamefont {Balakrishnan}, \citenamefont
  {Pomjakushina}, \citenamefont {Paulsen}, \citenamefont {Lhotel},
  \citenamefont {Keller}, \citenamefont {Frontzek}, \citenamefont {Capelli},
  \citenamefont {Zaharko}, \citenamefont {McClarty}, \citenamefont {Bramwell},
  \citenamefont {Henelius},\ and\ \citenamefont
  {Fennell}}]{PhysRevLett.121.067202}%
  \BibitemOpen
  \bibfield  {author} {\bibinfo {author} {\bibfnamefont {S.~R.}\ \bibnamefont
  {Giblin}}, \bibinfo {author} {\bibfnamefont {M.}~\bibnamefont
  {Twengstr\"om}}, \bibinfo {author} {\bibfnamefont {L.}~\bibnamefont {Bovo}},
  \bibinfo {author} {\bibfnamefont {M.}~\bibnamefont {Ruminy}}, \bibinfo
  {author} {\bibfnamefont {M.}~\bibnamefont {Bartkowiak}}, \bibinfo {author}
  {\bibfnamefont {P.}~\bibnamefont {Manuel}}, \bibinfo {author} {\bibfnamefont
  {J.~C.}\ \bibnamefont {Andresen}}, \bibinfo {author} {\bibfnamefont
  {D.}~\bibnamefont {Prabhakaran}}, \bibinfo {author} {\bibfnamefont
  {G.}~\bibnamefont {Balakrishnan}}, \bibinfo {author} {\bibfnamefont
  {E.}~\bibnamefont {Pomjakushina}}, \bibinfo {author} {\bibfnamefont
  {C.}~\bibnamefont {Paulsen}}, \bibinfo {author} {\bibfnamefont
  {E.}~\bibnamefont {Lhotel}}, \bibinfo {author} {\bibfnamefont
  {L.}~\bibnamefont {Keller}}, \bibinfo {author} {\bibfnamefont
  {M.}~\bibnamefont {Frontzek}}, \bibinfo {author} {\bibfnamefont {S.~C.}\
  \bibnamefont {Capelli}}, \bibinfo {author} {\bibfnamefont {O.}~\bibnamefont
  {Zaharko}}, \bibinfo {author} {\bibfnamefont {P.~A.}\ \bibnamefont
  {McClarty}}, \bibinfo {author} {\bibfnamefont {S.~T.}\ \bibnamefont
  {Bramwell}}, \bibinfo {author} {\bibfnamefont {P.}~\bibnamefont {Henelius}},
  \ and\ \bibinfo {author} {\bibfnamefont {T.}~\bibnamefont {Fennell}},\ }\href
  {\doibase 10.1103/PhysRevLett.121.067202} {\bibfield  {journal} {\bibinfo
  {journal} {Phys. Rev. Lett.}\ }\textbf {\bibinfo {volume} {121}},\ \bibinfo
  {pages} {067202} (\bibinfo {year} {2018})}\BibitemShut {NoStop}%
\bibitem [{\citenamefont {Diep}(2013)}]{doi:10.1142/8676}%
  \BibitemOpen
  \bibfield  {author} {\bibinfo {author} {\bibfnamefont {H.~T.}\ \bibnamefont
  {Diep}},\ }\href {\doibase 10.1142/8676} {\emph {\bibinfo {title} {Frustrated
  Spin Systems}}},\ \bibinfo {edition} {2nd}\ ed.\ (\bibinfo  {publisher}
  {World Scientific},\ \bibinfo {year} {2013})\BibitemShut {NoStop}%
\bibitem [{\citenamefont {Rosenkranz}\ \emph {et~al.}(2000)\citenamefont
  {Rosenkranz}, \citenamefont {Ramirez}, \citenamefont {Hayashi}, \citenamefont
  {Cava}, \citenamefont {Siddharthan},\ and\ \citenamefont
  {Shastry}}]{doi:10.1063/1.372565}%
  \BibitemOpen
  \bibfield  {author} {\bibinfo {author} {\bibfnamefont {S.}~\bibnamefont
  {Rosenkranz}}, \bibinfo {author} {\bibfnamefont {A.~P.}\ \bibnamefont
  {Ramirez}}, \bibinfo {author} {\bibfnamefont {A.}~\bibnamefont {Hayashi}},
  \bibinfo {author} {\bibfnamefont {R.~J.}\ \bibnamefont {Cava}}, \bibinfo
  {author} {\bibfnamefont {R.}~\bibnamefont {Siddharthan}}, \ and\ \bibinfo
  {author} {\bibfnamefont {B.~S.}\ \bibnamefont {Shastry}},\ }\href {\doibase
  10.1063/1.372565} {\bibfield  {journal} {\bibinfo  {journal} {J. Appl.
  Phys.}\ }\textbf {\bibinfo {volume} {87}},\ \bibinfo {pages} {5914} (\bibinfo
  {year} {2000})}\BibitemShut {NoStop}%
\bibitem [{\citenamefont {Fukazawa}\ \emph {et~al.}(2002)\citenamefont
  {Fukazawa}, \citenamefont {Melko}, \citenamefont {Higashinaka}, \citenamefont
  {Maeno},\ and\ \citenamefont {Gingras}}]{PhysRevB.65.054410}%
  \BibitemOpen
  \bibfield  {author} {\bibinfo {author} {\bibfnamefont {H.}~\bibnamefont
  {Fukazawa}}, \bibinfo {author} {\bibfnamefont {R.~G.}\ \bibnamefont {Melko}},
  \bibinfo {author} {\bibfnamefont {R.}~\bibnamefont {Higashinaka}}, \bibinfo
  {author} {\bibfnamefont {Y.}~\bibnamefont {Maeno}}, \ and\ \bibinfo {author}
  {\bibfnamefont {M.~J.~P.}\ \bibnamefont {Gingras}},\ }\href {\doibase
  10.1103/PhysRevB.65.054410} {\bibfield  {journal} {\bibinfo  {journal} {Phys.
  Rev. B}\ }\textbf {\bibinfo {volume} {65}},\ \bibinfo {pages} {054410}
  (\bibinfo {year} {2002})}\BibitemShut {NoStop}%
\bibitem [{\citenamefont {Flood}(1974)}]{doi:10.1063/1.1663909}%
  \BibitemOpen
  \bibfield  {author} {\bibinfo {author} {\bibfnamefont {D.~J.}\ \bibnamefont
  {Flood}},\ }\href {\doibase 10.1063/1.1663909} {\bibfield  {journal}
  {\bibinfo  {journal} {J. Appl. Phys.}\ }\textbf {\bibinfo {volume} {45}},\
  \bibinfo {pages} {4041} (\bibinfo {year} {1974})}\BibitemShut {NoStop}%
\bibitem [{\citenamefont {Jana}\ \emph {et~al.}(2002)\citenamefont {Jana},
  \citenamefont {Sengupta},\ and\ \citenamefont {Ghosh}}]{JANA20027}%
  \BibitemOpen
  \bibfield  {author} {\bibinfo {author} {\bibfnamefont {Y.}~\bibnamefont
  {Jana}}, \bibinfo {author} {\bibfnamefont {A.}~\bibnamefont {Sengupta}}, \
  and\ \bibinfo {author} {\bibfnamefont {D.}~\bibnamefont {Ghosh}},\ }\href
  {\doibase https://doi.org/10.1016/S0304-8853(01)00983-0} {\bibfield
  {journal} {\bibinfo  {journal} {J. Magn. Magn. Mater.}\ }\textbf {\bibinfo
  {volume} {248}},\ \bibinfo {pages} {7} (\bibinfo {year} {2002})}\BibitemShut
  {NoStop}%
\bibitem [{\citenamefont {Matsuhira}\ \emph {et~al.}(2000)\citenamefont
  {Matsuhira}, \citenamefont {Hinatsu}, \citenamefont {Tenya},\ and\
  \citenamefont {Sakakibara}}]{Matsuhira00:12}%
  \BibitemOpen
  \bibfield  {author} {\bibinfo {author} {\bibfnamefont {K.}~\bibnamefont
  {Matsuhira}}, \bibinfo {author} {\bibfnamefont {Y.}~\bibnamefont {Hinatsu}},
  \bibinfo {author} {\bibfnamefont {K.}~\bibnamefont {Tenya}}, \ and\ \bibinfo
  {author} {\bibfnamefont {T.}~\bibnamefont {Sakakibara}},\ }\href {\doibase
  10.1088/0953-8984/12/40/103} {\bibfield  {journal} {\bibinfo  {journal} {J.
  Condens. Matter Phys.}\ }\textbf {\bibinfo {volume} {12}},\ \bibinfo {pages}
  {L649} (\bibinfo {year} {2000})}\BibitemShut {NoStop}%
\bibitem [{\citenamefont {Fennell}\ \emph {et~al.}(2004)\citenamefont
  {Fennell}, \citenamefont {Petrenko}, \citenamefont {F\aa{}k}, \citenamefont
  {Bramwell}, \citenamefont {Enjalran}, \citenamefont {Yavors'kii},
  \citenamefont {Gingras}, \citenamefont {Melko},\ and\ \citenamefont
  {Balakrishnan}}]{PhysRevB.70.134408}%
  \BibitemOpen
  \bibfield  {author} {\bibinfo {author} {\bibfnamefont {T.}~\bibnamefont
  {Fennell}}, \bibinfo {author} {\bibfnamefont {O.~A.}\ \bibnamefont
  {Petrenko}}, \bibinfo {author} {\bibfnamefont {B.}~\bibnamefont {F\aa{}k}},
  \bibinfo {author} {\bibfnamefont {S.~T.}\ \bibnamefont {Bramwell}}, \bibinfo
  {author} {\bibfnamefont {M.}~\bibnamefont {Enjalran}}, \bibinfo {author}
  {\bibfnamefont {T.}~\bibnamefont {Yavors'kii}}, \bibinfo {author}
  {\bibfnamefont {M.~J.~P.}\ \bibnamefont {Gingras}}, \bibinfo {author}
  {\bibfnamefont {R.~G.}\ \bibnamefont {Melko}}, \ and\ \bibinfo {author}
  {\bibfnamefont {G.}~\bibnamefont {Balakrishnan}},\ }\href {\doibase
  10.1103/PhysRevB.70.134408} {\bibfield  {journal} {\bibinfo  {journal} {Phys.
  Rev. B}\ }\textbf {\bibinfo {volume} {70}},\ \bibinfo {pages} {134408}
  (\bibinfo {year} {2004})}\BibitemShut {NoStop}%
\bibitem [{\citenamefont {Kadowaki}\ \emph {et~al.}(2002)\citenamefont
  {Kadowaki}, \citenamefont {Ishii}, \citenamefont {Matsuhira},\ and\
  \citenamefont {Hinatsu}}]{PhysRevB.65.144421}%
  \BibitemOpen
  \bibfield  {author} {\bibinfo {author} {\bibfnamefont {H.}~\bibnamefont
  {Kadowaki}}, \bibinfo {author} {\bibfnamefont {Y.}~\bibnamefont {Ishii}},
  \bibinfo {author} {\bibfnamefont {K.}~\bibnamefont {Matsuhira}}, \ and\
  \bibinfo {author} {\bibfnamefont {Y.}~\bibnamefont {Hinatsu}},\ }\href
  {\doibase 10.1103/PhysRevB.65.144421} {\bibfield  {journal} {\bibinfo
  {journal} {Phys. Rev. B}\ }\textbf {\bibinfo {volume} {65}},\ \bibinfo
  {pages} {144421} (\bibinfo {year} {2002})}\BibitemShut {NoStop}%
\bibitem [{\citenamefont {Matsuhira}\ \emph {et~al.}(2001)\citenamefont
  {Matsuhira}, \citenamefont {Hinatsu},\ and\ \citenamefont
  {Sakakibara}}]{Matsuhira01:13}%
  \BibitemOpen
  \bibfield  {author} {\bibinfo {author} {\bibfnamefont {K.}~\bibnamefont
  {Matsuhira}}, \bibinfo {author} {\bibfnamefont {Y.}~\bibnamefont {Hinatsu}},
  \ and\ \bibinfo {author} {\bibfnamefont {T.}~\bibnamefont {Sakakibara}},\
  }\href {\doibase 10.1088/0953-8984/13/31/101} {\ \textbf {\bibinfo {volume}
  {13}},\ \bibinfo {pages} {L737} (\bibinfo {year} {2001})}\BibitemShut
  {NoStop}%
\bibitem [{\citenamefont {Snyder}\ \emph {et~al.}(2003)\citenamefont {Snyder},
  \citenamefont {Ueland}, \citenamefont {Slusky}, \citenamefont {Karunadasa},
  \citenamefont {Cava}, \citenamefont {Mizel},\ and\ \citenamefont
  {Schiffer}}]{PhysRevLett.91.107201}%
  \BibitemOpen
  \bibfield  {author} {\bibinfo {author} {\bibfnamefont {J.}~\bibnamefont
  {Snyder}}, \bibinfo {author} {\bibfnamefont {B.~G.}\ \bibnamefont {Ueland}},
  \bibinfo {author} {\bibfnamefont {J.~S.}\ \bibnamefont {Slusky}}, \bibinfo
  {author} {\bibfnamefont {H.}~\bibnamefont {Karunadasa}}, \bibinfo {author}
  {\bibfnamefont {R.~J.}\ \bibnamefont {Cava}}, \bibinfo {author}
  {\bibfnamefont {A.}~\bibnamefont {Mizel}}, \ and\ \bibinfo {author}
  {\bibfnamefont {P.}~\bibnamefont {Schiffer}},\ }\href {\doibase
  10.1103/PhysRevLett.91.107201} {\bibfield  {journal} {\bibinfo  {journal}
  {Phys. Rev. Lett.}\ }\textbf {\bibinfo {volume} {91}},\ \bibinfo {pages}
  {107201} (\bibinfo {year} {2003})}\BibitemShut {NoStop}%
\bibitem [{\citenamefont {Snyder}\ \emph
  {et~al.}(2001{\natexlab{b}})\citenamefont {Snyder}, \citenamefont {Slusky},
  \citenamefont {Cava},\ and\ \citenamefont {Schiffer}}]{snyder2001spin}%
  \BibitemOpen
  \bibfield  {author} {\bibinfo {author} {\bibfnamefont {J.}~\bibnamefont
  {Snyder}}, \bibinfo {author} {\bibfnamefont {J.~S.}\ \bibnamefont {Slusky}},
  \bibinfo {author} {\bibfnamefont {R.~J.}\ \bibnamefont {Cava}}, \ and\
  \bibinfo {author} {\bibfnamefont {P.}~\bibnamefont {Schiffer}},\ }\href
  {\doibase https://doi.org/10.1038/35092516} {\bibfield  {journal} {\bibinfo
  {journal} {Nature}\ }\textbf {\bibinfo {volume} {413}},\ \bibinfo {pages}
  {48} (\bibinfo {year} {2001}{\natexlab{b}})}\BibitemShut {NoStop}%
\bibitem [{\citenamefont {Ehlers}\ \emph {et~al.}(2002)\citenamefont {Ehlers},
  \citenamefont {Cornelius}, \citenamefont {c}, \citenamefont {Kajnakov},
  \citenamefont {Fennell}, \citenamefont {Bramwell},\ and\ \citenamefont
  {Gardner}}]{Ehlers02:15}%
  \BibitemOpen
  \bibfield  {author} {\bibinfo {author} {\bibfnamefont {G.}~\bibnamefont
  {Ehlers}}, \bibinfo {author} {\bibfnamefont {A.~L.}\ \bibnamefont
  {Cornelius}}, \bibinfo {author} {\bibfnamefont {M.~O.}\ \bibnamefont {c}},
  \bibinfo {author} {\bibfnamefont {M.}~\bibnamefont {Kajnakov}}, \bibinfo
  {author} {\bibfnamefont {T.}~\bibnamefont {Fennell}}, \bibinfo {author}
  {\bibfnamefont {S.~T.}\ \bibnamefont {Bramwell}}, \ and\ \bibinfo {author}
  {\bibfnamefont {J.~S.}\ \bibnamefont {Gardner}},\ }\href {\doibase
  10.1088/0953-8984/15/2/102} {\bibfield  {journal} {\bibinfo  {journal} {J.
  Condens. Matter Phys.}\ }\textbf {\bibinfo {volume} {15}},\ \bibinfo {pages}
  {L9} (\bibinfo {year} {2002})}\BibitemShut {NoStop}%
\bibitem [{\citenamefont {Zhou}\ \emph
  {et~al.}(2008{\natexlab{b}})\citenamefont {Zhou}, \citenamefont {Wiebe},
  \citenamefont {Janik}, \citenamefont {Balicas}, \citenamefont {Yo},
  \citenamefont {Qiu}, \citenamefont {Copley},\ and\ \citenamefont
  {Gardner}}]{PrSn_zhou2008dynamic}%
  \BibitemOpen
  \bibfield  {author} {\bibinfo {author} {\bibfnamefont {H.~D.}\ \bibnamefont
  {Zhou}}, \bibinfo {author} {\bibfnamefont {C.~R.}\ \bibnamefont {Wiebe}},
  \bibinfo {author} {\bibfnamefont {J.~A.}\ \bibnamefont {Janik}}, \bibinfo
  {author} {\bibfnamefont {L.}~\bibnamefont {Balicas}}, \bibinfo {author}
  {\bibfnamefont {Y.~J.}\ \bibnamefont {Yo}}, \bibinfo {author} {\bibfnamefont
  {Y.}~\bibnamefont {Qiu}}, \bibinfo {author} {\bibfnamefont {J.~R.~D.}\
  \bibnamefont {Copley}}, \ and\ \bibinfo {author} {\bibfnamefont {J.~S.}\
  \bibnamefont {Gardner}},\ }\href {\doibase 10.1103/PhysRevLett.101.227204}
  {\bibfield  {journal} {\bibinfo  {journal} {Phys. Rev. Lett.}\ }\textbf
  {\bibinfo {volume} {101}},\ \bibinfo {pages} {227204} (\bibinfo {year}
  {2008}{\natexlab{b}})}\BibitemShut {NoStop}%
\bibitem [{\citenamefont {Gardner}\ \emph {et~al.}(2011)\citenamefont
  {Gardner}, \citenamefont {Ehlers}, \citenamefont {Fouquet}, \citenamefont
  {Farago},\ and\ \citenamefont {Stewart}}]{Gardner_2011}%
  \BibitemOpen
  \bibfield  {author} {\bibinfo {author} {\bibfnamefont {J.~S.}\ \bibnamefont
  {Gardner}}, \bibinfo {author} {\bibfnamefont {G.}~\bibnamefont {Ehlers}},
  \bibinfo {author} {\bibfnamefont {P.}~\bibnamefont {Fouquet}}, \bibinfo
  {author} {\bibfnamefont {B.}~\bibnamefont {Farago}}, \ and\ \bibinfo {author}
  {\bibfnamefont {J.~R.}\ \bibnamefont {Stewart}},\ }\href {\doibase
  10.1088/0953-8984/23/16/164220} {\bibfield  {journal} {\bibinfo  {journal}
  {J. Condens. Matter Phys.}\ }\textbf {\bibinfo {volume} {23}},\ \bibinfo
  {pages} {164220} (\bibinfo {year} {2011})}\BibitemShut {NoStop}%
\bibitem [{\citenamefont {Ehlers}\ \emph {et~al.}(2004)\citenamefont {Ehlers},
  \citenamefont {Cornelius}, \citenamefont {Fennell}, \citenamefont {Koza},
  \citenamefont {Bramwell},\ and\ \citenamefont {Gardner}}]{Ehlers04:16}%
  \BibitemOpen
  \bibfield  {author} {\bibinfo {author} {\bibfnamefont {G.}~\bibnamefont
  {Ehlers}}, \bibinfo {author} {\bibfnamefont {A.~L.}\ \bibnamefont
  {Cornelius}}, \bibinfo {author} {\bibfnamefont {T.}~\bibnamefont {Fennell}},
  \bibinfo {author} {\bibfnamefont {M.}~\bibnamefont {Koza}}, \bibinfo {author}
  {\bibfnamefont {S.~T.}\ \bibnamefont {Bramwell}}, \ and\ \bibinfo {author}
  {\bibfnamefont {J.~S.}\ \bibnamefont {Gardner}},\ }\href {\doibase
  10.1088/0953-8984/16/11/010} {\bibfield  {journal} {\bibinfo  {journal} {J.
  Condens. Matter Phys.}\ }\textbf {\bibinfo {volume} {16}},\ \bibinfo {pages}
  {S635} (\bibinfo {year} {2004})}\BibitemShut {NoStop}%
\bibitem [{\citenamefont {Snyder}\ \emph {et~al.}(2002)\citenamefont {Snyder},
  \citenamefont {Slusky}, \citenamefont {Cava},\ and\ \citenamefont
  {Schiffer}}]{PhysRevB.66.064432}%
  \BibitemOpen
  \bibfield  {author} {\bibinfo {author} {\bibfnamefont {J.}~\bibnamefont
  {Snyder}}, \bibinfo {author} {\bibfnamefont {J.~S.}\ \bibnamefont {Slusky}},
  \bibinfo {author} {\bibfnamefont {R.~J.}\ \bibnamefont {Cava}}, \ and\
  \bibinfo {author} {\bibfnamefont {P.}~\bibnamefont {Schiffer}},\ }\href
  {\doibase 10.1103/PhysRevB.66.064432} {\bibfield  {journal} {\bibinfo
  {journal} {Phys. Rev. B}\ }\textbf {\bibinfo {volume} {66}},\ \bibinfo
  {pages} {064432} (\bibinfo {year} {2002})}\BibitemShut {NoStop}%
\bibitem [{\citenamefont {Matsuhira}\ \emph
  {et~al.}(2011{\natexlab{b}})\citenamefont {Matsuhira}, \citenamefont
  {Wakeshima}, \citenamefont {Hinatsu}, \citenamefont {Sekine}, \citenamefont
  {Paulsen}, \citenamefont {Sakakibara},\ and\ \citenamefont
  {Takagi}}]{Matsuhira_2011}%
  \BibitemOpen
  \bibfield  {author} {\bibinfo {author} {\bibfnamefont {K.}~\bibnamefont
  {Matsuhira}}, \bibinfo {author} {\bibfnamefont {M.}~\bibnamefont
  {Wakeshima}}, \bibinfo {author} {\bibfnamefont {Y.}~\bibnamefont {Hinatsu}},
  \bibinfo {author} {\bibfnamefont {C.}~\bibnamefont {Sekine}}, \bibinfo
  {author} {\bibfnamefont {C.}~\bibnamefont {Paulsen}}, \bibinfo {author}
  {\bibfnamefont {T.}~\bibnamefont {Sakakibara}}, \ and\ \bibinfo {author}
  {\bibfnamefont {S.}~\bibnamefont {Takagi}},\ }\href {\doibase
  10.1088/1742-6596/320/1/012050} {\bibfield  {journal} {\bibinfo  {journal}
  {J. Phys.: Conf. Ser.}\ }\textbf {\bibinfo {volume} {320}},\ \bibinfo {pages}
  {012050} (\bibinfo {year} {2011}{\natexlab{b}})}\BibitemShut {NoStop}%
\bibitem [{\citenamefont {Snyder}\ \emph
  {et~al.}(2004{\natexlab{b}})\citenamefont {Snyder}, \citenamefont {Ueland},
  \citenamefont {Slusky}, \citenamefont {Karunadasa}, \citenamefont {Cava},\
  and\ \citenamefont {Schiffer}}]{PhysRevB.69.064414}%
  \BibitemOpen
  \bibfield  {author} {\bibinfo {author} {\bibfnamefont {J.}~\bibnamefont
  {Snyder}}, \bibinfo {author} {\bibfnamefont {B.~G.}\ \bibnamefont {Ueland}},
  \bibinfo {author} {\bibfnamefont {J.~S.}\ \bibnamefont {Slusky}}, \bibinfo
  {author} {\bibfnamefont {H.}~\bibnamefont {Karunadasa}}, \bibinfo {author}
  {\bibfnamefont {R.~J.}\ \bibnamefont {Cava}}, \ and\ \bibinfo {author}
  {\bibfnamefont {P.}~\bibnamefont {Schiffer}},\ }\href {\doibase
  10.1103/PhysRevB.69.064414} {\bibfield  {journal} {\bibinfo  {journal} {Phys.
  Rev. B}\ }\textbf {\bibinfo {volume} {69}},\ \bibinfo {pages} {064414}
  (\bibinfo {year} {2004}{\natexlab{b}})}\BibitemShut {NoStop}%
\bibitem [{\citenamefont {Yaraskavitch}\ \emph {et~al.}(2012)\citenamefont
  {Yaraskavitch}, \citenamefont {Revell}, \citenamefont {Meng}, \citenamefont
  {Ross}, \citenamefont {Noad}, \citenamefont {Dabkowska}, \citenamefont
  {Gaulin},\ and\ \citenamefont {Kycia}}]{PhysRevB.85.020410}%
  \BibitemOpen
  \bibfield  {author} {\bibinfo {author} {\bibfnamefont {L.~R.}\ \bibnamefont
  {Yaraskavitch}}, \bibinfo {author} {\bibfnamefont {H.~M.}\ \bibnamefont
  {Revell}}, \bibinfo {author} {\bibfnamefont {S.}~\bibnamefont {Meng}},
  \bibinfo {author} {\bibfnamefont {K.~A.}\ \bibnamefont {Ross}}, \bibinfo
  {author} {\bibfnamefont {H.~M.~L.}\ \bibnamefont {Noad}}, \bibinfo {author}
  {\bibfnamefont {H.~A.}\ \bibnamefont {Dabkowska}}, \bibinfo {author}
  {\bibfnamefont {B.~D.}\ \bibnamefont {Gaulin}}, \ and\ \bibinfo {author}
  {\bibfnamefont {J.~B.}\ \bibnamefont {Kycia}},\ }\href {\doibase
  10.1103/PhysRevB.85.020410} {\bibfield  {journal} {\bibinfo  {journal} {Phys.
  Rev. B}\ }\textbf {\bibinfo {volume} {85}},\ \bibinfo {pages} {020410}
  (\bibinfo {year} {2012})}\BibitemShut {NoStop}%
\bibitem [{\citenamefont {Quilliam}\ \emph {et~al.}(2011)\citenamefont
  {Quilliam}, \citenamefont {Yaraskavitch}, \citenamefont {Dabkowska},
  \citenamefont {Gaulin},\ and\ \citenamefont {Kycia}}]{PhysRevB.83.094424}%
  \BibitemOpen
  \bibfield  {author} {\bibinfo {author} {\bibfnamefont {J.~A.}\ \bibnamefont
  {Quilliam}}, \bibinfo {author} {\bibfnamefont {L.~R.}\ \bibnamefont
  {Yaraskavitch}}, \bibinfo {author} {\bibfnamefont {H.~A.}\ \bibnamefont
  {Dabkowska}}, \bibinfo {author} {\bibfnamefont {B.~D.}\ \bibnamefont
  {Gaulin}}, \ and\ \bibinfo {author} {\bibfnamefont {J.~B.}\ \bibnamefont
  {Kycia}},\ }\href {\doibase 10.1103/PhysRevB.83.094424} {\bibfield  {journal}
  {\bibinfo  {journal} {Phys. Rev. B}\ }\textbf {\bibinfo {volume} {83}},\
  \bibinfo {pages} {094424} (\bibinfo {year} {2011})}\BibitemShut {NoStop}%
\bibitem [{\citenamefont {Lau}\ \emph {et~al.}(2006{\natexlab{a}})\citenamefont
  {Lau}, \citenamefont {Freitas}, \citenamefont {Ueland}, \citenamefont
  {Muegge}, \citenamefont {Duncan}, \citenamefont {Schiffer},\ and\
  \citenamefont {Cava}}]{lau2006zero}%
  \BibitemOpen
  \bibfield  {author} {\bibinfo {author} {\bibfnamefont {G.~C.}\ \bibnamefont
  {Lau}}, \bibinfo {author} {\bibfnamefont {R.~S.}\ \bibnamefont {Freitas}},
  \bibinfo {author} {\bibfnamefont {B.}~\bibnamefont {Ueland}}, \bibinfo
  {author} {\bibfnamefont {B.~D.}\ \bibnamefont {Muegge}}, \bibinfo {author}
  {\bibfnamefont {E.~L.}\ \bibnamefont {Duncan}}, \bibinfo {author}
  {\bibfnamefont {P.}~\bibnamefont {Schiffer}}, \ and\ \bibinfo {author}
  {\bibfnamefont {R.~J.}\ \bibnamefont {Cava}},\ }\href {\doibase
  https://doi.org/10.1038/nphys270} {\bibfield  {journal} {\bibinfo  {journal}
  {Nat. Phys.}\ }\textbf {\bibinfo {volume} {2}},\ \bibinfo {pages} {249}
  (\bibinfo {year} {2006}{\natexlab{a}})}\BibitemShut {NoStop}%
\bibitem [{\citenamefont {Aldus}\ \emph {et~al.}(2013)\citenamefont {Aldus},
  \citenamefont {Fennell}, \citenamefont {Deen}, \citenamefont {Ressouche},
  \citenamefont {Lau}, \citenamefont {Cava},\ and\ \citenamefont
  {Bramwell}}]{Aldus_2013}%
  \BibitemOpen
  \bibfield  {author} {\bibinfo {author} {\bibfnamefont {R.~J.}\ \bibnamefont
  {Aldus}}, \bibinfo {author} {\bibfnamefont {T.}~\bibnamefont {Fennell}},
  \bibinfo {author} {\bibfnamefont {P.~P.}\ \bibnamefont {Deen}}, \bibinfo
  {author} {\bibfnamefont {E.}~\bibnamefont {Ressouche}}, \bibinfo {author}
  {\bibfnamefont {G.~C.}\ \bibnamefont {Lau}}, \bibinfo {author} {\bibfnamefont
  {R.~J.}\ \bibnamefont {Cava}}, \ and\ \bibinfo {author} {\bibfnamefont
  {S.~T.}\ \bibnamefont {Bramwell}},\ }\href {\doibase
  10.1088/1367-2630/15/1/013022} {\bibfield  {journal} {\bibinfo  {journal}
  {New J. Phys.}\ }\textbf {\bibinfo {volume} {15}},\ \bibinfo {pages} {013022}
  (\bibinfo {year} {2013})}\BibitemShut {NoStop}%
\bibitem [{\citenamefont {Lau}\ \emph {et~al.}(2006{\natexlab{b}})\citenamefont
  {Lau}, \citenamefont {Muegge}, \citenamefont {McQueen}, \citenamefont
  {Duncan},\ and\ \citenamefont {Cava}}]{LAU20063126}%
  \BibitemOpen
  \bibfield  {author} {\bibinfo {author} {\bibfnamefont {G.}~\bibnamefont
  {Lau}}, \bibinfo {author} {\bibfnamefont {B.}~\bibnamefont {Muegge}},
  \bibinfo {author} {\bibfnamefont {T.}~\bibnamefont {McQueen}}, \bibinfo
  {author} {\bibfnamefont {E.}~\bibnamefont {Duncan}}, \ and\ \bibinfo {author}
  {\bibfnamefont {R.}~\bibnamefont {Cava}},\ }\href {\doibase
  https://doi.org/10.1016/j.jssc.2006.06.007} {\bibfield  {journal} {\bibinfo
  {journal} {J. Solid State Chem.}\ }\textbf {\bibinfo {volume} {179}},\
  \bibinfo {pages} {3126} (\bibinfo {year} {2006}{\natexlab{b}})}\BibitemShut
  {NoStop}%
\bibitem [{\citenamefont {Zhou}\ \emph {et~al.}(2007)\citenamefont {Zhou},
  \citenamefont {Wiebe}, \citenamefont {Jo}, \citenamefont {Balicas},
  \citenamefont {Qiu}, \citenamefont {Copley}, \citenamefont {Ehlers},
  \citenamefont {Fouquet},\ and\ \citenamefont {Gardner}}]{Zhou_2007}%
  \BibitemOpen
  \bibfield  {author} {\bibinfo {author} {\bibfnamefont {H.~D.}\ \bibnamefont
  {Zhou}}, \bibinfo {author} {\bibfnamefont {C.~R.}\ \bibnamefont {Wiebe}},
  \bibinfo {author} {\bibfnamefont {Y.~J.}\ \bibnamefont {Jo}}, \bibinfo
  {author} {\bibfnamefont {L.}~\bibnamefont {Balicas}}, \bibinfo {author}
  {\bibfnamefont {Y.}~\bibnamefont {Qiu}}, \bibinfo {author} {\bibfnamefont
  {J.~R.~D.}\ \bibnamefont {Copley}}, \bibinfo {author} {\bibfnamefont
  {G.}~\bibnamefont {Ehlers}}, \bibinfo {author} {\bibfnamefont
  {P.}~\bibnamefont {Fouquet}}, \ and\ \bibinfo {author} {\bibfnamefont
  {J.~S.}\ \bibnamefont {Gardner}},\ }\href {\doibase
  10.1088/0953-8984/19/34/342201} {\bibfield  {journal} {\bibinfo  {journal}
  {J. Condens. Matter Phys.}\ }\textbf {\bibinfo {volume} {19}},\ \bibinfo
  {pages} {342201} (\bibinfo {year} {2007})}\BibitemShut {NoStop}%
\bibitem [{\citenamefont {Ehlers}\ \emph {et~al.}(2008)\citenamefont {Ehlers},
  \citenamefont {Gardner}, \citenamefont {Qiu}, \citenamefont {Fouquet},
  \citenamefont {Wiebe}, \citenamefont {Balicas},\ and\ \citenamefont
  {Zhou}}]{PhysRevB.77.052404}%
  \BibitemOpen
  \bibfield  {author} {\bibinfo {author} {\bibfnamefont {G.}~\bibnamefont
  {Ehlers}}, \bibinfo {author} {\bibfnamefont {J.~S.}\ \bibnamefont {Gardner}},
  \bibinfo {author} {\bibfnamefont {Y.}~\bibnamefont {Qiu}}, \bibinfo {author}
  {\bibfnamefont {P.}~\bibnamefont {Fouquet}}, \bibinfo {author} {\bibfnamefont
  {C.~R.}\ \bibnamefont {Wiebe}}, \bibinfo {author} {\bibfnamefont
  {L.}~\bibnamefont {Balicas}}, \ and\ \bibinfo {author} {\bibfnamefont
  {H.~D.}\ \bibnamefont {Zhou}},\ }\href {\doibase 10.1103/PhysRevB.77.052404}
  {\bibfield  {journal} {\bibinfo  {journal} {Phys. Rev. B}\ }\textbf {\bibinfo
  {volume} {77}},\ \bibinfo {pages} {052404} (\bibinfo {year}
  {2008})}\BibitemShut {NoStop}%
\bibitem [{\citenamefont {Ueland}\ \emph {et~al.}(2008)\citenamefont {Ueland},
  \citenamefont {Lau}, \citenamefont {Freitas}, \citenamefont {Snyder},
  \citenamefont {Dahlberg}, \citenamefont {Muegge}, \citenamefont {Duncan},
  \citenamefont {Cava},\ and\ \citenamefont {Schiffer}}]{PhysRevB.77.144412}%
  \BibitemOpen
  \bibfield  {author} {\bibinfo {author} {\bibfnamefont {B.~G.}\ \bibnamefont
  {Ueland}}, \bibinfo {author} {\bibfnamefont {G.~C.}\ \bibnamefont {Lau}},
  \bibinfo {author} {\bibfnamefont {R.~S.}\ \bibnamefont {Freitas}}, \bibinfo
  {author} {\bibfnamefont {J.}~\bibnamefont {Snyder}}, \bibinfo {author}
  {\bibfnamefont {M.~L.}\ \bibnamefont {Dahlberg}}, \bibinfo {author}
  {\bibfnamefont {B.~D.}\ \bibnamefont {Muegge}}, \bibinfo {author}
  {\bibfnamefont {E.~L.}\ \bibnamefont {Duncan}}, \bibinfo {author}
  {\bibfnamefont {R.~J.}\ \bibnamefont {Cava}}, \ and\ \bibinfo {author}
  {\bibfnamefont {P.}~\bibnamefont {Schiffer}},\ }\href {\doibase
  10.1103/PhysRevB.77.144412} {\bibfield  {journal} {\bibinfo  {journal} {Phys.
  Rev. B}\ }\textbf {\bibinfo {volume} {77}},\ \bibinfo {pages} {144412}
  (\bibinfo {year} {2008})}\BibitemShut {NoStop}%
\bibitem [{\citenamefont {Ramon}\ \emph {et~al.}(2019)\citenamefont {Ramon},
  \citenamefont {Wang}, \citenamefont {Ishida}, \citenamefont {Bernardo},
  \citenamefont {Leite}, \citenamefont {Vichi}, \citenamefont {Gardner},\ and\
  \citenamefont {Freitas}}]{Ramon19:99}%
  \BibitemOpen
  \bibfield  {author} {\bibinfo {author} {\bibfnamefont {J.~G.~A.}\
  \bibnamefont {Ramon}}, \bibinfo {author} {\bibfnamefont {C.~W.}\ \bibnamefont
  {Wang}}, \bibinfo {author} {\bibfnamefont {L.}~\bibnamefont {Ishida}},
  \bibinfo {author} {\bibfnamefont {P.~L.}\ \bibnamefont {Bernardo}}, \bibinfo
  {author} {\bibfnamefont {M.~M.}\ \bibnamefont {Leite}}, \bibinfo {author}
  {\bibfnamefont {F.~M.}\ \bibnamefont {Vichi}}, \bibinfo {author}
  {\bibfnamefont {J.~S.}\ \bibnamefont {Gardner}}, \ and\ \bibinfo {author}
  {\bibfnamefont {R.~S.}\ \bibnamefont {Freitas}},\ }\href {\doibase
  10.1103/PhysRevB.99.214442} {\bibfield  {journal} {\bibinfo  {journal} {Phys.
  Rev. B}\ }\textbf {\bibinfo {volume} {99}},\ \bibinfo {pages} {214442}
  (\bibinfo {year} {2019})}\BibitemShut {NoStop}%
\bibitem [{\citenamefont {Liu}\ \emph {et~al.}(2014{\natexlab{b}})\citenamefont
  {Liu}, \citenamefont {Zou}, \citenamefont {Ling}, \citenamefont {Zhang},
  \citenamefont {Tong}, \citenamefont {Zhang},\ and\ \citenamefont
  {Zhang}}]{LIU2014107}%
  \BibitemOpen
  \bibfield  {author} {\bibinfo {author} {\bibfnamefont {H.}~\bibnamefont
  {Liu}}, \bibinfo {author} {\bibfnamefont {Y.}~\bibnamefont {Zou}}, \bibinfo
  {author} {\bibfnamefont {L.}~\bibnamefont {Ling}}, \bibinfo {author}
  {\bibfnamefont {L.}~\bibnamefont {Zhang}}, \bibinfo {author} {\bibfnamefont
  {W.}~\bibnamefont {Tong}}, \bibinfo {author} {\bibfnamefont {C.}~\bibnamefont
  {Zhang}}, \ and\ \bibinfo {author} {\bibfnamefont {Y.}~\bibnamefont
  {Zhang}},\ }\href {\doibase https://doi.org/10.1016/j.jmmm.2014.06.031}
  {\bibfield  {journal} {\bibinfo  {journal} {J. Magn. Magn. Mater.}\ }\textbf
  {\bibinfo {volume} {369}},\ \bibinfo {pages} {107} (\bibinfo {year}
  {2014}{\natexlab{b}})}\BibitemShut {NoStop}%
\bibitem [{\citenamefont {Nandi}\ \emph {et~al.}(2017)\citenamefont {Nandi},
  \citenamefont {Jana}, \citenamefont {Swarnakar}, \citenamefont {Alam},
  \citenamefont {Bag},\ and\ \citenamefont {Nath}}]{NANDI2017318}%
  \BibitemOpen
  \bibfield  {author} {\bibinfo {author} {\bibfnamefont {S.}~\bibnamefont
  {Nandi}}, \bibinfo {author} {\bibfnamefont {Y.}~\bibnamefont {Jana}},
  \bibinfo {author} {\bibfnamefont {D.}~\bibnamefont {Swarnakar}}, \bibinfo
  {author} {\bibfnamefont {J.}~\bibnamefont {Alam}}, \bibinfo {author}
  {\bibfnamefont {P.}~\bibnamefont {Bag}}, \ and\ \bibinfo {author}
  {\bibfnamefont {R.}~\bibnamefont {Nath}},\ }\href {\doibase
  https://doi.org/10.1016/j.jallcom.2017.04.160} {\bibfield  {journal}
  {\bibinfo  {journal} {J. Alloys Compd.}\ }\textbf {\bibinfo {volume} {714}},\
  \bibinfo {pages} {318} (\bibinfo {year} {2017})}\BibitemShut {NoStop}%
\bibitem [{\citenamefont {Chien}\ and\ \citenamefont
  {Sleight}(1978{\natexlab{b}})}]{chien78:18}%
  \BibitemOpen
  \bibfield  {author} {\bibinfo {author} {\bibfnamefont {C.~L.}\ \bibnamefont
  {Chien}}\ and\ \bibinfo {author} {\bibfnamefont {A.~W.}\ \bibnamefont
  {Sleight}},\ }\href {\doibase 10.1103/PhysRevB.18.2031} {\bibfield  {journal}
  {\bibinfo  {journal} {Phys. Rev. B}\ }\textbf {\bibinfo {volume} {18}},\
  \bibinfo {pages} {2031} (\bibinfo {year} {1978}{\natexlab{b}})}\BibitemShut
  {NoStop}%
\bibitem [{\citenamefont {Pal}\ \emph {et~al.}(2018)\citenamefont {Pal},
  \citenamefont {Singh}, \citenamefont {Ghosh},\ and\ \citenamefont
  {Chatterjee}}]{PAL18:462}%
  \BibitemOpen
  \bibfield  {author} {\bibinfo {author} {\bibfnamefont {A.}~\bibnamefont
  {Pal}}, \bibinfo {author} {\bibfnamefont {A.}~\bibnamefont {Singh}}, \bibinfo
  {author} {\bibfnamefont {A.~K.}\ \bibnamefont {Ghosh}}, \ and\ \bibinfo
  {author} {\bibfnamefont {S.}~\bibnamefont {Chatterjee}},\ }\href {\doibase
  https://doi.org/10.1016/j.jmmm.2018.04.060} {\bibfield  {journal} {\bibinfo
  {journal} {J. Magn. Magn. Mater.}\ }\textbf {\bibinfo {volume} {462}},\
  \bibinfo {pages} {1} (\bibinfo {year} {2018})}\BibitemShut {NoStop}%
\bibitem [{\citenamefont {Dasgupta}\ \emph {et~al.}(2007)\citenamefont
  {Dasgupta}, \citenamefont {Jana}, \citenamefont {{Nag Chattopadhyay}},
  \citenamefont {Higashinaka}, \citenamefont {Maeno},\ and\ \citenamefont
  {Ghosh}}]{DASGUPTA2007347}%
  \BibitemOpen
  \bibfield  {author} {\bibinfo {author} {\bibfnamefont {P.}~\bibnamefont
  {Dasgupta}}, \bibinfo {author} {\bibfnamefont {Y.}~\bibnamefont {Jana}},
  \bibinfo {author} {\bibfnamefont {A.}~\bibnamefont {{Nag Chattopadhyay}}},
  \bibinfo {author} {\bibfnamefont {R.}~\bibnamefont {Higashinaka}}, \bibinfo
  {author} {\bibfnamefont {Y.}~\bibnamefont {Maeno}}, \ and\ \bibinfo {author}
  {\bibfnamefont {D.}~\bibnamefont {Ghosh}},\ }\href {\doibase
  https://doi.org/10.1016/j.jpcs.2006.11.022} {\bibfield  {journal} {\bibinfo
  {journal} {J. Phys. Chem. Solids}\ }\textbf {\bibinfo {volume} {68}},\
  \bibinfo {pages} {347} (\bibinfo {year} {2007})}\BibitemShut {NoStop}%
\bibitem [{\citenamefont {Zinkin}\ \emph {et~al.}(1996)\citenamefont {Zinkin},
  \citenamefont {Harris}, \citenamefont {Tun}, \citenamefont {Cowley},\ and\
  \citenamefont {Wanklyn}}]{TmTi_zinkin1996lifting}%
  \BibitemOpen
  \bibfield  {author} {\bibinfo {author} {\bibfnamefont {M.~P.}\ \bibnamefont
  {Zinkin}}, \bibinfo {author} {\bibfnamefont {M.~J.}\ \bibnamefont {Harris}},
  \bibinfo {author} {\bibfnamefont {Z.}~\bibnamefont {Tun}}, \bibinfo {author}
  {\bibfnamefont {R.~A.}\ \bibnamefont {Cowley}}, \ and\ \bibinfo {author}
  {\bibfnamefont {B.~M.}\ \bibnamefont {Wanklyn}},\ }\href {\doibase
  https://doi.org/10.1088/0953-8984/8/2/007} {\bibfield  {journal} {\bibinfo
  {journal} {J. Condens. Matter Phys.}\ }\textbf {\bibinfo {volume} {8}},\
  \bibinfo {pages} {193} (\bibinfo {year} {1996})}\BibitemShut {NoStop}%
\bibitem [{\citenamefont {Chattopadhyay}\ \emph {et~al.}(2004)\citenamefont
  {Chattopadhyay}, \citenamefont {Dasgupta}, \citenamefont {Jana},\ and\
  \citenamefont {Ghosh}}]{CHATTOPADHYAY20046}%
  \BibitemOpen
  \bibfield  {author} {\bibinfo {author} {\bibfnamefont {A.}~\bibnamefont
  {Chattopadhyay}}, \bibinfo {author} {\bibfnamefont {P.}~\bibnamefont
  {Dasgupta}}, \bibinfo {author} {\bibfnamefont {Y.}~\bibnamefont {Jana}}, \
  and\ \bibinfo {author} {\bibfnamefont {D.}~\bibnamefont {Ghosh}},\ }\href
  {\doibase https://doi.org/10.1016/j.jallcom.2004.03.133} {\bibfield
  {journal} {\bibinfo  {journal} {J. Alloys Compd.}\ }\textbf {\bibinfo
  {volume} {384}},\ \bibinfo {pages} {6} (\bibinfo {year} {2004})}\BibitemShut
  {NoStop}%
\bibitem [{\citenamefont {Sheetal}\ and\ \citenamefont
  {Yadav}(2022)}]{SHEETAL2022169255}%
  \BibitemOpen
  \bibfield  {author} {\bibinfo {author} {\bibnamefont {Sheetal}}\ and\
  \bibinfo {author} {\bibfnamefont {C.}~\bibnamefont {Yadav}},\ }\href
  {\doibase https://doi.org/10.1016/j.jmmm.2022.169255} {\bibfield  {journal}
  {\bibinfo  {journal} {J. Magn. Magn. Mater.}\ }\textbf {\bibinfo {volume}
  {553}},\ \bibinfo {pages} {169255} (\bibinfo {year} {2022})}\BibitemShut
  {NoStop}%
\bibitem [{\citenamefont {Taira}\ \emph {et~al.}(2002)\citenamefont {Taira},
  \citenamefont {Wakeshima},\ and\ \citenamefont {Hinatsu}}]{Taira02:12}%
  \BibitemOpen
  \bibfield  {author} {\bibinfo {author} {\bibfnamefont {N.}~\bibnamefont
  {Taira}}, \bibinfo {author} {\bibfnamefont {M.}~\bibnamefont {Wakeshima}}, \
  and\ \bibinfo {author} {\bibfnamefont {Y.}~\bibnamefont {Hinatsu}},\ }\href
  {\doibase 10.1039/B110596P} {\bibfield  {journal} {\bibinfo  {journal} {J.
  Mater. Chem.}\ }\textbf {\bibinfo {volume} {12}},\ \bibinfo {pages} {1475}
  (\bibinfo {year} {2002})}\BibitemShut {NoStop}%
\bibitem [{\citenamefont {Lea}\ \emph {et~al.}(1962)\citenamefont {Lea},
  \citenamefont {Leask},\ and\ \citenamefont {Wolf}}]{Lea62:23}%
  \BibitemOpen
  \bibfield  {author} {\bibinfo {author} {\bibfnamefont {K.~R.}\ \bibnamefont
  {Lea}}, \bibinfo {author} {\bibfnamefont {M.~J.~M.}\ \bibnamefont {Leask}}, \
  and\ \bibinfo {author} {\bibfnamefont {W.}~\bibnamefont {Wolf}},\ }\href
  {\doibase https://doi.org/10.1016/0022-3697(62)90192-0} {\bibfield  {journal}
  {\bibinfo  {journal} {J. Phys. Chem. Solids}\ }\textbf {\bibinfo {volume}
  {23}},\ \bibinfo {pages} {1381} (\bibinfo {year} {1962})}\BibitemShut
  {NoStop}%
\bibitem [{\citenamefont {Greedan}(1992)}]{LandoltBornstein1992}%
  \BibitemOpen
  \bibfield  {author} {\bibinfo {author} {\bibfnamefont {J.~E.}\ \bibnamefont
  {Greedan}},\ }\href {\doibase 10.1007/10057685_18} {\emph {\bibinfo {title}
  {Oxides with Trirutile and Pyrochlore Structures, Landolt-B{\"{o}}rnstein -
  Group III Condensed Matter}}},\ edited by\ \bibinfo {editor} {\bibfnamefont
  {H.}~\bibnamefont {Wijn}},\ Vol.\ \bibinfo {volume} {27, Part g}\ (\bibinfo
  {publisher} {Springer-Verlag Berlin Heidelberg},\ \bibinfo {address}
  {Berlin},\ \bibinfo {year} {1992})\BibitemShut {NoStop}%
\bibitem [{\citenamefont {Sibille}\ \emph
  {et~al.}(2016{\natexlab{b}})\citenamefont {Sibille}, \citenamefont {Lhotel},
  \citenamefont {Hatnean}, \citenamefont {Balakrishnan}, \citenamefont
  {F{\aa}k}, \citenamefont {Gauthier}, \citenamefont {Fennell},\ and\
  \citenamefont {Kenzelmann}}]{PrHf_sibille2016candidate}%
  \BibitemOpen
  \bibfield  {author} {\bibinfo {author} {\bibfnamefont {R.}~\bibnamefont
  {Sibille}}, \bibinfo {author} {\bibfnamefont {E.}~\bibnamefont {Lhotel}},
  \bibinfo {author} {\bibfnamefont {M.~C.}\ \bibnamefont {Hatnean}}, \bibinfo
  {author} {\bibfnamefont {G.}~\bibnamefont {Balakrishnan}}, \bibinfo {author}
  {\bibfnamefont {B.}~\bibnamefont {F{\aa}k}}, \bibinfo {author} {\bibfnamefont
  {N.}~\bibnamefont {Gauthier}}, \bibinfo {author} {\bibfnamefont
  {T.}~\bibnamefont {Fennell}}, \ and\ \bibinfo {author} {\bibfnamefont
  {M.}~\bibnamefont {Kenzelmann}},\ }\href {\doibase
  10.1103/PhysRevB.94.024436} {\bibfield  {journal} {\bibinfo  {journal} {Phys.
  Rev. B}\ }\textbf {\bibinfo {volume} {94}},\ \bibinfo {pages} {024436}
  (\bibinfo {year} {2016}{\natexlab{b}})}\BibitemShut {NoStop}%
\bibitem [{\citenamefont {Anand}\ \emph {et~al.}(2016)\citenamefont {Anand},
  \citenamefont {Opherden}, \citenamefont {Xu}, \citenamefont {Adroja},
  \citenamefont {Islam}, \citenamefont {Herrmannsd\"orfer}, \citenamefont
  {Hornung}, \citenamefont {Sch\"onemann}, \citenamefont {Uhlarz},
  \citenamefont {Walker}, \citenamefont {Casati},\ and\ \citenamefont
  {Lake}}]{PrHf_anand2016physical}%
  \BibitemOpen
  \bibfield  {author} {\bibinfo {author} {\bibfnamefont {V.~K.}\ \bibnamefont
  {Anand}}, \bibinfo {author} {\bibfnamefont {L.}~\bibnamefont {Opherden}},
  \bibinfo {author} {\bibfnamefont {J.}~\bibnamefont {Xu}}, \bibinfo {author}
  {\bibfnamefont {D.~T.}\ \bibnamefont {Adroja}}, \bibinfo {author}
  {\bibfnamefont {A.~T. M.~N.}\ \bibnamefont {Islam}}, \bibinfo {author}
  {\bibfnamefont {T.}~\bibnamefont {Herrmannsd\"orfer}}, \bibinfo {author}
  {\bibfnamefont {J.}~\bibnamefont {Hornung}}, \bibinfo {author} {\bibfnamefont
  {R.}~\bibnamefont {Sch\"onemann}}, \bibinfo {author} {\bibfnamefont
  {M.}~\bibnamefont {Uhlarz}}, \bibinfo {author} {\bibfnamefont {H.~C.}\
  \bibnamefont {Walker}}, \bibinfo {author} {\bibfnamefont {N.}~\bibnamefont
  {Casati}}, \ and\ \bibinfo {author} {\bibfnamefont {B.}~\bibnamefont
  {Lake}},\ }\href {\doibase 10.1103/PhysRevB.94.144415} {\bibfield  {journal}
  {\bibinfo  {journal} {Phys. Rev. B}\ }\textbf {\bibinfo {volume} {94}},\
  \bibinfo {pages} {144415} (\bibinfo {year} {2016})}\BibitemShut {NoStop}%
\bibitem [{\citenamefont {Petit}\ \emph
  {et~al.}(2016{\natexlab{b}})\citenamefont {Petit}, \citenamefont {Lhotel},
  \citenamefont {Guitteny}, \citenamefont {Florea}, \citenamefont {Robert},
  \citenamefont {Bonville}, \citenamefont {Mirebeau}, \citenamefont {Ollivier},
  \citenamefont {Mutka}, \citenamefont {Ressouche}, \citenamefont {Decorse},
  \citenamefont {Ciomaga~Hatnean},\ and\ \citenamefont
  {Balakrishnan}}]{Pettit16:94}%
  \BibitemOpen
  \bibfield  {author} {\bibinfo {author} {\bibfnamefont {S.}~\bibnamefont
  {Petit}}, \bibinfo {author} {\bibfnamefont {E.}~\bibnamefont {Lhotel}},
  \bibinfo {author} {\bibfnamefont {S.}~\bibnamefont {Guitteny}}, \bibinfo
  {author} {\bibfnamefont {O.}~\bibnamefont {Florea}}, \bibinfo {author}
  {\bibfnamefont {J.}~\bibnamefont {Robert}}, \bibinfo {author} {\bibfnamefont
  {P.}~\bibnamefont {Bonville}}, \bibinfo {author} {\bibfnamefont
  {I.}~\bibnamefont {Mirebeau}}, \bibinfo {author} {\bibfnamefont
  {J.}~\bibnamefont {Ollivier}}, \bibinfo {author} {\bibfnamefont
  {H.}~\bibnamefont {Mutka}}, \bibinfo {author} {\bibfnamefont
  {E.}~\bibnamefont {Ressouche}}, \bibinfo {author} {\bibfnamefont
  {C.}~\bibnamefont {Decorse}}, \bibinfo {author} {\bibfnamefont
  {M.}~\bibnamefont {Ciomaga~Hatnean}}, \ and\ \bibinfo {author} {\bibfnamefont
  {G.}~\bibnamefont {Balakrishnan}},\ }\href {\doibase
  10.1103/PhysRevB.94.165153} {\bibfield  {journal} {\bibinfo  {journal} {Phys.
  Rev. B}\ }\textbf {\bibinfo {volume} {94}},\ \bibinfo {pages} {165153}
  (\bibinfo {year} {2016}{\natexlab{b}})}\BibitemShut {NoStop}%
\bibitem [{\citenamefont {Bonville}\ \emph {et~al.}(2016)\citenamefont
  {Bonville}, \citenamefont {Guitteny}, \citenamefont {Gukasov}, \citenamefont
  {Mirebeau}, \citenamefont {Petit}, \citenamefont {Decorse}, \citenamefont
  {Hatnean},\ and\ \citenamefont {Balakrishnan}}]{Bonville16:94}%
  \BibitemOpen
  \bibfield  {author} {\bibinfo {author} {\bibfnamefont {P.}~\bibnamefont
  {Bonville}}, \bibinfo {author} {\bibfnamefont {S.}~\bibnamefont {Guitteny}},
  \bibinfo {author} {\bibfnamefont {A.}~\bibnamefont {Gukasov}}, \bibinfo
  {author} {\bibfnamefont {I.}~\bibnamefont {Mirebeau}}, \bibinfo {author}
  {\bibfnamefont {S.}~\bibnamefont {Petit}}, \bibinfo {author} {\bibfnamefont
  {C.}~\bibnamefont {Decorse}}, \bibinfo {author} {\bibfnamefont {M.~C.}\
  \bibnamefont {Hatnean}}, \ and\ \bibinfo {author} {\bibfnamefont
  {G.}~\bibnamefont {Balakrishnan}},\ }\href {\doibase
  10.1103/PhysRevB.94.134428} {\bibfield  {journal} {\bibinfo  {journal} {Phys.
  Rev. B}\ }\textbf {\bibinfo {volume} {94}},\ \bibinfo {pages} {134428}
  (\bibinfo {year} {2016})}\BibitemShut {NoStop}%
\bibitem [{\citenamefont {Onoda}\ and\ \citenamefont
  {Tanaka}(2011)}]{Onoda11:83}%
  \BibitemOpen
  \bibfield  {author} {\bibinfo {author} {\bibfnamefont {S.}~\bibnamefont
  {Onoda}}\ and\ \bibinfo {author} {\bibfnamefont {Y.}~\bibnamefont {Tanaka}},\
  }\href {\doibase 10.1103/PhysRevB.83.094411} {\bibfield  {journal} {\bibinfo
  {journal} {Phys. Rev. B}\ }\textbf {\bibinfo {volume} {83}},\ \bibinfo
  {pages} {094411} (\bibinfo {year} {2011})}\BibitemShut {NoStop}%
\bibitem [{\citenamefont {Koohpayeh}\ \emph {et~al.}(2014)\citenamefont
  {Koohpayeh}, \citenamefont {Wen}, \citenamefont {Trump}, \citenamefont
  {Broholm},\ and\ \citenamefont {McQueen}}]{KOOHPAYEH2014291}%
  \BibitemOpen
  \bibfield  {author} {\bibinfo {author} {\bibfnamefont {S.}~\bibnamefont
  {Koohpayeh}}, \bibinfo {author} {\bibfnamefont {J.-J.}\ \bibnamefont {Wen}},
  \bibinfo {author} {\bibfnamefont {B.}~\bibnamefont {Trump}}, \bibinfo
  {author} {\bibfnamefont {C.}~\bibnamefont {Broholm}}, \ and\ \bibinfo
  {author} {\bibfnamefont {T.}~\bibnamefont {McQueen}},\ }\href {\doibase
  https://doi.org/10.1016/j.jcrysgro.2014.06.037} {\bibfield  {journal}
  {\bibinfo  {journal} {J. Cryst. Growth}\ }\textbf {\bibinfo {volume} {402}},\
  \bibinfo {pages} {291} (\bibinfo {year} {2014})}\BibitemShut {NoStop}%
\bibitem [{pri()}]{private}%
  \BibitemOpen
  \href@noop {} {}\bibinfo {howpublished} {Private Communication}\BibitemShut
  {NoStop}%
\bibitem [{MRL()}]{MRL}%
  \BibitemOpen
  \href@noop {} {}\bibinfo {note} {\url{https://www.mrfn.org/}}\BibitemShut
  {NoStop}%
\end{thebibliography}%

\end{document}